\documentclass[12pt, a4paper]{article}
\usepackage[T2A]{fontenc}
\usepackage[utf8]{inputenc}
\usepackage{colortbl}
\usepackage{cite}
\usepackage{mathtext}
\usepackage{amssymb,amsthm,amsmath}
\usepackage{bbm}
\usepackage[mathscr]{eucal}
\usepackage{array}
\usepackage[english]{babel}
\usepackage[title,titletoc]{appendix}
\usepackage{graphicx}
\usepackage{float}
\usepackage{upgreek}
\usepackage{mathabx}

\begin{document}
\author{\bf Yu.A.\,Markov$\!\,$\thanks{e-mail:markov@icc.ru},\,
M.A.\,Markova$\!\,$\thanks{e-mail:markova@icc.ru}}

\title{Canonical formulation of the plasmon\,–\,hard particle scattering process
in a quark-gluon plasma}
%
%
%
\date{\it\normalsize
\begin{itemize}
\item[]Matrosov Institute for System Dynamics and Control Theory, Siberian Branch, Russian Academy of Sciences, Irkutsk, 664033 Russia
\vspace{-0.3cm}
%
\end{itemize}}
%
\thispagestyle{empty}
\maketitle{}


\def\theequation{\arabic{section}.\arabic{equation}}
{
\[
\mbox{\bf Abstract}
\]

\noindent It is shown that the Hamiltonian formalism proposed previously in \cite{markov_2023} to describe the nonlinear dynamics of only {\it soft} fermionic and bosonic excitations contains much more information than  initially assumed. In this paper, we have demonstrated in detail that it also proved to be very appropriate and powerful in describing a wide range of other physical phenomena, including the scattering of colorless plasmons off {\it hard} thermal (or external) color-charged particles moving in a hot quark-gluon plasma. A generalization of the Poisson superbracket including both anticommuting variables for hard modes and normal variables of the soft Bose field, is presented for the case of continuous medium.  The corresponding Hamilton equations are defined, and the most general form of the third- and fourth-order interaction Hamiltonians is written out in terms of the normal boson field variables and hard momentum modes of the quark-gluon plasma. The canonical transformations involving both bosonic and hard mode degrees of freedom of the system under consideration, are discussed. The canonicity conditions for these transformations based on the Poisson superbracket, are derived. The most general structure of canonical transformations in the form of integro-power series up to sixth order in a new normal field variable and a new hard mode variable, is presented. For the hard momentum mode of quark-gluon plasma excitations, an ansatz separating the color and momentum degrees of freedom, is proposed. The question of approximation of the total effective scattering amplitude when the momenta of hard excitations are much larger than those of soft excitations of the plasma, is considered. The self-consistent system of Boltzmann-type kinetic equations taking into account the time evolution of the mean value of the color charge of the hard particle is obtained.

}

 \newpage

\section{Introduction}
\setcounter{equation}{0}

The present work is a direct continuation of our earlier work \cite{markov_2023}, where we have developed in detail the approach in the construction of the Hamiltonian formalism for the self-consistent description of the nonlinear scattering processes of soft collective excitations of both bosonic and fermionic types in the QGP. The use of the methods and of the results we received in \cite{markov_2023} allowed us to develop a somewhat different, more rigorous, as we think, approach to the problem posed in \cite{markov_I_2024}. Making use of just the same initial equations and relations (canonicity conditions, Poisson's superbracket, Hamilton's equations) written out in \cite{markov_2023} for soft collective modes of QGP excitations, we show step by step how one can derive from them the equations and relations describing qualitatively new physical phenomena and interaction processes. This, in turn, gives a deeper understanding of the kinetic equations themselves for soft bosonic and fermionic excitations obtained in \cite{markov_I_2024} and the possibility of using them to describe the hard momentum degrees of freedom of QGP.\\ 
\indent It should be noted at once that the kinetic equations and the equation of evolution of the color charge of a hard particle, which we derive in the present paper, do not coincide literally with the equations of the paper \cite{markov_I_2024}. In the current approach, new terms appear that sometimes qualitatively change the dynamics of the evolution of physical quantities. Moreover, in contrast to the results in \cite{markov_I_2024}, which are valid for arbitrary color group $SU(N_{c})$, here for a self-consistent description it is necessary to be restricted to the value $N_{c} = 3$ (not considering the ``trivial'' case $N_{c} = 2$).\\
%
%
\indent The paper is organized as follows. In section \ref{section_2}, a generalization of the Poisson superbracket including both the anticommuting variables for hard modes $(\xi^{\,i}_{{\bf p}^{\phantom\prime}},\,\xi^{\hspace{0.03cm}\ast\hspace{0.03cm}i}_{{\bf p}})$ and the normal variables $(a^{\phantom{\ast}\!\!a}_{\hspace{0.03cm}{\bf k}},\, a^{\,{\ast}\,a}_{\hspace{0.03cm}{\bf k}})$ for soft boson field to the case of a continuous medium is performed. The corresponding Hamilton equations
are defined and the general structure of the third- and fourth-order interaction Hamiltonians in the normal field variables and in the hard modes   of the hot quark-gluon plasma, is written out. In section \ref{section_3}, the canonical transformations including bosonic and hard mode degrees of freedom of the quark-gluon plasma are discussed. Two systems of canonicity conditions for these transformations, based on the Poisson superbracket are derived. The most general structure of canonical transformations in the form of integro-power series in the new normal field variables $(c^{\phantom{\ast}\!\!a}_{\hspace{0.03cm}{\bf k}},\,c^{\,\ast\,a}_{\hspace{0.03cm}{\bf k}})$ and new hard momentum mode variables $(\zeta^{\,i}_{{\bf p}^{\phantom\prime}},\,\zeta^{\hspace{0.03cm}\ast
\hspace{0.03cm}i}_{{\bf p}})$ up to the terms of sixth order is presented.\\
\indent In section \ref{section_4}, the problem of removing the “non-essential” third-order Hamiltonian $H^{(3)}$ is addressed. Explicit expressions for the coefficient functions in quadratic terms in $c^{\!\phantom{\ast}a}_{\hspace{0.03cm}{\bf k}}$ and $\zeta^{\,i}_{{\bf p}^{\phantom\prime}}$ of canonical transformations, are obtained. An explicit form of the complete effective amplitude $\mathscr{T}^{\,(2)\hspace{0.03cm} i\; i_{1}\, a_{1}\, a_{2}}_{\; {\bf p},\, {\bf p}_{1},\, {\bf k}_{1},\, {\bf k}_{2}}$ describing the elastic scattering process of plasmon off a hard color particle in leading tree-level order is given and the corresponding effective fourth-order Hamiltonian ${\mathcal H}^{(4)}_{g\hspace{0.02cm}G\hspace{0.02cm}\rightarrow\hspace{0.02cm} g\hspace{0.02cm}G}$, is written out. Section \ref{section_5} is concerned with the calculation of fourth- and sixth-order correlation functions in the new normal field variable $c^{\phantom{\ast}\!\!a}_{\hspace{0.03cm}{\bf k}}$ and the new hard mode variable $\zeta^{\,i}_{{\bf p}^{\phantom\prime}}$. The notions of the plasmon number density ${\mathcal N}^{\;a\hspace{0.03cm}a^{\prime}_{\phantom{1}}\!}_{\hspace{0.02cm}
{\bf k}}$\!, and of the number density of hard modes ${\mathfrak n}^{\hspace{0.03cm}i^{\hspace{0.02cm}\prime}\hspace{0.02cm}i}_{{\bf p}}$
are introduced. For the hard momentum modes of quark-gluon plasma excitations, we suggest an ansatz that separates the color and momentum degrees of freedom. On the basis of Hamilton’s equations of motion with the Poisson superbracket, a differential equation to which the fourth-order correlation function obeys, is defined.\\ 
\indent In section \ref{section_6} an approximate solution to the equation for the fourth-order correlator, accounting for the deviation of the four-point correlation function from the Gaussian approximation at a low  level of nonlinearity in interacting Bose-excitations is found. On the basis of this solution, a matrix kinetic equation for the number density of color plasmons describing the elastic scattering process of collective gluon excitations off a hard color-charged particle, is constructed. In section \ref{section_7} the question of approximation of effective subamplitudes $\mathscr{T}^{\,(2,\hspace{0.03cm}{\mathcal A})}_{{\bf p}_{1},\, {\bf p}_{2}, {\bf k}, {\bf k}_{1}}$ and $\mathscr{T}^{\,(2,\hspace{0.03cm}{\mathcal S})}_{{\bf p}_{1},\, {\bf p}_{2}, {\bf k}, {\bf k}_{1}}$ in the limit when the momenta of the hard excitations are much larger than the momenta of the soft plasma excitations, is considered. An approximate expression for the effective amplitude $\mathscr{T}^{\,(2,\hspace{0.03cm}{\mathcal A})}_{{\bf p}_{1},\, {\bf p}_{2}, {\bf k}, {\bf k}_{1}}$ is derived and a simple graphical interpretation of the individual terms in the effective amplitude, is provided.\\
\indent In section \ref{section_8} we consider an approximation of the matrix kinetic equation for soft gluon excitations in the limit of large hard excitation momenta. The color decomposition of the matrix function ${\mathcal N}^{\;a\hspace{0.03cm}a^{\prime}_{\phantom{1}}\!}_{\hspace{0.02cm}
{\bf k}}$ is written out and the first moment about color of the matrix kinetic equation defining a scalar kinetic equation for the colorless part $N^{\hspace{0.03cm}l}_{\bf k}$ of this decomposition, is calculated. 
Section \ref{section_9} is devoted to the determination of the second moment about color of the matrix kinetic equation. A special case of the $SU(3_{c})$ color group, is discussed. In section \ref{section_10} the derivation of the equation of motion for the expected value of the colorless charge ${\mathcal Q}$, is considered. In section \ref{section_11} the derivation of the equation of motion for the expected value of the color charge ${\mathcal Q}^{\hspace{0.03cm}a}$, is discussed. Nonlinear differential equations of first order for the colorless combinations of second ${\mathfrak q}_{2}(t)$ and third ${\mathfrak q}_{3}(t)$ orders with respect to the mean value $\langle{\mathcal Q}^{\hspace{0.03cm}a}\rangle$, are derived. It is shown that for the special case of the $SU(3_{c})$ color group, these two equations are completely self-consistent and their explicit analytical solutions, are obtained.\\ 
\indent In Appendix \ref{appendix_A} we provide the basic expressions for the effective three-plasmon vertex functions and the effective gluon propagator within the framework of the hard thermal loop approximation. In Appendix \ref{appendix_B}, all the necessary relations and traces of a product of generators in the defining representation of the $SU(N_{c})$ color group, are given. In Appendix \ref{appendix_C} the necessary traces of a product of generators in the adjoint representation of the $SU(N_{c})$ color group up to the fifth order as well as some useful relations between these generators are given.

\section{Interaction Hamiltonian of plasmons and hard particles}
\setcounter{equation}{0}
\label{section_2}

Let us consider the application of the general Zakharov theory \cite{zakharov_1971, zakharov_1974, zakharov_1985_1, zakharov_book_1992, krasitskii_1990, zakharov_1997} to a specific system, namely to a high-temperature quark-gluon plasma in the semiclassical approximation. The gauge field potentials describing the gluon field in the system are $N_c\times N_c$ matrices in the color space and are defined in terms of $A_{\mu}(x) = A_{\mu}^{a}(x)\, t^{a}$ with $N^{\hspace{0.03cm}2}_c - 1$ Hermitian generators $t^{a}$ of the color $SU(N_c)$ group in the fundamental representation\footnote{\,The color indices $a,\,b,\,c,\,\ldots$ run through values $1,2,\,\ldots\,,N^{\hspace{0.02cm}2}_{c}-1$, while the vector indices $\mu,\,\nu,\,\lambda,\,\ldots$ run through values $0,1,2,3$. Everywhere in this article, we imply summation over repeated indices and use the system of units with $\hbar = c = 1$.}.\\
\indent It is known that there exist two types of the physical soft gluon fields in an equilibrium hot quark-gluon plasma:  transverse- and longitudinal-polarized ones \cite{kalashnikov_1980}. For simplicity, we confine our analysis only to processes involving longitudinally polarized plasma excitations, which are known as {\it plasmons}. These excitations are a purely collective effect of the medium, which has no analogs in the conventional quantum field theory. Let us consider the gauge field potential in the form of the decomposition into plane waves \cite{blaizot_1994(1), hakim_book_2011}
\begin{equation}
A^{a}_{\mu}(x) = \int\!d\hspace{0.02cm}{\bf k}\left(\frac{Z_{l}({\bf k})}
{2\omega^{l}_{{\bf k}}}\right)^{\!\!1/2}\!\!
\left\{\epsilon^{\ \! l}_{\mu}({\bf k})\, a^{\phantom{\ast}\!\!a}_{{\bf k}}\ \!e^{-i\hspace{0.03cm}\omega^{l}_{{\bf k}}\hspace{0.02cm}t\hspace{0.03cm} +\hspace{0.03cm} i\hspace{0.03cm}{\bf k}\hspace{0.02cm}\cdot\hspace{0.02cm} {\bf x}}
+
\epsilon^{\ast\, l}_{\mu}({\bf k})\, a^{\ast\ \!\!a}_{{\bf k}}\ \!e^{\hspace{0.02cm}i\hspace{0.03cm}\omega^{l}_{{\bf k}}\hspace{0.02cm}t\hspace{0.03cm} -\hspace{0.03cm} i\hspace{0.03cm}{\bf k}\hspace{0.02cm}\cdot\hspace{0.02cm} {\bf x}}
\right\},
\label{eq:2q}
\end{equation}
where $\epsilon^{\ \! l}_{\mu} ({\bf k})$ is the polarization vector of a longitudinal mode  (${\bf k}$ is the wave vector). The asterisk $\ast$ denotes the complex conjugation. The factor $Z_{l}({\bf k})$ is the residue of the effective gluon propagator at the longitudinal pole. Finally, $\omega^{\ \! l}_{{\bf k}}$ is the dispersion relation of the longitudinal mode. We consider the amplitude  $a^{\phantom{\ast}\!\!a}_{{\bf k}}$ for longitudinal excitations as ordinary (complex) random function. The expectation value of the product of two bosonic amplitudes is
\begin{equation}
\bigl\langle\hspace{0.03cm}a^{\ast\hspace{0.03cm}a}_{{\bf k}}\hspace{0.03cm} a^{\phantom{\ast}\!\!b}_{{\bf k}^{\prime}}\bigr\rangle
=
\delta^{\hspace{0.03cm}a\hspace{0.02cm}b}\hspace{0.03cm}\delta({\bf k} - {\bf k}^{\prime})\hspace{0.05cm}{\mathcal N}^{\hspace{0.03cm}l}_{\bf k},
\label{eq:2w}
\end{equation}
where ${\mathcal N}^{\hspace{0.03cm}l}_{\bf k}$ is the number density of the longitudinal plasma waves. The dispersion relation $\omega^{\ \! l}_{{\bf k}}$ for plasmons satisfies the following dispersion equation \cite{kalashnikov_1980}:
\begin{equation}
{\rm Re}\ \!\varepsilon^{l}(\omega,{\bf k})=0\ \!,
\label{eq:2e}
\end{equation}
where
\[
\varepsilon^{l}(\omega,{\bf k})=1+\frac{3\omega^{2}_{pl}}{{\bf k}^{\ \! 2}}
\biggl[1-F\biggl(\frac{\omega}{|{\bf k}|^{2}}\biggr)\biggr],
\quad
F(x) = \frac{x}{2}\left[\hspace{0.03cm}\ln\left|\frac{1+x}{1-x}\right|-i\pi\theta(1-|x|)\right]
\]
is the longitudinal permittivity, $\omega_{\rm pl}^2 = g^2(2N_c+N_f)T^2/18$ is a plasma frequency squared, $T$ is the temperature of the system, $g$ is the strong interaction constant, and $N_{f}$ represents the number of flavors of massless quarks.\\
\indent As it was said already above, the amplitudes $a^{\phantom{\ast}\!\!a}_{{\bf k}}$ and $a^{\ast\ \!\!a}_{{\bf k}}$ in the expansion for the longitudinal mode of oscillations (\ref{eq:2q}) are usual (commuting) normal variables of the gauge field satisfying the Poisson superbracket relations
\begin{equation}
\bigl\{a^{\phantom{\ast}\!\!a}_{{\bf k}},\,a^{\phantom{\ast}\!\!b}_{{\bf k}^{\prime}}\bigr\}_{\rm SPB} = 0,
\quad\!
\bigl\{a^{\ast\ \!\!a}_{{\bf k}},\,a^{\ast\ \!\!b}_{{\bf k}^{\prime}}\bigr\}_{\rm SPB} = 0, 
\quad\!
\bigl\{a^{\phantom{\ast}\!\!a}_{{\bf k}},\,a^{\ast\ \!\!b}_{{\bf k}^{\prime}}\bigr\}_{\rm SPB}
=
\delta^{\hspace{0.02cm}a\hspace{0.02cm}b}\hspace{0.03cm} \delta({\bf k} - {\bf k}^{\prime}).
\label{eq:2r}
\end{equation}
From the other hand, in full analogy to our work \cite{markov_2023}, we consider the amplitudes $\xi^{\,i}_{{\bf p}}$ and $\xi^{\hspace{0.03cm}\ast\,i}_{{\bf p}}$ for {\it hard} momentum modes of excitations of a quark-gluon plasma as Grassmann-valued (anticommu\-ting) variables, the Poisson superbrackets $({\rm SPB})$ of which have the following standard form:
\begin{equation}
\hspace{0.04cm}
\bigl\{\xi^{\,i}_{{\bf p}^{\phantom\prime}},\hspace{0.03cm} \xi^{\,j}_{{\bf p}^{\prime}}\!\hspace{0.02cm}\bigr\}_{\rm SPB}\! = 0,
\quad
\bigl\{\xi^{\,\ast\, i}_{{\bf p}^{\phantom\prime}},\hspace{0.03cm} \xi^{\hspace{0.03cm}\ast\hspace{0.03cm}j}_{{\bf p}^{\prime}}\hspace{0.01cm} \bigr\}_{\rm SPB}\! = 0, 
\quad
\bigl\{\xi^{\,i}_{{\bf p}^{\phantom\prime}},\hspace{0.03cm} \xi^{\hspace{0.03cm}\ast\hspace{0.03cm}j}_{{\bf p}^{\prime}}\hspace{0.01cm} \bigr\}_{\rm SPB}\!
=
\delta^{\,i\hspace{0.01cm}j}\hspace{0.03cm}\delta({\bf p}\, -\, {\bf p}^{\prime}),
\label{eq:2t}
\end{equation}
here, $i,\,j = 1,\ldots,N_{c}$. For the case of a continuous media we take the following expression as the definition of the Poisson superbracket
\begin{equation}
\bigl\{F,\,G\bigr\}_{\rm SPB} 
\label{eq:2y}
\end{equation}
\[
=\!
\int\! d\hspace{0.02cm}{\bf k\hspace{0.01cm}}'\!\hspace{0.02cm}
\left\{\frac{\delta\hspace{0.01cm} F}{\delta\hspace{0.02cm} a^{\phantom{\ast}\!\!c}_{{\bf k}'}}
\,\frac{\delta\hspace{0.01cm}  G}{\delta\hspace{0.02cm} a^{\ast\ \!\!c}_{{\bf k}'}}
\,-\,
\frac{\delta\hspace{0.01cm}  F}{\delta\hspace{0.02cm} a^{\ast\ \!\!c}_{{\bf k}'}}\,
\frac{\delta\hspace{0.01cm}  G}{\delta\hspace{0.02cm} a^{\phantom{\ast}\!\!c}_{{\bf k}'}}\right\}
+
\int\! d{\bf p\hspace{0.01cm}}'\!\hspace{0.02cm}
\left\{\frac{\!\!\overleftarrow{\delta}\! F}{\,\delta\hspace{0.02cm} \xi^{\phantom{\ast}\!\!i}_{{\bf p}^{\prime}}}\,
\frac{\!\!\overrightarrow{\delta}\! G}{\,\delta\hspace{0.02cm} \xi^{\hspace{0.03cm} \ast\ \!\!i}_{{\bf p}^{\prime}}}
\,+\,(-1)^{P_{F} + P_{G}}\,
\frac{\!\!\overrightarrow{\delta}\! F}{\,\delta\hspace{0.02cm} \xi^{\hspace{0.03cm} \ast\ \!\!i}_{{\bf p}^{\prime}}}\,
\frac{\!\!\overleftarrow{\delta}\! G}{\,\delta\hspace{0.02cm} \xi^{\phantom{\ast}\!\!i}_{{\bf p}^{\prime}}}\right\}.
\]
Here, $\overrightarrow{\delta}\!/\delta\hspace{0.02cm} \xi^{\,*\hspace{0.03cm}i}_{{\bf p}}$ and $\overleftarrow{\delta}\!/\delta\hspace{0.02cm} \xi^{\,i}_{{\bf p}}$ are the right and left functional derivatives\footnote{\,In our notations of the right and left variational derivatives we follow the notations accepted for the right and left derivatives adopted in \cite{casalbuoni_1976_1, casalbuoni_1976_2, berezin_1987} and therefore, 
\[
\delta F = 
\int\! d\hspace{0.02cm}{\bf k\hspace{0.01cm}}'\!\hspace{0.02cm}
\left\{\frac{\delta F}{\delta\hspace{0.02cm} a^{\phantom{\ast}\!\!c}_{{\bf k}'}}\, 
\delta\hspace{0.02cm} a^{\phantom{\ast}\!\!c}_{{\bf k}'}
\,+\,
\frac{\delta  F}{\delta\hspace{0.02cm} a^{\ast\ \!\!c}_{{\bf k}'}}\,
\delta\hspace{0.02cm} a^{\ast\ \!\!c}_{{\bf k}'}\right\}
\,+\,
\int\! d{\bf p\hspace{0.01cm}}'\!\hspace{0.02cm}
\left\{\frac{\!\!\overleftarrow{\delta}\! F}{\,\delta\hspace{0.02cm} \xi^{\phantom{\ast}\!\!i}_{{\bf p}^{\prime}}}\,
\delta\hspace{0.02cm} \xi^{\phantom{\ast}\!\!i}_{{\bf p}^{\prime}}
\,+\,
\delta\hspace{0.02cm} \xi^{\,\ast\ \!\!i}_{{\bf p}^{\prime}}\,
\frac{\!\!\overrightarrow{\delta}\! F}{\,\delta\hspace{0.02cm} \xi^{\,\ast\ \!\!i}_{{\bf p}^{\prime}}}\right\}.
\]}, 
$P_{F}$ and $P_{G}$ designate Grassmann parity of the functions $F$ and $G$, correspondingly. For simplicity of notation the abbreviation ${\rm SPB}$ will be omitted, thereby suggesting that by the braces $\{\,,\}$ we always mean the Poisson superbrackets.\\
\indent Let us write the Hamilton equations for the functions  $a^{\phantom{\ast}\!\!a}_{{\bf k}}$, $\xi^{\,i}_{{\bf p}}$ and their complex conjugation 
\begin{equation}
\frac{\partial\hspace{0.02cm}a^{\phantom{\ast}\!\!a}_{{\bf k}}}{\partial\hspace{0.02cm} t}
=
-i\hspace{0.02cm}\bigl\{a^{\phantom{\ast}\!\!a}_{{\bf k}}, H\bigr\} \equiv  -i\,\frac{\delta H}{\delta\hspace{0.01cm} a^{\ast\ \!\!a}_{{\bf k}}},
\qquad
\frac{\partial\hspace{0.02cm}a^{\ast\ \!\!a}_{{\bf k}}}{\partial\hspace{0.02cm} t}
=
-i\bigl\{a^{\ast\ \!\!a}_{{\bf k}}, H\bigr\} \equiv  i\,\frac{\delta H}{\delta\hspace{0.01cm} a^{\phantom{\ast}\!\!a}_{{\bf k}}},
\label{eq:2u}
\end{equation}
\begin{equation}
\frac{\partial\hspace{0.02cm}\xi^{\,i}_{{\bf p}}}{\partial\hspace{0.02cm} t}
=
-i\left\{\xi^{\,i}_{\bf p}, H\right\} \equiv  -i\,\frac{\!\!\!\!\overrightarrow{\delta}\! H}{\,\delta\hspace{0.015cm} \xi^{\,*\hspace{0.03cm}i}_{{\bf p}}},
\qquad
\frac{\partial\hspace{0.02cm}\xi^{\,\ast\,i}_{{\bf p}}}{\!\!\partial\hspace{0.02cm} t}
=
-i\left\{\xi^{\,\ast\,i}_{\bf p}, H\right\} \equiv  i\,\frac{\!\!\overleftarrow{\delta}\! H}
{\,\delta\hspace{0.015cm} \xi^{\,i}_{{\bf p}}}.
\label{eq:2i}
\end{equation}
Here, the function $H$ represents a Hamiltonian for the system of plasmons and hard particles, which is equal to a sum $H =  H^{(0)} + H_{int}$, where
\begin{equation}
H^{(0)} =  \!\int\!d\hspace{0.02cm}{\bf k}\ \omega^{l}_{{\bf k}}\ \!
a^{\ast\hspace{0.03cm}a}_{{\bf k}} a^{ a}_{{\bf k}}
+
\int\!d{\bf p}\;\varepsilon_{\bf p}\ \!
\xi^{\,\ast\hspace{0.04cm}i}_{{\bf p}}\ \!\xi^{\,i}_{{\bf p}}\
\label{eq:2o}
\end{equation}
is the Hamiltonian of noninteracting plasmons and hard particles, ${H}_{int}$ is the interaction Hamiltonian, and $\varepsilon_{\bf p}$ is hard particle energy
\begin{equation}
\varepsilon_{\bf p}\simeq |{\bf p}|.
\label{eq:2p}
\end{equation}
\indent In the approximation of small amplitudes, the interaction Hamiltonian can be presented in the form of a formal integro-power series in the bosonic functions ${a}^{a}_{{\bf k}}$ and ${a}^{\ast\hspace{0.03cm} a}_{{\bf k}}$, and in the fermionic ones $\xi^{\,i}_{{\bf p}}$ and $\xi^{\,\ast\hspace{0.04cm}i}_{{\bf p}}$:
\[
H_{int} = H^{(3)} + H^{(4)} + \, \ldots\,\,,
\]
where the third-order interaction Hamiltonian has the following structure:
\begin{align}
H^{(3)} 
&= 
\int\!d\hspace{0.02cm}{\bf k}\, d\hspace{0.02cm}{\bf k}_{1}\hspace{0.03cm} d\hspace{0.02cm}{\bf k}_{2}\,
\Bigl\{\hspace{0.02cm}{\mathcal V}^{\; a\, a_{1}\, a_{2}}_{{\bf k},\, {\bf k}_{1},\, {\bf k}_{2}}\, a^{\ast\hspace{0.03cm}  a}_{{\bf k}}\,
a^{\,a_{1}}_{{\bf k}_{1}}\, a^{\,a_{2}}_{{\bf k}_{2}}
\,+\,
{\mathcal V}^{\,*\,a\, a_{1}\, a_{2}}_{{\bf k},\, {\bf k}_{1},\, {\bf k}_{2}}\, 
a^{\!\phantom{\ast}a}_{{\bf k}}\, a^{\ast\,a_{1}}_{{\bf k}_{1}}\hspace{0.03cm}  a^{\ast\, a_{2}}_{{\bf k}_{2}}
\Bigr\}\hspace{0.03cm}\delta({\bf k} - {\bf k}_{1} - {\bf k}_{2}) 
\label{eq:2a}\\[0.7ex]
+\,\frac{1}{3}&\int\!d\hspace{0.02cm}{\bf k}\, d\hspace{0.02cm}{\bf k}_{1}\hspace{0.03cm} d\hspace{0.02cm}{\bf k}_{2}\,
\Bigl\{\hspace{0.02cm}{\mathcal U}^{\; a\, a_{1}\, a_{2}}_{\,{\bf k},\, {\bf k}_{1},\, {\bf k}_{2}}\, a^{a}_{{\bf k}}\, a^{a_{1}}_{{\bf k}_{1}}\,
a^{a_{2}}_{{\bf k}_{2}}
\,+\,
{\mathcal U}^{\,*\,a\, a_{1}\, a_{2}}_{\; {\bf k},\,{\bf k}_{1},\, {\bf k}_{2}}\, a^{\ast\ \!a}_{{\bf k}}\,
a^{\ast\,a_{1}}_{{\bf k}_{1}}\, a^{\ast\,a_{2}}_{{\bf k}_{2}}
\Bigr\}\hspace{0.03cm}\delta({\bf k} + {\bf k}_{1} + {\bf k}_{2}) 
\notag\\[0.7ex]
+\!&\int\!d\hspace{0.02cm}{\bf k}\, d{\bf p}_{1}\, d{\bf p}_{2}\,
\Bigl\{\hspace{0.02cm}{\Phi}^{\;a\,i_{1}\hspace{0.03cm}i_{2}}_{\hspace{0.03cm} {\bf k},\, {\bf p}_{1},\, {\bf p}_{2}}\; {a}^{\,a}_{{\bf k}}\,
\xi^{\,\ast\,i_{1}}_{{\bf p}_{1}}\, \xi^{\; i_{2}}_{{\bf p}_{2}}\,
\delta({\bf k} - {\bf p}_{1} + {\bf p}_{2})
+
{\Phi}^{\,\ast\, a\, i_{2}\,  i_{1}}_{\hspace{0.03cm} {\bf k},\, {\bf p}_{2},\, {\bf p}_{1}}\, {a}^{\,\ast\, a}_{{\bf k}}\,
\xi^{\,\ast\, i_{1}}_{{\bf p}_{1}}\, \xi^{\; i_{2}}_{{\bf p}_{2}}\,
\delta({\bf k} + {\bf p}_{1} - {\bf p}_{2})
\Bigr\}
\notag
\end{align}
and, correspondingly, the fourth-order interaction Hamiltonian is
\begin{equation}
\begin{split}
H^{(4)}\! 
=
&\int\!d{\bf p}\, d{\bf p}_{1}\hspace{0.03cm}d\hspace{0.02cm}{\bf k}_{1}\hspace{0.03cm} d\hspace{0.02cm}{\bf k}_{2}\,
T^{\hspace{0.03cm} (2)\hspace{0.03cm} i\; i_{1}\, a_{1}\, a_{2}}_{\; {\bf p},\, {\bf p}_{1},\, {\bf k}_{1},\, {\bf k}_{2}}\,
\xi^{\,\ast\, i}_{{\bf p}} \xi^{\;i_{1}}_{{\bf p}_{1}}\, a^{\ast\ \!\!a_{1}}_{{\bf k}_{1}} a^{\hspace{0.03cm}a_{2}}_{{\bf k}_{2}}\,
\delta({\bf p} + {\bf k}_{1} - {\bf p}_{1} - {\bf k}_{2}),
\\[1ex]
+\;
\frac{1}{2}&\int\!d{\bf p}\, d{\bf p}_{1}\hspace{0.03cm} d{\bf p}_{2}\hspace{0.04cm}d{\bf p}_{3}\, 
T^{\,(2)\, i\, i_{1}\, i_{2}\, i_{3}}_{{\bf p},\, {\bf p}_{1},\, {\bf p}_{2},\, {\bf p}_{3}}\, 
\xi^{\,\ast\, i}_{{\bf p}}\hspace{0.04cm} \xi^{\,\ast\, i_{1}}_{{\bf p}_{1}}\hspace{0.03cm} \xi^{\;i_{2}}_{{\bf p}_{2}}\, \xi^{\;i_{3}}_{{\bf p}_{3}}\,
\delta({\bf p} + {\bf p}_{1} - {\bf p}_{2} - {\bf p}_{3}).
\end{split}
\label{eq:2s}
\end{equation}
As shown in \cite{markov_2023} other possible the third-order contributions in (\ref{eq:2a}) in the framework of the hard thermal loop (HTL) approximation are equal to zero. In the expression (\ref{eq:2s}) the first term describes plasmon\,--\,hard-particle scattering with the resonance condition 
\[
\left\{
\begin{array}{l}
{\bf k} + {\bf p} = {\bf k}_{1} + {\bf p}_{1},\\[5pt]
\omega^{l}_{{\bf k}} + \varepsilon_{\bf p} = \omega^{l}_{{\bf k}_{1}} + \varepsilon_{{\bf p}_{1}}.
\end{array}
\right.
\]
The second term is associated with the interaction of hard excitations among themselves. The expression (\ref{eq:2a}) is a direct analog of the third-order interaction Hamiltonian (2.14) from the paper \cite{markov_2023}, where to the substitutions
\begin{equation}
{\bf q} \Rightarrow {\bf p}, \qquad \omega_{\bf q}^{-} \Rightarrow \varepsilon_{\bf p},
\qquad 
{b}^{\!\phantom{\ast}\! i}_{{\bf q}} \Rightarrow {\xi}^{\!\phantom{\ast}\! b}_{{\bf p}},
\qquad
{b}^{\ast\ \!\!i}_{{\bf q}} \Rightarrow {\xi}^{\ast\ \!\!i}_{{\bf p}},
\label{eq:2d}
\end{equation}
one should add substitutions of three- and four-point coefficient functions
\begin{equation}
{\mathcal P}^{\;a_{1}\,i\;i_{1}}_{{\bf k}_{1},\, {\bf q},\, {\bf q}_{1}}
\Rightarrow
{\Phi}^{\;a\,i_{1}\hspace{0.03cm}i_{2}}_{{\bf k},\, {\bf p}_{1},\, {\bf p}_{2}},
\quad
T^{\,(2)\, i\, i_{1}\, a_{1}\, a_{2}}_{\ {\bf q},\, {\bf q}_{1},\, {\bf k}_{1},\, {\bf k}_{2}}
\Rightarrow
T^{\,(2)\, i\; i_{1}\, a_{1}\, a_{2}}_{\; {\bf p},\, {\bf p}_{1},\, {\bf k}_{1},\, {\bf k}_{2}}.
\label{eq:2f}
\end{equation}
\indent The vertex functions ${\mathcal V}^{\; a\, a_{1}\, a_{2}}_{{\bf k},\, {\bf k}_{1},\, {\bf k}_{2}}$ and ${\mathcal U}^{\; a\, a_{1}\, a_{2}}_{{\bf k},\,{\bf k}_{1},\,{\bf k}_{2}}$ satisfy the ``conditions of natural symmetry'', which specify that the integrals in Eqs.\,(\ref{eq:2a}) and (\ref{eq:2s}) are unaffected by relabeling of the dummy color indices and integration variables. These conditions have the following form:
\begin{align}
&{\mathcal V}^{\; a\, a_{1}\, a_{2}}_{{\bf k},\, {\bf k}_{1},\, {\bf k}_{2}} = {\mathcal V}^{\; a\, a_{2}\, a_{1}}_{{\bf k},\, {\bf k}_{2},\, {\bf k}_{1}},
\quad
{\mathcal U}^{\; a\, a_{1}\, a_{2}}_{{\bf k},\, {\bf k}_{1},\, {\bf k}_{2}} = {\mathcal U}^{\; a\, a_{2}\, a_{1}}_{{\bf k},\, {\bf k}_{2},\, {\bf k}_{1}}
= {\mathcal U}^{\, a_{1}\, a_{2}\, a}_{{\bf k}_{1},\, {\bf k}_{2},\, {\bf k}} ,\notag\\[1ex]
&T^{\,(2)\, i\, i_{1}\, i_{2}\, i_{3}}_{{\bf p},\, {\bf p}_{1},\, {\bf p}_{2},\, {\bf p}_{3}}
=
-\hspace{0.03cm}T^{\,(2)\, i_{1}\hspace{0.03cm}i\, i_{2}\, i_{3}}_{{\bf p}_{1},\,{\bf p},\,{\bf p}_{2},\, {\bf p}_{3}}
=
-\hspace{0.03cm}T^{\,(2)\, i\, i_{1}\, i_{3}\, i_{2}}_{{\bf p},\, {\bf p}_{1},\, {\bf p}_{3},\, {\bf p}_{2}}.
\notag
\end{align}
The real nature of the Hamiltonian (\ref{eq:2a}) is obvious. A  reality of the Hamiltonian (\ref{eq:2s}) entails a validity of additional relations for the vertex functions $T^{\hspace{0.03cm} (2)\hspace{0.03cm} i\; i_{1}\, a_{1}\, a_{2}}_{\; {\bf p},\, {\bf p}_{1},\, {\bf k}_{1},\, {\bf k}_{2}}$ and $T^{\,(2)\, i\, i_{1}\, i_{2}\, i_{3}}_{{\bf p},\, {\bf p}_{1},\, {\bf p}_{2},\, {\bf p}_{3}}$:
\begin{equation*}
T^{\hspace{0.03cm} (2)\hspace{0.03cm} i\; i_{1}\, a_{1}\, a_{2}}_{\; {\bf p},\, {\bf p}_{1},\, {\bf k}_{1},\, {\bf k}_{2}}
=
T^{\,\ast\hspace{0.03cm} (2)\, i_{1}\, i\, a_{2}\, a_{1}}_{\ {\bf p}_{1},\, {\bf p},\, {\bf k}_{2},\, {\bf k}_{1}},
\qquad
T^{\,(2)\, i\, i_{1}\, i_{2}\, i_{3}}_{{\bf p},\, {\bf p}_{1},\, {\bf p}_{2},\, {\bf p}_{3}}
=
T^{\,(2)\, i_{2}\, i_{3}\, i\, i_{1}}_{{\bf p}_{2},\,{\bf p}_{3},\, {\bf p},\, {\bf p}_{1}}.
\end{equation*}
The information about a concrete physical system, in our case about a hot quark-gluon plasma, is contained in the dispersion law $\omega^{l}_{{\bf k}}$ and in the form of the interaction vertex functions in the Hamiltonians $H^{(3)}$ and $H^{(4)}$. In particular, an explicit form of the three-point amplitudes ${\mathcal V}^{\; a\, a_{1}\, a_{2}}_{{\bf k},\, {\bf k}_{1},\, {\bf k}_{2}}$ and ${\mathcal U}^{\; a\, a_{1}\, a_{2}}_{{\bf k},\, {\bf k}_{1},\, {\bf k}_{2}}$ within the hard thermal loop approximation was obtained in \cite{markov_2020}. They have the following color and momentum structures: 
\begin{equation}
{\mathcal V}^{\ \! a\, a_{1}\hspace{0.03cm} a_{2}}_{{\bf k},\, {\bf k}_{1},\, {\bf k}_{2}}
=
f^{\hspace{0.03cm} a\, a_{1}\hspace{0.03cm} a_{2}\,}\hspace{0.02cm}{\mathcal V}_{\, {\bf k},\, {\bf k}_{1},\, {\bf k}_{2}},
\qquad
{\mathcal U}^{\ \! a\, a_{1}\hspace{0.03cm} a_{2}}_{{\bf k},\, {\bf k}_{1},\, {\bf k}_{2}}
=
f^{\hspace{0.03cm} a\, a_{1}\hspace{0.03cm} a_{2}\,}\hspace{0.02cm}{\mathcal U}_{\, {\bf k},\, {\bf k}_{1},\, {\bf k}_{2}},
\label{eq:2j}
\end{equation}
where the explicit form of the functions ${\mathcal V}_{\, {\bf k},\, {\bf k}_{1},\, {\bf k}_{2}}\!$ and ${\mathcal U}_{\, {\bf k},\, {\bf k}_{1},\, {\bf k}_{2}}$ is written out in  Appendix \ref{appendix_A}, Eqs.\,(\ref{ap:A1}) and (\ref{ap:A2}).

\section{Canonical transformations}
\setcounter{equation}{0}
\label{section_3}

Let us consider the transformation from the initial bosonic and fermionic variables $a^{a}_{\bf k}$ and $\xi^{\,i}_{\bf p}$ to the new bosonic and fermionic ones $c^{a}_{\hspace{0.02cm}\bf k}$ and $\zeta^{\,i}_{\bf p}$:
\begin{align}
&a^{a}_{\bf k} = a^{a}_{{\bf k}}(c^{a}_{\hspace{0.02cm}\bf k},\, c^{\ast\ \!\!a}_{{\hspace{0.02cm}\bf k}}\!,\, 
\zeta^{\,i}_{\bf p},\, \zeta^{\,\ast\ \!\!i}_{{\bf p}}),
\label{eq:3q}\\[0.8ex]
&\xi^{\,i}_{\bf p} = \xi^{\,i}_{{\bf p}}\hspace{0.02cm}(\hspace{0.02cm}c^{a}_{\hspace{0.02cm}\bf k},\, 
c^{\ast\ \!\!a}_{{\hspace{0.02cm}\bf k}}\!,\, \zeta^{\,i}_{\bf p},\,\zeta^{\,\ast\ \!\!i}_{{\bf p}}\hspace{0.02cm}).
\label{eq:3w}
\end{align}
We shall demand that the Hamilton equations in terms of new functions have the form (\ref{eq:2u}) and (\ref{eq:2i}) with the same Hamiltonian $H$. Straightforward but rather cumbersome calculations result in two systems of integral relations. The first of them has the following form:
\begin{subequations} 
\label{eq:3e}
\begin{align}
&\int\! d\hspace{0.02cm}{\bf k\hspace{0.01cm}}'\!\hspace{0.01cm}
\left\{\frac{\delta\hspace{0.01cm}  a^{\phantom{\ast}\!\!a}_{{\bf k}}}{\delta c^{\phantom{\ast}\!\!c}_{\hspace{0.02cm}{\bf k}'}}
\,\frac{\delta\hspace{0.01cm}  a^{\ast\ \!\!b}_{{\bf k}''}}{\delta c^{\ast\ \!\!c}_{\hspace{0.02cm}{\bf k}'}}
\,-\,
\frac{\delta\hspace{0.01cm}  a^{\phantom{\ast}\!\!a}_{{\bf k}}}
{\delta c^{\ast\ \!\!c}_{\hspace{0.02cm}{\bf k}'}}\,
\frac{\delta\hspace{0.01cm}  a^{\ast\ \!\!b}_{{\bf k}''}}
{\delta c^{\phantom{\ast}\!\!c}_{\hspace{0.02cm}{\bf k}'}}\right\}
+
\int\! d\hspace{0.02cm}{\bf p\hspace{0.01cm}}'\!\hspace{0.01cm}
\left\{\frac{\overleftarrow{\delta}\!\!\hspace{0.04cm} a^{\phantom{\ast}\!\!a}_{{\bf k}}}{\delta\hspace{0.01cm} \zeta^{\phantom{\ast}\!\!k}_{{\bf p}^{\prime}}}\,
\frac{\overrightarrow{\delta}\!\!\hspace{0.04cm} a^{\ast\ \!\! b}_{{\bf k}''}}{\delta\hspace{0.01cm} \zeta^{\, \ast\ \!\!k}_{{\bf p}^{\prime}}}
\,+\,
\frac{\overrightarrow{\delta}\!\!\hspace{0.04cm} a^{\phantom{\ast}\!\!a}_{{\bf k}}}{\delta\hspace{0.01cm} \zeta^{\, \ast\ \!\!k}_{{\bf p}^{\prime}}}\,
\frac{\overleftarrow{\delta}\!\!\hspace{0.04cm} a^{\ast\ \!\! b}_{{\bf k}''}}{\delta\hspace{0.01cm} \zeta^{\phantom{\ast}\!\!k}_{{\bf p}^{\prime}}}\right\}
\!=
\delta^{ab}\delta ({\bf k}-{\bf k}\!\ ''),
\label{eq:3ea}
\\[0.8ex]
&\int\! d\hspace{0.02cm}{\bf k\hspace{0.01cm}}'\!\hspace{0.01cm}\left\{\frac{\delta\hspace{0.02cm}  a^{\phantom{\ast}\!\!a}_{{\bf k}}}
{\delta c^{\phantom{\ast}\!\!c}_{\hspace{0.02cm}{\bf k}^{\prime}}}
\,\frac{\delta\hspace{0.02cm}  a^{\phantom{\ast}\!\!b}_{{\bf k}''}}
{\delta c^{\ast\ \!\!c}_{\hspace{0.02cm}{\bf k}^{\prime}}}
\,-\,
\frac{\delta\hspace{0.02cm}  a^{\phantom{\ast}\!\!a}_{{\bf k}}}
{\delta c^{\ast\ \!\!c}_{\hspace{0.02cm}{\bf k}^{\prime}}}\,
\frac{\delta\hspace{0.02cm}  a^{\phantom{\ast}\!\!b}_{{\bf k}''}}
{\delta c^{\phantom{\ast}\!\!c}_{\hspace{0.02cm}{\bf k}^{\prime}}}\right\}
+
\int\! d\hspace{0.02cm}{\bf p\hspace{0.01cm}}'\!\hspace{0.01cm}
\left\{\frac{\overleftarrow{\delta}\!\!\hspace{0.04cm} a^{\phantom{\ast}\!\!a}_{{\bf k}}}{\delta\hspace{0.01cm} \zeta^{\phantom{\ast}\!\!k}_{{\bf p}^{\prime}}}\,
\frac{\overrightarrow{\delta}\!\!\hspace{0.04cm} a^{b}_{{\bf k}''}}{\delta\hspace{0.01cm} \zeta^{\, \ast\ \!\!k}_{{\bf p}^{\prime}}}
\,+\,
\frac{\overrightarrow{\delta}\!\!\hspace{0.04cm} a^{\phantom{\ast}\!\!a}_{{\bf k}}}{\delta\hspace{0.01cm} \zeta^{\, \ast\ \!\!k}_{{\bf p}^{\prime}}}\,
\frac{\overleftarrow{\delta}\!\!\hspace{0.04cm} a^{b}_{{\bf k}''}}{\delta\hspace{0.01cm} \zeta^{\phantom{\ast}\!\!k}_{{\bf p}^{\prime}}}\right\} = 0,
\label{eq:3eb}
\\[0.8ex]
&\int\! d\hspace{0.02cm}{\bf k\hspace{0.01cm}}'\!\hspace{0.01cm}\left\{\frac{\delta\hspace{0.02cm}  a^{\phantom{\ast}\!\!a}_{{\bf k}}}
{\delta c^{\phantom{\ast}\!\!c}_{\hspace{0.02cm}{\bf k}^{\prime}}}
\,\frac{\delta\hspace{0.02cm}  \xi^{\,i}_{{\bf p}''}}
{\delta c^{\ast\ \!\!c}_{\hspace{0.02cm}{\bf k}^{\prime}}}
\,-\,
\frac{\delta\hspace{0.02cm}  a^{\phantom{\ast}\!\!a}_{{\bf k}}}
{\delta c^{\ast\ \!\!c}_{\hspace{0.02cm}{\bf k}^{\prime}}}\,
\frac{\delta\hspace{0.02cm}  \xi^{\,i}_{{\bf p}''}}
{\delta c^{\phantom{\ast}\!\!c}_{\hspace{0.02cm}{\bf k}^{\prime}}}\right\}
+
\int\! d\hspace{0.02cm}{\bf p\hspace{0.01cm}}'\!\hspace{0.01cm}
\left\{\frac{\overleftarrow{\delta}\!\!\hspace{0.04cm} a^{\phantom{\ast}\!\!a}_{{\bf k}}}{\delta\hspace{0.01cm} \zeta^{\phantom{\ast}\!\!k}_{{\bf p}^{\prime}}}\,
\frac{\overrightarrow{\delta}\! \xi^{\,i}_{{\bf p}''}}{\delta\hspace{0.01cm} \zeta^{\, \ast\ \!\!k}_{{\bf p}^{\prime}}}
\,-\,
\frac{\overrightarrow{\delta}\!\!\hspace{0.04cm} a^{\phantom{\ast}\!\!a}_{{\bf k}}}{\delta\hspace{0.01cm} \zeta^{\, \ast\ \!\!k}_{{\bf p}^{\prime}}}\,
\frac{\overleftarrow{\delta}\! \xi^{\,i}_{{\bf p}''}}{\delta\hspace{0.01cm} \zeta^{\phantom{\ast}\!\!k}_{{\bf p}^{\prime}}}\right\} = 0,
\label{eq:3ec}\\[0.8ex]
&\int\! d\hspace{0.02cm}{\bf k\hspace{0.01cm}}'\!\hspace{0.01cm}\left\{\frac{\delta\hspace{0.02cm}  a^{\phantom{\ast}\!\!a}_{{\bf k}}}
{\delta c^{\phantom{\ast}\!\!c}_{\hspace{0.02cm}{\bf k}^{\prime}}}
\,\frac{\delta\hspace{0.02cm} \xi^{\,\ast\hspace{0.02cm}i}_{{\bf p}''}}
{\delta c^{\ast\ \!\!c}_{\hspace{0.02cm}{\bf k}^{\prime}}}
\,-\,
\frac{\delta\hspace{0.02cm}  a^{\phantom{\ast}\!\!a}_{{\bf k}}}
{\delta c^{\ast\ \!\!c}_{\hspace{0.02cm}{\bf k}^{\prime}}}\,
\frac{\delta\hspace{0.02cm}  \xi^{\,\ast\hspace{0.02cm}i}_{{\bf p}''}}
{\delta c^{\phantom{\ast}\!\!c}_{\hspace{0.02cm}{\bf k}^{\prime}}}\right\}
+
\int\! d\hspace{0.02cm}{\bf p\hspace{0.01cm}}'\!\hspace{0.01cm}
\left\{\frac{\overleftarrow{\delta}\!\!\hspace{0.04cm} a^{\phantom{\ast}\!\!a}_{{\bf k}}}{\delta\hspace{0.01cm} \zeta^{\phantom{\ast}\!\!k}_{{\bf p}^{\prime}}}\,
\frac{\overrightarrow{\delta}\! \xi^{\,\ast\hspace{0.02cm}i}_{{\bf p}''}}{\delta\hspace{0.01cm} \zeta^{\, \ast\ \!\!k}_{{\bf p}^{\prime}}}
\,-\,
\frac{\overrightarrow{\delta}\!\!\hspace{0.04cm} a^{\phantom{\ast}\!\!a}_{{\bf k}}}{\delta\hspace{0.01cm} \zeta^{\, \ast\ \!\!k}_{{\bf p}^{\prime}}}\,
\frac{\overleftarrow{\delta}\! \xi^{\,\ast\hspace{0.02cm}i}_{{\bf p}''}}{\delta\hspace{0.01cm} \zeta^{\phantom{\ast}\!\!k}_{{\bf p}^{\prime}}}\right\} = 0.
\label{eq:3ed}
\end{align}
\end{subequations}
The second system is written in a similar way. These canonicity conditions can be presented in a very compact form if we make use of the definition of the Poisson superbracket (\ref{eq:2y}) and replace the variation variables by the new ones: $a^{a}_{\bf k}\rightarrow c^{\,a}_{\hspace{0.02cm}\bf k}$ and $\xi^{\,i}_{\bf p} \rightarrow \zeta^{\,i}_{\bf p}$. In this case the superbrackets for the original variables $a^{\,a}_{\bf k}$ and $\xi^{\,i}_{\bf p}$, Eqs.\,(\ref{eq:2r}) and (\ref{eq:2t}), turn to the canonicity conditions (\ref{eq:3e}), which impose certain restrictions on the functional dependencies (\ref{eq:3q}) and (\ref{eq:3w}). Let us present the right-hand sides of  (\ref{eq:3q}) and (\ref{eq:3w}) in the form of integro-power series in the normal variables $c^{\,a}_{\hspace{0.02cm}\bf k}$ and $\zeta^{\,i}_{\bf p}$. The most common dependence of the transformation (\ref{eq:3q}) up to cubic terms in $c^{\,a}_{\hspace{0.02cm}\bf k}$ and $\zeta^{\,i}_{\bf p}$ has the following form:
\begin{equation}
a^{\,a}_{{\bf k}} = c^{\,a}_{\hspace{0.02cm}{\bf k}}\,+
\label{eq:3t}
\end{equation}
\[
+ \int\!d\hspace{0.02cm}{\bf k}_{1}\hspace{0.02cm}d\hspace{0.02cm}{\bf k}_{2} 
\left[V^{(1)\, a\, a_{1}\, a_{2}}_{\ {\bf k},\, {\bf k}_{1},\, {\bf k}_{2}}\, c^{a_{1}}_{{\bf k}_{1}}\, c^{a_{2}}_{{\bf k}_{2}}
\,+\,
V^{(2)\, a\, a_{1}\, a_{2}}_{\ {\bf k},\, {\bf k}_{1},\, {\bf k}_{2}}\, c^{\ast\, a_{1}}_{{\bf k}_{1}}\,c^{\phantom{\ast}\!\!a_{2}}_{{\bf k}_{2}}
\,+\,
V^{(3)\ a\, a_{1}\, a_{2}}_{\ {\bf k},\, {\bf k}_{1},\, {\bf k}_{2}}\; 
c^{\ast\, a_{1}}_{{\bf k}_{1}} c^{\ast\, a_{2}}_{{\bf k}_{2}}\right] 
\]
\[
+\int\!d{\bf p}_{1}\hspace{0.02cm}d{\bf p}_{2}\hspace{0.03cm}
\left[
F^{(1)\, a\, i_{1}\, i_{2}}_{\ {\bf k},\, {\bf p}_{1},\, {\bf p}_{2}}\; \zeta^{\,i_{1}}_{{\bf p}_{1}} \zeta^{\,i_{2}}_{{\bf p}_{2}}
\,+\,
F^{(2)\, a\, i_{1}\, i_{2}}_{\ {\bf k},\, {\bf p}_{1},\, {\bf p}_{2}}\ \zeta^{\,\ast\, i_{1}}_{{\bf p}_{1}} \zeta^{\phantom{\ast}\!i_{2}}_{{\bf p}_{2}}
+
F^{(3)\, a\, i_{1}\, i_{2}}_{\ {\bf k},\, {\bf p}_{1},\, {\bf p}_{2}} \zeta^{\,\ast\hspace{0.04cm}i_{1}}_{{\bf p}_{1}} 
\zeta^{\,\ast\hspace{0.04cm}i_{2}}_{{\bf p}_{2}}\right] 
\hspace{0.6cm}
\]
\vspace{-0.5cm}
\begin{align}
\hspace{0.5cm}
+
\!\int\!d\hspace{0.02cm}{\bf k}_{1}\hspace{0.02cm} d{\bf p}_{1}\hspace{0.03cm}d{\bf p}_{2}\,
\Bigl[\hspace{0.03cm} 
&J^{\hspace{0.03cm}(1)\, a\, a_{1}\, i_{1}\, i_{2}}_{\ {\bf k},\, {\bf k}_{1},\, {\bf p}_{1},\, {\bf p}_{2}}\, c^{a_{1}}_{{\bf k}_{1}}\, \zeta^{\, i_{1}}_{{\bf p}_{1}} \zeta^{\, i_{2}}_{{\bf p}_{2}}
\,+\,
J^{\hspace{0.03cm}(2)\, a\, a_{1}\, i_{1}\, i_{2}}_{\ {\bf k},\, {\bf k}_{1},\, {\bf p}_{1},\, {\bf p}_{2}}\, c^{a_{1}}_{{\bf k}_{1}}\,
\zeta^{\,\ast\, i_{1}}_{{\bf p}_{1}} \zeta^{\phantom{\ast}\!\!i_{2}}_{{\bf p}_{2}}
\notag\\[0.5ex]
+\;
&J^{\hspace{0.03cm}(3)\, a\, a_{1}\, i_{1}\, i_{2}}_{\ {\bf k},\, {\bf k}_{1},\, {\bf p}_{1},\, {\bf p}_{2}}\ 
c^{a_{1}}_{{\bf k}_{1}}\, \zeta^{\,\ast\, i_{1}}_{{\bf p}_{1}} \zeta^{\,\ast\,i_{2}}_{{\bf p}_{2}}
\,+\,
J^{\hspace{0.03cm}(4)\, a\, a_{1}\, i_{1}\, i_{2}}_{\ {\bf k},\, {\bf k}_{1},\, {\bf p}_{1},\, {\bf p}_{2}}\, c^{\ast\, a_{1}}_{{\bf k}_{1}}\, \zeta^{\, i_{1}}_{{\bf p}_{1}}\, \zeta^{\, i_{2}}_{{\bf p}_{2}}
\notag\\[1ex]
+\;
&J^{\hspace{0.03cm}(5)\, a\, a_{1}\, i_{1}\, i_{2}}_{\ {\bf k},\, {\bf k}_{1},\, {\bf p}_{1},\, {\bf p}_{2}}\, 
c^{\ast\, a_{1}}_{{\bf k}_{1}} \zeta^{\,\ast\, i_{1}}_{{\bf p}_{1}}\hspace{0.03cm} \zeta^{\, i_{2}}_{{\bf p}_{2}}
\,+\,
J^{\hspace{0.03cm}(6)\, a\, a_{1}\, i_{1}\, i_{2}}_{\ {\bf k},\, {\bf k}_{1},\, {\bf p}_{1},\, {\bf p}_{2}}\, 
c^{\ast\, a_{1}}_{{\bf k}_{1}} \zeta^{\,\ast\, i_{1}}_{{\bf p}_{1}} \zeta^{\,\ast\, i_{2}}_{{\bf p}_{2}}\Bigr] +\,\ldots\;\ _{.}
\notag
\end{align}
Similarly, the most common dependence for the transformation (\ref{eq:3w}) up to cubic terms is
\begin{equation}
\xi^{\,i}_{{\bf p}} = \zeta^{\,i}_{{\bf p}}\,+
\label{eq:3y}
\end{equation}
\[
+\int\!d\hspace{0.02cm}{\bf k}_{1}\hspace{0.02cm}d{\bf p}_{1}\hspace{0.03cm}
\Bigl[\,
Q^{(1)\, i\, a_{1}\, i_{1}}_{\ {\bf p},\, {\bf k}_{1},\, {\bf p}_{1}}\, c^{a_{1}}_{{\bf k}_{1}}\, \zeta^{\,i_{1}}_{{\bf p}_{1}}
+\,
Q^{(2)\, i\, a_{1}\, i_{1}}_{\ {\bf p},\, {\bf k}_{1},\, {\bf p}_{1}}\, c^{a_{1}}_{{\bf k}_{1}}\, \zeta^{\, \ast\,i_{1}}_{{\bf p}_{1}}
+\,
Q^{(3)\, i\, a_{1}\, i_{1}}_{\ {\bf p},\, {\bf k}_{1},\, {\bf p}_{1}}\, c^{\ast\, a_{1}}_{{\bf k}_{1}}\hspace{0.03cm} \zeta^{\,i_{1}}_{{\bf p}_{1}}
+\,
Q^{(4)\, i\, a_{1}\, i_{1}}_{\ {\bf p},\, {\bf k}_{1},\, {\bf p}_{1}}\, c^{\ast\, a_{1}}_{{\bf k}_{1}}\hspace{0.03cm} 
\zeta^{\, \ast\,i_{1}}_{{\bf p}_{1}}\,\Bigr]
\]
\vspace{-0.3cm}
\begin{align}
+\!\int\!d\hspace{0.02cm}{\bf k}_{1}\hspace{0.02cm}d\hspace{0.02cm}{\bf k}_{2}\hspace{0.02cm}d{\bf p}_{1}\hspace{0.03cm}
\Bigl[&R^{\,(1)\, i\, a_{1}\, a_{2}\, i_{1}}_{\ {\bf p},\, {\bf k}_{1},\, {\bf k}_{2},\, {\bf p}_{1}}\, c^{a_{1}}_{{\bf k}_{1}}\, c^{a_{2}}_{{\bf k}_{2}}\, \zeta^{\,i_{1}}_{{\bf p}_{1}}
\,+\,
R^{\,(2)\, i\, a_{1}\, a_{2}\, i_{1}}_{\ {\bf p},\, {\bf k}_{1},\, {\bf k}_{2},\, {\bf p}_{1}}\, c^{\ast\, a_{1}}_{{\bf k}_{1}}\, c^{a_{2}}_{{\bf k}_{2}}\, \zeta^{\,i_{1}}_{{\bf p}_{1}} 
\notag\\[0.5ex]
+\, &R^{\,(3)\, i\, a_{1}\, a_{2}\, i_{1}}_{\ {\bf p},\, {\bf k}_{1},\, {\bf k}_{2},\, {\bf p}_{1}}\, c^{\ast\, a_{1}}_{{\bf k}_{1}}\hspace{0.03cm} c^{\ast\, a_{2}}_{{\bf k}_{2}}\hspace{0.03cm} \zeta^{\,i_{1}}_{{\bf p}_{1}} 
\,+\,
R^{\,(4)\, i\, a_{1}\, a_{2}\, i_{1}}_{\ {\bf p},\, {\bf k}_{1},\, {\bf k}_{2},\, {\bf p}_{1}}\, c^{a_{1}}_{{\bf k}_{1}}\, c^{a_{2}}_{{\bf k}_{2}}\, \zeta^{\, \ast\,i_{1}}_{{\bf p}_{1}}
\notag\\[1.7ex]
+\, &R^{\,(5)\, i\, a_{1}\, a_{2}\, i_{1}}_{\ {\bf p},\, {\bf k}_{1},\, {\bf k}_{2},\, {\bf p}_{1}}\, c^{\ast\, a_{1}}_{{\bf k}_{1}}\hspace{0.03cm} c^{a_{2}}_{{\bf k}_{2}}\, \zeta^{\, \ast\,i_{1}}_{{\bf p}_{1}}
\,+\,
R^{\,(6)\, i\, a_{1}\, a_{2}\, i_{1}}_{\ {\bf p},\, {\bf k}_{1},\, {\bf k}_{2},\, {\bf p}_{1}}\, c^{\ast\, a_{1}}_{{\bf k}_{1}}\, c^{\ast\, a_{2}}_{{\bf k}_{2}}\, \zeta^{\, \ast\,i_{1}}_{{\bf p}_{1}}\Bigr] 
\notag
\end{align}
\vspace{-0.7cm}
\begin{align}
\hspace{1.4cm}
+\!\int\!d{\bf p}_{1}\hspace{0.02cm}d{\bf p}_{2}\hspace{0.02cm}d{\bf p}_{3}\hspace{0.03cm}
\Bigl[&S^{\,(1)\, i\; i_{1}\, i_{2}\, i_{3}}_{\ {\bf p},\, {\bf p}_{1},\, {\bf p}_{2},\, {\bf p}_{3}}\, \zeta^{\,i_{1}}_{{\bf p}_{1}}\, \zeta^{\,i_{2}}_{{\bf p}_{2}}\, \zeta^{\,i_{3}}_{{\bf p}_{3}}
\,+\,
S^{\,(2)\, i\; i_{1}\, i_{2}\, i_{3}}_{\ {\bf p},\, {\bf p}_{1},\, {\bf p}_{2},\, {\bf p}_{3}}\, \zeta^{\,\ast\, i_{1}}_{{\bf p}_{1}}\hspace{0.03cm} \zeta^{\,i_{2}}_{{\bf p}_{2}}\, \zeta^{\,i_{3}}_{{\bf p}_{3}}
\notag\\[0.6ex]
+\,
&S^{\,(3)\, i\; i_{1}\, i_{2}\, i_{3}}_{\ {\bf p},\, {\bf p}_{1},\, {\bf p}_{2},\, {\bf p}_{3}}\, \zeta^{\,\ast\, i_{1}}_{{\bf p}_{1}}\hspace{0.03cm} \zeta^{\,\ast\, i_{2}}_{{\bf p}_{2}}\hspace{0.03cm} \zeta^{\,i_{3}}_{{\bf p}_{3}}
+
S^{\,(4)\, i\; i_{1}\, i_{2}\, i_{3}}_{\ {\bf p},\, {\bf p}_{1},\, {\bf p}_{2},\, {\bf p}_{3}}\, \zeta^{\,\ast\, i_{1}}_{{\bf p}_{1}}\hspace{0.03cm} \zeta^{\,\ast\, i_{2}}_{{\bf p}_{2}}\hspace{0.03cm} \zeta^{\,\ast\, i_{3}}_{{\bf p}_{3}}\Bigr] +\,\ldots\;\ _{.}
\notag
\end{align}
Note first of all that the coefficient functions $V^{(1)\, a\, a_{1}\, a_{2}}_{\ {\bf k},\, {\bf k}_{1},\, {\bf k}_{2}}$, $V^{(1)\, a\, a_{1}\, a_{2}}_{\ {\bf k},\, {\bf k}_{1},\, {\bf k}_{2}}$, $F^{(1,3)\, a\, i_{1}\, i_{2}}_{\ {\bf k},\, {\bf p}_{1},\, {\bf p}_{2}}$, $J^{\hspace{0.03cm}(1,3,4,6)\, a\, a_{1}\, i_{1}\, i_{2}}_{\ {\bf k},\, {\bf k}_{1},\, {\bf p}_{1},\, {\bf p}_{2}}$, $R^{\hspace{0.03cm}(1,3,4,6)\, i\, a_{1}\, a_{2}\, i_{1}}_{\ {\bf p},\, {\bf k}_{1},\, {\bf k}_{2},\, {\bf p}_{1}}$ and $S^{\hspace{0.03cm}(1,2,3,4)\, i\; i_{1}\, i_{2}\, i_{3}}_{\ {\bf p},\, {\bf p}_{1},\, {\bf p}_{2},\, {\bf p}_{3}}$ must satisfy the evident conditions of natural symmetry which we don't write out here.\\
\indent Further, substituting the expansions (\ref{eq:3t}) and  (\ref{eq:3y}) into the system of the canonicity conditions (\ref{eq:3e}), we obtain rather nontrivial integral relations connecting various coefficient functions among themselves. A complete list of the integral relations connecting the coefficient functions of the second and third orders can be written out in full analogy with the corresponding relations from the paper \cite{markov_2023}. These integral relations will not be needed in the present work, so we will not give them. Here, we provide only algebraic relations for the second-order coefficient functions:
\begin{equation}
\begin{split}
&V^{(2)\, a\, a_{1}\, a_{2}}_{\ {\bf k},\, {\bf k}_{1},\, {\bf k}_{2}} = -\hspace{0.01cm}2\hspace{0.03cm}V^{\,\ast\hspace{0.03cm}(1)\, a_{2}\, a_{1}\, a}_{\ {\bf k}_{2},\, {\bf k}_{1},\, {\bf k}},
\quad
V^{(3)\, a\, a_{1}\, a_{2}}_{\ {\bf k},\, {\bf k}_{1},\, {\bf k}_{2}} 
= 
V^{(3)\, a_{1}\, a\, a_{2}}_{\ {\bf k}_{1},\, {\bf k},\, 
{\bf k}_{2}},\\[1ex]
&Q^{(1)\,i_{1}\hspace{0.03cm}a\,i_{2}}_{\ {\bf p}_{1},\, {\bf k},\, {\bf p}_{2}} 
= 
-\hspace{0.01cm}F^{\,\ast\hspace{0.03cm} (2)\, a\, i_{2}\, i_{1}}_{\ {\bf k},\, {\bf p}_{2},\, {\bf p}_{1}},
\qquad
Q^{(2)\,i_{1}\hspace{0.03cm}a\,i_{2}}_{\ {\bf p}_{1},\, {\bf k},\, {\bf p}_{2}} 
= 
2\hspace{0.01cm} F^{\,\ast\hspace{0.03cm} (1)\, a\, i_{1}\, i_{2}}_{\ {\bf k},\, {\bf p}_{1},\, {\bf p}_{2}},
\\[1ex]
&Q^{(3)\,i_{1}\hspace{0.03cm}a\,i_{2}}_{\ {\bf p}_{1},\, {\bf k},\, {\bf p}_{2}} 
=  
F^{(2)\, a\, i_{1}\, i_{2}}_{\ {\bf k},\, {\bf p}_{1},\, {\bf p}_{2}},
\qquad\quad\,
Q^{(4)\,i_{1}\hspace{0.03cm}a\,i_{2}}_{\ {\bf p}_{1},\, {\bf k},\, {\bf p}_{2}} 
=  
2\hspace{0.01cm}F^{(3)\, a\, i_{1}\, i_{2}}_{\ {\bf k},\, {\bf p}_{1},\, {\bf p}_{2}}.
\end{split}
\label{eq:3p}
\end{equation}

\section{Eliminating ``non-essential '' Hamiltonian $H^{(3)}$. Effective fourth-order Hamiltonian}
\setcounter{equation}{0}
\label{section_4}

The next step in the construction of an effective theory is the procedure of eliminating the third-order interaction Hamiltonian $H^{(3)}$, Eq.\,(\ref{eq:2a}), upon switching from the original bosonic and fermionic functions $a^{\hspace{0.03cm}a}_{\bf k}$ and $\xi^{\,i}_{\bf p}$ to the new functions $c^{\,a}_{\bf k}$ and $\zeta^{\,i}_{\bf p}$ as a result of the canonical transformations (\ref{eq:3t}) and (\ref{eq:3y}). This elimination procedure is presented in detail in \cite{markov_2023}, so here we only give a brief description of the procedure and its final result, which follows from expressions (4.3) of \cite{markov_2023}, with appropriate substitutions (\ref{eq:2d}) and (\ref{eq:2f}).\\
\indent To achieve eliminating the third-order interaction Hamiltonian $H^{(3)}$, we substitute the expansions (\ref{eq:3t}) and (\ref{eq:3y}) into the free-field Hamiltonian $H^{(0)}$, Eq.\,(\ref{eq:2o}), and keep only the terms cubic in $c^{\,a}_{{\bf k}}$ and $\zeta^{\,i}_{\bf p}$. Then in the Hamiltonian $H^{(3)}$, Eq.\,(\ref{eq:2a}), we perform the replacements: $a^{\hspace{0.03cm}a}_{\bf k}\rightarrow c^{\,a}_{\bf k}$ and $\xi^{\,i}_{\bf p} \rightarrow \zeta^{\,i}_{\bf p}$. Adding the expression thus obtained to that which follows from the free-field Hamiltonian $H^{(0)}$, collecting similar terms and using the relations (\ref{eq:3p}), finally we obtain an explicit form of the coefficient functions in the quadratic part of the canonical transformations (\ref{eq:3t}) and (\ref{eq:3y}) that exclude the cubic terms in the interaction Hamiltonian:
\begin{equation}
\left\{\!\!\!
\begin{array}{ll}
&V^{(1)\,a\,a_{1}\hspace{0.03cm}a_{2}}_{\ {\bf k},\, {\bf k}_{1},\, {\bf k}_{2}} 
=
-\hspace{0.02cm}\displaystyle\frac{{\mathcal V}^{\,a\,a_{1}\hspace{0.03cm}a_{2}}_{\ {\bf k},\, {\bf k}_{1},\, {\bf k}_{2}}}
{\omega^{\,l}_{{\bf k}} - \omega^{\,l}_{{\bf k}_{1}} - \omega^{\,l}_{{\bf k}_{2}}}\,
\delta({\bf k}-{\bf k}_{1}-{\bf k}_{2}), 
\\[3ex]
&V^{(3)\,a\,a_{1}\hspace{0.03cm}a_{2}}_{\ {\bf k},\, {\bf k}_{1},\, {\bf k}_{2}}
= 
-\hspace{0.02cm}\displaystyle\frac{{\mathcal U}^{\hspace{0.03cm}*\,a\, a_{1}\hspace{0.03cm}a_{2}}_{\ {\bf k},\, {\bf k}_{1},\, {\bf k}_{2}}}{\omega^{\,l}_{{\bf k}} + \omega^{\,l}_{{\bf k}_{1}} + \omega^{\,l}_{{\bf k}_{2}}}\,
\delta({\bf k}+{\bf k}_{1}+{\bf k}_{2}),
\end{array}\
\right.
\label{eq:4q}
\end{equation}
\vspace{0.15cm}
\[
\left\{\!\!\!
\begin{array}{ll}
&F^{(1)\,a_{1}\hspace{0.03cm}i\,i_{1}}_{\;{\bf k}_1,\, {\bf p},\, {\bf p}_{1}}
\,=\,
F^{(3)\,a_{1}\hspace{0.03cm}i\,i_{1}}_{\;{\bf k}_1,\, {\bf p},\, {\bf p}_{1}}
\,=\,0,
\\[3ex]
&F^{(2)\,a_{1}\hspace{0.03cm}i\,i_{1}}_{\;{\bf k}_1,\, {\bf p},\, {\bf p}_{1}}
=
-\hspace{0.02cm}\displaystyle\frac{\Phi^{\hspace{0.03cm}\ast\,a_{1}\hspace{0.03cm}i_{1}\,i}_{\;{\bf k}_1,\, {\bf p}_{1},\, {\bf p}}}
{ \omega^{\,l}_{{\bf k}_{1}} - \varepsilon_{{\bf p}_{1}} + \varepsilon_{{\bf p}}}\,
\delta({\bf k}_{1} - {\bf p}_{1} + {\bf p}).
\end{array}\
\right.
\]
The coefficients $V^{(2)}$ and $Q^{(n)},\,n = 1,\,2,\,3,\,4$ are found from Eq.\,(\ref{eq:3p}). We have previously obtained the relations (\ref{eq:4q}) in \cite{markov_2020}. These expressions imply that due to specific character of the dispersion equations for soft bosonic excitations (\ref{eq:2e}) and for hard mode excitations (\ref{eq:2p}) in the hot quark-gluon plasma, the resonance conditions for three-wave processes with plasmons and  for Cherenkov radiation (or absorption) of plasmons by a hard particle
\[
\vspace{-0.3cm}
\left\{
\begin{array}{ll}
{\bf k} = {\bf k}_{1} + {\bf k}_{2}, \\[1.5ex]
\omega^{\,l}_{{\bf k}} = \omega^{\,l}_{{\bf k}_{1}} + \omega^{\,l}_{{\bf k}_{2}},
\end{array}
\right.
\quad
\left\{
\begin{array}{ll}
{\bf k} + {\bf k}_{1} + {\bf k}_{2} = 0, \\[1.5ex]
\omega^{\,l}_{{\bf k}} + \omega^{\,l}_{{\bf k}_{1}} + \omega^{\,l}_{{\bf k}_{2}} = 0,
\end{array}\
\right.
\quad
\left\{
\begin{array}{ll}
{\bf p}={\bf p}_{1} - {\bf k}_{1}, \\[1.5ex]
\varepsilon_{{\bf p}} = \varepsilon_{{\bf p}_{1}} - \omega^{\,l}_{{\bf k}_{1}}
\end{array}\
\right.
\]
have no solutions. In other words, the processes of emission or absorption of collective excitation by another collective excitation and by a hard particle that lie on the mass shells $\omega = \omega^{\,l}_{\bf k}$ and $\varepsilon = \varepsilon_{\bf p}$ are forbidden.\\
\indent Next we write out an explicit form of the effective fourth-order Hamiltonian, which describes the elastic scattering of plasmon off hard particle. In terms of the original variables $a^{a}_{\bf k}$ and $\xi^{\,i}_{\bf p}$, the Hamiltonian for the scattering process is defined by the first term on the right-hand side of (\ref{eq:2s}). In this term we make the substitution $a^{\,a}_{\bf k}\rightarrow c^{\,a}_{\bf k}$ and $\xi^{\,i}_{\bf p} \rightarrow \zeta^{\,i}_{\bf p}$. Further we define all similar terms of fourth-order product $c^{\,\ast\, a}_{{\bf k}}\hspace{0.03cm} c^{\;a_{1}}_{{\bf k}_{1}}\hspace{0.03cm} {\zeta}^{\ast\ \!\!i_{1}}_{{\bf p}_{1}}\hspace{0.03cm} {\zeta}^{\!\phantom{\ast}\! i_{2}}_{{\bf p}_{2}}$ from the free-field Hamiltonian $H^{(0)}$, Eq.\,(\ref{eq:2o}), and from the Hamiltonian $H^{(3)}$, Eq.\,(\ref{eq:2a}), to be arisen under the canonical transformations (\ref{eq:3t}) and (\ref{eq:3y}). Putting the pieces together, we result in the effective fourth-order Hamiltonian describing the elastic scattering process of plasmon off a hard color particle:  
\begin{equation}
{\mathcal H}^{(4)}_{g\hspace{0.02cm}G\hspace{0.02cm}\rightarrow\hspace{0.02cm} g\hspace{0.02cm}G} 
=
\int\!
\mathscr{T}^{\hspace{0.03cm} (2)\hspace{0.03cm} i\; i_{1}\, a_{1}\, a_{2}}_{\; {\bf p},\, {\bf p}_{1},\, {\bf k}_{1},\, {\bf k}_{2}}\,
\zeta^{\,\ast\, i}_{{\bf p}} \zeta^{\;i_{1}}_{{\bf p}_{1}}\, c^{\ast\ \!\!a_{1}}_{{\bf k}_{1}} c^{\hspace{0.03cm}a_{2}}_{{\bf k}_{2}}\,
\delta({\bf p} + {\bf k}_{1} - {\bf p}_{1} - {\bf k}_{2})\,
d{\bf p}\, d{\bf p}_{1}\hspace{0.03cm}d\hspace{0.02cm}{\bf k}_{1}\hspace{0.03cm} d\hspace{0.02cm}{\bf k}_{2},
\label{eq:4e}
\end{equation}
where the {\it complete effective amplitude} $\mathscr{T}^{\hspace{0.03cm} (2)\hspace{0.03cm} i\; i_{1}\, a_{1}\, a_{2}}_{\; {\bf p},\, {\bf p}_{1},\, {\bf k}_{1},\, {\bf k}_{2}}$ has the following structure:
\begin{equation}
\mathscr{T}^{\,(2)\hspace{0.03cm} i\; i_{1}\, a_{1}\, a_{2}}_{\; {\bf p},\, {\bf p}_{1},\, {\bf k}_{1},\, {\bf k}_{2}}
=
T^{\,(2)\hspace{0.03cm} i\; i_{1}\, a_{1}\, a_{2}}_{\; {\bf p},\, {\bf p}_{1},\, {\bf k}_{1},\, {\bf k}_{2}}
\label{eq:4r}
\vspace{-0.3cm}
\end{equation}
\begin{align}
-\,
\frac{1}{2}\,\Biggl[
\Biggl(&\frac{1}{\omega^{\hspace{0.02cm} l}_{{\bf k}_{2}} - \varepsilon_{{\bf p}} + \varepsilon_{{\bf p} - {\bf k}_{2}}} 
\,+\,
\frac{1}{\omega^{\hspace{0.02cm} l}_{{\bf k}_{1}} - \varepsilon_{{\bf p}_{1}} + \varepsilon_{{\bf p}_{1} - {\bf k}_{1}}} 
\Biggr)\hspace{0.02cm}
{\Phi}^{\, a_{2}\hspace{0.03cm} i\hspace{0.03cm}  j}_{{\bf k}_{2},\, {\bf p},\, {\bf p} - {\bf k}_{2}}\, 
{\Phi}^{\hspace{0.03cm}\ast\, a_{1}\hspace{0.03cm} i_{1}\hspace{0.03cm} j}_{{\bf k}_{1},\, {\bf p}_{1},\, {\bf p}_{1} - {\bf k}_{1}}
\notag\\[1.5ex]
-\,
\Biggl(&\frac{1}{\omega^{\hspace{0.02cm} l}_{{\bf k}_{2}} - \varepsilon_{{\bf k}_{2} +\hspace{0.03cm} {\bf p}_{1}} + 
\varepsilon_{{\bf p}_{1}}} 
+
\frac{1}{\omega^{\hspace{0.02cm} l}_{{\bf k}_{1}} - \varepsilon_{{\bf k}_{1} +\hspace{0.03cm} {\bf p}} + \varepsilon_{{\bf p}}} 
\Biggr)\hspace{0.02cm}
{\Phi}^{\, a_{2}\hspace{0.03cm} j\hspace{0.03cm} i_{1}}_{{\bf k}_{2},\, {\bf k}_{2} + {\bf p}_{1},\, {\bf p}_{1}}\, 
{\Phi}^{\hspace{0.03cm}\ast\, a_{1}\hspace{0.03cm} j\hspace{0.03cm} i}_{{\bf k}_{1},\, {\bf k}_{1} + {\bf p},\, {\bf p}}
\Biggr]
\notag\\[1.5ex]
+\,
\Biggl(
&\frac{1}{\omega^{\hspace{0.02cm} l}_{{\bf k}_{1}} - \omega^{\hspace{0.02cm} l}_{{\bf k}_{2}} - \omega^{\hspace{0.02cm} l}_{{\bf k}_{1} - {\bf k}_{2}}}
\,-\,
\frac{1}{\omega^{\hspace{0.02cm} l}_{{\bf p}_{1} - {\bf p}} - \varepsilon_{{\bf p}_{1}} + \varepsilon_{{\bf p}}}
\Biggr)\hspace{0.02cm}
{\mathcal V}^{\;a_{1}\hspace{0.03cm}a_{2}\, a}_{{\bf k}_{1}, {\bf k}_{2},\, {\bf k}_{1} - {{\bf k}_{2}}}\, 
{\Phi}^{\hspace{0.03cm}\ast\,a\,i_{1}\hspace{0.03cm}i}_{{\bf p}_{1} - {\bf p},\, {\bf p}_{1},\, {\bf p}}
\notag\\[1.5ex]
+\, 
\Biggl(
&\frac{1}{\omega^{\hspace{0.02cm} l}_{{\bf k}_{2}} - \omega^{\hspace{0.02cm} l}_{{\bf k}_{1}} - \omega^{\hspace{0.02cm} l}_{{\bf k}_{2} - {\bf k}_{1}}}
\,-\,
\frac{1}{\omega^{\hspace{0.02cm} l}_{{\bf p} - {\bf p}_{1}} - \varepsilon_{{\bf p}} + \varepsilon_{{\bf p}_{1}}}
\Biggr)\hspace{0.02cm}
{\Phi}^{\,a\, i\,i_{1}}_{{\bf p} - {\bf p}_{1},\, {\bf p},\, {\bf p}_{1}}\, 
{\mathcal V}^{\hspace{0.03cm}\ast\, a_{2}\hspace{0.03cm}a_{1}\hspace{0.03cm}a}_{{\bf k}_{2}, {\bf k}_{1},\, {\bf k}_{2} - {\bf k}_{1}}.
\notag
\end{align}
Hereinafter, the effective Hamiltonians will be de\-signated by the calligraphic letter ${\mathcal H}$, including also the Hamiltonian ${\mathcal H}^{(0)}$ for non-interacting plasmons and hard particles in the new variables:
\begin{equation*}
{\mathcal H}^{(0)} =  \!\int\!d\hspace{0.02cm}{\bf k}\;\omega^{l}_{{\bf k}}\ \!
c^{\ast\hspace{0.03cm}a}_{{\bf k}}\hspace{0.03cm} c^{\phantom{\ast}\!\! a}_{{\bf k}}
+
\int\!d{\bf p}\;\varepsilon^{\phantom{b}}_{\bf p}\hspace{0.04cm}
\zeta^{\,\ast\hspace{0.03cm}i}_{{\bf p}}\hspace{0.01cm} \zeta^{\,i}_{{\bf p}}.
\end{equation*}

\section{\bf Fourth-order correlation function for soft and hard excita\-tions}
\label{section_5}
\setcounter{equation}{0}

Let us consider the construction of a system of kinetic equations describing the elastic scattering process of plasmon off a hard particle. As the interaction Hamiltonian here, we take the effective Hamiltonian ${\mathcal H}^{(4)}_{gG\rightarrow gG}$, Eq.\,(\ref{eq:4e}). The equations of motion for the fermionic $\zeta^{\,i^{\hspace{0.03cm}\prime}}_{{\bf p}^{\prime}}, \,\zeta^{\hspace{0.03cm}\ast\hspace{0.03cm}i}_{{\bf p}}$ and bosonic $c^{\phantom{\hspace{0.03cm}\ast} \!\!a}_{{\bf k}}, \,c^{\hspace{0.03cm}\ast\,i}_{{\bf k}^{\prime}}$ normal variables are defined by the corresponding Hamilton equations. For the hard particle excitations we have
\begin{equation}
\frac{\partial\hspace{0.04cm}\zeta^{\,i^{\hspace{0.03cm}\prime}}_{{\bf p}^{\prime}}}{\partial\hspace{0.03cm} t}
=
-\hspace{0.03cm}i\hspace{0.04cm}\Bigl\{\zeta^{\,i^{\hspace{0.02cm}\prime}}_{{\bf p}^{\prime}}\hspace{0.03cm},\hspace{0.03cm} {\mathcal H}^{(0)\!} + {\mathcal H}^{(4)}_{g\hspace{0.02cm}G\rightarrow g\hspace{0.02cm}G}\Bigr\}
=
-\hspace{0.02cm}i\hspace{0.04cm}\varepsilon^{\,\phantom{i^{\hspace{0.02cm}\prime}}}_{{\bf p}^{\prime}}\,\zeta^{\,i^{\hspace{0.02cm}\prime}}_{{\bf p}^{\prime}}
\label{eq:5q}
\end{equation}
\[
-\; i\!\int\! 
\mathscr{T}^{\,(2)\, i^{\hspace{0.02cm}\prime}\, i_{1}\, a_{1}\, a_{2}}_{\ {\bf p}^{\prime},\, {\bf p}_{1},\, {\bf k}_{1},\, {\bf k}_{2}}\, 
\zeta^{\;i_{1}}_{{\bf p}_{1}}\, 
c^{\ast\ \!\!a_{1}}_{{\bf k}_{1}}\hspace{0.03cm} 
c^{\hspace{0.03cm}a_{2}}_{{\bf k}_{2}}\,
\delta({\bf p}^{\prime} + {\bf k}_1 - {\bf p}_{1} - {\bf k}_{2})\,
d{\bf p}_{1}\hspace{0.02cm} d{\bf k}_{1}\hspace{0.03cm} 
d{\bf k}_{2},
\]
\vspace{0.15cm}
\begin{equation}
\hspace{0.3cm}
\frac{\partial\hspace{0.04cm} \zeta^{\hspace{0.03cm}\ast\hspace{0.03cm}i}_{{\bf p}}}{\partial\hspace{0.03cm}t}
=
-\hspace{0.03cm}i\hspace{0.04cm}\Bigl\{\zeta^{\hspace{0.03cm}\ast
	\hspace{0.03cm}i}_{{\bf p}}\hspace{0.03cm},\hspace{0.03cm} {\mathcal H}^{(0)\!} + {\mathcal H}^{(4)}_{g\hspace{0.02cm}G\rightarrow g\hspace{0.02cm}G}\Bigr\}
=
i\hspace{0.04cm}\varepsilon^{\phantom{\ast i}\!\!}_{{\bf p}}\, \zeta^{\hspace{0.03cm}\ast\hspace{0.03cm}i}_{{\bf p}}
\label{eq:5w}
\end{equation}
\[
\hspace{0.6cm}
+\; i\!\int\! 
\mathscr{T}^{\,\ast\hspace{0.03cm}(2)\, i\, i_{1}\, a_{1}\, a_{2}}_{\ {\bf p},\, {\bf p}_{1},\; {\bf k}_{1},\, {\bf k}_{2}}\, 
\zeta^{\hspace{0.03cm}\ast\hspace{0.03cm}i_{1}}_{{\bf p}_{1}}\hspace{0.03cm} 
c^{\hspace{0.03cm} a_{1}}_{{\bf k}_{1}}\hspace{0.03cm} 
c^{\ast\ \!\!a_{2}}_{{\bf k}_{2}}\, 
\delta({\bf p} + {\bf k}_{1} - {\bf p}_{1} - {\bf k}_{2})\,
d{\bf p}_{1}\hspace{0.02cm} d{\bf k}_{1}\hspace{0.03cm} 
d{\bf k}_{2}.
\]
In the latter equation we have taken into account the symmetry condition for the complete scattering amplitude
\begin{equation}
\mathscr{T}^{\hspace{0.03cm} (2)\hspace{0.03cm} i\, i_{1}\, a_{1}\, a_{2}}_{\, {\bf p},\, {\bf p}_{1},\, {\bf k}_{1},\, {\bf k}_{2}}
=
\mathscr{T}^{\,\ast\hspace{0.03cm}(2)\, i_{1}\, i\, a_{2}\, a_{1}}_{\ {\bf p}_{1},\, {\bf p},\; {\bf k}_{2},\ 
{\bf k}_{1}}.
\label{eq:5e}
\end{equation}
This relation is a consequence of the requirement of the reality of the effective Hamiltonian ${\mathcal H}^{(4)}_{g\hspace{0.02cm}G\rightarrow g\hspace{0.02cm}G}$. Further, for soft Bose-excitations we define the second pair of the canonical equations of motions with the same Hamiltonian
\begin{equation}
\frac{\partial \hspace{0.02cm}c^{\phantom{\hspace{0.03cm}\ast} \!\!a^{\prime}}_{{\bf k}^{\prime}}}{\partial\hspace{0.03cm} t}
=
-\hspace{0.04cm}i\hspace{0.04cm}\Bigl\{c^{\phantom{\hspace{0.03cm}\ast} \!\!a^{\prime}}_{{\bf k}^{\prime}}\hspace{0.03cm},\hspace{0.03cm} {\mathcal H}^{(0)\!} + {\mathcal H}^{(4)}_{g\hspace{0.02cm}G\rightarrow g\hspace{0.02cm}G}\Bigr\}
=
- i\hspace{0.04cm}\omega^{\hspace{0.02cm}l}_{{\bf k}^{\prime}}\,c^{\phantom{\hspace{0.03cm}\ast} \!\!a^{\prime}}_{{\bf k}^{\prime}}
\label{eq:5r}
\end{equation}
\[
-\; i\!\int\! 
\mathscr{T}^{\,(2)\, i_{1}\, i_{2}\, a^{\prime}\; a_{1}}_{\, {\bf p}_{1},\, {\bf p}_{2},\, {\bf k}^{\prime},\, {\bf k}_{1}}\, 
\zeta^{\hspace{0.03cm}\ast\,i_{1}}_{{\bf p}_{1}}\hspace{0.03cm} 
\zeta^{\;i_{2}}_{{\bf p}_{2}}\hspace{0.03cm} 
c^{\hspace{0.03cm}a_{1}}_{{\bf k}_{1}}\,
\delta({\bf k}^{\prime} + {\bf p}_{1} - {\bf k}_{1} - {\bf p}_{2})\,
d{\bf p}_{1}\hspace{0.02cm} d{\bf p}_{2}\hspace{0.03cm} 
d{\bf k}_{1},
\]
\vspace{0.15cm}
\begin{equation}
\hspace{0.3cm}
\frac{\partial \hspace{0.02cm}c^{\hspace{0.03cm}\ast\,a}_{{\bf k}}}{\partial\hspace{0.03cm} t}
=
-\hspace{0.03cm}i\hspace{0.04cm}\Bigl\{c^{\hspace{0.03cm}\ast\,a}_{{\bf k}}\hspace{0.03cm},\hspace{0.03cm} {\mathcal H}^{(0)\!} + {\mathcal H}^{(4)}_{g\hspace{0.02cm}G\rightarrow g\hspace{0.02cm}G}\Bigr\}
=
i\hspace{0.04cm}\omega^{\hspace{0.02cm}l}_{{\bf k}}\, c^{\hspace{0.03cm}\ast\,a}_{{\bf k}}
\label{eq:5t}
\end{equation}
\[
\hspace{0.5cm}
-\; i\!\int\!
\mathscr{T}^{\,\ast\hspace{0.03cm}(2)\hspace{0.03cm} i_{1}\, i_{2}\, a\, a_{1}}_{\ {\bf p}_{1},\, {\bf p}_{2},\; {\bf k},\, {\bf k}_{1}}\, 
\zeta^{\;i_{1}}_{{\bf p}_{1}}\hspace{0.03cm} 
\zeta^{\hspace{0.03cm}\ast\,i_{2}}_{{\bf p}_{2}}\hspace{0.03cm} 
c^{\hspace{0.03cm}\ast\,a_{1}}_{{\bf k}_{1}}\, 
\delta({\bf k} + {\bf p}_{1} - {\bf k}_{1} - {\bf p}_{2})\,
d{\bf p}_{1}\hspace{0.02cm} d{\bf p}_{2}\hspace{0.03cm} 
d{\bf k}_{1}.
\]
In the case when an external gauge field is absent in the system, the exact equations (\ref{eq:5q}), (\ref{eq:5w}), (\ref{eq:5r}), and (\ref{eq:5t}) enable us to define the kinetic equations for the hard particle number density ${\mathfrak n}^{i\hspace{0.03cm} i^{\hspace{0.02cm}\prime}}_{{\bf p}}$ and for the plasmon number density  ${\mathcal N}^{\,a\hspace{0.03cm} a^{\prime}}_{\bf k}$. If the ensemble of interacting Bose-excitations at low nonlinearity level has  random phases, then it can be statistically described by introducing the bosonic correlation function \cite{markov_2020}:
\begin{equation}
\bigl\langle\hspace{0.03cm}c^{\ast\ \!\!a}_{\hspace{0.03cm}{\bf k}^{\phantom{\prime}}}\hspace{0.03cm} 
c^{\phantom{\ast}\!\!a^{\prime}}_{\hspace{0.03cm}{\bf k}^{\prime}}\hspace{0.03cm}\bigr\rangle
=
\delta({\bf k} - {\bf k}^{\prime})\hspace{0.04 cm}{\mathcal N}^{\,a\hspace{0.02cm}a^{\prime}}_{{\bf k}}.
\label{eq:5y}
\end{equation}
However, now we do not consider the spectral density ${\mathcal N}^{\,a\hspace{0.02cm}a^{\prime}}_{{\bf k}}$ to have a trivial diagonal structure in an effective color space (see Eq.\,(\ref{eq:2w})) as was the case in the previous paper \cite{markov_2020}. The color decomposition of ${\mathcal N}^{\,a\hspace{0.02cm}a^{\prime}}_{{\bf k}}$ will be presented below.\\
\indent For hard momentum modes of quark-gluon plasma excitations, we make use an ansatz separating the color and momentum degrees of freedom, namely we assume that 
\begin{equation}
\zeta^{\,i}_{{\bf p}} = \theta^{\,i}\hspace{0.03cm}\zeta_{{\bf p}}, 
\qquad 
\zeta^{\hspace{0.03cm}\ast\,i}_{{\bf p}} 
= \theta^{\hspace{0.03cm}\ast\,i}\hspace{0.03cm} 
\zeta^{\hspace{0.03cm}\ast}_{{\bf p}}.
\label{eq:5u}
\end{equation}
Here we have introduced a set of the Grassmann-valued color charges $\theta^{\hspace{0.03cm}\ast\ \!\!i}$ and $\theta^{\phantom{\ast}\!\!i}$ belonging to the defining representation of the $SU(N_{c})$ Lie algebra \cite{balachandran_1977, barducci_1977}. These color charges are in involution with respect to the conjugation operation $\ast$. The complex function $\zeta_{{\bf p}}$ is an usual commutative random function of the momentum variable ${\bf p}$. In the representation (\ref{eq:5u}) we have a complete decoupling of the color and momentum degrees of freedom. This is true only if we neglect the influence of soft collective excitations of the gauge field on the change of the momentum of a hard particle, i.e. the momentum of the particle is fixed and all interaction is carried out only through the color degree of freedom. For determination of the desired kinetic equations, it is necessary first to perform calculations exactly, without using any approximation. Only at the end of all calculations we must take into account the fact that the momentum of hard particles is much greater than the momentum of soft plasma excitations, i.e.,
\[
|{\bf p}_{1}|,\,|{\bf p}_{2}| \gg |{\bf k}|,\,|{\bf k}_{1}|,
\]   
and perform the corresponding approximations of the derived expressions. By virtue of the decomposition (\ref{eq:5u}), we can represent also the hard mode correlation function in the factorized form 
\[
\bigl\langle\,\!\zeta^{\hspace{0.03cm}\ast\,i}_{{\bf p}} \zeta^{\,i^{\hspace{0.02cm}\prime}}_{{\bf p}^{\prime}}\hspace{0.03cm}\bigr\rangle
=
\bigl\langle\,\!\zeta^{\hspace{0.03cm}\ast}_{\hspace{0.03cm} {\bf p}}\hspace{0.04cm}  \zeta^{\phantom{\ast}}_{\hspace{0.03cm}{\bf p}^{\prime}\!}\hspace{0.03cm}\bigr\rangle
\bigl\langle\,\!\theta^{\hspace{0.03cm}\ast\,i}\hspace{0.02cm}  \theta^{\,i^{\hspace{0.02cm}\prime}}\hspace{0.03cm}\bigr\rangle,
\]
where, in turn, we believe
\[
\bigl\langle\,\!\zeta^{\hspace{0.03cm}\ast}_{\hspace{0.03cm} {\bf p}}\hspace{0.04cm}  \zeta^{\phantom{\ast}}_{\hspace{0.03cm} {\bf p}^{\prime}}\hspace{0.01cm}\bigr\rangle
=
\delta({\bf p} - {\bf p}^{\prime})\hspace{0.04cm}n_{\bf p}.
\]
Thus, in full analogy with (\ref{eq:5y}) we can write
\begin{equation*}
\!\!\bigl\langle\,\!\zeta^{\hspace{0.03cm}\ast\,i}_{{\bf p}} \zeta^{\,i^{\hspace{0.02cm}\prime}}_{{\bf p}^{\prime}}\hspace{0.03cm}\bigr\rangle
=
\delta({\bf p} - {\bf p}^{\prime})\hspace{0.03cm}{\mathfrak n}^{i^{\hspace{0.02cm}\prime}\hspace{0.03cm}i}_{\bf p},
\end{equation*} 
where we have introduced the matrix function ${\mathfrak n}^{\hspace{0.03cm}i^{\hspace{0.02cm}\prime}\hspace{0.02cm}i}_{{\bf p}}$ setting by the definition
\begin{equation*}
{\mathfrak n}^{\hspace{0.03cm}i^{\hspace{0.02cm}\prime}\hspace{0.02cm}i}_{{\bf p}} 
\hspace{0.02cm}\stackrel{\text{\tiny def}}{=} \hspace{0.02cm}
n_{\bf p}\hspace{0.03cm} 
\bigl\langle\hspace{0.03cm}\theta^{\hspace{0.03cm}\ast\ \!\!i}\hspace{0.04cm} \theta^{\phantom{\ast}\!\!i^{\hspace{0.02cm}\prime}}\hspace{0.03cm}
\bigr\rangle.
\end{equation*}
We draw your attention to the arrangement of color indices on the left- and right-hand sides of the previous expression.\\
\indent Let us derive the kinetic equations for the number densities of hard excitations ${\mathfrak n}^{\hspace{0.03cm}i^{\hspace{0.02cm}\prime}\hspace{0.02cm}i}_{{\bf p}}$ and plasmons ${\mathcal N}^{\,a\hspace{0.03cm}a^{\prime}}_{{\bf k}}$ employing the Hamilton equations (\ref{eq:5q}), (\ref{eq:5w}), (\ref{eq:5r}) and (\ref{eq:5t}). Using precisely the same reasoning as in 
paper \cite{markov_2023}, we obtain matrix analog of the equations (10.7) and (10.8) in the above-mentioned work
\begin{equation}
\delta({\bf p} - {\bf p}\!\ ')\,
\frac{\partial\hspace{0.03cm} {\mathfrak n}^{\hspace{0.03cm}i^{\hspace{0.02cm}\prime}\hspace{0.02cm}i}_{{\bf p}}}{\partial\hspace{0.03cm} t}\
=
-\hspace{0.03cm}i\!\int\!d{\bf p}_{1}\hspace{0.02cm} d{\bf k}_{1}\hspace{0.03cm} d{\bf k}_{2}\,\times
\label{eq:5p}
\end{equation}
\[
\times\hspace{0.02cm} 
\biggl\{\mathscr{T}^{\hspace{0.03cm} (2)\hspace{0.03cm} i^{\hspace{0.02cm}\prime}\hspace{0.03cm} i_{1}\, a_{1}\, a_{2}}_{\, {\bf p}^{\prime},\, {\bf p}_{1},\, {\bf k}_{1},\, {\bf k}_{2}}\, 
I^{\, i\, i_{1}\, a_{1}\, a_{2}}_{{\bf p},\, {\bf p}_{1},\, {\bf k}_{1},\, {\bf k}_{2}}\,
\delta({\bf p}^{\prime} + {\bf k}_{1} - {\bf p}_{1} - {\bf k}_{2})
- 
\mathscr{T}^{\hspace{0.03cm}\ast\hspace{0.03cm}(2)\hspace{0.03cm} i\, i_{1}\, a_{1}\, a_{2}}_{\ {\bf p},\, {\bf p}_{1},\, {\bf k}_{1},\, {\bf k}_{2}}\, 
I^{\, i_{1}\hspace{0.03cm} i^{\hspace{0.02cm}\prime}\hspace{0.03cm} a_{2}\, a_{1}}_{{\bf p}_{1},\, {\bf p}^{\prime},\, {\bf k}_{2},\, {\bf k}_{1}}\,
\delta({\bf p} + {\bf k}_{1} - {\bf p}_{1} - {\bf k}_{2})\biggr\}
\]
and
\begin{equation}
\delta({\bf k} - {\bf k}\!\ ')\,
\frac{\partial\hspace{0.01cm} {\mathcal N}^{\,a\hspace{0.03cm}a^{\prime}}_{{\bf k}}}{\partial\hspace{0.03cm} t}\
=
-\hspace{0.03cm}i\!\int\!d{\bf p}_{1}\hspace{0.02cm} d{\bf p}_{2}\hspace{0.03cm} d{\bf k}_{1}\,\times
\label{eq:5a}
\end{equation}
\[
\times\hspace{0.03cm}
\biggl\{\mathscr{T}^{\,(2)\, i_{1}\, i_{2}\, a^{\prime}\, a_{1}}_{\, {\bf p}_{1},\, {\bf p}_{2},\, {\bf k}^{\prime},\, {\bf k}_{1}}\, 
I^{\;i_{1}\, i_{2}\, a\, a_{1}}_{{\bf p}_{1},\, {\bf p}_{2},\, {\bf k},\, {\bf k}_{1}}\,
\delta({\bf k}^{\prime} + {\bf p}_{1} - {\bf k}_{1} - {\bf p}_{2})
- 
\mathscr{T}^{\,\ast\hspace{0.03cm}(2)\, i_{1}\, i_{2}\, a\, a_{1}}_{\ {\bf p}_{1},\, {\bf p}_{2},\, {\bf k},\, {\bf k}_{1}}\, 
I^{\;i_{2}\, i_{1}\, a_{1}\, a^{\prime}}_{{\bf p}_{2},\, {\bf p}_{1},\, {\bf k}_{1},\, {\bf k}^{\prime}}\,
\delta({\bf k} + {\bf p}_{1} - {\bf k}_{1} - {\bf p}_{2})\biggr\},
\]
where
\[
I^{\, i\; i_{1}\, a_{1}\, a_{2}}_{{\bf p},\, {\bf p}_{1},\, {\bf k}_{1},\, {\bf k}_{2}}
=
\bigl\langle \zeta^{\hspace{0.03cm}\ast\,i}_{{\bf p}}\, 
\zeta^{\,i_{1}}_{{\bf p}_{1}}\, 
c^{\ast\ \!\!a_{1}}_{{\bf k}_{1}}\hspace{0.03cm} 
c^{\hspace{0.03cm}a_{2}}_{{\bf k}_{2}}\,\bigr\rangle
\]
is the four-point correlation function. By differentiating the correlation function $I^{\, i\ i_{1}\ a_{1}\ a_{2}}_{{\bf p},\, {\bf p}_{1},\, {\bf k}_{1},\, {\bf k}_{2}}$ with respect to $t$ with allowance made for (\ref{eq:5q}), (\ref{eq:5w}), (\ref{eq:5r}) and (\ref{eq:5t}), we derive the equation the right-hand side of which contains the six-order correlation functions of the variables $\zeta^{\hspace{0.03cm}\ast\, i}_{{\bf p}}\!,\, \zeta^{\phantom{\hspace{0.03cm}\ast}\!\!i}_{{\bf p}}$ and $c^{\phantom{\hspace{0.03cm}\ast} \!\!a}_{{\bf k}}, \,c^{\hspace{0.03cm}\ast\,a}_{{\bf k}}$:
\begin{equation}
\frac{\partial\hspace{0.01cm} I^{\, i\; i_{1}\, a_{1}\, a_{2}}_{{\bf p},\, {\bf p}_{1},\, {\bf k}_{1},\, {\bf k}_{2}}}{\partial\hspace{0.03cm} t}
=
i\hspace{0.03cm}\bigl[\,\varepsilon_{{\bf p}} + \omega^{\hspace{0.02cm}l}_{{\bf k}_{1}} - \varepsilon_{{\bf p}_{1}} -
\omega^{\hspace{0.02cm}l}_{{\bf k}_{2}}\bigr]
\, I^{\, i\; i_{1}\, a_{1}\, a_{2}}_{{\bf p},\, {\bf p}_{1},\, {\bf k}_{1},\, {\bf k}_{2}}\ +
\label{eq:5s}
\end{equation}
\begin{align}
+\; i\!&\int
\mathscr{T}^{\,\ast\hspace{0.03cm}(2)\, i\; i^{\hspace{0.02cm}\prime}_{1}\, a^{\prime}_{1}\, a^{\prime}_{2}}_{\ {\bf p},\, 
{\bf p}^{\prime}_{1},\, {\bf k}^{\prime}_{1},\, {\bf k}^{\prime}_{2}}\,
\bigl\langle
\zeta^{\hspace{0.04cm}\ast\,i^{\hspace{0.02cm}\prime}_{1}}_{{\bf p}^{\prime}_{1}} 
\zeta^{\,i^{\phantom{\prime}}_{1}}_{{\bf p}^{\phantom{\prime}}_{1}}\,
c^{\ast\ \!\!a^{\prime}_{2}}_{{\bf k}^{\prime}_{2}}\, 
c^{\hspace{0.03cm}a^{\prime}_{1}}_{{\bf k}^{\prime}_{1}}\,
c^{\ast\ \!\!a^{\phantom{\prime}}_{1}}_{{\bf k}^{\phantom{\prime}}_{1}}\hspace{0.03cm} 
c^{\hspace{0.03cm}a^{\phantom{\prime}}_{2}}_{{\bf k}^{\phantom{\prime}}_{2}}\,\bigr\rangle\, 
\delta({\bf p}^{\prime}_{1} + {\bf k}^{\prime}_{2} - {\bf p} - {\bf k}^{\prime}_{1})\,
d{\bf p}^{\prime}_{1}\hspace{0.03cm} d{\bf k}^{\prime}_{1}\hspace{0.03cm} d{\bf k}^{\prime}_{2}
\notag\\[1ex]
-\; i\!&\int 
\mathscr{T}^{\,(2)\, i_{1}\, i^{\hspace{0.02cm}\prime}_{1}\, a^{\prime}_{1}\, a^{\prime}_{2}}_{\ {\bf p}_{1},\, 
{\bf p}^{\prime}_{1},\, {\bf k}^{\prime}_{1},\, {\bf k}^{\prime}_{2}}\,
\bigl\langle
\zeta^{\hspace{0.04cm}\ast\,i^{\phantom{\prime}}}_{{\bf p}^{\phantom{\prime}}} 
\zeta^{\;i^{\hspace{0.02cm}\prime}_{1}}_{{\bf p}^{\prime}_{1}}\,
c^{\ast\ \!\!a^{\prime}_{1}}_{{\bf k}^{\prime}_{1}}\hspace{0.03cm} 
c^{\hspace{0.03cm}a^{\prime}_{2}}_{{\bf k}^{\prime}_{2}}\,
c^{\ast\ \!\!a^{\phantom{\prime}}_{1}}_{{\bf k}^{\phantom{\prime}}_{1}}\hspace{0.03cm} 
c^{\hspace{0.03cm}a^{\phantom{\prime}}_{2}}_{{\bf k}^{\phantom{\prime}}_{2}}\,\bigr\rangle\, 
\delta({\bf p}^{\phantom{\prime}}_{1} + {\bf k}^{\prime}_{1} - {\bf p}^{\prime}_{1} - {\bf k}^{\prime}_{2})\,
d{\bf p}^{\prime}_{1}\hspace{0.03cm} d{\bf k}^{\prime}_{1}\hspace{0.03cm} d{\bf k}^{\prime}_{2}
\notag\\[1ex]
+\; i\!&\int 
\mathscr{T}^{\,\ast\hspace{0.03cm}(2)\, i^{\hspace{0.02cm}\prime}_{2}\, i^{\hspace{0.02cm}\prime}_{1}\,a^{\phantom{\prime}}_{1}\,a^{\prime}_{1}}_{\ {\bf p}^{\prime}_{2},\, 
{\bf p}^{\prime}_{1},\, {\bf k}_{1},\, {\bf k}^{\prime}_{1}}\,
\bigl\langle
\zeta^{\hspace{0.04cm}\ast\,i^{\phantom{\prime}}\!\!}_{{\bf p}}\hspace{0.03cm}
\zeta^{\,i^{\phantom{\prime}}_{1}}_{{\bf p}^{\phantom{\prime}}_{1}}\hspace{0.03cm}
\zeta^{\hspace{0.04cm}\ast\,i^{\hspace{0.02cm}\prime}_{1}}_{{\bf p}^{\prime}_{1}}\hspace{0.03cm}
\zeta^{\phantom{\hspace{0.02cm}\ast}\!i^{\hspace{0.02cm}\prime}_{2}}_{{\bf p}^{\prime}_{2}}\,
c^{\ast\ \!\!a^{\prime}_{1}}_{{\bf k}^{\prime}_{1}}\hspace{0.03cm} 
c^{\hspace{0.03cm}a^{\phantom{\prime}}_{2}}_{{\bf k}^{\phantom{\prime}}_{2}}
\,\bigr\rangle\, 
\delta({\bf p}^{\prime}_{1} + {\bf k}^{\prime}_{1} - {\bf p}^{\prime}_{2} - {\bf k}^{\phantom{\prime}}_{1})\,
d{\bf p}^{\prime}_{1}\hspace{0.03cm} d{\bf p}^{\prime}_{2}\hspace{0.04cm} d{\bf k}^{\prime}_{1}
\notag\\[1ex]
-\; i\!&\int 
\mathscr{T}^{\,(2)\, i^{\hspace{0.02cm}\prime}_{1}\, i^{\hspace{0.02cm}\prime}_{2}\,a^{\phantom{\prime}}_{2}\; a^{\prime}_{1}}_{\ {\bf p}^{\prime}_{1},\, {\bf p}^{\prime}_{2},\, {\bf k}_{2},\, {\bf k}^{\prime}_{1}}\,
\bigl\langle
\zeta^{\hspace{0.04cm}\ast\,i^{\phantom{\prime}}\!\!}_{{\bf p}}\hspace{0.03cm}
\zeta^{\,i^{\phantom{\prime}}_{1}}_{{\bf p}^{\phantom{\prime}}_{1}}\,
\zeta^{\hspace{0.04cm}\ast\,i^{\hspace{0.02cm}\prime}_{1}}_{{\bf p}^{\prime}_{1}}\hspace{0.03cm}
\zeta^{\phantom{\hspace{0.03cm}\ast}\!i^{\hspace{0.02cm}\prime}_{2}}_{{\bf p}^{\prime}_{2}}
\,c^{\ast\ \!\!a^{\phantom{\prime}}_{1}}_{{\bf k}^{\phantom{\prime}}_{1}}\hspace{0.03cm} 
c^{\hspace{0.03cm}a^{\prime}_{1}}_{{\bf k}^{\prime}_{1}}
\,\bigr\rangle\, 
\delta({\bf p}^{\prime}_{1} + {\bf k}^{\phantom{\prime}}_{2} - {\bf p}^{\prime}_{2} - {\bf k}^{\prime}_{1})\,
d{\bf p}^{\prime}_{1}\hspace{0.03cm} d{\bf p}^{\prime}_{2}\hspace{0.04cm} d{\bf k}^{\prime}_{1}.
\notag
\end{align}
As in the pure fermionic case \cite{markov_2023}, we close the chain of equations by expressing the six-order correlation functions in terms of the pair correlation functions. We keep only those terms that give the proper contributions to the required kinetic equations:
\begin{equation}
\begin{split}
&\bigl\langle
\zeta^{\hspace{0.04cm}\ast\, i^{\hspace{0.02cm}\prime}_{1}}_{{\bf p}^{\prime}_{1}}\! 
\zeta^{\,i^{\phantom{\prime}}_{1}}_{{\bf p}^{\phantom{\prime}}_{1}}\,
c^{\ast\ \!\!a^{\prime}_{2}}_{{\bf k}^{\prime}_{2}}\hspace{0.03cm} 
c^{\hspace{0.03cm}a^{\prime}_{1}}_{{\bf k}^{\prime}_{1}}\,
c^{\ast\ \!\!a^{\phantom{\prime}}_{1}}_{{\bf k}^{\phantom{\prime}}_{1}}\hspace{0.03cm} 
c^{\hspace{0.03cm}a^{\phantom{\prime}}_{2}}_{{\bf k}^{\phantom{\prime}}_{2}}\bigr\rangle
\,\simeq\,
\delta({\bf p}^{\prime}_{1} - {\bf p}^{\phantom{\prime}}_{1})
\delta({\bf k}^{\prime}_{1} - {\bf k}^{\phantom{\prime}}_{1})
\delta({\bf k}^{\prime}_{2} - {\bf k}^{\phantom{\prime}}_{2})\,
{\mathfrak n}^{i^{\phantom{\prime}}_{1}i^{\hspace{0.02cm}\prime}_{1}}_{{\bf p}_{1}} 
{\mathcal N}^{\,a^{\phantom{\prime}}_{1}a^{\prime}_{1}}_{{\bf k}_{1}} 
{\mathcal N}^{\,a^{\prime}_{2}a^{\phantom{\prime}}_{2}}_{{\bf k}_{2}},
\\[1.5ex]
&\bigl\langle
\zeta^{\hspace{0.04cm}\ast\, i^{\phantom{\prime}}}_{{\bf p}^{\phantom{\prime}}} 
\zeta^{\,i^{\hspace{0.02cm}\prime}_{1}}_{{\bf p}^{\prime}_{1}}\,
c^{\ast\ \!\!a^{\prime}_{1}}_{{\bf k}^{\prime}_{1}}\hspace{0.03cm} 
c^{\hspace{0.03cm}a^{\prime}_{2}}_{{\bf k}^{\prime}_{2}}\,
c^{\ast\ \!\!a^{\phantom{\prime}}_{1}}_{{\bf k}^{\phantom{\prime}}_{1}}\hspace{0.02cm} 
c^{\hspace{0.03cm}a^{\phantom{\prime}}_{2}}_{{\bf k}^{\phantom{\prime}}_{2}}\,\bigr\rangle
\simeq\,
\delta({\bf p}^{\prime}_{1} - {\bf p})
\hspace{0.03cm}
\delta({\bf k}^{\prime}_{1} - {\bf k}^{\phantom{\prime}}_{2})
\hspace{0.03cm}
\delta({\bf k}^{\prime}_{2} - {\bf k}^{\phantom{\prime}}_{1})\,
{\mathfrak n}^{i^{\hspace{0.02cm}\prime}_{1}i^{\phantom{\prime}}}_{{\bf p}} 
{\mathcal N}^{\,a^{\phantom{\prime}}_{1}a^{\prime}_{2}}_{{\bf k}_{1}} 
{\mathcal N}^{\,a^{\prime}_{1}a^{\phantom{\prime}}_{2}}_{{\bf k}_{2}},
\\[1.5ex]
&\bigl\langle
\zeta^{\hspace{0.04cm}\ast\,i^{\phantom{\prime}}\!\!}_{{\bf p}}\hspace{0.03cm}
\zeta^{\,i^{\phantom{\prime}}_{1}}_{{\bf p}^{\phantom{\prime}}_{1}}\,
\zeta^{\hspace{0.04cm}\ast\,i^{\hspace{0.02cm}\prime}_{1}}_{{\bf p}^{\prime}_{1}}\hspace{0.03cm} 
\zeta^{\phantom{\hspace{0.03cm}\ast}\!i^{\hspace{0.02cm}\prime}_{2}}_{{\bf p}^{\prime}_{2}}\,
c^{\ast\ \!\!a^{\prime}_{1}}_{{\bf k}^{\prime}_{1}}\hspace{0.03cm} 
c^{\hspace{0.03cm}a^{\phantom{\prime}}_{2}}_{{\bf k}^{\phantom{\prime}}_{2}}
\hspace{0.03cm}\bigr\rangle
\simeq
-\hspace{0.03cm}
\delta({\bf p}^{\prime}_{2} - {\bf p})
\hspace{0.03cm}
\delta({\bf p}^{\prime}_{1} - {\bf p}_{1})
\hspace{0.03cm}
\delta({\bf k}^{\prime}_{1} - {\bf k}^{\phantom{\prime}}_{2})\,
{\mathfrak n}^{i^{\hspace{0.02cm}\prime}_{2}i}_{{\bf p}}\, 
{\mathfrak n}^{i_{1}i^{\hspace{0.02cm}\prime}_{1}}_{{\bf p}_{1}} 
{\mathcal N}^{\,a^{\prime}_{1}a^{\phantom{\prime}}_{2}}_{{\bf k}_{2}},
\\[1.5ex]
&\bigl\langle
\zeta^{\hspace{0.04cm}\ast\,i^{\phantom{\prime}}\!\!}_{{\bf p}}\hspace{0.03cm}
\zeta^{\,i^{\phantom{\prime}}_{1}}_{{\bf p}^{\phantom{\prime}}_{1}}\,
\zeta^{\hspace{0.04cm}\ast\,i^{\hspace{0.02cm}\prime}_{1}}_{{\bf p}^{\prime}_{1}}\hspace{0.03cm}
\zeta^{\phantom{\hspace{0.03cm}\ast}\!i^{\hspace{0.02cm}\prime}_{2}}_{{\bf p}^{\prime}_{2}}\,
c^{\ast\ \!\!a^{\phantom{\prime}}_{1}}_{{\bf k}^{\phantom{\prime}}_{1}}\hspace{0.03cm} 
c^{\hspace{0.03cm}a^{\prime}_{1}}_{{\bf k}^{\prime}_{1}}
\hspace{0.03cm}\bigr\rangle
\simeq
-\hspace{0.03cm}
\delta({\bf p}^{\prime}_{2} - {\bf p})
\hspace{0.03cm}
\delta({\bf p}^{\prime}_{1} - {\bf p}_{1})
\hspace{0.03cm}
\delta({\bf k}^{\prime}_{1} - {\bf k}^{\phantom{\prime}}_{1})\,
{\mathfrak n}^{i^{\hspace{0.02cm}\prime}_{2}i}_{{\bf p}}\, 
{\mathfrak n}^{i_{1}i^{\hspace{0.02cm}\prime}_{1}}_{{\bf p}_{1}} 
{\mathcal N}^{\,a^{\phantom{\prime}}_{1}a^{\prime}_{1}}_{{\bf k}_{1}}.
\end{split}
\label{eq:5d}
\end{equation}
In the interaction Hamiltonian (\ref{eq:2a}) we set for the three-point vertex function ${\Phi}^{\; a\, i_{1}\;  i_{2}}_{{\bf k},\, {\bf p}_{1},\, {\bf p}_{2}}$: 
\[
{\Phi}^{\;a\,i_{1}\hspace{0.03cm}i_{2}}_{{\bf k},\, {\bf p}_{1},\, {\bf p}_{2}} 
=
(t^{\hspace{0.03cm}a})^{\, i_{1}\hspace{0.03cm} i_{2}}\hspace{0.02cm}
{\Phi}_{{\bf k},\, {\bf p}_{1},\, {\bf p}_{2}}.
\]
Then, by taking into account the representation (\ref{eq:2j}) for the vertex function ${\mathcal V}^{\ \! a\, a_{1}\hspace{0.03cm} a_{2}}_{{\bf k},\, {\bf k}_{1},\, {\bf k}_{2}}$ the color structure of the complete effective amplitude $\mathscr{T}^{\hspace{0.03cm} (2)\hspace{0.03cm} i\; i_{1}\, a_{1}\, a_{2}}_{\; {\bf p},\, {\bf p}_{1},\, {\bf k}_{1},\, {\bf k}_{2}}$, Eq.\,(\ref{eq:4r}), looks like
\begin{equation}
\mathscr{T}^{\hspace{0.03cm} (2)\hspace{0.03cm} i\; i_{1}\, a_{1}\, a_{2}}_{\; {\bf p},\, {\bf p}_{1},\, {\bf k}_{1},\, {\bf k}_{2}}
=
[\,t^{a_{1}}\!,\hspace{0.03cm}t^{a_{2}}]^{\, i\hspace{0.03cm} i_{1}}\,
\mathscr{T}^{\,(2,\hspace{0.03cm}{\mathcal A})}_{\; {\bf p},\, {\bf p}_{1},\, {\bf k}_{1},\, {\bf k}_{2}}
+
\{t^{a_{1}}\!,\hspace{0.03cm}t^{a_{2}}\}^{\, i\hspace{0.03cm} i_{1}}\,
\mathscr{T}^{\,(2,\hspace{0.03cm}{\mathcal S})}_{\; {\bf p},\, {\bf p}_{1},\, {\bf k}_{1},\, {\bf k}_{2}},
\label{eq:5f}
\end{equation}
where the {\it effective subamplitudes} $\mathscr{T}^{\,(2,\hspace{0.03cm}{\mathcal A})}$ and $\mathscr{T}^{\,(2,\hspace{0.03cm}{\mathcal S})}$ have the following structures:
\begin{equation}
\mathscr{T}^{\,(2,\hspace{0.03cm}{\mathcal A})}_{\; {\bf p},\, {\bf p}_{1},\, {\bf k}_{1},\, {\bf k}_{2}}
=
T^{\,(2,\hspace{0.03cm}{\mathcal A})}_{\; {\bf p},\, {\bf p}_{1},\, {\bf k}_{1},\, {\bf k}_{2}}
\label{eq:5g}
\end{equation}
\begin{align}
+\,\frac{1}{4}\,\Biggl[
\Biggl(&\frac{1}{\omega^{\hspace{0.02cm} l}_{{\bf k}_{2}} - \varepsilon_{{\bf k}_{2} +\hspace{0.03cm} {\bf p}_{1}} + 
	\varepsilon_{{\bf p}_{1}}} 
+
\frac{1}{\omega^{\hspace{0.02cm} l}_{{\bf k}_{1}} - \varepsilon_{{\bf k}_{1} +\hspace{0.03cm} {\bf p}} + \varepsilon_{{\bf p}}} 
\Biggr)\hspace{0.02cm}
{\Phi}_{{\bf k}_{2},\, {\bf k}_{2} + {\bf p}_{1},\, {\bf p}_{1}}\, 
{\Phi}^{\hspace{0.03cm}\ast}_{{\bf k}_{1},\, {\bf k}_{1} + {\bf p},\, {\bf p}}
\notag\\[1.5ex]
+\,
\Biggl(&\frac{1}{\omega^{\hspace{0.02cm} l}_{{\bf k}_{2}} - \varepsilon_{{\bf p}} + \varepsilon_{{\bf p} - {\bf k}_{2}}} 
\,+\,
\frac{1}{\omega^{\hspace{0.02cm} l}_{{\bf k}_{1}} - \varepsilon_{{\bf p}_{1}} + \varepsilon_{{\bf p}_{1} - {\bf k}_{1}}} 
\Biggr)\hspace{0.02cm}
{\Phi}_{{\bf k}_{2},\, {\bf p},\, {\bf p} - {\bf k}_{2}}\, 
{\Phi}^{\hspace{0.03cm}\ast}_{{\bf k}_{1},\, {\bf p}_{1},\, {\bf p}_{1} - {\bf k}_{1}}\Biggr]
\notag\\[1.5ex]
-\,i\hspace{0.03cm}\Biggl[
\Biggl(
&\frac{1}{\omega^{\hspace{0.02cm} l}_{{\bf k}_{1}} - \omega^{\hspace{0.02cm} l}_{{\bf k}_{2}} - \omega^{\hspace{0.02cm} l}_{{\bf k}_{1} - {\bf k}_{2}}}
\,-\,
\frac{1}{\omega^{\hspace{0.02cm} l}_{{\bf p}_{1} - {\bf p}} - \varepsilon_{{\bf p}_{1}} + \varepsilon_{{\bf p}}}
\Biggr)\hspace{0.02cm}
{\mathcal V}_{{\bf k}_{1}, {\bf k}_{2},\, {\bf k}_{1} - {{\bf k}_{2}}}\, 
{\Phi}^{\hspace{0.03cm}\ast}_{{\bf p}_{1} - {\bf p},\, {\bf p}_{1},\, {\bf p}}
\notag\\[1.5ex]
-\, 
\Biggl(
&\frac{1}{\omega^{\hspace{0.02cm} l}_{{\bf k}_{2}} - \omega^{\hspace{0.02cm} l}_{{\bf k}_{1}} - \omega^{\hspace{0.02cm} l}_{{\bf k}_{2} - {\bf k}_{1}}}
\,-\,
\frac{1}{\omega^{\hspace{0.02cm} l}_{{\bf p} - {\bf p}_{1}} - \varepsilon_{{\bf p}} + \varepsilon_{{\bf p}_{1}}}
\Biggr)\hspace{0.02cm}
{\Phi}_{{\bf p} - {\bf p}_{1},\, {\bf p},\, {\bf p}_{1}}\, 
{\mathcal V}^{\hspace{0.03cm}\ast}_{{\bf k}_{2}, {\bf k}_{1},\, {\bf k}_{2} - {\bf k}_{1}}\Biggr],
\notag
\end{align}
\vspace{0.3cm}
\begin{equation}
\mathscr{T}^{\,(2,\hspace{0.03cm}{\mathcal S})}_{\; {\bf p},\, {\bf p}_{1},\, {\bf k}_{1},\, {\bf k}_{2}}
=
T^{\,(2,\hspace{0.03cm}{\mathcal S})}_{\; {\bf p},\, {\bf p}_{1},\, {\bf k}_{1},\, {\bf k}_{2}}
\label{eq:5h}
\vspace{-0.3cm}
\end{equation}
\begin{align}
	+\,\frac{1}{4}\,\Biggl[
	\Biggl(&\frac{1}{\omega^{\hspace{0.02cm} l}_{{\bf k}_{2}} - \varepsilon_{{\bf k}_{2} +\hspace{0.03cm} {\bf p}_{1}} + 
		\varepsilon_{{\bf p}_{1}}} 
	+
	\frac{1}{\omega^{\hspace{0.02cm} l}_{{\bf k}_{1}} - \varepsilon_{{\bf k}_{1} +\hspace{0.03cm} {\bf p}} + \varepsilon_{{\bf p}}} 
	\Biggr)\hspace{0.02cm}
	{\Phi}_{{\bf k}_{2},\, {\bf k}_{2} + {\bf p}_{1},\, {\bf p}_{1}}\, 
	{\Phi}^{\hspace{0.03cm}\ast}_{{\bf k}_{1},\, {\bf k}_{1} + {\bf p},\, {\bf p}}
	\notag\\[1.5ex]
	-\,
	\Biggl(&\frac{1}{\omega^{\hspace{0.02cm} l}_{{\bf k}_{2}} - \varepsilon_{{\bf p}} + \varepsilon_{{\bf p} - {\bf k}_{2}}} 
	\,+\,
	\frac{1}{\omega^{\hspace{0.02cm} l}_{{\bf k}_{1}} - \varepsilon_{{\bf p}_{1}} + \varepsilon_{{\bf p}_{1} - {\bf k}_{1}}} 
	\Biggr)\hspace{0.02cm}
	{\Phi}_{{\bf k}_{2},\, {\bf p},\, {\bf p} - {\bf k}_{2}}\, 
	{\Phi}^{\hspace{0.03cm}\ast}_{{\bf k}_{1},\, {\bf p}_{1},\, {\bf p}_{1} - {\bf k}_{1}}\Biggl].
\notag
\end{align}
We should have put the imaginary unit $i$ before the first term on the right-hand side of (\ref{eq:5f}), but we didn't do that. From the decomposition (\ref{eq:5f}) and the realness condition (\ref{eq:5e}) the symmetry properties for the effective subamplitudes $\mathscr{T}^{\,(2,\hspace{0.03cm}{\mathcal A})}$ and $\mathscr{T}^{\,(2,\hspace{0.03cm}{\mathcal S})}$ follow:
\begin{equation}
\mathscr{T}^{\,(2,\hspace{0.03cm}{\mathcal A})}_{\; {\bf p},\, {\bf p}_{1},\, {\bf k}_{1},\, {\bf k}_{2}}
=
\mathscr{T}^{\,\ast\,(2,\hspace{0.03cm}{\mathcal A})}_{\; {\bf p}_{1},\, {\bf p},\, {\bf k}_{2},\, {\bf k}_{1}},
\qquad
\mathscr{T}^{\,(2,\hspace{0.03cm}{\mathcal S})}_{\; {\bf p},\, {\bf p}_{1},\, {\bf k}_{1},\, {\bf k}_{2}}
=
\mathscr{T}^{\,\ast\,(2,\hspace{0.03cm}{\mathcal S})}_{\; {\bf p}_{1},\, {\bf p},\, {\bf k}_{2},\, {\bf k}_{1}}.
\label{eq:5j}
\end{equation}
\indent In section \ref{section_7} we show that in the limit $|{\bf p}|,\,|{\bf p}_{1}| \gg |{\bf k}_{1}|,\,|{\bf k}_{2}|$ the following approximate equality for the effective subamplitude $\mathscr{T}^{\,(2,\hspace{0.03cm}{\mathcal S})}_{\; {\bf p},\, {\bf p}_{1},\, {\bf k}_{1},\, {\bf k}_{2}}$ will be true 
\[
\mathscr{T}^{\,(2,\hspace{0.03cm}{\mathcal S})}_{\; {\bf p},\, {\bf p}_{1},\, {\bf k}_{1},\, {\bf k}_{2}} \simeq 0,
\]
so that in the future in the color decomposition (\ref{eq:5f}) we leave only the contribution with subamplitude $\mathscr{T}^{\,(2,\hspace{0.03cm}{\mathcal A})}_{\; {\bf p},\, {\bf p}_{1},\, {\bf k}_{1},\, {\bf k}_{2}}$, i.e., we set  
\begin{equation}
	\mathscr{T}^{\hspace{0.03cm} (2)\hspace{0.03cm} i\; i_{1}\, a_{1}\, a_{2}}_{\; {\bf p},\, {\bf p}_{1},\, {\bf k}_{1},\, {\bf k}_{2}}
	\,\simeq\,
	i\hspace{0.02cm}f^{\hspace{0.03cm} a_{1}\hspace{0.03cm} a_{2}\,e}\hspace{0.02cm}
	(\hspace{0.02cm}t^{\,e\hspace{0.02cm}})^{\hspace{0.03cm}i\hspace{0.03cm} i_{1}}\,
	\mathscr{T}^{\,(2,\hspace{0.03cm}{\mathcal A})}_{\; {\bf p},\, {\bf p}_{1},\, {\bf k}_{1},\, {\bf k}_{2}}
	\equiv 
	-\hspace{0.03cm}\bigl(\hspace{0.02cm}T^{\,e\hspace{0.02cm}}\bigr)^{a_{1}\hspace{0.03cm} a_{2}}\hspace{0.03cm}(\hspace{0.02cm}t^{\,e\hspace{0.02cm}})^{\hspace{0.03cm} i\hspace{0.03cm} i_{1}}\hspace{0.03cm}
	\mathscr{T}^{\,(2,\hspace{0.03cm}{\mathcal A})}_{\; {\bf p},\, {\bf p}_{1},\, {\bf k}_{1},\, {\bf k}_{2}},
	\label{eq:5k}
\end{equation}
where $\bigl(T^{\,a}\bigr)^{\hspace{0.01cm}b\hspace{0.03cm}c} \equiv -i\hspace{0.03cm} f^{\hspace{0.03cm}a\hspace{0.02cm}b\hspace{0.03cm}c}$.
For convenience of further considerations, let us also write out an expression for the conjugate amplitude:
\begin{equation}
\mathscr{T}^{\hspace{0.03cm} \ast\hspace{0.03cm}(2)\hspace{0.03cm} i\; i_{1}\, a_{1}\, a_{2}}_{\; {\bf p},\, {\bf p}_{1},\, {\bf k}_{1},\, {\bf k}_{2}}
\hspace{0.03cm}\simeq\hspace{0.03cm}
\bigl(\hspace{0.02cm}T^{\,e\hspace{0.02cm}}\bigr)^{a_{1}\hspace{0.03cm} a_{2}}\hspace{0.03cm}(\hspace{0.02cm}t^{\,e\hspace{0.02cm}})^{\hspace{0.03cm} i_{1}\hspace{0.03cm} i}\,
\mathscr{T}^{\hspace{0.03cm} \ast\hspace{0.03cm}(2,\hspace{0.03cm}{\mathcal A})}_{\; {\bf p},\, {\bf p}_{1},\, {\bf k}_{1},\, {\bf k}_{2}}.
\label{eq:5l}
\end{equation}
\indent Substituting the expressions (\ref{eq:5d}), (\ref{eq:5k}) and (\ref{eq:5l}) into the right-hand side of (\ref{eq:5s}) and considering the symmetry condition (\ref{eq:5e}) for the scattering amplitude, instead of (\ref{eq:5s}) we derive the equation for the fourth-order correlation function 
\begin{equation}
\frac{\partial\hspace{0.01cm} I^{\, i\; i_{1}\, a_{1}\, a_{2}}_{{\bf p},\, {\bf p}_{1},\, {\bf k}_{1},\, {\bf k}_{2}}}{\partial\hspace{0.03cm} t}
=
i\hspace{0.03cm}\bigl[\,\varepsilon_{{\bf p}} + \omega^{\hspace{0.02cm}l}_{{\bf k}_{1}} - \varepsilon_{{\bf p}_{1}} -
\omega^{\hspace{0.02cm}l}_{{\bf k}_{2}}\bigr]
\, I^{\, i\; i_{1}\, a_{1}\, a_{2}}_{{\bf p},\, {\bf p}_{1},\, {\bf k}_{1},\, {\bf k}_{2}}
\label{eq:5z}
\end{equation}
\[
-\,i\,
\mathscr{T}^{\hspace{0.03cm} \ast\hspace{0.03cm}(2,\hspace{0.03cm}{\mathcal A})}_{\; {\bf p},\, {\bf p}_{1},\, {\bf k}_{1},\, {\bf k}_{2}}\, 
\Bigl\{-
\bigl({\mathfrak n}_{{\bf p}_{1}}t^{\,e}\hspace{0.03cm}\bigr)^{i_{1} i} \bigl({\mathcal N}_{{\bf k}_{1}} T^{\,e} {\mathcal N}_{{\bf k}_{2}}\bigr)^{a_{1}\hspace{0.02cm} a_{2}} 
+\,
\bigl(t^{\,e}\hspace{0.03cm}{\mathfrak n}_{{\bf p}}\bigr)^{i_{1} i} \bigl({\mathcal N}_{{\bf k}_{1}} T^{\,e} {\mathcal N}_{{\bf k}_{2}}\bigr)^{a_{1}\hspace{0.02cm} a_{2}} 
\]
\[
+\,
\bigl({\mathfrak n}_{{\bf p}_{1}} t^{\,e}\hspace{0.03cm}
{\mathfrak n}_{{\bf p}}\bigr)^{i_{1}i} \bigl(T^{\,e} {\mathcal N}_{{\bf k}_{2}}\bigr)^{a_{1}\hspace{0.02cm} a_{2}}  
-\,
\bigl({\mathfrak n}_{{\bf p}_{1}}\hspace{0.03cm}t^{\,e}
{\mathfrak n}_{{\bf p}}\bigr)^{i_{1}i} \bigl({\mathcal N}_{{\bf k}_{1}} T^{\,e}\bigr)^{a_{1}\hspace{0.02cm} a_{2}}\Bigr\}\,
\delta({\bf p} + {\bf k}_{1} - {\bf p}_{1} - {\bf k}_{2}).
\]

\section{\bf Kinetic equation for soft gluon excitations}
\label{section_6}
\setcounter{equation}{0}

The self-consistent equations (\ref{eq:5p}), (\ref{eq:5a}) and (\ref{eq:5z}) determine, in principle, the time evolution of number densities of the hard particles ${\mathfrak n}^{\hspace{0.03cm}i\hspace{0.04cm}i^{\prime}}_{\hspace{0.03cm}{\bf p}}$ and soft plasmons ${\mathcal N}^{\,a\hspace{0.03cm}a^{\prime}}_{{\bf k}}$. However, we introduce one more simplification: in 
Eq.\,(\ref{eq:5z}), we disregard the term with the time derivative as compared to the term containing the difference in the eigenfrequencies of wave packets and hard particle energies. Instead of equation (\ref{eq:5z}), we have
\begin{equation}
I^{\, i\; i_{1}\, a_{1}\, a_{2}}_{{\bf p},\, {\bf p}_{1},\, {\bf k}_{1},\, {\bf k}_{2}} \simeq
\delta({\bf p} - {\bf p}_{1})\hspace{0.03cm} \delta({\bf k}_{1} - {\bf k}_{2})\,
{\mathfrak n}^{\hspace{0.03cm}i_{1}\hspace{0.02cm}i}_{\hspace{0.03cm}{\bf p}} {\mathcal N}^{\,a_{1}\hspace{0.02cm}a_{2}}_{{\bf k}_{1}}
\vspace{0.15cm}
\label{eq:6q}
\end{equation}
\[
+\; \frac{1}{\Delta\hspace{0.02cm}\omega_{\hspace{0.03cm}{\bf p},\,{\bf p}_{1},\,{\bf k}_{1},\,{\bf k}_{2}} - i\hspace{0.03cm}0}\
\mathscr{T}^{\hspace{0.03cm} \ast\hspace{0.03cm}(2,\hspace{0.03cm}{\mathcal A})}_{\; {\bf p},\, {\bf p}_{1},\, {\bf k}_{1},\, {\bf k}_{2}}\, 
\Bigl\{-
\bigl({\mathfrak n}_{\hspace{0.03cm}{\bf p}_{1}}t^{\,e}\bigr)^{i_{1} i} \bigl({\mathcal N}_{{\bf k}_{1}} T^{\,e} {\mathcal N}_{{\bf k}_{2}}\bigr)^{a_{1}\, a_{2}} 
+\,
\bigl(t^{\,e}\hspace{0.03cm}{\mathfrak n}_{\hspace{0.03cm}{\bf p}}\bigr)^{i_{1} i} \bigl({\mathcal N}_{{\bf k}_{1}} T^{\,e} {\mathcal N}_{{\bf k}_{2}}\bigr)^{a_{1}\, a_{2}} 
\]
\[
+\,
\bigl({\mathfrak n}_{\hspace{0.03cm}{\bf p}_{1}} t^{\,e}\hspace{0.03cm} {\mathfrak n}_{\hspace{0.03cm}{\bf p}}\bigr)^{i_{1}i} \bigl(T^{\,e} {\mathcal N}_{{\bf k}_{2}}\bigr)^{a_{1}\, a_{2}}  
-\,
\bigl({\mathfrak n}_{\hspace{0.03cm}{\bf p}_{1}} t^{\,e}\hspace{0.03cm} {\mathfrak n}_{\hspace{0.03cm}{\bf p}}\bigr)^{i_{1}i} \bigl({\mathcal N}_{{\bf k}_{1}} T^{\,e}\hspace{0.03cm}\bigr)^{a_{1}\, a_{2}}\Bigr\}
\,\delta({\bf p} + {\bf k}_{1} - {\bf p}_{1} - {\bf k}_{2}),
\]
where now the resonance frequency difference is
\begin{equation}
\Delta\hspace{0.02cm}\omega_{\hspace{0.03cm}{\bf p},\,{\bf p}_{1},\,{\bf k}_{1},\,{\bf k}_{2}} \equiv
\varepsilon_{{\bf p}} + \omega^{l}_{{\bf k}_{1}} - \varepsilon_{{\bf p}_{1}} - \omega^{l}_{{\bf k}_{2}}.
\label{eq:6w}
\end{equation}
The first term on the right-hand side of (\ref{eq:6q}), which corresponds to completely uncorrelated waves (Gaussian fluctuations) is the solution to the homogeneous equation for the fourth-order correlation function
$ I^{\, i\; i_{1}\, a_{1}\, a_{2}}_{{\bf p},\, {\bf p}_{1},\, {\bf k}_{1},\, {\bf k}_{2}}$. The second term determines the deviation of the four-point correlator from the Gaussian approximation for a low nonlinearity level of interacting waves.\\
\indent We substitute the first term from (\ref{eq:6q}) into the right-hand side of Eq.\,(\ref{eq:5a}) for ${\mathcal N}^{\,a\hspace{0.03cm}a^{\prime}}_{{\bf k}}$. As a result we obtain
\begin{equation}
-\hspace{0.03cm}i\hspace{0.04cm}\delta({\bf k} - {\bf k}^{\prime}) \hspace{0.03cm}\!\int\!d{\bf p}\, 
{\rm tr}\hspace{0.03cm}\bigl({\mathfrak n}_{\hspace{0.03cm}{\bf p}}\hspace{0.03cm}t^{\,e}\bigr)
\Bigl\{\!\bigl({\mathcal N}^{\phantom{I}\!}_{{\bf k}}\hspace{0.03cm} T^{\,e}\bigr)^{a\hspace{0.03cm}a^{\prime}}\hspace{0.03cm}
\mathscr{T}^{\,(2,\hspace{0.03cm}{\mathcal A})}_{{\bf p},\, {\bf p}, {\bf k}, {\bf k}}
-\,
\bigl(T^{\,e} {\mathcal N}^{\phantom{I}}_{{\bf k}} \bigr)^{a\hspace{0.03cm}a^{\prime}}\hspace{0.03cm}
\mathscr{T}^{\hspace{0.03cm} \ast\hspace{0.03cm}(2,\hspace{0.03cm}{\mathcal A})}_{{\bf p},\, {\bf p}, {\bf k}, {\bf k}}\Bigr\}.
\label{eq:6e}
\end{equation}
Further, we substitute the second term from (\ref{eq:6q}) into the right-hand side of Eq.\,(\ref{eq:5a}). Simple algebraic transformations, in view of the symmetry condition (\ref{eq:5e}), lead us to
\begin{equation}
i\hspace{0.04cm}\delta({\bf k} - {\bf k}^{\prime})
\!\int\!d{\bf p}_{1}\hspace{0.02cm} d{\bf p}_{2}\hspace{0.03cm} d{\bf k}_{1}\,
\biggl(\bigl|\mathscr{T}^{\,(2,\hspace{0.03cm}{\mathcal A})}_{{\bf p}_{1},\, {\bf p}_{2}, {\bf k}, {\bf k}_{1}}\bigr|^{\hspace{0.03cm}2}
\label{eq:6r}
\end{equation}
\[
\times
\left\{\frac{1}{\Delta\hspace{0.02cm}\omega_{\hspace{0.03cm}{\bf p}_{1},\, {\bf p}_{2},\,{\bf k},\,{\bf k}_{1}}\!- i\hspace{0.03cm}0}
\Bigl[\hspace{0.03cm} {\rm tr}\hspace{0.03cm}\bigl(t^{\,d}\hspace{0.03cm}
{\mathfrak n}_{\hspace{0.03cm}{\bf p}_{2}}t^{\,e}\bigr)
\bigl({\mathcal N}_{{\bf k}} T^{\,e} {\mathcal N}_{{\bf k}_{1}} T^{\,d}\bigr)^{a\hspace{0.03cm}a^{\prime}}
\right.
\!\!-
{\rm tr}\hspace{0.03cm}\bigl(t^{\,d}\hspace{0.03cm}t^{\,e}\hspace{0.03cm} {\mathfrak n}_{\hspace{0.03cm}{\bf p}_{1}}\bigr)
\bigl({\mathcal N}_{{\bf k}} T^{\,e} {\mathcal N}_{{\bf k}_{1}} T^{\,d}\hspace{0.03cm}\bigr)^{a\hspace{0.03cm}a^{\prime}}
\]
\[
-\,
{\rm tr}\hspace{0.03cm}\bigl(t^{\,d} {\mathfrak n}_{\hspace{0.03cm}{\bf p}_{2}}t^{\,e}\hspace{0.03cm}{\mathfrak n}_{\hspace{0.03cm}{\bf p}_{1}}\bigr)
\bigl(T^{\,e} {\mathcal N}_{{\bf k}_{1}} T^{\,d}\hspace{0.03cm}\bigr)^{a\hspace{0.03cm}a^{\prime}}
\!+\,
{\rm tr}\hspace{0.03cm}\bigl(t^{\,d}\hspace{0.03cm} 
{\mathfrak n}_{\hspace{0.03cm}{\bf p}_{2}}t^{\,e}\hspace{0.03cm}{\mathfrak n}_{\hspace{0.03cm}{\bf p}_{1}}\bigr)
\bigl({\mathcal N}_{{\bf k}} T^{\,e}T^{\,d}\hspace{0.03cm}\bigr)^{a\hspace{0.03cm}a^{\prime}}
\Bigr]\!\biggr\}\hspace{0.03cm}
\delta({\bf k} - {\bf k}_{1} + {\bf p}_{1} - {\bf p}_{2})
\vspace{0.15cm}
\]
\[
-\hspace{0.03cm}\bigl|\mathscr{T}^{\,(2,\hspace{0.03cm}{\mathcal A})}_{{\bf p}_{1},\,{\bf p}_{2},\,{\bf k},\,{\bf k}_{1}}\bigr|^{\hspace{0.03cm}2}\!
\left\{\frac{1}{\Delta\hspace{0.02cm}\omega_{\hspace{0.03cm}{\bf p}_{2},\, {\bf p}_{1}, {\bf k}_{1}, {\bf k}} - i\hspace{0.03cm}0}
\Bigl[\hspace{0.03cm}{\rm tr}\hspace{0.03cm}\bigl(t^{\,d}{\mathfrak n}_{\hspace{0.03cm}{\bf p}_{1}}t^{\,e}\bigr)
\bigl(T^{\,d}{\mathcal N}_{{\bf k}_{1}} T^{\,e} {\mathcal N}_{{\bf k}}\bigr)^{a\hspace{0.03cm}a^{\prime}}
\right.\!
\!-\hspace{0.04cm}
{\rm tr}\hspace{0.03cm}\bigl(t^{\,d}\hspace{0.03cm}t^{\,e} {\mathfrak n}_{\hspace{0.03cm}{\bf p}_{2}}\bigr)
\bigl(T^{\,d}{\mathcal N}_{{\bf k}_{1}} T^{\,e} {\mathcal N}_{{\bf k}}\bigr)^{\!a\hspace{0.03cm}a^{\prime}}
\]
\[
-\,
{\rm tr}\hspace{0.03cm}\bigl(t^{\,d} {\mathfrak n}_{\hspace{0.03cm}{\bf p}_{1}}t^{\,e} {\mathfrak n}_{\hspace{0.03cm}{\bf p}_{2}}\bigr)
\bigl(T^{\,d}T^{\,e} {\mathcal N}_{{\bf k}}\hspace{0.03cm}\bigr)^{a\hspace{0.03cm}a^{\prime}}
+\,
{\rm tr}\hspace{0.03cm}\bigl(t^{\,d}\hspace{0.03cm}
{\mathfrak n}_{\hspace{0.03cm}{\bf p}_{1}}t^{\,e}\hspace{0.03cm} 
{\mathfrak n}_{\hspace{0.03cm}{\bf p}_{2}}\bigr)
\bigl(T^{\,d}{\mathcal N}_{{\bf k}_{1}} T^{\,e}\hspace{0.03cm}\bigr)^{a\hspace{0.03cm}a^{\prime}}
\Bigr]\hspace{0.03cm}\!\biggr\}\hspace{0.03cm}
\delta({\bf k} - {\bf k}_{1} + {\bf p}_{1} - {\bf p}_{2})\biggr).
\]
We consider the equality
\[
\frac{1}{\Delta\hspace{0.02cm}\omega_{{\bf p}_{2},\,{\bf p}_{1},\,{\bf k}_{1},\,{\bf k}} -\hspace{0.03cm} 
i\hspace{0.03cm}0}
\,=\,
-\frac{1}{\Delta\hspace{0.02cm}\omega_{{\bf p}_{1},\,{\bf p}_{2},\,{\bf k},\,{\bf k}_{1}} +\hspace{0.03cm} i\hspace{0.03cm}0}.
\]
to be evident by virtue of the definition (\ref{eq:6w}). Taking into account the obtained expressions (\ref{eq:6e}) and (\ref{eq:6r}), changing, where necessary, the dummy color summation indices and reducing the factor $\delta({\bf k} - {\bf k}^{\prime})$, we get the  the following kinetic equation for the plasmon number density ${\mathcal N}^{\,a\hspace{0.03cm}a^{\prime}}_{{\bf k}}$, instead of (\ref{eq:5a}):
\begin{equation}
\frac{\partial\hspace{0.03cm} {\mathcal N}^{\,a\hspace{0.03cm}a^{\prime}}_{{\bf k}}}{\partial\hspace{0.03cm} t}
=
-\hspace{0.03cm}i\hspace{0.03cm}\!\int\!d\hspace{0.02cm}{\bf p}\, 
{\rm tr}\hspace{0.03cm}\bigl({\mathfrak n}_{\hspace{0.03cm}{\bf p}}\hspace{0.03cm}t^{\,e}\bigr)
\Bigl\{\!\bigl({\mathcal N}^{\phantom{I}\!}_{{\bf k}}\hspace{0.03cm} T^{\,e}\bigr)^{a\hspace{0.03cm}a^{\prime}}\hspace{0.03cm}
\mathscr{T}^{\,(2,\hspace{0.03cm}{\mathcal A})}_{{\bf p},\, {\bf p}, {\bf k}, {\bf k}}
-\,
\bigl(T^{\,e} {\mathcal N}^{\phantom{I}}_{{\bf k}} \bigr)^{a\hspace{0.03cm}a^{\prime}}\hspace{0.03cm}
\mathscr{T}^{\,\ast\hspace{0.03cm}(2,\hspace{0.03cm}{\mathcal A})}_{{\bf p},\, {\bf p}, {\bf k}, {\bf k}}\Bigr\}
\label{eq:6t}
\end{equation}
\[
+\,
i\hspace{0.04cm}\!\int\!d{\bf p}_{1}\hspace{0.02cm} d{\bf p}_{2}\hspace{0.03cm} d{\bf k}_{1}\,
\delta({\bf k} - {\bf k}_{1} + {\bf p}_{1} - {\bf p}_{2})\,
\bigl|\mathscr{T}^{\,(2,\hspace{0.03cm}{\mathcal A})}_{{\bf p}_{1},\, {\bf p}_{2}, {\bf k}, {\bf k}_{1}}\bigr|^{2}
\]
\[
\times
\left\{\frac{1}{\Delta\hspace{0.02cm}\omega_{{\bf p}_{1},\,{\bf p}_{2},\,{\bf k},\,{\bf k}_{1}}\! - i\hspace{0.03cm}0}\,
\Bigl(\Bigl[\,{\rm tr}\hspace{0.03cm}\bigl(t^{\,e}\hspace{0.03cm}t^{\,d}\hspace{0.03cm} {\mathfrak n}_{\hspace{0.03cm}{\bf p}_{2}}\bigr)
\right.
\!\!-\,
{\rm tr}\hspace{0.03cm}\bigl(t^{\,d}\hspace{0.03cm}t^{\,e}\hspace{0.03cm} {\mathfrak n}_{\hspace{0.03cm}{\bf p}_{1}}\bigr)\Bigr]
\bigl({\mathcal N}^{\phantom{d}}_{{\bf k}} T^{\,e} {\mathcal N}^{\phantom{d}}_{{\bf k}_{1}} T^{\,d}\hspace{0.03cm}\bigr)^{a\hspace{0.03cm}a^{\prime}}
\]
\[
-\,
\Bigl[
\bigl(T^{\,e} {\mathcal N}^{\phantom{d}}_{{\bf k}_{1}} T^{\,d}\hspace{0.03cm}\bigr)^{a\hspace{0.03cm}a^{\prime}}
\!-\,
\bigl({\mathcal N}^{\phantom{d}}_{{\bf k}} T^{\,e}T^{\,d}\hspace{0.03cm}\bigr)^{a\hspace{0.03cm}a^{\prime}}
\Bigr]
{\rm tr}\hspace{0.03cm}\bigl(t^{\,d}\hspace{0.03cm} {\mathfrak n}_{\hspace{0.03cm}{\bf p}_{2}}t^{\,e}\hspace{0.03cm} 
{\mathfrak n}_{\hspace{0.03cm}{\bf p}_{1}}\bigr)\!
\Bigr)
\vspace{0.15cm}
\]
\[
-\,
\frac{1}{\Delta\hspace{0.02cm}\omega_{{\bf p}_{1},\,{\bf p}_{2},\,{\bf k},\,{\bf k}_{1}}\! + i\hspace{0.03cm}0}\,
\Bigl(\Bigl[\,{\rm tr}\hspace{0.03cm}\bigl(t^{\,e}\hspace{0.03cm}t^{\,d}\hspace{0.03cm} {\mathfrak n}_{\hspace{0.03cm}{\bf p}_{2}}\bigr)
\!\!-\,
{\rm tr}\hspace{0.03cm}\bigl(t^{\,d}\hspace{0.03cm}t^{\,e}\hspace{0.03cm} {\mathfrak n}_{\hspace{0.03cm}{\bf p}_{1}}\bigr)\Bigr]
\bigl(T^{\,e}{\mathcal N}_{{\bf k}_{1}} T^{\,d} {\mathcal N}_{{\bf k}}\bigr)^{a\hspace{0.03cm}a^{\prime}}
\]
\[
-\,\Bigl[
\bigl(T^{\,e}{\mathcal N}^{\phantom{d}}_{{\bf k}_{1}} T^{\,d}\hspace{0.03cm}\bigr)^{a\hspace{0.03cm}a^{\prime}}
-\,
\bigl(T^{\,e}T^{\,d} {\mathcal N}^{\phantom{d}}_{{\bf k}}\hspace{0.03cm}\bigr)^{a\hspace{0.03cm}a^{\prime}}
\Bigr]
{\rm tr}\hspace{0.03cm}\bigl(t^{\,d}\hspace{0.03cm} {\mathfrak n}_{\hspace{0.03cm}{\bf p}_{2}}t^{\,e}\hspace{0.03cm} 
{\mathfrak n}_{\hspace{0.03cm}{\bf p}_{1}}\bigr)\Bigr)\hspace{0.03cm}\!\biggr\}.
\]
In contrast to our previous works \cite{markov_2020, markov_2023}, where the plasmon number density matrix ${\mathcal N}^{\,a\hspace{0.03cm}a^{\prime}}_{{\bf k}}$ was chosen as the unit diagonal matrix in color space (as well as the matrix function ${\mathfrak n}^{\hspace{0.03cm}i^{\hspace{0.02cm}\prime}\hspace{0.02cm}i}_{\bf p}\,$), the required difference 
\begin{equation}
\frac{1}{\Delta\hspace{0.02cm}\omega_{{\bf p}_{1},\,{\bf p}_{2},\,{\bf k},\,{\bf k}_{1}}\! - i\hspace{0.03cm}0}
\,-\,
\frac{1}{\Delta\hspace{0.02cm}\omega_{{\bf p}_{1},\,{\bf p}_{2},\,{\bf k},\,{\bf k}_{1}}\! + i\hspace{0.03cm}0}\;
\big(\!\equiv2\hspace{0.02cm}\pi\hspace{0.03cm}i\hspace{0.04cm}
\delta(\Delta\hspace{0.03cm}\omega_{{\bf p}_{1},\,{\bf p}_{2},\,{\bf k},\,{\bf k}_{1}})\bigr).
\label{eq:6y}
\end{equation}
is literally not collected here. In the kinetic equation (\ref{eq:6t}) we have nontrivial arrangements of color matrices in the fundamental $t^{\,a}$ and the adjoint $T^{\hspace{0.03cm}a}$ representations, and also the matrix densities of the number of plasmons ${\mathcal N}_{\bf k}$ and hard particles ${\mathfrak n}_{\hspace{0.03cm}{\bf p}}$. It is necessary to calculate the available traces in advance.

\section{\bf Approximation of the effective amplitude $\mathscr{T}^{\hspace{0.03cm} (2)\hspace{0.03cm} i\; i_{1}\, a_{1}\, a_{2}}_{\; {\bf p},\, {\bf p}_{1},\, {\bf k}_{1},\, {\bf k}_{2}}$}
\label{section_7}
\setcounter{equation}{0}

\begin{figure}[hbtp]
	\begin{center}
		\includegraphics[width=1\textwidth]{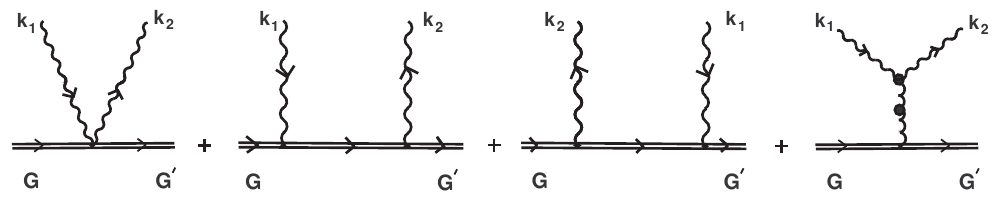}
	\end{center}
	\vspace{-0.5cm}
\caption{\small The effective amplitude $\mathscr{T}^{\,(2,\hspace{0.03cm}{\cal A})}_{\; {\bf p},\, {\bf p},\, {\bf k}_{1},\, {\bf k}_{2}}$ for the elastic scattering process of plasmon off a hard color particle. The blob stands for HTL-resummation and the double line denotes the hard  particle}
\label{fig1}
\end{figure}

Let us consider approximation of the effective subamplitudes  $\mathscr{T}^{\,(2,\hspace{0.03cm}{\mathcal A})}$ and $\mathscr{T}^{\,(2,\hspace{0.03cm}{\mathcal S})}$, Eqs.\,(\ref{eq:5g}) and (\ref{eq:5h}), in the limit 
\begin{equation}
|{\bf p}|,\,|{\bf p}_{1}| \gg |{\bf k}_{1}|,\,|{\bf k}_{2}|.
\label{eq:7q}
\end{equation}
As a preliminary step, by virtue of the momentum conservation law in (\ref{eq:4e}), we rewrite the expressions (\ref{eq:5g}) and (\ref{eq:5h}) setting 
\[
{\bf p}_{1}  = {\bf p} + {\bf k}_{1} - {\bf k}_{2} 
\equiv 
{\bf p} + \Delta{\bf k}.
\]
Then, for example, for the first effective amplitude $\mathscr{T}^{\,(2,\hspace{0.03cm}{\cal A})}$ we have 
\begin{equation}
	\mathscr{T}^{\,(2,\hspace{0.03cm}{\cal A})}_{\; {\bf p},\, {\bf p}_{1},\, {\bf k}_{1},\, {\bf k}_{2}}
	=
	T^{\,(2,\hspace{0.03cm}{\cal A})}_{\; {\bf p},\, {\bf p} + \Delta{\bf k},\, {\bf k}_{1},\, {\bf k}_{2}}
	\label{eq:7w}
\end{equation}
\begin{align}
	+\,\frac{1}{4}\,\Biggl[
	\Biggl(&\frac{1}{\omega^{\hspace{0.02cm} l}_{{\bf k}_{2}} - \varepsilon_{{\bf p} + {\bf k}_{1}} + 
		\varepsilon_{{\bf p} + \Delta{\bf k}}} 
	+
	\frac{1}{\omega^{\hspace{0.02cm} l}_{{\bf k}_{1}} - \varepsilon_{{\bf p}+ {\bf k}_{1}} + \varepsilon_{{\bf p}}} 
	\Biggr)\hspace{0.02cm}
	{\Phi}_{\,{\bf k}_{2},\, {\bf p}\hspace{0.03cm} + {\bf k}_{1},\, {\bf p} + \Delta{\bf k}}\, 
	{\Phi}^{\hspace{0.03cm}\ast}_{\,{\bf k}_{1},\, {\bf p}+ {\bf k}_{1},\, {\bf p}}
	\notag\\[1.5ex]
	+\,
	\Biggl(&\frac{1}{\omega^{\hspace{0.02cm} l}_{{\bf k}_{2}} - \varepsilon_{{\bf p}} + \varepsilon_{{\bf p} - {\bf k}_{2}}} 
	\,+\,
	\frac{1}{\omega^{\hspace{0.02cm} l}_{{\bf k}_{1}} - \varepsilon_{{\bf p} + \Delta{\bf k}} + \varepsilon_{{\bf p} - {\bf k}_{2}}} 
	\Biggr)\hspace{0.02cm}
	{\Phi}_{{\bf k}_{2},\, {\bf p},\, {\bf p} - {\bf k}_{2}}\, 
	{\Phi}^{\hspace{0.03cm}\ast}_{{\bf k}_{1},\,{\bf p} + \Delta{\bf k},\, {\bf p} - {\bf k}_{2}}\Biggr]
	\notag\\[1.5ex]
	-\,i\hspace{0.03cm}\Biggl[
	\Biggl(
	&\frac{1}{\omega^{\hspace{0.02cm} l}_{{\bf k}_{1}} - \omega^{\hspace{0.02cm} l}_{{\bf k}_{2}} - \omega^{\hspace{0.02cm} l}_{{\bf k}_{1} - {\bf k}_{2}}}
	\,-\,
	\frac{1}{\omega^{\hspace{0.02cm} l}_{{\bf k}_{1} - {\bf k}_{2}} - \varepsilon_{{\bf p} + \Delta{\bf k}} + \varepsilon_{{\bf p}}}
	\Biggr)\hspace{0.02cm}
	{\mathcal V}_{\,{\bf k}_{1}, {\bf k}_{2},\, {\bf k}_{1} - {\bf k}_{2}}\, 
	{\Phi}^{\hspace{0.03cm}\ast}_{\,{\bf k}_{1} - {\bf k}_{2},\, {\bf p} + \Delta{\bf k},\, {\bf p}}
	\notag\\[1.5ex]
	-\,\Biggl(
	&\frac{1}{\omega^{\hspace{0.02cm} l}_{{\bf k}_{2}} - \omega^{\hspace{0.02cm} l}_{{\bf k}_{1}} - \omega^{\hspace{0.02cm} l}_{{\bf k}_{2} - {\bf k}_{1}}}
	\,-\,
	\frac{1}{\omega^{\hspace{0.02cm} l}_{{\bf k}_{2} - {\bf k}_{1}} - \varepsilon_{{\bf p}} + \varepsilon_{{\bf p} + \Delta{\bf k}}}
	\Biggr)\hspace{0.02cm}
	{\Phi}_{\,{\bf k}_{2} - {\bf k}_{1},\, {\bf p},\, {\bf p} + \Delta{\bf k}}\, 
	{\mathcal V}^{\hspace{0.03cm}\ast}_{\,{\bf k}_{2}, {\bf k}_{1},\, {\bf k}_{2} - {\bf k}_{1}}\Biggr].
	\notag
\end{align}
In the limiting case (\ref{eq:7q}) for the expressions in the denominators on the right-hand side of (\ref{eq:7w}) we get
\[
\varepsilon_{{\bf p} + \Delta{\bf k}} - \varepsilon_{{\bf p} + {\bf k}_{1}}  
 \simeq - {\bf v}\cdot {\bf k}_{2}, 
\qquad
\varepsilon_{{\bf p} +\hspace{0.03cm} {\bf k}_{1}} - \varepsilon_{{\bf p}} \simeq {\bf v}\cdot {\bf k}_{1}
\]
etc. Here, we have denoted ${\bf v} = \partial\hspace{0.03cm} \varepsilon_{\bf p}/\partial\hspace{0.03cm} {\bf p}$. In light of these expansions we finally find an approximate expression for the effective amplitude $\mathscr{T}^{\,(2,\hspace{0.03cm}{\cal A})}$:
\begin{equation}
\mathscr{T}^{\,(2,\hspace{0.03cm}{\cal A})}_{\; {\bf p},\, {\bf p},\, {\bf k}_{1},\, {\bf k}_{2}}
=
T^{\,(2,\hspace{0.03cm}{\cal A})}_{\; {\bf p},\, {\bf p},\, {\bf k}_{1},\, {\bf k}_{2}}
+
\frac{1}{2}\,\Biggl(
\frac{1}{\omega^{\hspace{0.02cm} l}_{{\bf k}_{1}} - {\bf v}\cdot {\bf k}_{1}}  
+
\frac{1}{\omega^{\hspace{0.02cm} l}_{{\bf k}_{2}} - {\bf v}\cdot {\bf k}_{2}} 
\Biggr)\hspace{0.02cm}
{\Phi}^{\hspace{0.03cm}\ast}_{\,{\bf k}_{1},\, {\bf p},\, {\bf p}}\,
{\Phi}_{\,{\bf k}_{2},\, {\bf p},\, {\bf p}}. 
\label{eq:7r}
\end{equation}
\begin{align}
-\,i\hspace{0.03cm}\Biggl[\Biggl(
&\frac{1}{\omega^{\hspace{0.02cm} l}_{{\bf k}_{1}} - \omega^{\hspace{0.02cm} l}_{{\bf k}_{2}} - \omega^{\hspace{0.02cm} l}_{{\bf k}_{1} - {\bf k}_{2}}}
\,-\,
\frac{1}{\omega^{\hspace{0.02cm} l}_{{\bf k}_{1} - {\bf k}_{2}} - {\bf v}\cdot ({\bf k}_{1} - {\bf k}_{2})}
\Biggr)\hspace{0.02cm}
{\mathcal V}_{\,{\bf k}_{1}, {\bf k}_{2},\, {\bf k}_{1} - {\bf k}_{2}}\, 
{\Phi}^{\hspace{0.03cm}\ast}_{\,{\bf k}_{1} - {\bf k}_{2},\, {\bf p},\, {\bf p}}
\notag\\[1.5ex]
-\,\Biggl(
&\frac{1}{\omega^{\hspace{0.02cm} l}_{{\bf k}_{2}} - \omega^{\hspace{0.02cm} l}_{{\bf k}_{1}} - \omega^{\hspace{0.02cm} l}_{{\bf k}_{2} - {\bf k}_{1}}}
\,-\,
\frac{1}{\omega^{\hspace{0.02cm} l}_{{\bf k}_{2} - {\bf k}_{1}} - {\bf v}\cdot ({\bf k}_{2} - {\bf k}_{1})}
\Biggr)\hspace{0.02cm}
{\mathcal V}^{\hspace{0.03cm}\ast}_{\,{\bf k}_{2}, {\bf k}_{1},\, {\bf k}_{2} - {\bf k}_{1}}\,
{\Phi}_{\,{\bf k}_{2} - {\bf k}_{1},\, {\bf p},\, {\bf p}}
\Biggr].
\notag
\end{align}
\indent Fig.\,\ref{fig1} gives the diagrammatic interpretation of different terms in the effective amplitude  
$\mathscr{T}^{\,(2,\hspace{0.03cm}{\cal A})}_{\; {\bf p},\, {\bf p},\, {\bf k}_{1},\, {\bf k}_{2}}$.
The first graph represents a direct interaction of two plasmons with hard test particle induced by the amplitude $T^{\,(2,\hspace{0.03cm}{\cal A})}_{\; {\bf p},\, {\bf p},\, {\bf k}_{1},\, {\bf k}_{2}}$ in the general expression (\ref{eq:7r}). The second and third graphs describe the Compton scattering of soft boson excitations off a hard particle. In the effective amplitude (\ref{eq:7r}) they correspond to the term with product of the elementary interaction vertices of soft boson excitations with the hard test color-charged particle, namely ${\Phi}^{\hspace{0.03cm}\ast}_{\,{\bf k}_{1},\, {\bf p},\, {\bf p}}$ and ${\Phi}_{\,{\bf k}_{2},\, {\bf p},\, {\bf p}}$. The remaining graph is connected with the interaction of hard particle with plasmon and of three plasmons among themselves generated by the amplitudes ${\Phi}^{\hspace{0.03cm}\ast}_{\,{\bf k}_{1} - {\bf k}_{2},\, {\bf p},\, {\bf p}}$ and ${\mathcal V}_{\,{\bf k}_{1}, {\bf k}_{2},\, {\bf k}_{1} - {\bf k}_{2}}$ with intermediate ``virtual'' oscillation.\\
\indent Similar reasoning for the second effective subamplitude
$\mathscr{T}^{\,(2,\hspace{0.03cm}{\cal S})}_{\; {\bf p},\, {\bf p}_{1},\, {\bf k}_{1},\, {\bf k}_{2}}$ (\ref{eq:5h}) lead us to the following expression:
\[
\mathscr{T}^{\,(2,\hspace{0.03cm}{\cal S})}_{\; {\bf p},\, {\bf p},\, {\bf k}_{1},\, {\bf k}_{2}}
=
T^{\,(2,\hspace{0.03cm}{\cal S})}_{\; {\bf p},\, {\bf p},\, {\bf k}_{1},\, {\bf k}_{2}}
\]
\begin{align}
	+\,
	\frac{1}{4}\,\Biggl[
	\Biggl(&\frac{1}{\omega^{\hspace{0.02cm} l}_{{\bf k}_{2}} - {\bf v}\cdot {\bf k}_{2}} 
	\,+\,
	\frac{1}{\omega^{\hspace{0.02cm} l}_{{\bf k}_{1}} - {\bf v}\cdot {\bf k}_{1}} 
	\Biggr)\hspace{0.02cm}
	{\Phi}^{\hspace{0.03cm}\ast}_{\,{\bf k}_{1},\, {\bf p},\, {\bf p}}\,
	{\Phi}_{\,{\bf k}_{2},\, {\bf p},\, {\bf p}} 
	\notag\\[1.5ex]
	-\,
	\Biggl(&\frac{1}{\omega^{\hspace{0.02cm} l}_{{\bf k}_{2}} - {\bf v}\cdot {\bf k}_{2}} 
	\,+\,
	\frac{1}{\omega^{\hspace{0.02cm} l}_{{\bf k}_{1}} - {\bf v}\cdot {\bf k}_{1}} 
	\Biggr)\hspace{0.02cm}
	{\Phi}^{\hspace{0.03cm}\ast}_{\,{\bf k}_{1},\, {\bf p},\, {\bf p}}\,
	{\Phi}_{\,{\bf k}_{2},\, {\bf p},\, {\bf p}} 
	\Biggr].
\notag
\end{align}
In the limit (\ref{eq:7q}) the terms with the product $\Phi^{\ast}\hspace{0.02cm}\Phi$ exactly reduce each other and thus
\[
\mathscr{T}^{\,(2,\hspace{0.03cm}{\mathcal S})}_{\; {\bf p},\, {\bf p},\, {\bf k}_{1},\, {\bf k}_{2}} \simeq 0
\]
within the framework of these approximations, as already mentioned in the section \ref{section_5}. The complete effective amplitude $\mathscr{T}^{\hspace{0.03cm} (2)\,i\; i_{1}\, a_{1}\, a_{2}}_{\; {\bf p},\, {\bf p}_{1},\, {\bf k}_{1},\, {\bf k}_{2}}$, Eq.\,(\ref{eq:4r}), in this approximation has the simple color structure 
(\ref{eq:5k}), which, in turn, allows us to write the effective fourth-order Hamiltonian, Eq.\,(\ref{eq:4e}), describing the elastic scattering process of plasmon off a hard color particle as follows:
\[
	{\mathcal H}^{(4)}_{g\hspace{0.02cm}G\hspace{0.02cm}\rightarrow\hspace{0.02cm} g\hspace{0.02cm}G} 
	=
	i\hspace{0.02cm}f^{\hspace{0.03cm} a_{1}\hspace{0.03cm} a_{2}\hspace{0.03cm} a_{3}}
	\hspace{0.03cm}\biggl(\int\!|\hspace{0.035cm}\zeta_{\bf p}|^{\hspace{0.035cm}2}\,{\bf p}^{2}\hspace{0.03cm}
	d\hspace{0.02cm}|{\bf p}|\biggr)\!
	\int\!d\hspace{0.04cm}\Omega_{\hspace{0.035cm}{\bf v}}\!
	\int\!d\hspace{0.02cm}{\bf k}_{1}\hspace{0.03cm} d\hspace{0.02cm}{\bf k}_{2}\;
	\mathscr{T}^{\,(2,\hspace{0.03cm}{\mathcal A})}_{\; {\bf p},\, {\bf p},\, {\bf k}_{1},\, {\bf k}_{2}}\hspace{0.03cm}
	 c^{\ast\ \!\!a_{1}}_{{\bf k}_{1}} c^{\hspace{0.03cm}a_{2}}_{{\bf k}_{2}}\,\mathcal{Q}^{\hspace{0.03cm}a_{3}},
 \]
where $d\hspace{0.035cm}\Omega_{\hspace{0.035cm}{\bf v}}$ is a differential solid angle with respect to the velocity direction ${\bf v}$, and the classical (commuting) color charge $\mathcal{Q}^{\hspace{0.03cm}a}$ on the right-hand side is defined as
\begin{equation}
\mathcal{Q}^{\hspace{0.03cm}a}
\equiv
\theta^{\,\ast\ \!\!i}\hspace{0.01cm}(t^{\,a})^{i\hspace{0.03cm}j}
\hspace{0.01cm}\theta^{\phantom{\ast}\!\!j}.
\label{eq:7u}
\end{equation}

\section{Approximation of the kinetic equation (\ref{eq:6t}). The first moment with respect to color}
\label{section_8}
\setcounter{equation}{0}

Let us now turn to the approximation of the original kinetic equation (\ref{eq:6t}). In the second term, we perform integration over $d{\bf p}_{2}$, which gives us ${\bf p}_{2} = {\bf k} - {\bf k}_{1} + {\bf p}_{1}$ and consider the approximation $|{\bf p}_{1}| \gg |{\bf k}|,\, |{\bf k}_{1}|$. By using the definition of the color charge (\ref{eq:7u}), for the trace in the first term on the right-hand side of (\ref{eq:6t}) we have
\[
{\rm tr}\hspace{0.03cm}\bigl({\mathfrak n}_{\hspace{0.03cm}{\bf p}}t^{\,e}\bigr) = {\mathfrak n}^{i\hspace{0.02cm}j}_{\hspace{0.03cm}{\bf p}}\bigl(t^{\,e}\bigr)^{j\hspace{0.02cm}i}
=
n_{\hspace{0.03cm}{\bf p}}\hspace{0.03cm}\bigl\langle\hspace{0.03cm}\theta^{\ast\ \!\!j}\hspace{0.03cm} \theta^{\phantom{\ast}\!\!i}\hspace{0.03cm}\bigr\rangle \bigl(t^{\,e}\bigr)^{j\hspace{0.02cm}i} 
=
n_{\hspace{0.03cm}{\bf p}}\hspace{0.03cm}\bigl\langle\hspace{0.03cm}\mathcal{Q}^{\hspace{0.03cm}e}
\hspace{0.03cm}\bigr\rangle.
\]
Here, $n_{\hspace{0.03cm}{\bf p}}$ is an ordinary scalar function of the momentum ${\bf p}$ of a hard particle. Then in the second contribution on the right-hand side (\ref{eq:6t}) we have for the difference of traces 
\[
{\rm tr}\hspace{0.03cm}\bigl(t^{\,e}\hspace{0.03cm}t^{\,d}\hspace{0.03cm}{\mathfrak n}_{\hspace{0.03cm}{\bf p}_{1} + \Delta{\bf k}}\bigr)
-
{\rm tr}\hspace{0.03cm}\bigl(t^{\,d}\hspace{0.03cm}t^{\,e}\hspace{0.03cm}{\mathfrak n}_{\hspace{0.03cm}{\bf p}_{1}}\bigr)
=
{\rm tr}\hspace{0.03cm}\bigl(\bigl[\hspace{0.03cm}t^{\,e\!},t^{\,d}\hspace{0.03cm}\bigr] {\mathfrak n}_{\hspace{0.03cm}{\bf p}_{1}}\bigr)
+
{\rm tr}\hspace{0.03cm}\Bigl(t^{\,e}\hspace{0.03cm}t^{\,d}\,\frac{\partial\hspace{0.04cm} {\mathfrak n}_{\hspace{0.03cm}{\bf p}_{1}}}{\partial\hspace{0.03cm} {\bf p}_{1}}\cdot\Delta{\bf k}\Bigr) +\,\ldots,
\]
where we have designated $\Delta{\bf k} \equiv {\bf k} - {\bf k}_{1}$. In the abelian case the first term on the right-hand side here is equal to zero and it is necessary to take into account the next term of the expansion that is linear in $\Delta{\bf k}$. This takes place in the theory of weak wave turbulence for ordinary electron-ion  plasma (see, for example, \cite{kovrizhnykh_1965 }). Thus in the leading (zero) order in $\Delta{\bf k}$ for the non-Abelian case we have for the difference of traces:
\begin{equation}
{\rm tr}\hspace{0.03cm}\bigl(t^{\,e}\hspace{0.03cm}t^{\,d}\hspace{0.03cm} {\mathfrak n}_{\hspace{0.03cm}{\bf p}_{1} + \Delta{\bf k}}\bigr)
-
{\rm tr}\hspace{0.03cm}\bigl(t^{\,d}\hspace{0.03cm}t^{\,e}\hspace{0.03cm} {\mathfrak n}_{\hspace{0.03cm}{\bf p}_{1}}\bigr)
\simeq
{\rm tr}\hspace{0.03cm}\bigl(\bigl[\hspace{0.03cm}t^{\,e\!},\hspace{0.03cm}t^{\,d}\hspace{0.03cm}\bigr] {\mathfrak n}_{\hspace{0.03cm}{\bf p}_{1}}\bigr)
\equiv
i f^{\hspace{0.03cm}e\hspace{0.02cm}d\hspace{0.0cm}f}\bigl\langle\hspace{0.01cm}\mathcal{Q}^{\hspace{0.03cm}f}\hspace{0.03cm}\bigr\rangle n_{\hspace{0.03cm}{\bf p}_{1}}.
\label{eq:8q}
\end{equation}
\indent Let us consider further the more complex trace
\[
{\rm tr}\hspace{0.03cm}\bigl(t^{\,d}\hspace{0.03cm} {\mathfrak n}_{\hspace{0.03cm}{\bf p}_{1} + \Delta{\bf k}}\hspace{0.03cm}t^{\,e}\hspace{0.03cm} 
{\mathfrak n}_{\hspace{0.03cm}{\bf p}_{1}}\bigr)
=
{\rm tr}\hspace{0.03cm}\bigl(t^{\,d}\hspace{0.03cm} {\mathfrak n}_{\hspace{0.03cm}{\bf p}_{1}}t^{\,e}\hspace{0.03cm} 
{\mathfrak n}_{\hspace{0.03cm}{\bf p}_{1}}\bigr)
+
{\rm tr}\hspace{0.03cm}\Bigl(t^{\,d}\, 
\frac{\partial\hspace{0.04cm} {\mathfrak n}_{\hspace{0.03cm}{\bf p}_{1}}}{\partial\hspace{0.03cm} {\bf p}_{1}}
\cdot\Delta{\bf k}\,t^{\,e}\hspace{0.03cm} {\mathfrak n}_{\hspace{0.03cm}{\bf p}_{1}}\Bigr) +\,\ldots
\]
\[ 
=
\biggl\{n^{2}_{{\bf p}_{1}}\hspace{0.03cm} + \frac{1}{2}\,\frac{\partial\hspace{0.04cm} {n}^{2}_{{\bf p}_{1}}}{\partial\hspace{0.03cm} |{\bf p}_{1}|}\,({\bf v}_{1}\cdot\Delta{\bf k}) \,+\,\ldots\, \biggr\}\hspace{0.02cm}
\bigl[
(t^{\,d}\hspace{0.03cm})^{\hspace{0.03cm}j_{1}\hspace{0.02cm}i_{2}} 
\hspace{0.03cm}
\bigl\langle\hspace{0.03cm}\theta^{\hspace{0.03cm}\ast\ \!\!i_{1}}\hspace{0.03cm} \theta^{\phantom{\ast}\!\!i_{2}}\hspace{0.03cm}\bigr\rangle\hspace{0.03cm} 
(t^{\,e}\hspace{0.03cm})^{\hspace{0.03cm}i_{1}\hspace{0.02cm}j_{2}} 
\hspace{0.03cm}
\bigl\langle\hspace{0.03cm}\theta^{\hspace{0.03cm}\ast\ \!\!j_{1}}\hspace{0.03cm} \theta^{\phantom{\ast}\!\!j_{2}}\hspace{0.03cm}\bigr\rangle\bigr].  
\]
Here, unlike (\ref{eq:8q}), we cannot immediately present this expression in terms of the product of two commutative color charges $\mathcal{Q}^{\,d}$ and $\mathcal{Q}^{\,e}$. Let us rewrite the kinetic equation (\ref{eq:6t}) once more, leaving only zero order in $\Delta{\bf k}$ and assuming that the effective amplitude $\mathscr{T}^{\,(2,\hspace{0.03cm}{\mathcal A})}$ depends {\it only} on the velocity ${\bf v} = {\bf p}/|{\bf p}|$:
\[
\frac{\partial\hspace{0.03cm} {\mathcal N}^{\,a\hspace{0.03cm}a^{\prime}}_{{\bf k}}}{\!\!\partial\hspace{0.03cm} t}
=
-\hspace{0.03cm}i\,\biggl(\int\!n_{\bf p}\,{\bf p}^{2}\hspace{0.03cm}
d\hspace{0.02cm}|{\bf p}|\biggr)\!
\int\!d\hspace{0.04cm}\Omega_{\hspace{0.035cm}{\bf v}}\hspace{0.03cm} 
\Bigl\{\!\bigl({\mathcal N}_{{\bf k}}\hspace{0.03cm} T^{\,e}\bigr)^{a\hspace{0.03cm}a^{\prime}}\hspace{0.03cm}
\mathscr{T}^{\,(2,\hspace{0.03cm}{\mathcal A})}_{{\bf k},\,{\bf k}}({\bf v})
-\,
\bigl(T^{\,e} {\mathcal N}_{{\bf k}} \bigr)^{a\hspace{0.03cm}a^{\prime}}\hspace{0.03cm}
\mathscr{T}^{\,\ast\,(2,\hspace{0.03cm}{\mathcal A})}_{{\bf k},\,{\bf k}}({\bf v})\Bigr\}\hspace{0.03cm}
\bigl\langle\hspace{0.01cm}\mathcal{Q}^{\hspace{0.03cm}e}\hspace{0.03cm}\bigr\rangle
\]
\[
+\;
i\,\biggl(\int\!n_{\bf p}\,{\bf p}^{2}\hspace{0.03cm}
d\hspace{0.02cm}|{\bf p}|\biggr)\!
\int\!d\hspace{0.04cm}\Omega_{\hspace{0.035cm}{\bf v}}\!
\int\!d{\bf k}_{1}\,
\bigl|\hspace{0.02cm}\mathscr{T}^{\,(2,\hspace{0.03cm}{\mathcal A})}_{{\bf k},\,{\bf k}_{1}}({\bf v})\hspace{0.02cm}\bigr|^{2}\,
\bigl(T^{\,f}\bigr)^{d\hspace{0.02cm} e}\bigl\langle\hspace{0.01cm}\mathcal{Q}^{\hspace{0.03cm}f}
\hspace{0.03cm}\bigr\rangle
\]
\begin{equation}
\times
\left(\frac{\bigl({\mathcal N}_{{\bf k}}\hspace{0.03cm} T^{\,e} {\mathcal N}_{{\bf k}_{1}} T^{\,d}\hspace{0.03cm}\bigr)^{a\hspace{0.03cm}a^{\prime}}}{\Delta\hspace{0.02cm}\omega_{{\bf p},\, {\bf p}, {\bf k}, {\bf k}_{1}} -\, i\hspace{0.03cm}0} 
\,-\,
\frac{\bigl(T^{\,e} {\mathcal N}_{{\bf k}_{1}} T^{\,d}{\mathcal N}_{{\bf k}} \hspace{0.03cm}\bigr)^{a\hspace{0.03cm}a^{\prime}}}{\Delta\hspace{0.02cm}\omega_{{\bf p},\, {\bf p}, {\bf k}, {\bf k}_{1}} +\, i\hspace{0.03cm}0}\, 
\right)
\vspace{0.3cm}
\label{eq:8w}
\end{equation}
\[
-\;
i\,\biggl(\int\!n^{2}_{\bf p}\,{\bf p}^{2}\hspace{0.03cm}
d\hspace{0.02cm}|{\bf p}|\biggr)\!
\int\!d\hspace{0.04cm}\Omega_{\hspace{0.035cm}{\bf v}}\!
\int\!d{\bf k}_{1}\,
\bigl|\hspace{0.02cm}\mathscr{T}^{\,(2,\hspace{0.03cm}{\mathcal A})}_{{\bf k},\,{\bf k}_{1}}({\bf v})\hspace{0.02cm}\bigr|^{2}\,
\bigl[
(t^{\,d}\hspace{0.03cm})^{\hspace{0.03cm}j_{1}\hspace{0.02cm}i_{2}} 
\hspace{0.03cm}
\bigl\langle\hspace{0.03cm}\theta^{\hspace{0.03cm}\ast\ \!\!i_{1}}\hspace{0.03cm} \theta^{\phantom{\ast}\!\!i_{2}}\hspace{0.03cm}\bigr\rangle\hspace{0.03cm} 
(t^{\,e}\hspace{0.03cm})^{\hspace{0.03cm}i_{1}\hspace{0.02cm}j_{2}} 
\hspace{0.03cm}
\bigl\langle\hspace{0.03cm}\theta^{\hspace{0.03cm}\ast\ \!\!j_{1}}\hspace{0.03cm} \theta^{\phantom{\ast}\!\!j_{2}}\hspace{0.03cm}\bigr\rangle\bigr] 
\]
\[
\times
\left(\frac{\bigl(T^{\,e} {\mathcal N}_{{\bf k}_{1}} T^{\,d}\hspace{0.03cm}\bigr)^{a\hspace{0.03cm}a^{\prime}} 
- 
\bigl({\mathcal N}_{{\bf k}}\hspace{0.03cm} T^{\,e} T^{\,d}\hspace{0.03cm}\bigr)^{a\hspace{0.03cm}a^{\prime}}}
{\Delta\hspace{0.02cm}\omega_{{\bf p},\, {\bf p}, {\bf k}, {\bf k}_{1}} -\, i\hspace{0.03cm}0} 
\,-\,
\frac{\bigl(T^{\,e} {\mathcal N}_{{\bf k}_{1}} T^{\,d} \hspace{0.03cm}\bigr)^{a\hspace{0.03cm}a^{\prime}} 
- 
\bigl(T^{\,e} T^{\,d}{\mathcal N}_{{\bf k}} \hspace{0.03cm}\bigr)^{a\hspace{0.03cm}a^{\prime}}}
{\Delta\hspace{0.02cm}\omega_{{\bf p},\, {\bf p}, {\bf k}, {\bf k}_{1}} +\, i\hspace{0.03cm}0}\, 
\right),
\]
where we have replaced the integration variable ${\bf p}_{1}$ by ${\bf p}$ and supposed
\[
\mathscr{T}^{\,(2,\hspace{0.03cm}{\mathcal A})}_{{\bf p},\,{\bf p}, {\bf k}, {\bf k}_{1}}
\equiv
\mathscr{T}^{\,(2,\hspace{0.03cm}{\mathcal A})}_{\ {\bf k},\,{\bf k}_{1}}({\bf v}).
\]
Further, the resonance frequency difference (\ref{eq:6w}) in the expression (\ref{eq:8w}) is approximated as
\[
\Delta\hspace{0.02cm}\omega_{{\bf p},\, {\bf p}, {\bf k}, {\bf k}_{1}}
\simeq
\omega^{l}_{{\bf k}} - \omega^{l}_{{\bf k}_{1}} - {\mathbf v}\cdot({\bf k} - {\bf k}_{1}).
\]
Consider the following color decomposition of the matrix function ${\mathcal N}^{\,a\hspace{0.03cm}a^{\prime}}_{\bf k}$:
\begin{equation}
{\mathcal N}^{\,a\hspace{0.02cm}a^{\prime}}_{\bf k} = \delta^{\,a\hspace{0.02cm}a^{\prime}}\! 
N^{\hspace{0.03cm}l}_{\bf k} 
\hspace{0.03cm}+\hspace{0.03cm}
\bigl(T^{\,c}\bigr)^{a\hspace{0.02cm}a^{\prime}}\!\bigl\langle\hspace{0.01cm}\mathcal{Q}^{\hspace{0.03cm}c}\hspace{0.03cm}\bigr\rangle\hspace{0.03cm} W^{\hspace{0.03cm}l}_{\bf k}.
\label{eq:8r}
\end{equation}
We take the trace of the left and right-hand sides of (\ref{eq:8w}) with respect to color indices, i.e., we set $a = a^{\prime}$ and sum over $a$. Using the explicit representation (\ref{eq:8r}) and the formulae for the traces of the product of two and three color matrices in the adjoint representation from Appendix \ref{appendix_C}, Eqs.\,(\ref{ap:C4}) and (\ref{ap:C5}), we easily find for the trace on the left-hand side and for the traces in the first and third summands on the right-hand side of (\ref{eq:8w})
\[
{\rm tr}\,{\mathcal N}_{\bf k} = (N^{2}_{c} - 1)\hspace{0.03cm}N^{\hspace{0.03cm}l}_{\bf k} \equiv d_{A}\hspace{0.03cm}N^{\hspace{0.03cm}l}_{\bf k},
\qquad
{\rm tr}\hspace{0.03cm}\bigl(T^{\,e} {\mathcal N}_{{\bf k}} \bigr) = N_{c}\hspace{0.04cm}\bigl\langle\hspace{0.01cm}\mathcal{Q}^{\hspace{0.03cm}e}\hspace{0.03cm}\bigr\rangle\hspace{0.03cm} W^{\hspace{0.03cm}l}_{\bf k},
\]
\[
{\rm tr}\bigl[\bigl(T^{\,e} {\mathcal N}_{{\bf k}_{1}} T^{\,d}\hspace{0.03cm}\bigr) 
- 
\bigl({\mathcal N}_{{\bf k}} T^{\,e} T^{\,d}\hspace{0.03cm}\bigr)\bigr]
=
{\rm tr}\bigl[\bigl(T^{\,e} {\mathcal N}_{{\bf k}_{1}} T^{\,d}\hspace{0.03cm}\bigr) 
- 
\bigl(T^{\,e} T^{\,d} {\mathcal N}_{{\bf k}}\hspace{0.03cm}\bigr)\bigr]
\]
\[
=
\delta^{e\hspace{0.03cm}d}N_{c}\bigl(N^{\hspace{0.03cm}l}_{{\bf k}_{1}} - N^{\hspace{0.03cm}l}_{\bf k}\bigr)
+
\frac{1}{2}\,N_{c}\bigl(T^{\,c}\bigr)^{e\hspace{0.03cm}d}\bigl(W^{\hspace{0.03cm}l}_{{\bf k}_{1}} + W^{\hspace{0.03cm}l}_{\bf k}\hspace{0.03cm}\bigr)
\bigl\langle\hspace{0.01cm}\mathcal{Q}^{\hspace{0.03cm}c}\hspace{0.03cm}\bigr\rangle.
\]
The trace in the second term in (\ref{eq:8w}) has a slightly more complicated structure and requires the use of the formula for the trace of the product of four matrices (\ref{ap:C6}). Here, after contracting with $\bigl(T^{\,f}\bigr)^{d\hspace{0.03cm}e}$ we finally have
\[
\bigl(T^{\,f}\bigr)^{d\hspace{0.03cm}e}\hspace{0.03cm}{\rm tr}\hspace{0.02cm}
\bigl({\mathcal N}_{{\bf k}}\hspace{0.03cm} T^{\,e} {\mathcal N}_{{\bf k}_{1}} T^{\,d}\hspace{0.03cm}\bigr)
=
-\hspace{0.03cm}\frac{1}{2}\,N^{\hspace{0.02cm}2}_{c}\hspace{0.03cm}\bigl\langle\hspace{0.01cm}\mathcal{Q}^{\hspace{0.03cm}f}\hspace{0.03cm}\bigr\rangle
\bigl(W^{\hspace{0.03cm}l}_{\bf k}\hspace{0.02cm}N^{\hspace{0.03cm}l}_{{\bf k}_{1}} 
-
N^{\hspace{0.03cm}l}_{\bf k}\hspace{0.03cm}W^{\hspace{0.03cm}l}_{{\bf k}_{1}}\bigr).
\] 
In obtaining this expression we used the symmetry property (\ref{ap:C9}). This allowed us to easily eliminate the term with the product $W^{\hspace{0.03cm}l}_{\bf k}\,W^{\hspace{0.03cm}l}_{{\bf k}_{1}}$. Taking into account the obtained expressions for the color traces, we can now write out the first moment about color for equation (\ref{eq:8w})
\begin{equation}
d_{A}\,\frac{\partial\hspace{0.02cm} N^{\hspace{0.03cm}l}_{\bf k}}{\!\!\partial\hspace{0.03cm} t}
=
2\hspace{0.02cm}N_{c}\biggl(\int\!n_{\bf p}\,{\bf p}^{2}\hspace{0.03cm}
d\hspace{0.02cm}|{\bf p}|\biggr)\!
\int\!d\hspace{0.04cm}\Omega_{\hspace{0.035cm}{\bf v}}
\,{\rm Im}\hspace{0.03cm}\mathscr{T}^{\,(2,\hspace{0.03cm}{\mathcal A})}_{{\bf k},\,{\bf k}}({\bf v})
\hspace{0.03cm}W^{\hspace{0.03cm}l}_{\bf k}\,
\bigl\langle\hspace{0.01cm}\mathcal{Q}^{\hspace{0.03cm}e}\hspace{0.03cm}\bigr\rangle
\bigl\langle\hspace{0.01cm}\mathcal{Q}^{\hspace{0.03cm}e}\hspace{0.03cm}\bigr\rangle
\label{eq:8t}
\end{equation}
\[
+\,
\frac{1}{2}\,N^{\hspace{0.02cm}2}_{c}\biggl(\int\!n_{\bf p}\,{\bf p}^{2}\hspace{0.03cm}
d\hspace{0.02cm}|{\bf p}|\biggr)\!
\int\!d\hspace{0.04cm}\Omega_{\hspace{0.035cm}{\bf v}}\!
\int\!d{\bf k}_{1}\,
\bigl|\hspace{0.02cm}\mathscr{T}^{\,(2,\hspace{0.03cm}{\mathcal A})}_{{\bf k},\,{\bf k}_{1}}({\bf v})\hspace{0.02cm}\bigr|^{2}\,
\bigl(W^{\hspace{0.03cm}l}_{\bf k}\hspace{0.02cm}N^{\hspace{0.03cm}l}_{{\bf k}_{1}} 
-
N^{\hspace{0.03cm}l}_{\bf k}\hspace{0.03cm}W^{\hspace{0.03cm}l}_{{\bf k}_{1}}\bigr)
\bigl\langle\hspace{0.01cm}\mathcal{Q}^{\hspace{0.03cm}e}\hspace{0.03cm}\bigr\rangle
\bigl\langle\hspace{0.01cm}\mathcal{Q}^{\hspace{0.03cm}e}\hspace{0.03cm}\bigr\rangle
\]
\[
\times\,
(2\pi)\,\delta(\omega^{l}_{{\bf k}} - \omega^{l}_{{\bf k}_{1}} - {\mathbf v}\cdot({\bf k} - {\bf k}_{1}))
\vspace{0.3cm}
\]
\[
-\,
N_{c}\,\biggl(\int\!n^{2}_{\bf p}\,{\bf p}^{2}\hspace{0.03cm}
d\hspace{0.02cm}|{\bf p}|\biggr)\!
\int\!d\hspace{0.04cm}\Omega_{\hspace{0.035cm}{\bf v}}\!
\int\!d{\bf k}_{1}\,
\bigl|\hspace{0.02cm}\mathscr{T}^{\,(2,\hspace{0.03cm}{\mathcal A})}_{{\bf k},\,{\bf k}_{1}}({\bf v})\hspace{0.02cm}\bigr|^{2}\,
\Bigl\{\delta^{e\hspace{0.03cm}d}\bigl(N^{\hspace{0.03cm}l}_{{\bf k}} - N^{\hspace{0.03cm}l}_{{\bf k}_{1}}\bigr)
+
\frac{i}{2}\,f^{\,e\hspace{0.03cm}d\hspace{0.03cm}c}\hspace{0.03cm}
\bigl(W^{\hspace{0.03cm}l}_{{\bf k}} + W^{\hspace{0.03cm}l}_{{\bf k}_{1}}\bigr)
\bigl\langle\hspace{0.01cm}\mathcal{Q}^{\hspace{0.03cm}c}\hspace{0.03cm}\bigr\rangle\!\Bigr\} 
\]
\[
\times\,
\bigl[
(t^{\,d}\hspace{0.03cm})^{\hspace{0.03cm}j_{1}\hspace{0.02cm}i_{2}} 
\hspace{0.03cm}
\bigl\langle\hspace{0.03cm}\theta^{\hspace{0.03cm}\ast\ \!\!i_{1}}\hspace{0.03cm} \theta^{\phantom{\ast}\!\!i_{2}}\hspace{0.03cm}\bigr\rangle\hspace{0.03cm} 
(t^{\,e}\hspace{0.03cm})^{\hspace{0.03cm}i_{1}\hspace{0.02cm}j_{2}} 
\hspace{0.03cm}
\bigl\langle\hspace{0.03cm}\theta^{\hspace{0.03cm}\ast\ \!\!j_{1}}\hspace{0.03cm} \theta^{\phantom{\ast}\!\!j_{2}}\hspace{0.03cm}\bigr\rangle\bigr] 
(2\pi)\,\delta(\omega^{l}_{{\bf k}} - \omega^{l}_{{\bf k}_{1}} - {\mathbf v}\cdot({\bf k} - {\bf k}_{1})).
\]
Here, we have taken into account the Sokhotsky formula (\ref{eq:6y}). We note that the expectation value of the color charge enters the first and the second terms on the right-hand side in the colorless quadratic combination $\bigl\langle\hspace{0.01cm}\mathcal{Q}^{\hspace{0.03cm}e}\hspace{0.03cm}\bigr\rangle
\bigl\langle\hspace{0.01cm}\mathcal{Q}^{\hspace{0.03cm}e}\hspace{0.03cm}\bigr\rangle$. Furthermore, the last term in braces in (\ref{eq:8t}) contains the imaginary part proportional to the sum $\bigl(W^{\hspace{0.03cm}l}_{{\bf k}} + W^{\hspace{0.03cm}l}_{{\bf k}_{1}}\bigr)$. However, it is easy to see that this contribution vanishes. Indeed, let us introduce the notation
\begin{equation}
Z^{d\hspace{0.02cm}e} 
\equiv
\bigl[
(t^{\,d}\hspace{0.03cm})^{\hspace{0.03cm}j_{1}\hspace{0.02cm}i_{2}} 
\hspace{0.03cm}
\bigl\langle\hspace{0.03cm}\theta^{\hspace{0.03cm}\ast\ \!\!i_{1}}\hspace{0.03cm} \theta^{\phantom{\ast}\!\!i_{2}}\hspace{0.03cm}\bigr\rangle\hspace{0.03cm} 
(t^{\,e}\hspace{0.03cm})^{\hspace{0.03cm}i_{1}\hspace{0.02cm}j_{2}} 
\hspace{0.03cm}
\bigl\langle\hspace{0.03cm}\theta^{\hspace{0.03cm}\ast\ \!\!j_{1}}\hspace{0.03cm} \theta^{\phantom{\ast}\!\!j_{2}}\hspace{0.03cm}\bigr\rangle\bigr].
\label{eq:8y}
\end{equation}
The symmetry property with respect to color indices $d$ and $e$ follows from the structure of this expression
\begin{equation}
Z^{\hspace{0.02cm}d\hspace{0.02cm}e} = Z^{\hspace{0.02cm}e\hspace{0.02cm}d},
\label{eq:8u}
\end{equation} 
whence it immediately follows  
\begin{equation}
f^{\,e\hspace{0.03cm}d\hspace{0.03cm}c}\hspace{0.03cm}Z^{\hspace{0.02cm}d\hspace{0.02cm}e} = 0.
\label{eq:8i}
\end{equation}
Let us consider the first term in braces in (\ref{eq:8t}) containing the difference  
$\bigl(N^{\hspace{0.03cm}l}_{{\bf k}} - N^{\hspace{0.03cm}l}_{{\bf k}_{1}}\bigr)$. Here, we have the contraction of the form 
\begin{equation}
\delta^{\,e\hspace{0.03cm}d}\hspace{0.03cm}Z^{\hspace{0.02cm}d\hspace{0.02cm}e}
=
\bigl[
(t^{\,e}\hspace{0.03cm})^{\hspace{0.03cm}j_{1}\hspace{0.02cm}i_{2}} 
\hspace{0.03cm}
\bigl\langle\hspace{0.03cm}\theta^{\hspace{0.03cm}\ast\ \!\!i_{1}}\hspace{0.03cm} \theta^{\phantom{\ast}\!\!i_{2}}\hspace{0.03cm}\bigr\rangle\hspace{0.03cm} 
(t^{\,e}\hspace{0.03cm})^{\hspace{0.03cm}i_{1}\hspace{0.02cm}j_{2}} 
\hspace{0.03cm}
\bigl\langle\hspace{0.03cm}\theta^{\hspace{0.03cm}\ast\ \!\!j_{1}}\hspace{0.03cm} \theta^{\phantom{\ast}\!\!j_{2}}\hspace{0.03cm}\bigr\rangle\bigr].
\label{eq:8o}
\end{equation}
To disentangle this expression, it is necessary to use the Fierz identity for the $t^{\,a}$ matrices, Eq.\,(\ref{ap:B3_b}). In this case we have
\[
(t^{\,e})^{j_{1}i_{2}}(t^{\,e})^{i_{1}j_{2}} 
= 
\biggl(\frac{N^{\hspace{0.02cm}2}_{c} - 4}{2\hspace{0.02cm}N^{\hspace{0.02cm}2}_{c}}\biggr)\hspace{0.02cm} \delta^{\,i_{1}i_{2}}\hspace{0.02cm}\delta^{\,j_{1}j_{2}}	
-
\frac{1}{N_{c}}\,(t^{\,e})^{\hspace{0.03cm}i_{1}i_{2}}(t^{\,e})^{\hspace{0.03cm}j_{1}j_{2}}
\]
and therefore instead of (\ref{eq:8o}) we obtain at once
\begin{equation}
\delta^{\,e\hspace{0.03cm}d}\hspace{0.02cm}Z^{\hspace{0.02cm}d\hspace{0.02cm}e}
=
\biggl(\frac{N^{\hspace{0.02cm}2}_{c} - 4}{2\hspace{0.02cm}N^{\hspace{0.02cm}2}_{c}}\biggr)\hspace{0.02cm} \bigl\langle\hspace{0.01cm}{\mathcal Q}\hspace{0.03cm}\bigr\rangle^{\!2}	
-
\frac{1}{N_{c}}\,\bigl\langle\hspace{0.01cm}\mathcal{Q}^{\hspace{0.03cm}e}
\hspace{0.03cm}\bigr\rangle\bigl\langle\hspace{0.01cm}
\mathcal{Q}^{\hspace{0.03cm}e}\hspace{0.03cm}\bigr\rangle.
\label{eq:8a}
\end{equation}
Here we have introduced a notation for the mean value of the commutative ``colorless'' charge
\[
\bigl\langle\hspace{0.01cm}{\mathcal Q}\hspace{0.03cm}\bigr\rangle
\equiv
\bigl\langle\hspace{0.03cm}\theta^{\hspace{0.03cm}\ast\ \!\!i}\hspace{0.03cm} \theta^{\phantom{\ast}\!\!i}\hspace{0.04cm}
\bigr\rangle.
\]
We see that it is impossible in this case to reduce the expression (\ref{eq:8o}) only to a quadratic combination of color charges $\bigl\langle\hspace{0.01cm}\mathcal{Q}^{\hspace{0.03cm}e}\hspace{0.03cm}\bigr\rangle\bigl\langle\hspace{0.01cm}\mathcal{Q}^{\hspace{0.03cm}e}\hspace{0.03cm}\bigr\rangle$. The square of the mean value of the colorless Grassmann charges combination $\langle\hspace{0.03cm}\theta^{\hspace{0.03cm}\ast\ \!\!i}\hspace{0.03cm} \theta^{\phantom{\ast}\!\!i}\hspace{0.03cm}\bigr\rangle$ inevitably appears. Substituting the expression (\ref{eq:8a}) into (\ref{eq:8t}) we find finally the kinetic equation for the colorless part of the plasmon number density $N^{\hspace{0.03cm}l}_{\bf k}$:
\[
	d_{A}\,\frac{\partial\hspace{0.02cm} N^{\hspace{0.03cm}l}_{\bf k}}{\!\!\partial\hspace{0.03cm} t}
	=
	2\hspace{0.02cm}N_{c}\biggl(\int\!n_{\bf p}\,{\bf p}^{2}\hspace{0.03cm}
	d\hspace{0.02cm}|{\bf p}|\biggr)\!
	\int\!d\hspace{0.04cm}\Omega_{\hspace{0.035cm}{\bf v}}\,
	{\rm Im}\hspace{0.03cm}\mathscr{T}^{\,(2,\hspace{0.03cm}{\mathcal A})}_{{\bf k},\,{\bf k}}({\bf v})
	\hspace{0.03cm}W^{\hspace{0.03cm}l}_{\bf k}\,
	\bigl\langle\hspace{0.01cm}\mathcal{Q}^{\hspace{0.03cm}e}\hspace{0.03cm}
	\bigr\rangle
	\bigl\langle\hspace{0.01cm}\mathcal{Q}^{\hspace{0.03cm}e}\hspace{0.03cm}
	\bigr\rangle
	\label{eq:8s}
\]
%
\[
+\;
\frac{1}{2}\,N^{\hspace{0.02cm}2}_{c}\biggl(\int\!n_{\bf p}\,{\bf p}^{2}\hspace{0.03cm}
d\hspace{0.02cm}|{\bf p}|\biggr)\!
\int\!d\hspace{0.04cm}\Omega_{\hspace{0.035cm}{\bf v}}\!
\int\!d{\bf k}_{1}\,
\bigl|\hspace{0.02cm}\mathscr{T}^{\,(2,\hspace{0.03cm}{\mathcal A})}_{{\bf k},\,{\bf k}_{1}}({\bf v})\hspace{0.02cm}\bigr|^{\hspace{0.02cm} 2}\,
\bigl(W^{\hspace{0.03cm}l}_{\bf k}\hspace{0.02cm}N^{\hspace{0.03cm}l}_{{\bf k}_{1}} 
-
N^{\hspace{0.03cm}l}_{\bf k}\hspace{0.03cm}W^{\hspace{0.03cm}l}_{{\bf k}_{1}}\bigr)
\bigl\langle\hspace{0.01cm}\mathcal{Q}^{\hspace{0.03cm}e}\hspace{0.03cm}\bigr\rangle
\bigl\langle\hspace{0.01cm}\mathcal{Q}^{\hspace{0.03cm}e}\hspace{0.03cm}\bigr\rangle
\vspace{0.1cm}
\]
\[
\times\,
(2\pi)\,\delta(\omega^{l}_{{\bf k}} - \omega^{l}_{{\bf k}_{1}} - {\mathbf v}\cdot({\bf k} - {\bf k}_{1}))
\vspace{0.3cm}
\]
\[
-\,
\biggl(\int\!n^{2}_{\bf p}\,{\bf p}^{2}\hspace{0.03cm}
d\hspace{0.02cm}|{\bf p}|\biggr)\!
\int\!d\hspace{0.04cm}\Omega_{\hspace{0.035cm}{\bf v}}\!
\int\!d{\bf k}_{1}\,
\bigl|\hspace{0.02cm}\mathscr{T}^{\,(2,\hspace{0.03cm}{\mathcal A})}_{{\bf k},\,{\bf k}_{1}}({\bf v})\hspace{0.02cm}\bigr|^{\hspace{0.02cm}2}\hspace{0.03cm}
\bigl(N^{\hspace{0.03cm}l}_{{\bf k}} - N^{\hspace{0.03cm}l}_{{\bf k}_{1}}\bigr)
\biggl\{\biggl(\frac{N^{\hspace{0.02cm}2}_{c} - 4} {2\hspace{0.02cm}N_{c}}\biggr)\hspace{0.02cm}
\bigl\langle\hspace{0.01cm}{\mathcal Q}\hspace{0.03cm}
\bigr\rangle^{\!2}
\hspace{0.03cm}-\hspace{0.03cm}
\bigl\langle\hspace{0.01cm}\mathcal{Q}^{\hspace{0.03cm}e}
\hspace{0.03cm}\bigr\rangle\bigl\langle\hspace{0.01cm}
\mathcal{Q}^{\hspace{0.03cm}e}\hspace{0.03cm}\bigr\rangle\!\hspace{0.02cm}
\biggr\} 
\vspace{0.1cm}
\]
\[
\hspace{0.08cm}
\times\,
(2\pi)\,\delta(\omega^{l}_{{\bf k}} - \omega^{l}_{{\bf k}_{1}} - {\mathbf v}\cdot({\bf k} - {\bf k}_{1})).
\]

\section{The second moment with respect to color}
\label{section_9}
\setcounter{equation}{0}

Let us return to our original equation (\ref{eq:8w}). Now let us contract the left- and right-hand sides of this equation with the color matrix $\bigl(T^{\hspace{0.03cm}s}\bigr)^{a^{\prime}\hspace{0.01cm}a}$. As a result, we find
\[
\frac{\partial\,{\rm tr}\hspace{0.03cm}\bigl(T^{\,s} {\mathcal N}_{{\bf k}} \bigr)}{\!\!\partial\hspace{0.03cm}t}
=
-\hspace{0.03cm}i\,\biggl(\int\!n_{\bf p}\,{\bf p}^{2}\hspace{0.03cm}
d\hspace{0.02cm}|{\bf p}|\biggr)\!
\int\!d\hspace{0.04cm}\Omega_{\hspace{0.035cm}{\bf v}}
\, 
\Bigl\{{\rm tr}\hspace{0.03cm}\bigl( T^{\,e}T^{\,s} {\mathcal N}_{{\bf k}}\hspace{0.02cm}\bigr)\hspace{0.03cm}
\mathscr{T}^{\,(2,\hspace{0.03cm}{\mathcal A})}_{{\bf k},\,{\bf k}}({\bf v})
-\,
{\rm tr}\hspace{0.03cm}\bigl(T^{\,s}T^{\,e} {\mathcal N}_{{\bf k}} \bigr)\hspace{0.03cm}
\mathscr{T}^{\,\ast\,(2,\hspace{0.03cm}{\mathcal A})}_{{\bf k},\,{\bf k}}({\bf v})\Bigr\}\hspace{0.03cm}
\bigl\langle\hspace{0.01cm}\mathcal{Q}^{\hspace{0.03cm}e}\hspace{0.03cm}
\bigr\rangle
\]
\begin{equation}
+\;
i\,\biggl(\int\!n_{\bf p}\,{\bf p}^{2}\hspace{0.03cm}
d\hspace{0.02cm}|{\bf p}|\biggr)\!
\int\!d\hspace{0.04cm}\Omega_{\hspace{0.035cm}{\bf v}}\!
\int\!d\hspace{0.03cm}{\bf k}_{1}\,
\bigl|\hspace{0.02cm}\mathscr{T}^{\,(2,\hspace{0.03cm}{\mathcal A})}_{{\bf k},\,{\bf k}_{1}}({\bf v})\hspace{0.02cm}\bigr|^{\hspace{0.03cm}2}\,
\bigl(T^{\,f}\bigr)^{d\hspace{0.02cm} e}\bigl\langle\hspace{0.01cm}\mathcal{Q}^{\hspace{0.03cm}f}\hspace{0.03cm}
\bigr\rangle
\label{eq:9q}
\end{equation}
\[
\times
\left(\frac{{\rm tr}\hspace{0.03cm}\bigl(T^{\,d}\hspace{0.03cm}T^{\,s}{\mathcal N}_{{\bf k}}\hspace{0.03cm} T^{\,e} {\mathcal N}_{{\bf k}_{1}} \hspace{0.03cm}\bigr)}{\Delta\hspace{0.02cm}\omega_{{\bf p},\, {\bf p}, {\bf k}, {\bf k}_{1}} -\, i\hspace{0.03cm}0} 
\,-\,
\frac{{\rm tr}\hspace{0.03cm}\bigl(T^{\,s}T^{\,e} {\mathcal N}_{{\bf k}_{1}} T^{\,d}{\mathcal N}_{{\bf k}} \hspace{0.03cm}\bigr)}{\Delta\hspace{0.02cm}\omega_{{\bf p},\, {\bf p}, {\bf k}, {\bf k}_{1}} +\, i\hspace{0.03cm}0}\, 
\right)
\vspace{0.3cm}
\]
\[
-\;
i\,\biggl(\int\!n^{2}_{\bf p}\,{\bf p}^{2}\hspace{0.03cm}
d\hspace{0.02cm}|{\bf p}|\biggr)\!
\int\!d\hspace{0.04cm}\Omega_{\hspace{0.035cm}{\bf v}}\!
\int\!d\hspace{0.03cm}{\bf k}_{1}\,
\bigl|\hspace{0.02cm}\mathscr{T}^{\,(2,\hspace{0.03cm}{\mathcal A})}_{{\bf k},\,{\bf k}_{1}}({\bf v})\hspace{0.02cm}\bigr|^{\hspace{0.03cm} 2}\,
\bigl[
(t^{\,d}\hspace{0.03cm})^{\hspace{0.03cm}j_{1}\hspace{0.02cm}i_{2}} 
\hspace{0.03cm}
\bigl\langle\hspace{0.03cm}\theta^{\hspace{0.03cm}\ast\ \!\!i_{1}}\hspace{0.03cm} \theta^{\phantom{\ast}\!\!i_{2}}\hspace{0.03cm}\bigr\rangle\hspace{0.03cm} 
(t^{\,e}\hspace{0.03cm})^{\hspace{0.03cm}i_{1}\hspace{0.02cm}j_{2}} 
\hspace{0.03cm}
\bigl\langle\hspace{0.03cm}\theta^{\hspace{0.03cm}\ast\ \!\!j_{1}}\hspace{0.03cm} \theta^{\phantom{\ast}\!\!j_{2}}\hspace{0.03cm}\bigr\rangle\bigr] 
\]
\[
\times
\left(\frac{{\rm tr}\hspace{0.03cm}\bigl(T^{\,d} \hspace{0.03cm}T^{\,s} \hspace{0.03cm}T^{\,e} {\mathcal N}_{{\bf k}_{1}} \hspace{0.03cm}\bigr) 
- 
{\rm tr}\hspace{0.03cm}\bigl(T^{\,e}  \hspace{0.03cm}T^{\,d}  \hspace{0.03cm}T^{\,s}{\mathcal N}_{{\bf k}}\hspace{0.03cm}\bigr)}
{\Delta\hspace{0.02cm}\omega_{{\bf p},\, {\bf p}, {\bf k}, {\bf k}_{1}} -\, i\hspace{0.03cm}0} 
\,-\,
\frac{{\rm tr}\hspace{0.03cm}\bigl(T^{\,d} \hspace{0.03cm}T^{\,s} \hspace{0.03cm}T^{\,e} {\mathcal N}_{{\bf k}_{1}} \hspace{0.03cm}\bigr) 
- 
{\rm tr}\hspace{0.03cm}\bigl(T^{\,e} \hspace{0.03cm}T^{\,d}{\mathcal N}_{{\bf k}}\hspace{0.03cm}T^{\,s} \hspace{0.02cm}\bigr)}
{\Delta\hspace{0.02cm}\omega_{{\bf p},\, {\bf p}, {\bf k}, {\bf k}_{1}} +\, i\hspace{0.03cm}0}\, 
\right).
\]
We consider the trace on the left-hand side and the traces in the first term on the right-hand side of Eq.\,(\ref{eq:9q}). With allowance made for the color decomposition (\ref{eq:8r}), simple calculations give 
\begin{equation}
{\rm tr}\hspace{0.03cm}\bigl(T^{\,s} {\mathcal N}_{{\bf k}} \bigr) = N_{c}\hspace{0.03cm}\bigl\langle\hspace{0.01cm}\mathcal{Q}^{\hspace{0.03cm}s}\hspace{0.03cm}\bigr\rangle\hspace{0.03cm} W^{\hspace{0.03cm}l}_{\bf k},
\label{eq:9w}
\end{equation}
\[
{\rm tr}\bigl(T^{\,e}T^{\,s}{\mathcal N}_{{\bf k}}\hspace{0.03cm}\bigr) 
=
\delta^{e\hspace{0.03cm}s}N_{c}\hspace{0.03cm}N^{\hspace{0.03cm}l}_{\bf k}
+
\frac{i}{2}\,N_{c}\hspace{0.03cm}f^{\,e\hspace{0.03cm} s\hspace{0.03cm}c}\hspace{0.03cm}\bigl\langle\hspace{0.01cm}\mathcal{Q}^{\hspace{0.03cm}c}\hspace{0.03cm}\bigr\rangle\hspace{0.03cm} W^{\hspace{0.03cm}l}_{\bf k},
\quad
{\rm tr}\bigl(T^{\,s} T^{\,e} {\mathcal N}_{{\bf k}} \hspace{0.03cm}\bigr) 
=
\delta^{e\hspace{0.03cm}s}N_{c}\hspace{0.03cm}N^{\hspace{0.03cm}l}_{\bf k}
-
\frac{i}{2}\,N_{c}\hspace{0.03cm}f^{\,e\hspace{0.03cm} s\hspace{0.03cm}c}\hspace{0.03cm}\bigl\langle\hspace{0.01cm}
\mathcal{Q}^{\hspace{0.03cm}c}\hspace{0.03cm}\bigr\rangle\hspace{0.03cm} W^{\hspace{0.03cm}l}_{\bf k}.
\]
The imaginary part in the last two expressions will turn to zero under contraction with the color charge $\bigl\langle\hspace{0.01cm}\mathcal{Q}^{\hspace{0.03cm}e}\hspace{0.03cm}
\bigr\rangle$ and as a result the expression in braces in the first term in (\ref{eq:9q}) may be cast in the following way:
\begin{equation}
\Bigl\{{\rm tr}\hspace{0.03cm}\bigl(T^{\,e}T^{\,s}{\mathcal N}_{{\bf k}}\hspace{0.02cm}\bigr)\hspace{0.03cm}
\mathscr{T}^{\,(2,\hspace{0.03cm}{\mathcal A})}_{{\bf k},\,{\bf k}}({\bf v})
-\,
{\rm tr}\hspace{0.03cm}\bigl(T^{\,s}T^{\,e} {\mathcal N}_{{\bf k}} \bigr)\hspace{0.03cm}
\mathscr{T}^{\,\ast\,(2,\hspace{0.03cm}{\mathcal A})}_{{\bf k},\,{\bf k}}({\bf v})\Bigr\}\hspace{0.03cm}
\bigl\langle\hspace{0.01cm}\mathcal{Q}^{\hspace{0.03cm}e}\hspace{0.03cm}\bigr\rangle
=
2\hspace{0.03cm}i\hspace{0.02cm}N_{c}\,
{\rm Im}\hspace{0.03cm}\mathscr{T}^{\,(2,\hspace{0.03cm}{\mathcal A})}_{{\bf k},\,{\bf k}}({\bf v})
\hspace{0.03cm}N^{\hspace{0.03cm}l}_{\bf k}\,
\bigl\langle\hspace{0.01cm}\mathcal{Q}^{\hspace{0.03cm}s}\hspace{0.03cm}\bigr\rangle.
\label{eq:9e}
\end{equation}
\indent Let us further consider more nontrivial traces in the second term in (\ref{eq:9q}). For the first trace, taking into account the decomposition (\ref{eq:8r}), we find the starting expression for the subsequent analysis
\begin{equation}
\begin{split}	
&{\rm tr}\hspace{0.03cm}\bigl(T^{\,d}\hspace{0.03cm}T^{\,s}{\mathcal N}_{{\bf k}}\hspace{0.03cm} T^{\,e} {\mathcal N}_{{\bf k}_{1}} \hspace{0.03cm}\bigr)
=
{\rm tr}\hspace{0.03cm}\bigl(T^{\,d}\hspace{0.02cm} T^{\,s}\hspace{0.02cm} T^{\,e}\hspace{0.03cm}\bigr)
N^{\hspace{0.03cm}l}_{{\bf k}}\hspace{0.01cm}N^{\hspace{0.03cm}l}_{{\bf k}_{1}}
+
{\rm tr}\hspace{0.03cm}\bigl(T^{\,d}\hspace{0.02cm} T^{\,s}\hspace{0.02cm} T^{\,c}\hspace{0.02cm} T^{\,e}\hspace{0.03cm}\bigr)
\bigl\langle\hspace{0.01cm}\mathcal{Q}^{\hspace{0.03cm}c}\hspace{0.03cm}\bigr\rangle
W^{\hspace{0.03cm}l}_{{\bf k}}\hspace{0.03cm}N^{\hspace{0.03cm}l}_{{\bf k}_{1}}\\[1.5ex]
&+\,
{\rm tr}\hspace{0.03cm}\bigl(T^{\,d}\hspace{0.02cm} T^{\,s}\hspace{0.02cm} T^{\,e}\hspace{0.02cm}T^{\,c}\hspace{0.03cm}\bigr)
\bigl\langle\hspace{0.01cm}\mathcal{Q}^{\hspace{0.03cm}c}\hspace{0.03cm}\bigr\rangle
N^{\hspace{0.03cm}l}_{{\bf k}}\hspace{0.03cm}W^{\hspace{0.03cm}l}_{{\bf k}_{1}}
+
{\rm tr}\hspace{0.03cm}\bigl(T^{\,d}\hspace{0.02cm} T^{\,s}\hspace{0.02cm} T^{\,c}\hspace{0.02cm} 
T^{\,e}\hspace{0.02cm} T^{\,c^{\prime}}\hspace{0.03cm}\bigr)
\bigl\langle\hspace{0.01cm}\mathcal{Q}^{\hspace{0.03cm}c}\hspace{0.03cm}\bigr\rangle
\bigl\langle\hspace{0.01cm}\mathcal{Q}^{\hspace{0.03cm}c^{\prime}}\hspace{0.03cm}\bigr\rangle
W^{\hspace{0.03cm}l}_{{\bf k}}\hspace{0.03cm}W^{\hspace{0.03cm}l}_{{\bf k}_{1}}.
\end{split}
\label{eq:9r}
\end{equation}
For the traces of three and four generators in the adjoint representation of $SU(N_{c})$ we make use of the corresponding formulae (\ref{ap:C5}) and (\ref{ap:C6}) given in Appendix \ref{appendix_C}. If we contract the expressions obtained in this way with $\bigl(T^{\,f}\bigr)^{d\hspace{0.02cm} e}\bigl\langle\hspace{0.01cm}\mathcal{Q}^{\hspace{0.03cm}f}\hspace{0.03cm}\bigr\rangle$, as it takes place in the original equation (\ref{eq:9q}), then we get, instead of (\ref{eq:9r}),
\[  
\bigl(T^{\,f}\bigr)^{d\hspace{0.02cm} e}\bigl\langle\hspace{0.01cm}\mathcal{Q}^{\hspace{0.03cm}f}\hspace{0.03cm}\bigr\rangle\,
{\rm tr}\hspace{0.03cm}\bigl(T^{\,d}\hspace{0.03cm}T^{\,s}{\mathcal N}_{{\bf k}}\hspace{0.03cm} T^{\,e} {\mathcal N}_{{\bf k}_{1}} \hspace{0.03cm}\bigr)
=
-\hspace{0.03cm}\frac{1}{2}\,\bigl\langle\hspace{0.01cm}\mathcal{Q}^{\hspace{0.03cm}s}\hspace{0.03cm}\bigr\rangle\hspace{0.03cm} N^{\hspace{0.02cm} 2}_{c} N^{\hspace{0.03cm}l}_{{\bf k}}\hspace{0.01cm}N^{\hspace{0.03cm}l}_{{\bf k}_{1}}
\]
\[
-\,\biggl\{\frac{1}{2}\,i\,f^{\,c\hspace{0.03cm} s\hspace{0.03cm}f} 
+
\frac{1}{4}\,N_{c}\hspace{0.03cm}\Bigl({\rm tr}\hspace{0.03cm}\bigl(T^{\,f}\hspace{0.02cm} T^{\,s}\hspace{0.02cm} T^{\,c}\hspace{0.03cm}\bigr)
-
{\rm tr}\hspace{0.03cm}\bigl(T^{\,f} D^{\,s} D^{\,c}\hspace{0.03cm}\bigr)\Bigr)\biggr\}
\hspace{0.03cm}
\bigl\langle\hspace{0.01cm}\mathcal{Q}^{\hspace{0.03cm}f}\hspace{0.03cm}\bigr\rangle
\hspace{0.03cm}
\bigl\langle\hspace{0.01cm}\mathcal{Q}^{\hspace{0.03cm}c}\hspace{0.03cm}\bigr\rangle
\hspace{0.04cm}
W^{\hspace{0.03cm}l}_{{\bf k}}\hspace{0.03cm}N^{\hspace{0.03cm}l}_{{\bf k}_{1}}
\]
\[
+\, \bigl(T^{\,f}\bigr)^{d\hspace{0.02cm} e}\hspace{0.03cm}
{\rm tr}\hspace{0.03cm}\bigl(T^{\,d}\hspace{0.02cm} T^{\,s}\hspace{0.02cm} T^{\,c}\hspace{0.02cm} 
T^{\,e}\hspace{0.02cm} T^{\,c^{\prime}}\hspace{0.03cm}\bigr)
\bigl\langle\hspace{0.01cm}\mathcal{Q}^{\hspace{0.03cm}f}\hspace{0.03cm}\bigr\rangle
\hspace{0.03cm}
\bigl\langle\hspace{0.01cm}\mathcal{Q}^{\hspace{0.03cm}c}\hspace{0.03cm}\bigr\rangle
\hspace{0.03cm}
\bigl\langle\hspace{0.01cm}\mathcal{Q}^{\hspace{0.03cm}c^{\prime}}\hspace{0.03cm}\bigr\rangle
\hspace{0.04cm}
W^{\hspace{0.03cm}l}_{{\bf k}}\hspace{0.05cm}W^{\hspace{0.03cm}l}_{{\bf k}_{1}}.
\]
For the third trace on the right-hand side of (\ref{eq:9r}) we have used the symmetry property (\ref{ap:C9}), by virtue of which it turns to zero. Further, from the formulae (\ref{ap:C5}) for third-order traces we have ${\rm tr}\hspace{0.03cm}\bigl(T^{\,f}\hspace{0.02cm} T^{\,s}\hspace{0.02cm} T^{\,c}\hspace{0.03cm}\bigr) \sim {\rm tr}\hspace{0.03cm}\bigl(T^{\,f} D^{\,s} D^{\,c}\hspace{0.03cm}\bigr) \sim f^{\hspace{0.03cm}f\hspace{0.02cm} s\hspace{0.03cm} c}$ and therefore the second term proportional the product $W^{\hspace{0.03cm}l}_{{\bf k}}\hspace{0.04cm}N^{\hspace{0.03cm}l}_{{\bf k}_{1}}$ also turns to zero by virtue of its contraction with the product $\bigl\langle\hspace{0.01cm}\mathcal{Q}^{\hspace{0.03cm}f}\hspace{0.03cm}\bigr\rangle
\bigl\langle\hspace{0.01cm}\mathcal{Q}^{\hspace{0.03cm}c}\hspace{0.03cm}\bigr\rangle$ symmetric on the color indices $f$ and $c$. We end up here with
\begin{equation}
\bigl(T^{\,f}\bigr)^{d\hspace{0.02cm} e}\bigl\langle\hspace{0.01cm}\mathcal{Q}^{\hspace{0.03cm}f}\hspace{0.03cm}\bigr\rangle\,
{\rm tr}\hspace{0.03cm}\bigl(T^{\,d}\hspace{0.03cm}T^{\,s}{\mathcal N}_{{\bf k}}\hspace{0.03cm} T^{\,e} {\mathcal N}_{{\bf k}_{1}} \hspace{0.03cm}\bigr)
=
-\hspace{0.03cm}\frac{1}{2}\,\bigl\langle\hspace{0.01cm}\mathcal{Q}^{\hspace{0.03cm}s}\hspace{0.03cm}\bigr\rangle\hspace{0.03cm} N^{\hspace{0.02cm} 2}_{c} N^{\hspace{0.03cm}l}_{{\bf k}}\hspace{0.01cm}N^{\hspace{0.03cm}l}_{{\bf k}_{1}}
\label{eq:9t}
\end{equation}
\[
+\, \bigl(T^{\,f}\bigr)^{d\hspace{0.02cm} e}\hspace{0.03cm}
{\rm tr}\hspace{0.03cm}\bigl(T^{\,d}\hspace{0.02cm} T^{\,s}\hspace{0.02cm} T^{\,c}\hspace{0.02cm} 
T^{\,e}\hspace{0.02cm} T^{\,c^{\prime}}\hspace{0.03cm}\bigr)
\bigl\langle\hspace{0.01cm}\mathcal{Q}^{\hspace{0.03cm}f}\hspace{0.03cm}\bigr\rangle
\bigl\langle\hspace{0.01cm}\mathcal{Q}^{\hspace{0.03cm}c}\hspace{0.03cm}\bigr\rangle
\bigl\langle\hspace{0.01cm}\mathcal{Q}^{\hspace{0.03cm}c^{\prime}}\hspace{0.03cm}\bigr\rangle
W^{\hspace{0.03cm}l}_{{\bf k}}\hspace{0.03cm}W^{\hspace{0.03cm}l}_{{\bf k}_{1}}.
\]
We just need to determine the contribution with the trace of five generators. This can be done directly using the general formula (\ref{ap:C12}). Here, however, we choose another somewhat simpler way, using the fact that this trace is contracted with the matrix $\bigl(T^{\,f}\bigr)^{d\hspace{0.02cm} e}$.\\
\indent Let us rewrite the contraction as follows:
\[
\bigl(T^{\,f}\bigr)^{d\hspace{0.02cm} e}\hspace{0.03cm}
{\rm tr}\hspace{0.03cm}\bigl(T^{\,d}\hspace{0.02cm} T^{\,s}\hspace{0.02cm} T^{\,c}\hspace{0.02cm} 
T^{\,e}\hspace{0.02cm} T^{\,c^{\prime}}\hspace{0.03cm}\bigr)
=
\bigl(T^{\,f}\bigr)^{d\hspace{0.02cm} e}\hspace{0.03cm}
{\rm tr}\hspace{0.03cm}\bigl(T^{\,s}\hspace{0.02cm} T^{\,c}\hspace{0.02cm} T^{\,e}\hspace{0.02cm} 
T^{\,c^{\prime}}\hspace{0.02cm} T^{\,d}\hspace{0.03cm}\bigr)
\equiv
\bigl(T^{\,f}\bigr)^{d\hspace{0.02cm} e}\hspace{0.03cm}
(T^{\,s}\hspace{0.02cm} T^{\,c}\hspace{0.02cm}\bigr)^{a\hspace{0.02cm}b} 
\bigl(T^{\,e}\hspace{0.02cm} T^{\,c^{\prime}}\hspace{0.02cm} T^{\,d}\hspace{0.03cm}\bigr)^{b\hspace{0.02cm}a}.
\]
Further, we can write
\[
\bigl(T^{\,f}\bigr)^{d\hspace{0.02cm} e}\hspace{0.03cm} \bigl(T^{\,e}\hspace{0.02cm} T^{\,c^{\prime}}\hspace{0.02cm} T^{\,d}\hspace{0.03cm}\bigr)^{b\hspace{0.02cm}a}
=
{\rm tr}\hspace{0.03cm}\bigl(T^{\,f}\hspace{0.02cm} T^{\,b}\hspace{0.02cm} T^{\,c^{\prime}}\hspace{0.02cm} 
T^{\,a}\hspace{0.03cm}\bigr)
\]
\[
=
\Bigl(\delta^{\hspace{0.02cm}f\hspace{0.02cm}b}\hspace{0.03cm}\delta^{\hspace{0.02cm}c^{\prime}a}
+
\delta^{\hspace{0.02cm}f\hspace{0.02cm}a}\hspace{0.03cm}\delta^{\hspace{0.02cm}c^{\prime}b}
+
\frac{1}{4}\,N_{c}\hspace{0.02cm}\Bigl[\bigl\{D^{\hspace{0.03cm}f},D^{\,c^{\prime}}\bigr\}^{b\hspace{0.02cm}a}
-
d^{\hspace{0.02cm}f\hspace{0.02cm}c^{\prime}\lambda}\hspace{0.03cm}\bigl(D^{\hspace{0.03cm}\lambda}\bigr)^{b\hspace{0.02cm}a}
\Bigr]\Bigr).
\]
Here, we have used the formula (\ref{ap:C6}) for the fourth-order trace. Let us contract the obtained expression with $(T^{\,s}\hspace{0.02cm} T^{\,c}\hspace{0.02cm}\bigr)^{a\hspace{0.02cm}b}$. Finally, we get
\begin{equation}
\bigl(T^{\,f}\bigr)^{d\hspace{0.02cm} e}\hspace{0.03cm}
{\rm tr}\hspace{0.03cm}\bigl(T^{\,d}\hspace{0.02cm} T^{\,s}\hspace{0.02cm} T^{\,c}\hspace{0.02cm} 
T^{\,e}\hspace{0.02cm} T^{\,c^{\prime}}\hspace{0.03cm}\bigr)
\label{eq:9y}
\end{equation}
\[
=
\bigl\{T^{c^{\prime}},T^{f}\bigr\}^{s\hspace{0.02cm}c}
+
\frac{1}{4}\,N_{c}\hspace{0.02cm}\Bigl[{\rm tr}\hspace{0.03cm}\bigl(T^{\,s}\hspace{0.02cm} T^{\,c}\hspace{0.02cm}\bigl\{D^{f},D^{c^{\prime}}\bigr\}\bigr)
-\,
d^{\hspace{0.02cm}f\hspace{0.02cm}c^{\prime}\lambda}\hspace{0.03cm}
{\rm tr}\hspace{0.03cm}\bigl(T^{\,s}\hspace{0.02cm} T^{\,c}\hspace{0.02cm}D^{\lambda}\bigr)
\Bigr],
\]
where in the last term we can immediately put ${\rm tr}\hspace{0.03cm}\bigl(T^{\,s}\hspace{0.02cm} T^{\,c}\hspace{0.02cm}D^{\lambda}\bigr) = \frac{1}{2}\hspace{0.03cm}N_{c}\hspace{0.03cm}d^{\hspace{0.03cm}s\hspace{0.03cm}c\hspace{0.03cm}\lambda}$. We write the fourth-order trace on the right-hand side of (\ref{eq:9y}) using the representation (\ref{ap:C7}) and as a result it is equal to 
\[
{\rm tr}\hspace{0.03cm}\bigl(T^{\,s}\hspace{0.02cm} T^{\,c}\hspace{0.02cm}\bigl\{D^{f},D^{c^{\prime}}\bigr\}\bigr)
=
\biggl(\frac{N^{\hspace{0.02cm}2}_{c} - 4}{N^{\hspace{0.02cm}2}_{c}}\biggr)
\bigl(2\hspace{0.03cm}\delta^{\hspace{0.02cm}s\hspace{0.02cm}c}\delta^{\hspace{0.02cm}f\hspace{0.02cm}c^{\prime}}
\!-
\delta^{\hspace{0.02cm}s\hspace{0.02cm}f}\delta^{\hspace{0.02cm}c\hspace{0.02cm}c^{\prime}}
\!-
\delta^{\hspace{0.02cm}s\hspace{0.02cm}c^{\prime}}\delta^{\hspace{0.02cm}c\hspace{0.02cm}f}\bigr)
\]
\[
+\,
\biggl(\frac{N^{\hspace{0.02cm}2}_{c} - 8}{4\hspace{0.02cm}N_{c}}\biggr)
\bigl(2\hspace{0.03cm}d^{\hspace{0.03cm}s\hspace{0.03cm}c\hspace{0.03cm}\lambda}
\hspace{0.03cm}d^{\hspace{0.03cm}f\hspace{0.03cm}c^{\prime}\hspace{0.03cm}\lambda}
\!-
d^{\hspace{0.03cm}s\hspace{0.03cm}f\hspace{0.03cm}\lambda}
\hspace{0.03cm}d^{\hspace{0.03cm}c\hspace{0.03cm}c^{\prime}\hspace{0.03cm}\lambda}
\!-
d^{\hspace{0.03cm}s\hspace{0.03cm}c^{\prime}\hspace{0.03cm}\lambda}
\hspace{0.03cm}d^{\hspace{0.03cm}c\hspace{0.03cm}f\hspace{0.03cm}\lambda}\bigr)
\,+\,
\frac{1}{4}\,N_{c}\hspace{0.03cm}
\bigl(d^{\hspace{0.03cm}s\hspace{0.03cm}c^{\prime}\hspace{0.03cm}\lambda}
\hspace{0.03cm}d^{\hspace{0.03cm}c\hspace{0.03cm}f\hspace{0.03cm}\lambda}
\!+
d^{\hspace{0.03cm}s\hspace{0.03cm}f\hspace{0.03cm}\lambda}
\hspace{0.03cm}d^{\hspace{0.03cm}c\hspace{0.03cm}c^{\prime}\hspace{0.03cm}\lambda}\bigr).
\]
According to (\ref{eq:9t}), the expression (\ref{eq:9y}) must be contracted with $\bigl\langle\hspace{0.01cm}\mathcal{Q}^{\hspace{0.03cm}f}\hspace{0.03cm}\bigr\rangle
\hspace{0.03cm}
\bigl\langle\hspace{0.01cm}\mathcal{Q}^{\hspace{0.03cm}c}\hspace{0.03cm}\bigr\rangle
\hspace{0.03cm}
\bigl\langle\hspace{0.01cm}\mathcal{Q}^{\hspace{0.03cm}c^{\prime}}\hspace{0.03cm}\bigr\rangle$. For the first term on the right-hand side of (\ref{eq:9y}) we have the trivial equality
\[
\bigl\{T^{\,c^{\prime}}\!,T^{\,f}\bigr\}^{s\hspace{0.02cm}c}
\bigl\langle\hspace{0.01cm}\mathcal{Q}^{\hspace{0.03cm}f}\hspace{0.03cm}\bigr\rangle
\hspace{0.03cm}
\bigl\langle\hspace{0.01cm}\mathcal{Q}^{\hspace{0.03cm}c}\hspace{0.03cm}\bigr\rangle
\hspace{0.03cm}
\bigl\langle\hspace{0.01cm}\mathcal{Q}^{\hspace{0.03cm}c^{\prime}}\hspace{0.03cm}\bigr\rangle
= 0.
\]
The contraction with the remaining terms in (\ref{eq:9y}) gives us
\[
\biggl(\frac{1}{8}\,N^{\hspace{0.02cm}2}_{c} - \frac{1}{8}\,N^{\hspace{0.02cm}2}_{c}\biggr) 
d^{\hspace{0.03cm}s\hspace{0.03cm}c\hspace{0.03cm}\lambda}
\hspace{0.03cm}d^{\hspace{0.03cm}f\hspace{0.03cm}c^{\prime}\hspace{0.03cm}\lambda}
\bigl\langle\hspace{0.01cm}\mathcal{Q}^{\hspace{0.03cm}f}\hspace{0.03cm}\bigr\rangle
\hspace{0.03cm}
\bigl\langle\hspace{0.01cm}\mathcal{Q}^{\hspace{0.03cm}c}\hspace{0.03cm}\bigr\rangle
\hspace{0.03cm}
\bigl\langle\hspace{0.01cm}\mathcal{Q}^{\hspace{0.03cm}c^{\prime}}\hspace{0.03cm}\bigr\rangle
+
\frac{1}{4}\,N_{c}\hspace{0.03cm}\biggl(\frac{N^{\hspace{0.02cm}2}_{c} - 4}{N^{\hspace{0.02cm}2}_{c}}\biggr)
(2 - 2)
\bigl\langle\hspace{0.01cm}\mathcal{Q}^{\hspace{0.03cm}s}\hspace{0.03cm}\bigr\rangle
\hspace{0.03cm}
\bigl\langle\hspace{0.01cm}\mathcal{Q}^{\hspace{0.03cm}c}\hspace{0.03cm}\bigr\rangle
\hspace{0.03cm}
\bigl\langle\hspace{0.01cm}\mathcal{Q}^{\hspace{0.03cm}c}\hspace{0.03cm}\bigr\rangle
\equiv 0.
\]
Thus the coefficient before the product $W_{{\bf k}}W_{{\bf k}_{1}}$ in (\ref{eq:9t}) is exactly zero. We independently verify this rather unexpected result for the special case $N_{c} = 3$ by directly computing the trace of the product of five matrices $T^{\,a}$.\\
\indent For the trace ${\rm tr}\hspace{0.03cm}\bigl(T^{\,s}T^{\,e} {\mathcal N}_{{\bf k}_{1}} T^{\,d}{\mathcal N}_{{\bf k}} \hspace{0.03cm}\bigr)$ in the second term in (\ref{eq:9q}) we get similar result. In the end, for the expression in parentheses in the second term in (\ref{eq:9q}), taking into account Sokhotsky's formula (\ref{eq:6y}), we obtain finally
\begin{equation}
\bigl(T^{\,f}\bigr)^{d\hspace{0.02cm} e}\bigl\langle\hspace{0.01cm}\mathcal{Q}^{\hspace{0.03cm}f}\hspace{0.03cm}\bigr\rangle\!
\left(\frac{{\rm tr}\hspace{0.03cm}\bigl(T^{\,s}{\mathcal N}_{{\bf k}}\hspace{0.03cm} T^{\,e} {\mathcal N}_{{\bf k}_{1}} T^{\,d}\hspace{0.03cm}\bigr)}{\Delta\hspace{0.02cm}\omega_{{\bf p},\, {\bf p}, {\bf k}, {\bf k}_{1}} -\, i\hspace{0.03cm}0} 
\,-\,
\frac{{\rm tr}\hspace{0.03cm}\bigl(T^{\,s}T^{\,e} {\mathcal N}_{{\bf k}_{1}} T^{\,d}{\mathcal N}_{{\bf k}} \hspace{0.03cm}\bigr)}{\Delta\hspace{0.02cm}\omega_{{\bf p},\, {\bf p}, {\bf k}, {\bf k}_{1}} +\, i\hspace{0.03cm}0}\, 
\right)
\label{eq:9u}
\end{equation}
\[
=
-\hspace{0.03cm}\frac{1}{2}\,i\hspace{0.03cm}N^{\hspace{0.02cm} 2}_{c} N_{{\bf k}}N_{{\bf k}_{1}}
\bigl\langle\hspace{0.01cm}\mathcal{Q}^{\hspace{0.03cm}s}\hspace{0.03cm}\bigr\rangle\hspace{0.03cm}
(2\pi)\,\delta(\omega^{l}_{{\bf k}} - \omega^{l}_{{\bf k}_{1}} - {\mathbf v}\cdot({\bf k} - {\bf k}_{1})).
\]
\indent Let us now consider the traces in the last contribution on the right-hand side of the original equation (\ref{eq:9q}). Here in the last trace ${\rm tr}\hspace{0.03cm}\bigl(T^{\,e} \hspace{0.03cm}T^{\,d}{\mathcal N}_{{\bf k}}\hspace{0.03cm}T^{\,s}\hspace{0.02cm}\bigr)$ in the expression in parentheses, we see a certain asymmetry in the arrangement of the matrix $T^{\,s}$ under the sign of the trace in comparison to the other similar traces. Therefore, as a first step, by taking into account the decomposition (\ref{eq:8r}), we transform this trace as follows:
\[
{\rm tr}\hspace{0.03cm}\bigl(T^{\,e} \hspace{0.03cm}T^{\,d}{\mathcal N}_{{\bf k}}\hspace{0.03cm}T^{\,s} \hspace{0.02cm}\bigr)
=
{\rm tr}\hspace{0.03cm}\bigl(T^{\,e} \hspace{0.03cm}T^{\,d}\hspace{0.03cm}T^{\,s}\hspace{0.03cm}{\mathcal N}_{{\bf k}} \hspace{0.02cm}\bigr)  
+
{\rm tr}\hspace{0.03cm}\bigl(T^{\,e} \hspace{0.03cm}T^{\,d}\hspace{0.03cm}\bigl[\hspace{0.03cm}{\mathcal N}_{{\bf k}},\hspace{0.03cm}T^{\,s}\bigr] \hspace{0.02cm}\bigr)  
\]
\[
=
{\rm tr}\hspace{0.03cm}\bigl(T^{\,e} \hspace{0.03cm}T^{\,d}\hspace{0.03cm}T^{\,s}\hspace{0.03cm}{\mathcal N}_{{\bf k}} \hspace{0.02cm}\bigr)  
+
i\hspace{0.03cm}f^{\hspace{0.03cm}c\hspace{0.02cm} s\hspace{0.03cm} \lambda}
{\rm tr}\hspace{0.03cm}\bigl(T^{\,e} \hspace{0.03cm}T^{\,d}\hspace{0.03cm}T^{\,\lambda}\hspace{0.03cm}\bigr) 
\bigl\langle\hspace{0.01cm}\mathcal{Q}^{\hspace{0.03cm}c}\hspace{0.03cm}\bigr\rangle\hspace{0.03cm}
W^{\,l}_{{\bf k}}
=
{\rm tr}\hspace{0.03cm}\bigl(T^{\,e} \hspace{0.03cm}T^{\,d}\hspace{0.03cm}T^{\,s}\hspace{0.03cm}{\mathcal N}_{{\bf k}} \hspace{0.02cm}\bigr)  
-
\frac{1}{2}\,f^{\hspace{0.03cm}c\hspace{0.02cm} s\hspace{0.03cm} \lambda}
f^{\hspace{0.03cm}e\hspace{0.03cm}d\hspace{0.03cm}\lambda}\hspace{0.02cm} N_{c}
\bigl\langle\hspace{0.01cm}\mathcal{Q}^{\hspace{0.03cm}c}\hspace{0.03cm}\bigr\rangle\hspace{0.03cm}
W^{\,l}_{{\bf k}}.
\]
The last term here contains the antisymmetric structural constant $f^{\hspace{0.03cm}e\hspace{0.03cm} d\hspace{0.03cm} \lambda}$ and so it can be discarded by virtue of the relation (\ref{eq:8i}). Given this fact and using Sokhotsky's formula (\ref{eq:6y}), the last line in equation (\ref{eq:9q}) can be rewritten as follows:
\begin{equation}
\frac{{\rm tr}\hspace{0.03cm}\bigl(T^{\,d} \hspace{0.03cm}T^{\,s} \hspace{0.03cm}T^{\,e} {\mathcal N}_{{\bf k}_{1}} \hspace{0.03cm}\bigr) 
- 
{\rm tr}\hspace{0.03cm}\bigl(T^{\,e}  \hspace{0.03cm}T^{\,d}  \hspace{0.03cm}T^{\,s}{\mathcal N}_{{\bf k}}\hspace{0.03cm}\bigr)}
{\Delta\hspace{0.02cm}\omega_{{\bf p},\, {\bf p}, {\bf k}, {\bf k}_{1}} -\, i\hspace{0.03cm}0} 
\,-\,
\frac{{\rm tr}\hspace{0.03cm}\bigl(T^{\,d} \hspace{0.03cm}T^{\,s} \hspace{0.03cm}T^{\,e} {\mathcal N}_{{\bf k}_{1}} \hspace{0.03cm}\bigr) 
- 
{\rm tr}\hspace{0.03cm}\bigl(T^{\,e} \hspace{0.03cm}T^{\,d}\hspace{0.03cm}T^{\,s}\hspace{0.03cm}{\mathcal N}_{{\bf k}} \hspace{0.02cm}\bigr)}
{\Delta\hspace{0.02cm}\omega_{{\bf p},\, {\bf p}, {\bf k}, {\bf k}_{1}} +\, i\hspace{0.03cm}0}
\label{eq:9i}
\end{equation}
\[
=
i\hspace{0.04cm}\bigl[\hspace{0.03cm}{\rm tr}\hspace{0.03cm}\bigl(T^{\,d} \hspace{0.03cm}T^{\,s} \hspace{0.03cm}T^{\,e} {\mathcal N}_{{\bf k}_{1}} \hspace{0.03cm}\bigr) 
- 
{\rm tr}\hspace{0.03cm}\bigl(T^{\,e}  \hspace{0.03cm}T^{\,d}  \hspace{0.03cm}T^{\,s}{\mathcal N}_{{\bf k}}\hspace{0.03cm}\bigr)\bigr]\hspace{0.03cm}
(2\pi)\hspace{0.03cm}\delta(\omega^{l}_{{\bf k}} - \omega^{l}_{{\bf k}_{1}} - {\mathbf v}\cdot({\bf k} - {\bf k}_{1})).
\]
Then, considering the color decomposition (\ref{eq:8r}), we transform the second trace on the right-hand side (\ref{eq:9i}) as follows:
\[
{\rm tr}\hspace{0.03cm}\bigl(T^{\,e} \hspace{0.03cm}T^{\,d} \hspace{0.03cm}T^{\,s} {\mathcal N}_{{\bf k}} \hspace{0.03cm}\bigr) 
\equiv
{\rm tr}\hspace{0.03cm}\bigl(T^{\,e} \hspace{0.03cm}T^{\,s} \hspace{0.03cm}T^{\,d} {\mathcal N}_{{\bf k}} \hspace{0.03cm}\bigr) 
+
{\rm tr}\hspace{0.03cm}\bigl(T^{\,e} \hspace{0.03cm}\bigl[\hspace{0.03cm}T^{\,d}, \hspace{0.03cm}T^{\,s}\bigr] {\mathcal N}_{{\bf k}} \hspace{0.03cm}\bigr) 
\]
\[
=
{\rm tr}\hspace{0.03cm}\bigl(T^{\,e} \hspace{0.03cm}T^{\,s} \hspace{0.03cm}T^{\,d} {\mathcal N}_{{\bf k}} \hspace{0.03cm}\bigr) 
+
\frac{1}{2}\,
N_{c}\hspace{0.03cm}\bigl(T^{\,d} \hspace{0.03cm}T^{\,e}\bigr)^{s\hspace{0.03cm}c}
\bigl\langle\hspace{0.01cm}\mathcal{Q}^{\hspace{0.03cm}c}\hspace{0.03cm}\bigr\rangle\hspace{0.03cm}
W^{\,l}_{{\bf k}}
\sim
{\rm tr}\hspace{0.03cm}\bigl(T^{\,e} \hspace{0.03cm}T^{\,s} \hspace{0.03cm}T^{\,d} {\mathcal N}_{{\bf k}} \hspace{0.03cm}\bigr) 
+
\frac{1}{4}\,N_{c}\hspace{0.03cm}\bigl\{T^{\,d}, \hspace{0.03cm}T^{\,e}\bigr\}^{s\hspace{0.03cm}c}\hspace{0.01cm}
\bigl\langle\hspace{0.01cm}\mathcal{Q}^{\hspace{0.03cm}c}\hspace{0.03cm}\bigr\rangle\hspace{0.03cm}
W^{\,l}_{{\bf k}}.
\]
In the final step here, we have taken into account that in the equation (\ref{eq:9q}) this trace is contracted with the factor $Z^{d\hspace{0.03cm}e}$ symmetric in indices $d$ and $e$ as defined by (\ref{eq:8u}). The advantage of choosing a trace with this arrangement of the matrices $T^{\,d}$ and $T^{\,e}$ is the automatic symmetry of the fourth-order traces (see below) over the permutation of the indices $d$ and $e$, as is the case for the factor $Z^{\hspace{0.03cm}d\hspace{0.03cm}e}$. Taking into account the relation above, the difference of traces on the right-hand side of (\ref{eq:9i}) takes then the following form
\begin{equation}
{\rm tr}\hspace{0.03cm}\bigl(T^{\,d} \hspace{0.03cm}T^{\,s} \hspace{0.03cm}T^{\,e} {\mathcal N}_{{\bf k}_{1}} \hspace{0.03cm}\bigr) 
- 
{\rm tr}\hspace{0.03cm}\bigl(T^{\,e}  \hspace{0.03cm}T^{\,d}  \hspace{0.03cm}T^{\,s}{\mathcal N}_{{\bf k}}\hspace{0.03cm}\bigr)
\label{eq:9o}
\end{equation}
\[
=
{\rm tr}\hspace{0.03cm}\bigl(T^{\,d} \hspace{0.03cm}T^{\,s} \hspace{0.03cm}T^{\,e} {\mathcal N}_{{\bf k}_{1}} \hspace{0.03cm}\bigr) 
- 
{\rm tr}\hspace{0.03cm}\bigl(T^{\,e} \hspace{0.03cm}T^{\,s} \hspace{0.03cm}T^{\,d} {\mathcal N}_{{\bf k}} \hspace{0.03cm}\bigr) 
-
\frac{1}{4}\,N_{c}\hspace{0.03cm}\bigl\{T^{\,d}, \hspace{0.03cm}T^{\,e}\bigr\}^{s\hspace{0.03cm}c}\hspace{0.01cm}
\bigl\langle\hspace{0.01cm}\mathcal{Q}^{\hspace{0.03cm}c}\hspace{0.03cm}\bigr\rangle\hspace{0.03cm}
W^{\,l}_{{\bf k}},
\]
where, in turn, taking into account the decomposition (\ref{eq:8r}) and the formulae for the traces of the third and fourth orders (\ref{ap:C5}) and (\ref{ap:C6}), we have
\begin{equation}
{\rm tr}\hspace{0.03cm}\bigl(T^{\,d} \hspace{0.03cm}T^{\,s} \hspace{0.03cm}T^{\,e} {\mathcal N}_{{\bf k}_{1}} \hspace{0.03cm}\bigr) 
- 
{\rm tr}\hspace{0.03cm}\bigl(T^{\,e} \hspace{0.03cm}T^{\,s} \hspace{0.03cm}T^{\,d} {\mathcal N}_{{\bf k}} \hspace{0.03cm}\bigr) 
\label{eq:9p}
\end{equation}
\[
=
{\rm tr}\hspace{0.03cm}\bigl(T^{\,d} \hspace{0.03cm}T^{\,s} \hspace{0.03cm}T^{\,e} \hspace{0.03cm}\bigr)
N^{\hspace{0.03cm}l}_{{\bf k}_{1}} 
- 
{\rm tr}\hspace{0.03cm}\bigl(T^{\,e}  \hspace{0.03cm}T^{\,s}  \hspace{0.03cm}T^{\,d}\hspace{0.03cm}\bigr)
N^{\hspace{0.03cm}l}_{{\bf k}}
+
{\rm tr}\hspace{0.03cm}\bigl(T^{\,d} \hspace{0.03cm}T^{\,s} \hspace{0.03cm}T^{\,e} \hspace{0.03cm}T^{\,c}\bigr)
\bigl\langle\hspace{0.01cm}\mathcal{Q}^{\hspace{0.03cm}c}\hspace{0.03cm}\bigr\rangle\hspace{0.03cm}
W^{\,l}_{{\bf k}_{1}}
-
{\rm tr}\hspace{0.03cm}\bigl(T^{\,e} \hspace{0.03cm}T^{\,s} \hspace{0.03cm}T^{\,d} \hspace{0.03cm}T^{\,c}\bigr)
\bigl\langle\hspace{0.01cm}\mathcal{Q}^{\hspace{0.03cm}c}\hspace{0.03cm}\bigr\rangle\hspace{0.03cm}
W^{\,l}_{{\bf k}}
\]
\[
=
\frac{i}{2}\,N_{c}\hspace{0.03cm} f^{\hspace{0.04cm}e\hspace{0.02cm} d\hspace{0.03cm} s}\hspace{0.03cm}
\bigl(N^{\,l}_{{\bf k}_{1}} + N^{\,l}_{{\bf k}}\hspace{0.03cm}\bigr)
-
\Bigl(\delta^{\hspace{0.02cm}e\hspace{0.02cm}s}\hspace{0.03cm}\delta^{\hspace{0.02cm}d\hspace{0.03cm}c}
+
\delta^{\hspace{0.03cm}e\hspace{0.02cm}c}\hspace{0.03cm}\delta^{\hspace{0.03cm}d\hspace{0.03cm}s}
+
\frac{1}{4}\,N_{c}\hspace{0.02cm}\Bigl[\bigl\{D^{\,e},D^{\,d}\bigr\}^{s\hspace{0.02cm}c}
-
d^{\hspace{0.03cm}e\hspace{0.02cm}d\lambda}\hspace{0.03cm}\bigl(D^{\,\lambda}\bigr)^{s\hspace{0.02cm}c}
\Bigr]\Bigr)
\bigl\langle\hspace{0.01cm}\mathcal{Q}^{\hspace{0.03cm}c}\hspace{0.03cm}\bigr\rangle\hspace{0.03cm}
\bigl(W^{\,l}_{{\bf k}} - W^{\,l}_{{\bf k}_{1}}\bigr).
\]
Here, the first (imaginary) term on the right-hand side containing the sum of the colorless part of the plasmon number density $N^{\,l}_{{\bf k}}$ turns to zero when contracted with the factor $Z^{\hspace{0.03cm}d\hspace{0.03cm}e}$. The second term when using a different representation of the fourth-order trace of the matrices $T^{\,a}$, Eq.\,(\ref{ap:C11}), can be represented in a slightly different form, simpler for further transformations
\[
-\Bigl(\delta^{\hspace{0.02cm}e\hspace{0.02cm}d}\hspace{0.03cm}\delta^{\hspace{0.02cm}s\hspace{0.03cm}c}
+
\frac{1}{2}\,\bigl(
\delta^{\hspace{0.02cm}e\hspace{0.02cm}s}\hspace{0.03cm}\delta^{\hspace{0.02cm}d\hspace{0.03cm}c}
+
\delta^{\hspace{0.03cm}e\hspace{0.02cm}c}\hspace{0.03cm}\delta^{\hspace{0.03cm}d\hspace{0.03cm}s}
\bigr)
-
\frac{1}{4}\,N_{c}\hspace{0.02cm}\Bigl[\bigl\{T^{\,e},T^{\,d}\bigr\}^{s\hspace{0.02cm}c}
-
d^{\hspace{0.03cm}e\hspace{0.02cm}d\lambda}\hspace{0.03cm}\bigl(D^{\,\lambda}\bigr)^{s\hspace{0.02cm}c}
\Bigr]\Bigr)
\bigl\langle\hspace{0.01cm}\mathcal{Q}^{\hspace{0.03cm}c}\hspace{0.03cm}\bigr\rangle\hspace{0.03cm}
\bigl(W^{\,l}_{{\bf k}} - W^{\,l}_{{\bf k}_{1}}\bigr).
\]
\indent In view of all the expressions (\ref{eq:9w}), (\ref{eq:9e}), (\ref{eq:9u}), (\ref{eq:9i}), (\ref{eq:9o}) and (\ref{eq:9p}) obtained above, the kinetic equation (\ref{eq:9q}) for the color part $W^{\,l}_{{\bf k}}$ of the plasmon number density takes the following form:
\begin{equation}
N_{c}\hspace{0.03cm}\frac{\partial\hspace{0.03cm}  \bigl(
\bigl\langle\hspace{0.01cm}\mathcal{Q}^{\hspace{0.03cm}s}\hspace{0.03cm}\bigr\rangle
\hspace{0.02cm} W^{\hspace{0.03cm}l}_{\bf k}\hspace{0.03cm} \bigr)}{\!\!\partial\hspace{0.03cm} t}
=
2\hspace{0.02cm}N_{c}\hspace{0.03cm}\biggl(\int\!n_{\bf p}\,{\bf p}^{2}\hspace{0.03cm}
d\hspace{0.02cm}|{\bf p}|\biggr)\!
\int\!d\hspace{0.04cm}\Omega_{\hspace{0.035cm}{\bf v}}
\, 
{\rm Im}\hspace{0.03cm}\mathscr{T}^{\,(2,\hspace{0.03cm}{\mathcal A})}_{{\bf k},\,{\bf k}}({\bf v})
\hspace{0.03cm}N^{\hspace{0.03cm}l}_{\bf k}\,
\bigl\langle\hspace{0.01cm}\mathcal{Q}^{\hspace{0.03cm}s}
\hspace{0.03cm}\bigr\rangle
\label{eq:9a}
\end{equation}
\[
+\,
\frac{1}{2}\,N^{\hspace{0.02cm}2}_{c}\hspace{0.03cm}\biggl(\int\!n_{\bf p}\,{\bf p}^{2}\hspace{0.03cm}
d\hspace{0.02cm}|{\bf p}|\biggr)\!
\int\!d\hspace{0.04cm}\Omega_{\hspace{0.035cm}{\bf v}}\!
\int\!d{\bf k}_{1}\,
\bigl|\hspace{0.02cm}\mathscr{T}^{\,(2,\hspace{0.03cm}{\mathcal A})}_{{\bf k},\,{\bf k}_{1}}({\bf v})\hspace{0.02cm}\bigr|^{\hspace{0.03cm}2}\,
N^{\,l}_{{\bf k}}\hspace{0.02cm}N^{\,l}_{{\bf k}_{1}}
\bigl\langle\hspace{0.01cm}\mathcal{Q}^{\hspace{0.03cm}s}\hspace{0.03cm}\bigr\rangle\hspace{0.03cm}
(2\pi)\,\delta(\omega^{l}_{{\bf k}} - \omega^{l}_{{\bf k}_{1}} - {\mathbf v}\cdot({\bf k} - {\bf k}_{1}))
\vspace{0.1cm}
\]
\[
-\,
\biggl(\int\!n^{2}_{\bf p}\,{\bf p}^{2}\hspace{0.03cm}
d\hspace{0.02cm}|{\bf p}|\biggr)\!
\int\!d\hspace{0.04cm}\Omega_{\hspace{0.035cm}{\bf v}}\!
\int\!d{\bf k}_{1}\,
\bigl|\hspace{0.02cm}\mathscr{T}^{\,(2,\hspace{0.03cm}{\mathcal A})}_{{\bf k},\,{\bf k}_{1}}({\bf v})\hspace{0.02cm}\bigr|^{\hspace{0.03cm}2}\,
\bigl[
(t^{\,d}\hspace{0.03cm})^{\hspace{0.03cm}j_{1}\hspace{0.02cm}i_{2}} 
\hspace{0.03cm}
\bigl\langle\hspace{0.03cm}\theta^{\hspace{0.03cm}\ast\ \!\!i_{1}}\hspace{0.03cm} \theta^{\phantom{\ast}\!\!i_{2}}\hspace{0.03cm}\bigr\rangle\hspace{0.03cm} 
(t^{\,e}\hspace{0.03cm})^{\hspace{0.03cm}i_{1}\hspace{0.02cm}j_{2}} 
\hspace{0.03cm}
\bigl\langle\hspace{0.03cm}\theta^{\hspace{0.03cm}\ast\ \!\!j_{1}}\hspace{0.03cm} \theta^{\phantom{\ast}\!\!j_{2}}\hspace{0.03cm}\bigr\rangle\bigr]
\]
\[
\times\biggl[
\Bigl(\delta^{\hspace{0.02cm}e\hspace{0.02cm}d}\hspace{0.03cm}\delta^{\hspace{0.02cm}s\hspace{0.03cm}c}
+
\frac{1}{2}\,\bigl(
\delta^{\hspace{0.02cm}e\hspace{0.02cm}s}\hspace{0.03cm}\delta^{\hspace{0.02cm}d\hspace{0.03cm}c}
+
\delta^{\hspace{0.03cm}e\hspace{0.02cm}c}\hspace{0.03cm}\delta^{\hspace{0.03cm}d\hspace{0.03cm}s}
\bigr)
-
\frac{1}{4}\,N_{c}\hspace{0.02cm}\Bigl[\bigl\{T^{\,e},T^{\,d}\bigr\}^{s\hspace{0.02cm}c}
-\,
d^{\hspace{0.03cm}e\hspace{0.02cm}d\hspace{0.02cm}\lambda}\hspace{0.02 cm}
\bigl(D^{\,\lambda}\bigr)^{s\hspace{0.02cm}c}
\Bigr]\Bigr)
\bigl\langle\hspace{0.01cm}\mathcal{Q}^{\hspace{0.03cm}c}\hspace{0.03cm}\bigr\rangle\hspace{0.03cm}
\bigl(W^{\,l}_{{\bf k}} - W^{\,l}_{{\bf k}_{1}}\bigr)
\]
\[
+\,
\frac{1}{4}\,N_{c}\hspace{0.03cm}\bigl\{T^{\,d}, \hspace{0.03cm}T^{\,e}\bigr\}^{s\hspace{0.03cm}c}\hspace{0.01cm}
\bigl\langle\hspace{0.01cm}\mathcal{Q}^{\hspace{0.03cm}c}\hspace{0.03cm}
\bigr\rangle\hspace{0.03cm}
W^{\,l}_{{\bf k}}\hspace{0.03cm}\biggr]
\hspace{0.03cm}
(2\pi)\hspace{0.03cm}\delta(\omega^{l}_{{\bf k}} - \omega^{l}_{{\bf k}_{1}} - {\mathbf v}\cdot({\bf k} - {\bf k}_{1})).
\]
Just below we will show that the third term on the right-hand side of (\ref{eq:9a}) containing the color structure $Z^{\,d\hspace{0.02cm}e}$, Eq.\,(\ref{eq:8y}), cannot be reduced to a function only of the averaged classical colorless and color charges $\bigl\langle\hspace{0.01cm}{\mathcal Q}\hspace{0.03cm}\bigr\rangle$ and $\bigl\langle\hspace{0.01cm}\mathcal{Q}^{\hspace{0.03cm}s}\hspace{0.03cm}\bigr\rangle$ for an arbitrary value  $N_{c}$. In addition, there is evident asymmetry with respect to the functions $W^{\,l}_{{\bf k}}$ and $W^{\,l}_{{\bf k}_{1}}$.\\
\indent We consider separately the terms in braces in the next to the last line in (\ref{eq:9a}), when contracting them with $Z^{\,d\hspace{0.02cm}e}$.
For the first term, allowing for (\ref{eq:8a}) we have
\[
\delta^{\hspace{0.02cm}e\hspace{0.02cm}d}\hspace{0.03cm}
\delta^{\hspace{0.02cm}s\hspace{0.03cm}c}
Z^{\,d\hspace{0.02cm}e} 
=
\delta^{\hspace{0.02cm}s\hspace{0.03cm}c}\hspace{0.02cm}
\biggl[
\biggl(\frac{N^{\hspace{0.02cm}2}_{c} - 4}{2\hspace{0.02cm}N^{\hspace{0.02cm}2}_{c}}\biggr)\hspace{0.02cm} \bigl\langle\hspace{0.01cm}{\mathcal Q}\hspace{0.03cm}\bigr\rangle^{\!2}	
-
\frac{1}{N_{c}}\,\bigl\langle\hspace{0.01cm}\mathcal{Q}^{\hspace{0.03cm}e}
\hspace{0.03cm}\bigr\rangle\bigl\langle\hspace{0.01cm}
\mathcal{Q}^{\hspace{0.03cm}e}\hspace{0.03cm}\bigr\rangle
\biggr].
\]
Then using the relation (\ref{ap:B8}) from Appendix \ref{appendix_B}, we find for the third term
\begin{equation}
\bigl\{T^{\,e},T^{\,d}\bigr\}^{s\hspace{0.02cm}c}
\hspace{0.02cm}  
Z^{\,d\hspace{0.02cm}e} 
=
\frac{1}{N_{c}}\,\delta^{\hspace{0.02cm}s\hspace{0.03cm}c}
\bigl\langle\hspace{0.01cm}{\mathcal Q}\hspace{0.03cm}\bigr\rangle^{\!2}
+
\bigl(D^{\,\lambda}\bigr)^{s\hspace{0.02cm}c}
\bigl\langle\hspace{0.01cm}
\mathcal{Q}^{\hspace{0.03cm}\lambda}\hspace{0.03cm}\bigr\rangle
\bigl\langle\hspace{0.01cm}{\mathcal Q}\hspace{0.03cm}\bigr\rangle
-
2\,\bigl\langle\hspace{0.01cm}
\mathcal{Q}^{\hspace{0.03cm}s}\hspace{0.03cm}
\bigr\rangle\bigl\langle\hspace{0.01cm}
\mathcal{Q}^{\hspace{0.03cm}c}\hspace{0.03cm}\bigr\rangle.
\label{eq:9s}
\end{equation}
In the end, for the last term, by virtue of the relation (\ref{ap:B6_b}), we have
\[
d^{\,e\hspace{0.03cm}d\hspace{0.03cm}\lambda}\hspace{0.03cm}
\bigl(D^{\,\lambda}\bigr)^{s\hspace{0.02cm}c}
Z^{\,d\hspace{0.02cm}e} 
=
\biggl(\frac{N^{\hspace{0.02cm}2}_{c} - 4}{N^{\hspace{0.02cm}2}_{c}}\biggr)
\bigl(D^{\,\lambda}\bigr)^{s\hspace{0.02cm}c}
\bigl\langle\hspace{0.01cm}
\mathcal{Q}^{\hspace{0.03cm}\lambda}\hspace{0.03cm}\bigr\rangle
\bigl\langle\hspace{0.01cm}{\mathcal Q}\hspace{0.03cm}\bigr\rangle
-
\frac{2}{N_{c}}\,
\bigl(D^{\,\lambda}\bigr)^{s\hspace{0.02cm}c}
d^{\,e\hspace{0.02cm}d\hspace{0.02cm}\lambda}
\bigl\langle\hspace{0.01cm}
\mathcal{Q}^{\hspace{0.03cm}e}\hspace{0.03cm}
\bigr\rangle\bigl\langle\hspace{0.01cm}
\mathcal{Q}^{\hspace{0.03cm}d}\hspace{0.03cm}\bigr\rangle.
\] 
Collecting all the calculations above, we finally obtain for the expression in braces in (\ref{eq:9a})
\begin{equation}
\Bigl\{\delta^{\hspace{0.02cm}e\hspace{0.02cm}d}\hspace{0.03cm}\delta^{\hspace{0.02cm}s\hspace{0.03cm}c}
+
\frac{1}{2}\,\bigl(
\delta^{\hspace{0.02cm}e\hspace{0.02cm}s}\hspace{0.03cm}\delta^{\hspace{0.02cm}d\hspace{0.03cm}c}
+
\delta^{\hspace{0.03cm}e\hspace{0.02cm}c}\hspace{0.03cm}\delta^{\hspace{0.03cm}d\hspace{0.03cm}s}
\bigr)
-
\frac{1}{8}\,N_{c}\hspace{0.02cm}\Bigl[\bigl\{T^{\,e},T^{\,d}\bigr\}^{s\hspace{0.02cm}c}
-
2\hspace{0.03cm}d^{\hspace{0.03cm}e\hspace{0.02cm}d\hspace{0.02cm}\lambda}
\hspace{0.03cm}\bigl(D^{\,\lambda}\bigr)^{s\hspace{0.02cm}c}
\Bigr]\Bigr\}
\vspace{0.1cm}
\label{eq:9d}
\end{equation}
\[
\times\,\bigl[
(t^{\,d}\hspace{0.03cm})^{\hspace{0.03cm}j_{1}\hspace{0.02cm}i_{2}} 
\hspace{0.03cm}
\bigl\langle\hspace{0.03cm}\theta^{\hspace{0.03cm}\ast\ \!\!i_{1}}\hspace{0.03cm} \theta^{\phantom{\ast}\!\!i_{2}}\hspace{0.03cm}\bigr\rangle\hspace{0.03cm} 
(t^{\,e}\hspace{0.03cm})^{\hspace{0.03cm}i_{1}\hspace{0.02cm}j_{2}} 
\hspace{0.03cm}
\bigl\langle\hspace{0.03cm}\theta^{\hspace{0.03cm}\ast\ \!\!j_{1}}\hspace{0.03cm} \theta^{\phantom{\ast}\!\!j_{2}}\hspace{0.03cm}\bigr\rangle\bigr]
\vspace{0.2cm}
\]
\[
=
\delta^{\hspace{0.02cm}s\hspace{0.03cm}c}\hspace{0.02cm}
\biggl\{
\biggl[
\biggl(\frac{N^{\hspace{0.02cm}2}_{c} - 4}{2\hspace{0.02cm}N^{\hspace{0.02cm}2}_{c}}\biggr)
-
\frac{1}{8}\hspace{0.03cm}\biggr]\hspace{0.02cm} \bigl\langle\hspace{0.01cm}{\mathcal Q}\hspace{0.03cm}\bigr\rangle^{\!2}	
-
\frac{1}{N_{c}}\,\bigl\langle\hspace{0.01cm}\mathcal{Q}^{\hspace{0.03cm}e}
\hspace{0.03cm}\bigr\rangle\bigl\langle\hspace{0.01cm}
\mathcal{Q}^{\hspace{0.03cm}e}\hspace{0.03cm}\bigr\rangle
\biggr\}
+
\biggl(\frac{N^{\hspace{0.02cm}2}_{c} - 8} {8\hspace{0.02cm}N_{c}}\biggr)
\hspace{0.03cm}
\bigl(D^{\,\lambda}\bigr)^{s\hspace{0.02cm}c}
\bigl\langle\hspace{0.01cm}
\mathcal{Q}^{\hspace{0.03cm}\lambda}\hspace{0.03cm}\bigr\rangle
\bigl\langle\hspace{0.01cm}{\mathcal Q}\hspace{0.03cm}\bigr\rangle
+
\frac{1}{4}\,N_{c}\hspace{0.03cm}
\bigl\langle\hspace{0.01cm}
\mathcal{Q}^{\,s}\hspace{0.03cm}
\bigr\rangle\bigl\langle\hspace{0.01cm}
\mathcal{Q}^{\,c}\hspace{0.03cm}\bigr\rangle
\vspace{0.1cm}
\]
\[
-\,
\frac{1}{2}\,
\bigl(D^{\,\lambda}\bigr)^{s\hspace{0.02cm}c}
d^{\,e\hspace{0.02cm}d\hspace{0.02cm}\lambda}
\bigl\langle\hspace{0.01cm}
\mathcal{Q}^{\hspace{0.03cm}e}\hspace{0.03cm}
\bigr\rangle\bigl\langle\hspace{0.01cm}
\mathcal{Q}^{\hspace{0.03cm}d}\hspace{0.03cm}\bigr\rangle
+
\frac{1}{2}\,
	\Bigl[(t^{\,s}\hspace{0.02cm})^{i_{1}\hspace{0.03cm}j_{2}}
	(t^{\,c}\hspace{0.03cm})^{j_{1}\hspace{0.03cm}i_{2}}
	+
	(t^{\,c}\hspace{0.02cm})^{\hspace{0.03cm}i_{1}\hspace{0.03cm}j_{2}}
	(t^{\,s}\hspace{0.03cm})^{\hspace{0.03cm}j_{1}\hspace{0.03cm}i_{2}}\Bigr]
\bigl\langle\hspace{0.03cm}\theta^{\hspace{0.03cm}\ast\ \!\!i_{1}}\hspace{0.03cm} \theta^{\phantom{\ast}\!\!i_{2}}\hspace{0.03cm}\bigr\rangle\hspace{0.03cm} 
\bigl\langle\hspace{0.03cm}\theta^{\hspace{0.03cm}\ast\ \!\!j_{1}}\hspace{0.03cm} \theta^{\phantom{\ast}\!\!j_{2}}\hspace{0.03cm}\bigr\rangle.
\]
We see that here there remains only one ``twisted'' term associated with the second color structure in round brackets (\ref{eq:9a}), namely with
\[
\frac{1}{2}\,\bigl(
\delta^{\hspace{0.02cm}e\hspace{0.02cm}s}\hspace{0.03cm}\delta^{\hspace{0.02cm}d\hspace{0.03cm}c}
+
\delta^{\hspace{0.03cm}e\hspace{0.02cm}c}\hspace{0.03cm}\delta^{\hspace{0.03cm}d\hspace{0.03cm}s}
\bigr).
\]
It generally does not allow to reduce the expression (\ref{eq:9d}) to a combination of the colorless $\bigl\langle\hspace{0.01cm}{\mathcal Q}\hspace{0.03cm}\bigr\rangle$ and color $\bigl\langle\hspace{0.01cm}
\mathcal{Q}^{\,s}\hspace{0.03cm}\bigr\rangle$ charges. This can be done only for the special case $N_{c} = 3$. Here we can use the relation (\ref{ap:B9}) for the summand in the last line (\ref{eq:9d}), which gives us
\[
\Bigl[(t^{\,s}\hspace{0.02cm})^{i_{1}\hspace{0.03cm}j_{2}}
(t^{\,c}\hspace{0.03cm})^{j_{1}\hspace{0.03cm}i_{2}}
+
(t^{\,c}\hspace{0.02cm})^{\hspace{0.03cm}i_{1}\hspace{0.03cm}j_{2}}
(t^{\,s}\hspace{0.03cm})^{\hspace{0.03cm}j_{1}\hspace{0.03cm}i_{2}}\Bigr]
\bigl\langle\hspace{0.03cm}\theta^{\hspace{0.03cm}\ast\ \!\!i_{1}}\hspace{0.03cm}
\theta^{\phantom{\ast}\!\!i_{2}}\hspace{0.03cm}\bigr\rangle\hspace{0.03cm} 
\bigl\langle\hspace{0.03cm}\theta^{\hspace{0.03cm}\ast\ \!\!j_{1}}\hspace{0.03cm} \theta^{\phantom{\ast}\!\!j_{2}}\hspace{0.03cm}\bigr\rangle
=
2\hspace{0.03cm}
\bigl\langle\hspace{0.01cm}
\mathcal{Q}^{\,s}\hspace{0.03cm}
\bigr\rangle\bigl\langle\hspace{0.01cm}
\mathcal{Q}^{\,c}\hspace{0.03cm}\bigr\rangle
\]
\[
+\,
\delta^{\hspace{0.03cm}s\hspace{0.03cm}c}\hspace{0.04cm}
\biggl\{\frac{1}{9}\,
\bigl\langle\hspace{0.01cm}{\mathcal Q}\hspace{0.03cm}\bigr\rangle^{\!2}	
-
\frac{1}{3}\,
\langle\hspace{0.01cm}\mathcal{Q}^{\hspace{0.03cm}e}
\hspace{0.03cm}\bigr\rangle\bigl\langle\hspace{0.01cm}
\mathcal{Q}^{\hspace{0.03cm}e}\hspace{0.03cm}\bigr\rangle\!
\bigg\}
+
\frac{2}{3}\,
\bigl(D^{\,\lambda}\bigr)^{s\hspace{0.02cm}c}
\bigl\langle\hspace{0.01cm}
\mathcal{Q}^{\hspace{0.03cm}\lambda}\hspace{0.03cm}\bigr\rangle
\bigl\langle\hspace{0.01cm}{\mathcal Q}\hspace{0.03cm}\bigr\rangle
-
2\hspace{0.03cm}
\bigl(D^{\,\lambda}\bigr)^{s\hspace{0.02cm}c}
d^{\,e\hspace{0.02cm}d\hspace{0.02cm}\lambda}
\bigl\langle\hspace{0.01cm}
\mathcal{Q}^{\hspace{0.03cm}e}\hspace{0.03cm}
\bigr\rangle\bigl\langle\hspace{0.01cm}
\mathcal{Q}^{\hspace{0.03cm}d}\hspace{0.03cm}\bigr\rangle.
\]
Considering this relation for the given particular value of $N_{c}$ we find instead of (\ref{eq:9d})
\begin{equation}
	\Bigl(\delta^{\hspace{0.02cm}e\hspace{0.02cm}d}\hspace{0.03cm}\delta^{\hspace{0.02cm}s\hspace{0.03cm}c}
	+
	\frac{1}{2}\,\bigl(
	\delta^{\hspace{0.02cm}e\hspace{0.02cm}s}\hspace{0.03cm}\delta^{\hspace{0.02cm}d\hspace{0.03cm}c}
	+
	\delta^{\hspace{0.03cm}e\hspace{0.02cm}c}\hspace{0.03cm}\delta^{\hspace{0.03cm}d\hspace{0.03cm}s}
	\bigr)
	-
	\frac{1}{8}\,N_{c}\hspace{0.02cm}\Bigl[\bigl\{T^{\,e},T^{\,d}\bigr\}^{s\hspace{0.02cm}c}
	-
	2\hspace{0.03cm}d^{\hspace{0.03cm}e\hspace{0.02cm}d\hspace{0.02cm}\lambda}
	\hspace{0.03cm}\bigl(D^{\,\lambda}\bigr)^{s\hspace{0.02cm}c}
	\Bigr]\Bigr)
	\vspace{0.1cm}
	\label{eq:9f}
\end{equation}
\[
\times\,\bigl[
(t^{\,d}\hspace{0.03cm})^{\hspace{0.03cm}j_{1}\hspace{0.02cm}i_{2}} 
\hspace{0.03cm}
\bigl\langle\hspace{0.03cm}\theta^{\hspace{0.03cm}\ast\ \!\!i_{1}}\hspace{0.03cm} \theta^{\phantom{\ast}\!\!i_{2}}\hspace{0.03cm}\bigr\rangle\hspace{0.03cm} 
(t^{\,e}\hspace{0.03cm})^{\hspace{0.03cm}i_{1}\hspace{0.02cm}j_{2}} 
\hspace{0.03cm}
\bigl\langle\hspace{0.03cm}\theta^{\hspace{0.03cm}\ast\ \!\!j_{1}}\hspace{0.03cm} \theta^{\phantom{\ast}\!\!j_{2}}\hspace{0.03cm}\bigr\rangle\bigr]
\big|_{N_{c}\hspace{0.03cm}=\hspace{0.03cm}3}
\vspace{0.1cm}
\]
\[
=
\delta^{\hspace{0.02cm}s\hspace{0.03cm}c}\hspace{0.02cm}
\biggl\{
\biggl(\frac{1}{3} - \frac{1}{8}\biggr)\hspace{0.03cm} \bigl\langle\hspace{0.01cm}{\mathcal Q}\hspace{0.03cm}\bigr\rangle^{\!2}	
-
\frac{1}{2}\,\bigl\langle\hspace{0.01cm}\mathcal{Q}^{\hspace{0.03cm}e}
\hspace{0.03cm}\bigr\rangle\bigl\langle\hspace{0.01cm}
\mathcal{Q}^{\hspace{0.03cm}e}\hspace{0.03cm}\bigr\rangle
\biggr\}
+
\frac{3}{8}\,
\bigl(D^{\,\lambda}\bigr)^{s\hspace{0.02cm}c}
\bigl\langle\hspace{0.01cm}
\mathcal{Q}^{\hspace{0.03cm}\lambda}\hspace{0.03cm}\bigr\rangle
\bigl\langle\hspace{0.01cm}{\mathcal Q}\hspace{0.03cm}\bigr\rangle
+
\frac{7}{4}\,
\bigl\langle\hspace{0.01cm}\mathcal{Q}^{\,s}\hspace{0.03cm}
\bigr\rangle\bigl\langle\hspace{0.01cm}\mathcal{Q}^{\,c}\hspace{0.03cm}
\bigr\rangle
\vspace{0.1cm}
\]
\[
-\,
\frac{3}{2}\,
\bigl(D^{\,\lambda}\bigr)^{s\hspace{0.03cm}c}
d^{\,e\hspace{0.02cm}d\hspace{0.02cm}\lambda}
\bigl\langle\hspace{0.01cm}
\mathcal{Q}^{\hspace{0.03cm}e}\hspace{0.03cm}
\bigr\rangle\bigl\langle\hspace{0.01cm}
\mathcal{Q}^{\hspace{0.03cm}d}\hspace{0.03cm}\bigr\rangle.
\]
Let us substitute (\ref{eq:9f}) and (\ref{eq:9s}) into the right-hand side of the kinetic equation (\ref{eq:9a}). Reducing the left- and right-hand sides by the factor $N_{c} = 3$, we find here finally for this particular value
\begin{equation}
	\bigl\langle\hspace{0.01cm}\mathcal{Q}^{\hspace{0.03cm}s}\hspace{0.03cm}\bigr\rangle\hspace{0.03cm}
\frac{\partial\hspace{0.03cm}  		
		\hspace{0.02cm} W^{\hspace{0.03cm}l}_{\bf k}\hspace{0.03cm} }{\!\!\partial\hspace{0.03cm}t}
\hspace{0.03cm}+\hspace{0.03cm}
W^{\hspace{0.03cm}l}_{\bf k}\,
	\frac{d\hspace{0.03cm}  		
	\bigl\langle\hspace{0.01cm}\mathcal{Q}^{\hspace{0.03cm}s}\hspace{0.03cm}\bigr\rangle 
 }{d\hspace{0.03cm} t}\,	
	=
	2\hspace{0.03cm}\biggl(\int\!n_{\bf p}\,{\bf p}^{2}\hspace{0.03cm}
	d\hspace{0.02cm}|{\bf p}|\biggr)\!
	\int\!d\hspace{0.04cm}\Omega_{\hspace{0.035cm}{\bf v}}\, 
	{\rm Im}\hspace{0.03cm}\mathscr{T}^{\,(2,\hspace{0.03cm}{\mathcal A})}_{{\bf k},\,{\bf k}}({\bf v})
	\hspace{0.03cm}N^{\hspace{0.03cm}l}_{\bf k}\,
	\bigl\langle\hspace{0.01cm}\mathcal{Q}^{\hspace{0.03cm}s}\hspace{0.03cm}\bigr\rangle
	\label{eq:9g}
\end{equation}
\[
+\,
\frac{3}{2}\,\biggl(\int\!n_{\bf p}\,{\bf p}^{2}\hspace{0.03cm}
d\hspace{0.02cm}|{\bf p}|\biggr)\!
\int\!d\hspace{0.04cm}\Omega_{\hspace{0.035cm}{\bf v}}\!
\int\!d{\bf k}_{1}\,
\bigl|\hspace{0.02cm}\mathscr{T}^{\,(2,\hspace{0.03cm}{\mathcal A})}_{{\bf k},\,{\bf k}_{1}}({\bf v})\hspace{0.02cm}\bigr|^{\hspace{0.03cm}2}\,
N^{\,l}_{{\bf k}}\hspace{0.02cm}N^{\,l}_{{\bf k}_{1}}
\bigl\langle\hspace{0.01cm}\mathcal{Q}^{\hspace{0.03cm}s}\hspace{0.03cm}\bigr\rangle
\]
\[
\times\hspace{0.03cm}
(2\pi)\,\delta(\omega^{l}_{{\bf k}} - \omega^{l}_{{\bf k}_{1}} - {\mathbf v}\cdot({\bf k} - {\bf k}_{1}))
\vspace{0.1cm}
\]
\[
-\,
\frac{1}{3}\,
\biggl(\int\!n^{2}_{\bf p}\,{\bf p}^{2}\hspace{0.03cm}
d\hspace{0.02cm}|{\bf p}|\biggr)\!
\int\!d\hspace{0.04cm}\Omega_{\hspace{0.035cm}{\bf v}}\!
\int\!d{\bf k}_{1}\,
\bigl|\hspace{0.02cm}\mathscr{T}^{\,(2,\hspace{0.03cm}{\mathcal A})}_{{\bf k},\,{\bf k}_{1}}({\bf v})\hspace{0.02cm}\bigr|^{\hspace{0.03cm}2}\,
\biggl[
	\biggl\{
	\delta^{\hspace{0.02cm}s\hspace{0.03cm}c}\hspace{0.02cm}
	\biggl(
	\frac{1}{12}\hspace{0.03cm} \bigl\langle\hspace{0.01cm}{\mathcal Q}\hspace{0.03cm}\bigr\rangle^{\!2}	
	-
	\frac{1}{2}\,\bigl\langle\hspace{0.01cm}\mathcal{Q}^{\hspace{0.03cm}e}
	\hspace{0.03cm}\bigr\rangle\bigl\langle\hspace{0.01cm}
	\mathcal{Q}^{\hspace{0.03cm}e}\hspace{0.03cm}\bigr\rangle
	\biggr)
\]
\begin{align}
&+
	\frac{5}{2}\,
	\bigl\langle\hspace{0.01cm}\mathcal{Q}^{\,s}\hspace{0.03cm}
	\bigr\rangle\bigl\langle\hspace{0.01cm}\mathcal{Q}^{\,c}\hspace{0.03cm}
	\bigr\rangle
	-
	\frac{3}{2}\,
	\bigl(D^{\,\lambda}\bigr)^{s\hspace{0.03cm}c}
	d^{\,e\hspace{0.02cm}d\hspace{0.02cm}\lambda}
	\bigl\langle\hspace{0.01cm}
	\mathcal{Q}^{\hspace{0.03cm}e}\hspace{0.03cm}
	\bigr\rangle\bigl\langle\hspace{0.01cm}
	\mathcal{Q}^{\hspace{0.03cm}d}\hspace{0.03cm}\bigr\rangle
	\!\hspace{0.03cm}\biggr\}
\bigl\langle\hspace{0.01cm}\mathcal{Q}^{\hspace{0.03cm}c}\hspace{0.03cm}\bigr\rangle\hspace{0.03cm}
\bigl(W^{\,l}_{{\bf k}} - W^{\,l}_{{\bf k}_{1}}\bigr)
\notag\\[0.6ex]
&+
\frac{3}{4}\,\biggl\{\frac{1}{3}\,\delta^{\hspace{0.02cm}s\hspace{0.03cm}c}
	\bigl\langle\hspace{0.01cm}{\mathcal Q}\hspace{0.03cm}\bigr\rangle^{\!2}
	+
	\bigl(D^{\,\lambda}\bigr)^{s\hspace{0.02cm}c}
	\bigl\langle\hspace{0.01cm}
	\mathcal{Q}^{\hspace{0.03cm}\lambda}\hspace{0.03cm}\bigr\rangle
	\bigl\langle\hspace{0.01cm}{\mathcal Q}\hspace{0.03cm}\bigr\rangle
	-
	2\,\bigl\langle\hspace{0.01cm}
	\mathcal{Q}^{\hspace{0.03cm}s}\hspace{0.03cm}
	\bigr\rangle\bigl\langle\hspace{0.01cm}
	\mathcal{Q}^{\hspace{0.03cm}c}\hspace{0.03cm}\bigr\rangle\biggr\}
	\hspace{0.01cm}
\bigl\langle\hspace{0.03cm}\mathcal{Q}^{\hspace{0.03cm}c}
\hspace{0.03cm}\bigr\rangle\hspace{0.03cm}
W^{\,l}_{{\bf k}}\hspace{0.03cm}\biggr]
\notag
\end{align}
\[
\times\hspace{0.03cm}
(2\pi)\hspace{0.03cm}\delta(\omega^{l}_{{\bf k}} - \omega^{l}_{{\bf k}_{1}} - {\mathbf v}\cdot({\bf k} - {\bf k}_{1})).
\]

\section{\bf Equation for the averaged colorless charge $\langle\hspace{0.01cm}\mathcal{Q}\hspace{0.03cm}\rangle$}
\label{section_10}
\setcounter{equation}{0}

In this and next sections, we analyze the kinetic equation for the hard particle number density ${\mathfrak n}^{\hspace{0.03cm}i^{\prime}\hspace{0.02cm}i}_{{\bf p}}$ defined by (\ref{eq:5p}) in the approximation $|{\bf p}|, |{\bf p}_{1}| \gg |{\bf k}_{1}|,\, |{\bf k}_{2 }|$. Let us write out the original equation here once more 
\[
\delta({\bf p} - {\bf p}\!\ ')\,
\frac{\partial\hspace{0.03cm} {\mathfrak n}^{\hspace{0.03cm}i^{\prime}\hspace{0.02cm}i}_{{\bf p}}}{\partial\hspace{0.03cm} t}\
=
-\hspace{0.03cm}i\!\int\!d{\bf p}_{1}\hspace{0.02cm} d{\bf k}_{1}\hspace{0.03cm} d{\bf k}_{2}
\]
\[
\times\,
\biggl\{\mathscr{T}^{\hspace{0.03cm} (2)\hspace{0.03cm} i^{\prime}\, i_{1}\, a_{1}\, a_{2}}_{\, {\bf p}^{\prime},\, {\bf p}_{1},\, {\bf k}_{1},\, {\bf k}_{2}}\, 
I^{\, i\, i_{1}\, a_{1}\, a_{2}}_{{\bf p},\, {\bf p}_{1},\, {\bf k}_{1},\, {\bf k}_{2}}\hspace{0.03cm}
\delta({\bf p}^{\prime} + {\bf k}_{1} - {\bf p}_{1} - {\bf k}_{2})
\hspace{0.03cm}-\hspace{0.03cm}
\mathscr{T}^{\hspace{0.03cm}\ast\hspace{0.03cm}(2)\hspace{0.03cm} i\, i_{1}\, a_{1}\, a_{2}}_{\ {\bf p},\, {\bf p}_{1},\, {\bf k}_{1},\, {\bf k}_{2}}\, 
I^{\, i_{1}\, i^{\prime}\, a_{2}\, a_{1}}_{{\bf p}_{1},\, {\bf p}^{\prime},\, {\bf k}_{2},\, {\bf k}_{1}}\hspace{0.03cm}
\delta({\bf p} + {\bf k}_{1} - {\bf p}_{1} - {\bf k}_{2})\biggr\}.
\]
As the fourth-order correlation function $I^{\,i\,i_{1}\, a_{1}\, a_{2}}_{{\bf p},\,{\bf p}_{1},\,{\bf k}_{1},\,{\bf k}_{2}}$ we take the expression (\ref{eq:6q}). Following the same line of the reasoning as in section \ref{section_6}, in this case we arrive at the following matrix kinetic equation supplementing Eq.\,(\ref{eq:8w}): 
\begin{equation}
\frac{\partial\hspace{0.04cm} {\mathfrak n}^{\hspace{0.03cm}i^{\prime}\hspace{0.02cm}i}_{{\bf p}}}{\partial\hspace{0.03cm}t}
=
-\hspace{0.03cm}i\!\int\!d\hspace{0.02cm}{\bf k}\, 
{\rm tr}\hspace{0.03cm}\bigl({\mathcal N}_{{\bf k}}T^{\,d}\hspace{0.03cm}\bigr)
\Bigl\{\mathscr{T}^{\,(2,\hspace{0.03cm}{\mathcal A})}_{{\bf p},\, {\bf p}, {\bf k}, {\bf k}}\hspace{0.03cm}
\bigl(t^{\,d}\hspace{0.03cm}{\mathfrak n}_{{\bf p}}\bigr)^{i^{\prime}\hspace{0.03cm}i}\hspace{0.03cm}
-\,
\mathscr{T}^{\,\ast\hspace{0.03cm}(2,\hspace{0.03cm}{\mathcal A})}_{{\bf p},\, {\bf p}, {\bf k}, {\bf k}}
\bigl({\mathfrak n}_{{\bf p}}\hspace{0.03cm}t^{\,d}\hspace{0.03cm}\bigr)^{i^{\prime}\hspace{0.02cm}i}
\hspace{0.03cm}\Bigr\}
\label{eq:10q}
\end{equation}
\[
+\,
i\!\int\!d{\bf p}_{1}\hspace{0.02cm}d{\bf k}_{1}\hspace{0.03cm}
d{\bf k}_{2}\,
\bigl|\mathscr{T}^{\,(2,\hspace{0.03cm}{\mathcal A})}_{{\bf p},\, {\bf p}_{1}, {\bf k}_{1}, {\bf k}_{2}}\bigr|^{2}
(2\pi)^{3}\hspace{0.03cm} \delta({\bf p} - {\bf p}_{1} + {\bf k}_{1} - {\bf k}_{2})
\]
\[
\times
\biggl\{\frac{1}{\Delta\hspace{0.02cm}\omega_{{\bf p},\, {\bf p}_{1}, {\bf k}_{1}, {\bf k}_{2}} -\, i\hspace{0.03cm}0}\,
\Bigl(\Bigl[\bigl(t^{\,d}\hspace{0.03cm}{\mathfrak n}_{{\bf p}_{1}}t^{\,e}\hspace{0.03cm}\bigr)^{i^{\prime}\hspace{0.03cm}i}
-
\bigl(t^{\,d}\hspace{0.03cm}t^{\,e}\hspace{0.03cm}{\mathfrak n}_{{\bf p}}\hspace{0.03cm}\bigr)^{i^{\prime}\hspace{0.03cm}i}\hspace{0.03cm}\Bigr]
{\rm tr}\hspace{0.03cm}\bigl(T^{\,d}{\mathcal N}_{{\bf k}_{1}} T^{\,e} {\mathcal N}_{{\bf k}_{2}} \hspace{0.03cm}\bigr)
\]
\[
-\,
\bigl(t^{\,d}\hspace{0.03cm} {\mathfrak n}_{{\bf p}_{1}}t^{\,e}\hspace{0.03cm} {\mathfrak n}_{{\bf p}}\bigr)^{i^{\prime}\hspace{0.01cm}i}
\Bigl[\,{\rm tr}\hspace{0.03cm}
\bigl(T^{\,d}T^{\,e} {\mathcal N}_{{\bf k}_{2}} \hspace{0.03cm}\bigr)
\!-\,
{\rm tr}\hspace{0.03cm}\bigl(T^{\,d}{\mathcal N}_{{\bf k}_{1}} T^{\,e}\hspace{0.03cm}\bigr)\Bigr]\Bigr)
\vspace{0.15cm}
\]
\[
-\,
\frac{1}{\Delta\hspace{0.02cm}\omega_{{\bf p},\, {\bf p}_{1}, {\bf k}_{1}, {\bf k}_{2}} +\, i\hspace{0.03cm}0}\,
\Bigl(\Bigl[\bigl(t^{\,e}\hspace{0.03cm}{\mathfrak n}_{{\bf p}_{1}}t^{\,d}\hspace{0.03cm}\bigr)^{i^{\prime}\hspace{0.03cm}i}
-
\bigl({\mathfrak n}_{{\bf p}}t^{\,e}\hspace{0.03cm}t^{\,d}\hspace{0.03cm}\bigr)^{i^{\prime}\hspace{0.03cm}i}\hspace{0.03cm}\Bigr]
{\rm tr}\hspace{0.03cm}\bigl(T^{\,d}{\mathcal N}_{{\bf k}_{2}} T^{\,e} {\mathcal N}_{{\bf k}_{1}} \hspace{0.03cm}\bigr)
\]
\[
-\,
\bigl({\mathfrak n}_{{\bf p}}\hspace{0.03cm}t^{\,e}\hspace{0.03cm} 
{\mathfrak n}_{{\bf p}_{1}}t^{\,d}
\hspace{0.03cm}\bigr)^{i^{\prime}\hspace{0.01cm}i}
\Bigl[\,{\rm tr}\hspace{0.03cm}
\bigl({\mathcal N}_{{\bf k}_{2}}T^{\,e}T^{\,d} \hspace{0.03cm}\bigr)
\!-\,
{\rm tr}\hspace{0.03cm}\bigl(T^{\,e}{\mathcal N}_{{\bf k}_{1}} T^{\,d}\hspace{0.03cm}\bigr)\Bigr]\Bigr)\!\biggr\}.
\vspace{0.15cm}
\]
Let us consider an approximation of this equation. The first step is to integrate over ${\bf p}_{1}$ in the second term on the right-hand side of (\ref{eq:10q}). This gives us ${\bf p}_{1} = {\bf p} + \Delta {\bf k}$, where $\Delta {\bf k} \equiv {\bf k}_{1} - {\bf k}_{2}$. We are interested in the approximation $|{\bf p}| \gg |{\bf k}_{1}|,\, |{\bf k}_{2}|$. We compute the trace of the left- and right-hand sides over color indices, i.e. we set $i = i^{\hspace{0.03cm}\prime}$ and sum over $i$. Taking into account that 
\[
{\rm tr}\hspace{0.03cm}(\hspace{0.03cm}
{\mathfrak n}_{{\bf p}}\hspace{0.03cm})
=
{\mathfrak n}^{\hspace{0.03cm}i\hspace{0.03cm}i}_{\hspace{0.03cm}{\bf p}} 
= 
n_{\hspace{0.03cm}{\bf p}}\hspace{0.03cm} 
\bigl\langle\hspace{0.03cm}\theta^{\hspace{0.03cm}\ast\ \!\!i}\hspace{0.04cm} \theta^{\phantom{\ast}\!\!i}
\hspace{0.03cm}\bigr\rangle
\equiv
n_{\hspace{0.03cm}{\bf p}}\hspace{0.03cm}\bigl\langle\hspace{0.01cm}\mathcal{Q}\hspace{0.03cm}\bigr\rangle,
\]
we find instead of (\ref{eq:10q})
\begin{equation}
n_{\bf p}\,
\frac{d\hspace{0.04cm} \langle\hspace{0.01cm}\mathcal{Q}\hspace{0.03cm}\rangle}{d\hspace{0.03cm} t}
=
-\hspace{0.03cm}i\hspace{0.03cm}\!\int\!d\hspace{0.02cm}{\bf k}\, 
{\rm tr}\hspace{0.03cm}\bigl({\mathcal N}_{{\bf k}}\hspace{0.03cm}T^{\,d}\hspace{0.03cm}\bigr)
\hspace{0.03cm}
{\rm tr}\hspace{0.03cm}\bigl(\hspace{0.03cm}t^{\,d}\hspace{0.03cm}
{\mathfrak n}_{{\bf p}}\hspace{0.03cm}\bigr)\hspace{0.03cm}
\Bigl\{\hspace{0.03cm}\mathscr{T}^{\,(2,\hspace{0.03cm}{\mathcal A})}_{{\bf k},\,{\bf k}}({\bf v})\hspace{0.03cm}
\bigr)
-\,
\mathscr{T}^{\,\ast\,(2,\hspace{0.03cm}{\mathcal A})}_{{\bf k},\,{\bf k}}({\bf v})\Bigr\}
\label{eq:10w}
\end{equation}
\[
+\,
i\hspace{0.04cm} \!\int\!d{\bf k}_{1}\hspace{0.03cm}d{\bf k}_{2}\,
\bigl|\hspace{0.02cm}\mathscr{T}^{\,(2,\hspace{0.03cm}{\mathcal A})}_{{\bf k}_{1},\,{\bf k}_{2}}({\bf v})\hspace{0.02cm}\bigr|^{2}
\left\{\frac{1}{\Delta\hspace{0.02cm}\omega_{\hspace{0.03cm}{\bf p},\, 
{\bf p}, {\bf k}_{1}, {\bf k}_{2}} -\, i\hspace{0.03cm}0}\,
\right.
\hspace{2.5cm}
\]
\[
\times
\biggl({\rm tr}\hspace{0.03cm}
\bigl(\bigl[t^{\,e},\hspace{0.03cm}t^{\,d}\hspace{0.03cm}\bigr]{\mathfrak n}_{{\bf p}}\hspace{0.03cm}\bigr)
\hspace{0.03cm}
{\rm tr}\hspace{0.03cm}\bigl(T^{\,d}{\mathcal N}_{{\bf k}_{1}} T^{\,e} {\mathcal N}_{{\bf k}_{2}} \hspace{0.03cm}\bigr)
-
{\rm tr}\hspace{0.03cm}
\bigl(\hspace{0.03cm}t^{\,d}\hspace{0.03cm} {\mathfrak n}_{{\bf p}}\hspace{0.03cm}t^{\,e}\hspace{0.03cm} {\mathfrak n}_{{\bf p}}\bigr)
\Bigl[\,{\rm tr}\hspace{0.03cm}
\bigl(T^{\,d}T^{\,e} {\mathcal N}_{{\bf k}_{2}}\hspace{0.03cm}\bigr)
\!-\,
{\rm tr}\hspace{0.03cm}\bigl(T^{\,e}T^{\,d}{\mathcal N}_{{\bf k}_{1}}\hspace{0.03cm}\bigr)\Bigr]\biggr)
\]
\[
\hspace{2cm}
-\,
\frac{1}{\Delta\hspace{0.02cm}\omega_{\hspace{0.03cm}{\bf p},\, {\bf p}, {\bf k}_{1}, {\bf k}_{2}} +\, i\hspace{0.03cm}0}\,
\]
\[
\times
\biggl({\rm tr}\hspace{0.03cm}
\bigl(\bigl[t^{\,e},\hspace{0.03cm}t^{\,d}\hspace{0.03cm}\bigr]{\mathfrak n}_{{\bf p}}\hspace{0.03cm}\bigr)
\hspace{0.03cm}
{\rm tr}\hspace{0.03cm}\bigl(T^{\,d}{\mathcal N}_{{\bf k}_{1}} T^{\,e} {\mathcal N}_{{\bf k}_{2}} \hspace{0.03cm}\bigr)
-
{\rm tr}\hspace{0.03cm}
\bigl(\hspace{0.03cm}t^{\,d}\hspace{0.03cm} {\mathfrak n}_{{\bf p}}\hspace{0.03cm}t^{\,e}\hspace{0.03cm}{\mathfrak n}_{{\bf p}}\bigr)
\Bigl[\,{\rm tr}\hspace{0.03cm}
\bigl(T^{\,d}\hspace{0.03cm}T^{\,e} {\mathcal N}_{{\bf k}_{2}} \hspace{0.03cm}\bigr)
\!-\,
{\rm tr}\hspace{0.03cm}\bigl(T^{\,e}\hspace{0.03cm}
T^{\,d}{\mathcal N}_{{\bf k}_{1}}\hspace{0.03cm}\bigr)\Bigr]\biggr)\!\biggr\}.
\]
Within the approximations used in this paper, we have assumed that the function ${n}_{{\bf p}}$ is independent of time.\\
\indent We analyze the first term on the right-hand side of Eq.\,(\ref{eq:10w}). Considering the traces
\begin{equation}
	{\rm tr}\hspace{0.03cm}\bigl(t^{\,d}\hspace{0.03cm}
	{\mathfrak n}_{{\bf p}}\bigr)
	=
	(t^{\,d})^{i\hspace{0.04cm}i^{\hspace{0.03cm}\prime}} {\mathfrak n}^{\hspace{0.03cm}i^{\hspace{0.03cm}\prime}
		\hspace{0.03cm}i}_{\hspace{0.03cm}{\bf p}} 
	= 
	n_{\hspace{0.03cm}{\bf p}}\hspace{0.03cm} (t^{\,d})^{i\hspace{0.04cm}i^{\hspace{0.03cm}\prime}}
	\bigl\langle\hspace{0.03cm}\theta^{\hspace{0.03cm}\ast\ \!\!i}\hspace{0.04cm} \theta^{\phantom{\ast}\!\!i^{\hspace{0.03cm}\prime}}
	\hspace{0.03cm}\bigr\rangle
	\equiv
	n_{\hspace{0.03cm}{\bf p}}\hspace{0.03cm}\bigl\langle\hspace{0.01cm}
	\mathcal{Q}^{\hspace{0.04cm}d}\hspace{0.03cm}\bigr\rangle
	\label{eq:10e}
\end{equation}
and 
\[
{\rm tr}\hspace{0.03cm}\bigl({\mathcal N}_{{\bf k}}\hspace{0.03cm}T^{\,d}\hspace{0.03cm}\bigr)
=
N_{c}\hspace{0.03cm}
\bigl\langle\hspace{0.01cm}{\mathcal Q}^{\hspace{0.03cm}d}\hspace{0.03cm}
\bigr\rangle
\hspace{0.03cm}W^{l}_{\bf k},
\]
it is not difficult to see that the integrand in the first term on the right-hand side of (\ref{eq:10w}) can be represented in the following form:
\begin{equation*}
\begin{split}
&{\rm tr}\hspace{0.03cm}\bigl({\mathcal N}_{{\bf k}}\hspace{0.03cm}T^{\,d}\hspace{0.03cm}\bigr)
	\hspace{0.03cm}
	{\rm tr}\hspace{0.03cm}\bigl(\hspace{0.03cm}t^{\,d}\hspace{0.03cm}
	{\mathfrak n}_{{\bf p}}\hspace{0.03cm}\bigr)\hspace{0.03cm}
	\Bigl\{\hspace{0.03cm}\mathscr{T}^{\,(2,\hspace{0.03cm}{\mathcal A})}_{{\bf k},\,{\bf k}}({\bf v})\hspace{0.03cm}
	\bigr)
	-\,
	\mathscr{T}^{\,\ast\,(2,\hspace{0.03cm}{\mathcal A})}_{{\bf k},\,{\bf k}}({\bf v})\Bigr\} \\[1.3ex]
&=
2\hspace{0.01cm}i\hspace{0.02cm}n_{\hspace{0.03cm}{\bf p}}\hspace{0.03cm}N_{c}\,
	{\rm Im}\hspace{0.03cm}\mathscr{T}^{\,(2,\hspace{0.03cm}{\mathcal A})}_{{\bf k},\,{\bf k}}({\bf v})
	\hspace{0.04cm}
	\hspace{0.03cm}W^{\hspace{0.03cm}l}_{\bf k}
	\hspace{0.03cm}\bigl\langle\hspace{0.01cm}\mathcal{Q}^{\,d}
	\hspace{0.03cm}\bigr\rangle
	\hspace{0.03cm}\bigl\langle\hspace{0.01cm}{\mathcal Q}^{\,d}\hspace{0.03cm}
	\bigr\rangle.
\end{split}
\end{equation*}
\indent Let us proceed to analyze the traces in the second term on the right-hand side of (\ref{eq:10w}). Given that 
\[
{\rm tr}\hspace{0.03cm}
\bigl(\bigl[t^{\,e},\hspace{0.03cm}t^{\,d}\hspace{0.03cm}\bigr]{\mathfrak n}_{{\bf p}}\hspace{0.03cm}\bigr)
=
i\hspace{0.03cm}
f^{\hspace{0.03cm}e\hspace{0.02cm}d\hspace{0.03cm}\kappa}
\hspace{0.03cm}\bigl\langle\hspace{0.01cm}{\mathcal Q}^{\hspace{0.03cm}\kappa}\hspace{0.03cm}
\bigr\rangle\hspace{0.03cm}n_{\bf p},
\]
we trivially find
\begin{equation*}
{\rm tr}\hspace{0.03cm}
\bigl(\bigl[t^{\,e},\hspace{0.03cm}t^{\,d}\hspace{0.03cm}\bigr]{\mathfrak n}_{{\bf p}}\hspace{0.03cm}\bigr)
\hspace{0.03cm}
{\rm tr}\hspace{0.03cm}\bigl(T^{\,d}{\mathcal N}_{{\bf k}_{1}} T^{\,e} {\mathcal N}_{{\bf k}_{2}} \hspace{0.03cm}\bigr)
\end{equation*}
\[
=
i\hspace{0.03cm}n_{\bf p}\hspace{0.03cm}
f^{\hspace{0.03cm}e\hspace{0.02cm}d\hspace{0.03cm}\kappa}
\hspace{0.03cm}\bigl\langle\hspace{0.01cm}{\mathcal Q}^{\hspace{0.03cm}\kappa}\hspace{0.03cm}
\bigr\rangle\hspace{0.03cm}
\Bigl\{\delta^{\hspace{0.03cm}e\hspace{0.03cm} d}N_{c}\hspace{0.03cm}  N_{{\bf k}_{1}}N_{{\bf k}_{2}}
+
\frac{i}{2}\,N_{c}\hspace{0.03cm} f^{\hspace{0.03cm}c\hspace{0.03cm}e\hspace{0.03cm}d}\hspace{0.03cm}
\bigl\langle\hspace{0.01cm}\mathcal{Q}^{\hspace{0.03cm}c}\hspace{0.03cm}\bigr\rangle\hspace{0.03cm}
\bigl(W_{{\bf k}_{1}}N_{{\bf k}_{2}} - N_{{\bf k}_{1}}W_{{\bf k}_{2}}\bigr)
\]
\[
+\,
\Bigl(\delta^{\hspace{0.02cm}c\hspace{0.02cm}e}\delta^{\hspace{0.02cm}\rho\hspace{0.02cm} d}
+
\delta^{\hspace{0.02cm}c\hspace{0.02cm}d}\delta^{\hspace{0.02cm}e\hspace{0.02cm}\rho}
+
\frac{1}{4}\,N_{c}\hspace{0.02cm}\Bigl[\bigl\{D^{c},D^{\rho}\bigr\}^{e\hspace{0.02cm}d}
-
d^{\hspace{0.03cm}c\hspace{0.03cm}\rho\hspace{0.03cm}\lambda}\bigl(D^{\lambda}\bigr)^{e\hspace{0.02cm}d}
\hspace{0.03cm}\Bigr]\Bigr)
\bigl\langle\hspace{0.01cm}\mathcal{Q}^{\hspace{0.03cm}c}\hspace{0.03cm}
\bigr\rangle
\bigl\langle\hspace{0.01cm}\mathcal{Q}^{\hspace{0.03cm}\rho}\hspace{0.03cm}
\bigr\rangle
W_{{\bf k}_{1}}W_{{\bf k}_{2}}\Bigr\}
\]
\[
=
-\hspace{0.03cm}\frac{1}{2}\,n_{\bf p}\hspace{0.03cm}N^{\hspace{0.03cm}2}_{c}
\hspace{0.03cm}
\bigl(W_{{\bf k}_{1}}N_{{\bf k}_{2}} - N_{{\bf k}_{1}}W_{{\bf k}_{2}}\bigr)
\bigl\langle\hspace{0.01cm}{\mathcal Q}^{\hspace{0.03cm}e}\hspace{0.03cm}
\bigr\rangle
\bigl\langle\hspace{0.01cm}{\mathcal Q}^{\hspace{0.03cm}e}\hspace{0.03cm}
\bigr\rangle.
\]
Here, we have used the representation (\ref{eq:8r}) for the matrix function ${\mathcal N}_{{\bf k}}$ and the formulae for traces (\ref{ap:C4})\,--\,(\ref{ap:C6}).\\
\indent Let us consider the other trace in the second term in (\ref{eq:10w}), which differ in color structure. By virtue of the decomposition (\ref{eq:8r}) and the traces (\ref{ap:C4}) and (\ref{ap:C5}), it can be represented as follows:
\begin{equation}
{\rm tr}\hspace{0.03cm}
\bigl(\hspace{0.03cm}t^{\,d}\hspace{0.03cm} {\mathfrak n}_{{\bf p}}\hspace{0.03cm}t^{\,e}\hspace{0.03cm} {\mathfrak n}_{{\bf p}}\bigr)
\Bigl[\,{\rm tr}\hspace{0.03cm}
\bigl(T^{\,d}T^{\,e} {\mathcal N}_{{\bf k}_{2}} \hspace{0.03cm}\bigr)
-
{\rm tr}\hspace{0.03cm}\bigl(T^{\,e}T^{\,d}{\mathcal N}_{{\bf k}_{1}}\hspace{0.01cm}\bigr)\Bigr]
\label{eq:10y}
\end{equation}
\[
=
{\rm tr}\hspace{0.03cm}
\bigl(\hspace{0.03cm}t^{\,d}\hspace{0.03cm} {\mathfrak n}_{{\bf p}}\hspace{0.03cm}t^{\,e}\hspace{0.03cm} {\mathfrak n}_{{\bf p}}\bigr)
\Bigl[\hspace{0.03cm} \delta^{\hspace{0.03cm}d\hspace{0.03cm}e}\hspace{0.03cm}N_{c}\hspace{0.03cm}
\bigl(N_{{\bf k}_{2}} - N_{{\bf k}_{1}}\hspace{0.01cm}\bigr)
+
\frac{1}{2}\,
i\hspace{0.03cm}N_{c}\hspace{0.03cm} f^{\hspace{0.03cm}d\hspace{0.02cm}e\hspace{0.03cm}c}
\hspace{0.03cm}\bigl(W_{{\bf k}_{2}} + W_{{\bf k}_{1}}\hspace{0.01cm}\bigr) 
\bigl\langle\hspace{0.01cm}\mathcal{Q}^{\hspace{0.03cm}c}\hspace{0.03cm}\bigr\rangle\Bigr].
\]
We examine the contribution proportional to the unit color  matrix  $\delta^{\hspace{0.03cm}d\hspace{0.03cm}e}$. Taking into account the relation (\ref{ap:B5}), we have the following chain of transformations
\begin{equation}
{\rm tr}\hspace{0.03cm}
\bigl(\hspace{0.03cm}t^{\,e}\hspace{0.03cm} {\mathfrak n}_{{\bf p}}\hspace{0.03cm}t^{\,e}\hspace{0.03cm}{\mathfrak n}_{{\bf p}}\bigr)
=
\frac{1}{2}\,{\rm tr}\hspace{0.03cm}\bigl(\hspace{0.03cm}{\mathfrak n}_{{\bf p}}\bigr)\hspace{0.03cm}
{\rm tr}\hspace{0.03cm}\bigl({\mathfrak n}_{{\bf p}}\bigr)
-
\frac{1}{2\hspace{0.02cm}N_{c}}\,{\rm tr}\hspace{0.03cm}\bigl({\mathfrak n}^{\hspace{0.02cm}2}_{{\bf p}}\hspace{0.03cm}\bigr)
\label{eq:10u}
\end{equation}
\[
=
\frac{1}{2}\,
(n_{\bf p})^{2}\hspace{0.03cm}
\hspace{0.03cm}\Bigl\{
\hspace{0.03cm}\bigl\langle\hspace{0.01cm}\mathcal{Q}\hspace{0.03cm}
\bigr\rangle^{2}
-
\frac{1}{N_{c}}\,
\delta^{\hspace{0.03cm}i_{1}\hspace{0.03cm}j_{2}}
\delta^{\hspace{0.03cm}j_{1}\hspace{0.03cm}i_{2}}
\hspace{0.03cm}
\bigl\langle\hspace{0.03cm}\theta^{\,\ast\hspace{0.04cm}i_{1}}\hspace{0.04cm}
\theta^{\hspace{0.04cm}i_{2}}\hspace{0.03cm}\bigr\rangle
\hspace{0.03cm}
\bigl\langle\hspace{0.03cm}\theta^{\,\ast\hspace{0.04cm}j_{1}}\hspace{0.04cm}
\theta^{\hspace{0.04cm}j_{2}}\hspace{0.03cm}\bigr\rangle
\Bigr\}.
\]
Further, for the color factor $\delta^{\hspace{0.03cm}j_{1}i_{2}}\hspace{0.03cm}
\delta^{\hspace{0.03cm}i_{1}\hspace{0.03cm}j_{2}}$ in the second term in (\ref{eq:10u}) we make use of identity (\ref{ap:B4}). Considering this identity, we find instead of (\ref{eq:10u})
\begin{equation*}
	{\rm tr}\hspace{0.03cm}
	\bigl(\hspace{0.03cm}t^{\,e}\hspace{0.03cm} {\mathfrak n}_{{\bf p}}\hspace{0.03cm}t^{\,e}\hspace{0.03cm}{\mathfrak n}_{{\bf p}}\bigr)
	=
	\frac{1}{2}\,
	(n_{\bf p})^{2}\hspace{0.03cm}
	\hspace{0.03cm}\biggl\{
	\biggl(\frac{N^{\hspace{0.02cm}2}_{c} - 1} {N^{\hspace{0.02cm}2}_{c}}\biggr)
	\bigl\langle\hspace{0.01cm}\mathcal{Q}\hspace{0.03cm}\bigr\rangle^{2}
	-
	\frac{2}{N_{c}}\,
		\langle\hspace{0.01cm}\mathcal{Q}^{\hspace{0.03cm}e}\hspace{0.03cm}
		\bigr\rangle
		\langle\hspace{0.01cm}\mathcal{Q}^{\hspace{0.03cm}e}\hspace{0.03cm}
		\bigr\rangle
		\!\hspace{0.03cm}\biggr\}.
\end{equation*}
The term in (\ref{eq:10y}) with the antisymmetric structure constants $f^{\hspace{0.03cm}d\hspace{0.02cm}e\hspace{0.03cm}c}$ will give us zero contribution due to the symmetry of the trace ${\rm tr}\hspace{0.03cm}
\bigl(\hspace{0.03cm}t^{\,d}\hspace{0.03cm} {\mathfrak n}_{{\bf p}}\hspace{0.03cm}t^{\,e}\hspace{0.03cm}{\mathfrak n}_{{\bf p}}\bigr)$ with respect to the permutation of indices $d$ and $e$. Thus we finally obtain for (\ref{eq:10y})
\begin{equation*}
	{\rm tr}\hspace{0.03cm}
	\bigl(\hspace{0.03cm}t^{\,d}\hspace{0.03cm} {\mathfrak n}_{{\bf p}}\hspace{0.03cm}t^{\,e}\hspace{0.03cm} {\mathfrak n}_{{\bf p}}\bigr)
	\Bigl[\,{\rm tr}\hspace{0.03cm}
	\bigl(T^{\,d}T^{\,e} {\mathcal N}_{{\bf k}_{2}} \hspace{0.03cm}\bigr)
	-
	{\rm tr}\hspace{0.03cm}\bigl(T^{\,e}T^{\,d}{\mathcal N}_{{\bf k}_{1}}\hspace{0.01cm}\bigr)\Bigr]
\end{equation*}
\[
=
\frac{1}{2}\,
(n_{\bf p})^{2}\hspace{0.03cm}N_{c}
\hspace{0.03cm}\biggl\{
\biggl(\frac{N^{\hspace{0.02cm}2}_{c} - 1} {N^{\hspace{0.02cm}2}_{c}}\biggr)
\bigl\langle\hspace{0.01cm}\mathcal{Q}\hspace{0.03cm}\bigr\rangle^{2}
-
\frac{2}{N_{c}}\,
	\langle\hspace{0.01cm}\mathcal{Q}^{\hspace{0.03cm}e}\hspace{0.03cm}
	\bigr\rangle
	\langle\hspace{0.01cm}\mathcal{Q}^{\hspace{0.03cm}e}\hspace{0.03cm}
	\bigr\rangle
\!\hspace{0.03cm}\biggr\}
\bigl(N_{{\bf k}_{2}} - N_{{\bf k}_{1}}\hspace{0.01cm}\bigr).
\]
\indent Taking into account all the above calculations, Sokhotsky's formula (\ref{eq:6y}) and reducing the left and right-hand sides by the common multiplier $n_{\bf p}$, we find instead of (\ref{eq:10w}) the following equation for the averaged colorless charge  $\bigl\langle\hspace{0.01cm}\mathcal{Q}\hspace{0.03cm}\bigr\rangle$:
\begin{equation}
	\frac{d\hspace{0.04cm} \langle\hspace{0.01cm}\mathcal{Q}\hspace{0.03cm}\rangle}{d\hspace{0.03cm} t}
	=
	2\hspace{0.02cm}N_{c}\hspace{0.04cm}{\mathfrak q}_{2}(t)\!
	\!\int\!d\hspace{0.02cm}{\bf k}\, 
		{\rm Im}\hspace{0.03cm}\mathscr{T}^{\,(2,\hspace{0.03cm}{\mathcal A})}_{{\bf k},\,{\bf k}}({\bf v})
	\hspace{0.04cm}
	W^{\hspace{0.03cm}l}_{\bf k}
\label{eq:10p}
\end{equation}
\[
+\,\frac{1}{2}\,N^{\hspace{0.03cm}2}_{c}
\hspace{0.04cm}{\mathfrak q}_{2}(t)\!\!\int\!d{\bf k}_{1}\hspace{0.03cm} d{\bf k}_{2}\,
\bigl|\hspace{0.02cm}\mathscr{T}^{\,(2,\hspace{0.03cm}{\mathcal A})}_{{\bf k}_{1},\,{\bf k}_{2}}({\bf v})\hspace{0.02cm}\bigr|^{\hspace{0.03cm}2}\hspace{0.03cm}
\bigl(W_{{\bf k}_{1}}N_{{\bf k}_{2}} - N_{{\bf k}_{1}}W_{{\bf k}_{2}}\bigr)
\hspace{0.03cm}
(2\pi)\,\delta(\omega^{l}_{{\bf k}_{1}} - \omega^{l}_{{\bf k}_{2}} - {\mathbf v}\cdot({\bf k}_{1} - {\bf k}_{2}))
\hspace{0.7cm}
\]
\vspace{0.1cm}
\[
-\,n_{\bf p}\!\int\!d{\bf k}_{1}\hspace{0.03cm}d{\bf k}_{2}\,
\bigl|\hspace{0.02cm}\mathscr{T}^{\,(2,\hspace{0.03cm}{\mathcal A})}_{{\bf k}_{1},\,{\bf k}_{2}}({\bf v})\hspace{0.02cm}\bigr|^{\hspace{0.03cm}2}
\hspace{0.03cm}
\hspace{0.03cm}\biggl\{
\biggl(\frac{N^{\hspace{0.02cm}2}_{c} - 1} {2\hspace{0.02cm}N_{c}}\biggr)
\bigl\langle\hspace{0.01cm}\mathcal{Q}\hspace{0.03cm}\bigr\rangle^{2}
-\,
{\mathfrak q}_{2}(t)\!\hspace{0.03cm}\biggr\}
\bigl(N_{{\bf k}_{1}} - N_{{\bf k}_{2}}\hspace{0.01cm}\bigr)
\vspace{0.1cm}
\]
\[
\times\hspace{0.03cm}
(2\pi)\,\delta(\omega^{l}_{{\bf k}_{1}} - \omega^{l}_{{\bf k}_{2}} - {\mathbf v}\cdot({\bf k}_{1} - {\bf k}_{2})).
\]
Here, we have introduced the shorthand notation for the colorless quadratic combination of the averaged color charge
\begin{equation}
{\mathfrak q}_{2}(t) \equiv \langle\hspace{0.01cm}\mathcal{Q}^{\hspace{0.03cm}e}\hspace{0.03cm}
\bigr\rangle
\hspace{0.03cm}
\langle\hspace{0.01cm}\mathcal{Q}^{\hspace{0.03cm}e}\hspace{0.03cm}
\bigr\rangle.
\label{eq:10a}
\end{equation}
\indent Let us analyze the right-hand side of the obtained equation (\ref{eq:10p}). The amplitude modulus square $\bigl|\hspace{0.02cm}\mathscr{T}^{\,(2,\hspace{0.03cm}{\mathcal A})}_{{\bf k}_{1},\,{\bf k}_{2}}({\bf v})\hspace{0.02cm}\bigr|^{2}$, due to the first property in (\ref{eq:5j}), is an even function with respect to the permutation ${\bf k}_{1} \rightleftarrows {\bf k}_{2}$. The resonance condition 
\[
\delta(\omega^{l}_{{\bf k}_{1}} - \omega^{l}_{{\bf k}_{2}} - {\mathbf v}\cdot({\bf k}_{1} - {\bf k}_{2}))
\] 
is also even with respect to the same permutation. Thus, we can see that the last two terms in (\ref{eq:10p}) have odd the functions $\bigl(W_{{\bf k}_{1}}N_{{\bf k}_{2}} - N_{{\bf k}_{1}}W_{{\bf k}_{2}}\bigr)$ and $\bigl(N_{{\bf k}_{1}} - N_{{\bf k}_{2}}\hspace{0.01cm}\bigr)$, and therefore they are equal to zero, which leaves us with
\[
\frac{d\hspace{0.04cm} \langle\hspace{0.01cm}\mathcal{Q}\hspace{0.03cm}\rangle}{d\hspace{0.03cm} t}
=
2\hspace{0.03cm}N_{c}\hspace{0.04cm}{\mathfrak q}_{2}(t)\!
\!\int\!d\hspace{0.02cm}{\bf k}\, 
W^{\hspace{0.03cm}l}_{\bf k}\,
{\rm Im}\hspace{0.03cm}\mathscr{T}^{\,(2,\hspace{0.03cm}{\mathcal A})}_{{\bf k},\,{\bf k}}({\bf v}).
\]
Further, let us take into account that the remaining term on the right-hand side is actually related to the collisionless (Landau) damping of the wave oscillations. Therefore the expression ${\rm Im}\hspace{0.03cm}\mathscr{T}^{\,(2,\hspace{0.03cm} {\mathcal A})}_{{\bf k},\,{\bf k}}({\bf v})$ must contain a $\delta$-function which reflects the corresponding conservation laws for energy and momentum:
\[
{\rm Im}\hspace{0.03cm}\mathscr{T}^{\,(2,\hspace{0.03cm}{\mathcal A})}_{{\bf k},\,{\bf k}}({\bf v})
\sim
\int\!\frac{d\vspace{0.4cm}\Omega_{{\bf v}^{\prime}}}{4\pi}\,
w_{{\bf v}^{\prime}}({\bf v}, {\bf k})\hspace{0.03cm}
(2\pi)\hspace{0.03cm}\delta(\omega^{l}_{{\bf k}} - {\mathbf v}^{\prime}\cdot{\bf k}),
\]
where the probability $w_{{\bf v}^{\prime}}({\bf v}, {\bf k})$ for the Landau damping process can be determined using explicit expressions for the scattering amplitude (\ref{eq:7r}), the three-point amplitude ${\mathcal V}_{{\bf k},\, {\bf k}_{1},\, {\bf k}_{2}}$, Eq.\,(\ref{ap:A1}), and the HTL-correction $\delta\hspace{0.025cm} \Gamma^{\mu\hspace{0.02cm}\nu\rho}(k, k_{1}, k_{2})$, Eq.\,(\ref{ap:A6}). However, as is well known, the {\it linear} Landau damping is kinematically forbidden in a hot quark-gluon plasma and therefore, this term can be setting zero and thus finally we obtain 
\[
\frac{d\hspace{0.04cm} \langle\hspace{0.01cm}\mathcal{Q}\hspace{0.03cm}\rangle}{d\hspace{0.03cm} t}
= 0,
\]
i.e., 
\begin{equation}
\langle\hspace{0.01cm}\mathcal{Q}\hspace{0.03cm}\rangle = {const}.
\label{eq:10g}
\end{equation}

\section{\bf Equation for the averaged color charge $\langle\hspace{0.01cm}\mathcal{Q}^{\hspace{0.03cm}s}\hspace{0.03cm}\rangle$}
\label{section_11}
\setcounter{equation}{0}

We now turn our attention to the derivation of the equation of motion for the color charge $\langle\hspace{0.01cm}\mathcal{Q}^{\hspace{0.03cm}s}\hspace{0.03cm}\rangle$.
For this purpose, we contract the left and right-hand sides of (\ref{eq:10q}) with the matrix $(t^{\,s})^{i\hspace{0.04cm}i^{\prime}}$.
Taking into account the trace (\ref{eq:10e}), we find in this case instead of (\ref{eq:10q})
\begin{equation}
n_{\bf p}\,
\frac{d\hspace{0.04cm} \langle\hspace{0.01cm}\mathcal{Q}^{\hspace{0.03cm}s}\hspace{0.03cm}\rangle}{d\hspace{0.03cm} t}
=
-\hspace{0.03cm}i\hspace{0.03cm}\!\int\!d\hspace{0.02cm}{\bf k}\, 
{\rm tr}\hspace{0.03cm}\bigl({\mathcal N}_{{\bf k}}\hspace{0.03cm}T^{\,d}\hspace{0.03cm}\bigr)
\Bigl\{\hspace{0.03cm}\mathscr{T}^{\,(2,\hspace{0.03cm}{\mathcal A})}_{{\bf k},\,{\bf k}}({\bf v})\hspace{0.03cm}
{\rm tr}\hspace{0.03cm}\bigl(t^{\,s}\hspace{0.03cm}t^{\,d}\hspace{0.03cm}
{\mathfrak n}_{{\bf p}}\bigr)
-\,
\mathscr{T}^{\,\ast\,(2,\hspace{0.03cm}{\mathcal A})}_{{\bf k},\,{\bf k}}({\bf v})\hspace{0.03cm}{\rm tr}\hspace{0.03cm}
\bigl(t^{\,d}\hspace{0.03cm}t^{\,s}\hspace{0.03cm}{\mathfrak n}_{{\bf p}}\bigr)\!
\Bigr\}
\label{eq:11w}
\end{equation}
\[
+\,
i\hspace{0.04cm} \!\int\!d{\bf k}_{1}\hspace{0.03cm}d{\bf k}_{2}\,
\bigl|\hspace{0.02cm}\mathscr{T}^{\,(2,\hspace{0.03cm}{\mathcal A})}_{{\bf k}_{1},\,{\bf k}_{2}}({\bf v})\hspace{0.02cm}\bigr|^{2}
\]
\[
\times\hspace{0.03cm}
\biggl\{\frac{1}{\Delta\hspace{0.02cm}\omega_{\hspace{0.03cm}{\bf p},\, {\bf p}, {\bf k}_{1}, {\bf k}_{2}} -\, i\hspace{0.03cm}0}\,
\Bigl(\Bigl[{\rm tr}\hspace{0.03cm}
\bigl(\hspace{0.03cm}t^{\,e}\hspace{0.03cm}t^{\,s}\hspace{0.03cm}t^{\,d}
\hspace{0.03cm}{\mathfrak n}_{{\bf p}}\hspace{0.03cm}\bigr)
-
{\rm tr}\hspace{0.03cm}
\bigl(\hspace{0.03cm}t^{\,s}\hspace{0.03cm}t^{\,d}\hspace{0.03cm}t^{\,e}
\hspace{0.03cm}{\mathfrak n}_{{\bf  p}}\hspace{0.03cm}\bigr)
\hspace{0.03cm}\Bigr]
{\rm tr}\hspace{0.03cm}\bigl(T^{\,d}{\mathcal N}_{{\bf k}_{1}} T^{\,e} {\mathcal N}_{{\bf k}_{2}} \hspace{0.03cm}\bigr)
\]
\[
-\;
{\rm tr}\hspace{0.03cm}
\bigl(t^{\,s}\hspace{0.03cm}t^{\,d}\hspace{0.03cm} {\mathfrak n}_{{\bf p}}\hspace{0.03cm}t^{\,e}\hspace{0.03cm} {\mathfrak n}_{{\bf p}}\bigr)
\Bigl[\,{\rm tr}\hspace{0.03cm}
\bigl(T^{\,d}T^{\,e} {\mathcal N}_{{\bf k}_{2}}\hspace{0.03cm}\bigr)
\!-\,
{\rm tr}\hspace{0.03cm}\bigl(T^{\,e}T^{\,d}{\mathcal N}_{{\bf k}_{1}}\hspace{0.03cm}\bigr)\Bigr]\Bigr)
\vspace{0.15cm}
\]
\[
-\,
\frac{1}{\Delta\hspace{0.02cm}\omega_{\hspace{0.03cm}{\bf p},\, {\bf p}, {\bf k}_{1}, {\bf k}_{2}} +\, i\hspace{0.03cm}0}\,
\Bigl(\Bigl[{\rm tr}\hspace{0.03cm}
\bigl(t^{\,e}\hspace{0.03cm}t^{\,s}\hspace{0.03cm}t^{\,d}\hspace{0.03cm}{\mathfrak n}_{{\bf p}}\hspace{0.03cm}\bigr)
-
{\rm tr}\hspace{0.03cm}
\bigl(t^{\,d}\hspace{0.03cm}t^{\,e}\hspace{0.03cm}t^{\,s}\hspace{0.03cm}{\mathfrak n}_{{\bf p}}\hspace{0.03cm}\bigr)\hspace{0.03cm}\Bigr]
{\rm tr}\hspace{0.03cm}\bigl(T^{\,d}{\mathcal N}_{{\bf k}_{1}} T^{\,e} {\mathcal N}_{{\bf k}_{2}} \hspace{0.03cm}\bigr)
\]
\[
-\,
{\rm tr}\hspace{0.03cm}
\bigl(t^{\,e}\hspace{0.01cm}t^{\,s}\hspace{0.03cm} {\mathfrak n}_{{\bf p}}\hspace{0.03cm}t^{\,d}\hspace{0.03cm}{\mathfrak n}_{{\bf p}}\bigr)
\Bigl[\,{\rm tr}\hspace{0.03cm}
\bigl(T^{\,d}\hspace{0.03cm}T^{\,e} {\mathcal N}_{{\bf k}_{2}} \hspace{0.03cm}\bigr)
\!-\,
{\rm tr}\hspace{0.03cm}\bigl(T^{\,e}\hspace{0.03cm}
T^{\,d}{\mathcal N}_{{\bf k}_{1}}\hspace{0.03cm}\bigr)\Bigr]\Bigr)\biggr\}.
\]
\indent Let us analyze the first term on the right-hand side of Eq.\,(\ref{eq:11w}). Using the formula (\ref{ap:B1}) for the first trace in this term we have
\begin{equation}
{\rm tr}\hspace{0.03cm}\bigl(t^{\,s}\hspace{0.03cm}t^{\,d}\hspace{0.03cm}
{\mathfrak n}_{{\bf p}}\bigr)
=
n_{\bf p}\hspace{0.03cm} \bigl(t^{\,s}\hspace{0.03cm}t^{\,d}\hspace{0.03cm}\bigr)^{\hspace{0.01cm}i\hspace{0.03cm}i^{\hspace{0.03cm}\prime}}\!
\bigl\langle\hspace{0.03cm}\theta^{\hspace{0.03cm}\ast\ \!\!i}\hspace{0.04cm}\theta^{\phantom{\ast}\!\!i^{\prime}}\hspace{0.03cm}\bigr\rangle
=
\biggl\{
\frac{1}{2\hspace{0.03cm}N_{c}}\,\delta^{\hspace{0.03cm}s\hspace{0.03cm}d}
\hspace{0.03cm}\bigl\langle\hspace{0.01cm}{\mathcal Q}\hspace{0.03cm}
\bigr\rangle
+
\frac{1}{2}\,\bigl(d^{\hspace{0.03cm}s\hspace{0.02cm}d\hspace{0.03cm}e}
+
i\hspace{0.03cm}
f^{\hspace{0.03cm}s\hspace{0.02cm}d\hspace{0.03cm}e}\hspace{0.03cm}\bigr)
\hspace{0.03cm}\bigl\langle\hspace{0.01cm}{\mathcal Q}^{\hspace{0.03cm}e}\hspace{0.03cm}
\bigr\rangle
\biggr\}\hspace{0.03cm}n_{\bf p}.
\label{eq:11e}
\end{equation}
The second trace ${\rm tr}\hspace{0.03cm}\bigl(t^{\,d}\hspace{0.03cm}t^{\,s}\hspace{0.03cm}
{\mathfrak n}_{{\bf p}}\bigr)$ trivially follows from (\ref{eq:11e}) by rearranging the indices $s\rightleftharpoons d$.\\
\indent Further taking into account the already known equality
\[
{\rm tr}\hspace{0.03cm}\bigl({\mathcal N}_{{\bf k}}\hspace{0.03cm}T^{\,d}\hspace{0.03cm}\bigr)
=
N_{c}\hspace{0.03cm}
\bigl\langle\hspace{0.01cm}{\mathcal Q}^{\hspace{0.03cm}d}\hspace{0.03cm}
\bigr\rangle
\hspace{0.03cm}W^{l}_{\bf k},
\]
it is easy to see that the integrand in the first contribution to (\ref{eq:11w}) can be represented in the following form:
\begin{equation*}
\begin{split}
&{\rm tr}\hspace{0.03cm}\bigl({\mathcal N}_{{\bf k}}\hspace{0.03cm}T^{\,d}\hspace{0.03cm}\bigr)
	\Bigl\{\hspace{0.03cm}\mathscr{T}^{\,(2,\hspace{0.03cm}{\mathcal A})}_{{\bf k},\,{\bf k}}({\bf v})\hspace{0.04cm}
	{\rm tr}\hspace{0.03cm}\bigl(t^{\,s}\hspace{0.03cm}t^{\,d}\hspace{0.03cm}
	{\mathfrak n}_{{\bf p}}\bigr)
	-\,
	\mathscr{T}^{\,\ast\,(2,\hspace{0.03cm}{\mathcal A})}_{{\bf k},\,{\bf k}}({\bf v})\hspace{0.04cm}{\rm tr}\hspace{0.03cm}
	\bigl(t^{\,d}\hspace{0.03cm}t^{\,s}\hspace{0.03cm}{\mathfrak n}_{{\bf p}}\bigr)\!
	\Bigr\}\\[1.3ex]
&=
i\hspace{0.03cm}n_{\bf p}\hspace{0.04cm}
{\rm Im}\hspace{0.03cm}\mathscr{T}^{\,(2,\hspace{0.03cm}{\mathcal A})}_{{\bf k},\,{\bf k}}({\bf v})
\hspace{0.03cm}W^{\hspace{0.03cm}l}_{\bf k}\,
\Bigl\{
\bigl\langle\hspace{0.01cm}\mathcal{Q}\hspace{0.03cm}\bigr\rangle
\bigl\langle\hspace{0.01cm}\mathcal{Q}^{\hspace{0.03cm}s}\hspace{0.03cm}\bigr\rangle
+
N_{c}\hspace{0.04cm}d^{\hspace{0.03cm}s\hspace{0.02cm}d\hspace{0.03cm}e}
\hspace{0.02cm}
\bigl\langle\hspace{0.01cm}\mathcal{Q}^{\hspace{0.03cm}d}\hspace{0.03cm}
\bigr\rangle
\bigl\langle\hspace{0.01cm}\mathcal{Q}^{\hspace{0.03cm}e}\hspace{0.03cm}
\bigr\rangle\Bigr\}.
\end{split}
\end{equation*}
\indent We proceed to the analysis of the traces in the second term on the right-hand side (\ref{eq:11w}). Our first step is to consider the following expression
\begin{equation}
\Bigl[{\rm tr}\hspace{0.03cm}
\bigl(t^{\,e}\hspace{0.03cm}t^{\,s}\hspace{0.03cm}t^{\,d}\hspace{0.03cm}
{\mathfrak n}_{{\bf p}}\hspace{0.03cm}\bigr)
-
{\rm tr}\hspace{0.03cm}
\bigl(t^{\,s}\hspace{0.03cm}t^{\,d}\hspace{0.03cm}t^{\,e}\hspace{0.03cm}
{\mathfrak n}_{{\bf p}}\hspace{0.03cm}\bigr)\hspace{0.03cm}\Bigr]
{\rm tr}\hspace{0.03cm}\bigl(T^{\,d}{\mathcal N}_{{\bf k}_{1}} T^{\,e} {\mathcal N}_{{\bf k}_{2}} \hspace{0.03cm}\bigr)
\label{eq:11t}
\end{equation}
\[
=
\Bigl[{\rm tr}\hspace{0.03cm}
\bigl(t^{\,e}\hspace{0.03cm}t^{\,s}\hspace{0.03cm}t^{\,d}\hspace{0.03cm}
{\mathfrak n}_{{\bf p}}\hspace{0.03cm}\bigr)
-
{\rm tr}\hspace{0.03cm}
\bigl(t^{\,s}t^{\,d}\hspace{0.03cm}t^{\,e}\hspace{0.03cm}{\mathfrak n}_{{\bf p}}\hspace{0.03cm}\bigr)\hspace{0.03cm}\Bigr]
\Bigl\{\delta^{\hspace{0.03cm}e\hspace{0.03cm} d}N_{c}\hspace{0.03cm}  N_{{\bf k}_{1}}N_{{\bf k}_{2}}
+
\frac{i}{2}\,N_{c}\hspace{0.03cm} f^{\hspace{0.03cm}c\hspace{0.03cm}e\hspace{0.03cm}d}\hspace{0.03cm}
\bigl\langle\hspace{0.01cm}\mathcal{Q}^{\hspace{0.03cm}c}\hspace{0.03cm}\bigr\rangle\hspace{0.03cm}
\bigl(W^{\,l}_{{\bf k}_{1}}N^{\,l}_{{\bf k}_{2}} - N^{\,l}_{{\bf k}_{1}}W^{\,l}_{{\bf k}_{2}}\bigr)
\]
\[
+\,
\Bigl(\delta^{\hspace{0.02cm}c\hspace{0.02cm}e}\delta^{\hspace{0.02cm}\rho\hspace{0.02cm} d}
+
\delta^{\hspace{0.02cm}c\hspace{0.02cm}d}\delta^{\hspace{0.02cm}e\hspace{0.02cm}\rho}
+
\frac{1}{4}\,N_{c}\hspace{0.02cm}\Bigl[\bigl\{D^{c},D^{\rho}\bigr\}^{e\hspace{0.02cm}d}
-
d^{\hspace{0.02cm}c\hspace{0.02cm}\rho\lambda}\bigl(D^{\lambda}\bigr)^{e\hspace{0.02cm}d}
\hspace{0.03cm}\Bigr]\Bigr)
\bigl\langle\hspace{0.01cm}\mathcal{Q}^{\hspace{0.03cm}c}\hspace{0.03cm}\bigr\rangle
\bigl\langle\hspace{0.01cm}\mathcal{Q}^{\hspace{0.03cm}\rho}\hspace{0.03cm}\bigr\rangle
W^{\,l}_{{\bf k}_{1}}W^{\,l}_{{\bf k}_{2}}\Bigr\}.
\]
Here, we have used the representation (\ref{eq:8r}) for the matrix function ${\mathcal N}_{{\bf k}}$ and the formulae for the traces (\ref{ap:C4})\,--\,(\ref{ap:C6}). We examine the term in braces with the simplest color structure $\delta^{\hspace{0.03cm}e\hspace{0.03cm}d}$. With allowance made for the relations (\ref{ap:B2}), the difference of traces in the square brackets in this case will be equal to
\[
{\rm tr}\hspace{0.03cm}
\bigl(t^{\,e}t^{\,s}t^{\,e}\hspace{0.03cm}{\mathfrak n}_{{\bf p}}\hspace{0.03cm}\bigr)
-
{\rm tr}\hspace{0.03cm}
\bigl(t^{\,s}t^{\,e}\hspace{0.03cm}t^{\,e}\hspace{0.03cm}{\mathfrak n}_{{\bf p}}\hspace{0.03cm}\bigr)
=
-\frac{1}{2}\,N_{c}\hspace{0.03cm}{\rm tr}\hspace{0.03cm}\bigl(t^{\,s}\hspace{0.03cm}{\mathfrak n}_{{\bf p}}\bigr)
=
-\frac{1}{2}\,n_{\bf p}\hspace{0.03cm}N_{c}\hspace{0.03cm} \bigl\langle\hspace{0.01cm}\mathcal{Q}^{\hspace{0.03cm}s}\hspace{0.03cm}\bigr\rangle.
\]
Thus, the term with $\delta^{\hspace{0.03cm}e\hspace{0.03cm}d}$ takes the form 
\[
-\frac{1}{2}\,n_{\bf p}\hspace{0.03cm}N^{\hspace{0.03cm}2}_{c}\hspace{0.03cm} \bigl\langle\hspace{0.01cm}\mathcal{Q}^{\hspace{0.03cm}s}\hspace{0.03cm}
\bigr\rangle\hspace{0.03cm} N^{\,l}_{{\bf k}_{1}}N^{\,l}_{{\bf k}_{2}}.
\]
\indent Next, we consider the term mixed in $W_{{\bf k}}$ and $N_{{\bf k}}$,  containing the antisymmetric structure constants $f^{\hspace{0.03cm}c\hspace{0.03cm}e\hspace{0.03cm}d}$. 
In this case, it is more convenient to represent the difference of traces in the square brackets as follows: 
\begin{equation}
{\rm tr}\hspace{0.03cm}
\bigl(t^{\,e}t^{\,s}t^{\,d}\hspace{0.03cm}{\mathfrak n}_{{\bf p}}\hspace{0.03cm}\bigr)
-
{\rm tr}\hspace{0.03cm}
\bigl(t^{\,s}t^{\,d}\hspace{0.03cm}t^{\,e}\hspace{0.03cm}
{\mathfrak n}_{{\bf p}}\hspace{0.03cm}\bigr)
\equiv
{\rm tr}\hspace{0.03cm}
\bigl(\bigl[t^{\,e},t^{\,s}\bigr]t^{\,d}\hspace{0.03cm}{\mathfrak n}_{{\bf p}}\hspace{0.03cm}\bigr)
+
{\rm tr}\hspace{0.03cm}
\bigl(t^{\,s}\bigl[t^{\,e},\hspace{0.03cm}t^{\,d}\hspace{0.03cm}\bigr]
\hspace{0.03cm}{\mathfrak n}_{{\bf p}}\hspace{0.03cm}\bigr)
\label{eq:11y}
\end{equation}
\[
=
\frac{1}{2}\,\Bigl[\bigl(T^{\,s}D^{\,\kappa}\bigr)^{e\hspace{0.03cm}d}
-\bigl(T^{\,s}\hspace{0.03cm}T^{\,\kappa}\bigr)^{e\hspace{0.03cm}d}\hspace{0.03cm}\Bigr]
n_{\bf p}\hspace{0.03cm}
\bigl\langle\hspace{0.01cm}\mathcal{Q}^{\hspace{0.03cm}\kappa}\hspace{0.03cm}\bigr\rangle
+
\frac{1}{2}\,if^{\hspace{0.03cm}e\hspace{0.03cm}d\hspace{0.03cm}\lambda}
\bigl(d^{\hspace{0.03cm}s\hspace{0.02cm}\lambda\hspace{0.03cm}\kappa}
+
i\hspace{0.03cm}
f^{\hspace{0.03cm}s\hspace{0.02cm}\lambda\hspace{0.03cm}\kappa}
\hspace{0.03cm}\bigr)n_{\bf p}\hspace{0.03cm}
\bigl\langle\hspace{0.01cm}\mathcal{Q}^{\hspace{0.03cm}\kappa}
\hspace{0.03cm}\bigr\rangle.
\]
Here, we used the equality (\ref{eq:11e}). The contribution with the ``colorless'' charge $\bigl\langle\hspace{0.01cm}{\mathcal Q}\hspace{0.03cm}
\bigr\rangle$ is reduced. If we contract this expression with  $f^{\hspace{0.03cm}c\hspace{0.03cm}e\hspace{0.03cm}d} = -i\hspace{0.03cm}\bigl(T^{\,c}\bigr)^{d\hspace{0.03cm}e}$ and employ the formulae for third-order traces (\ref{ap:C5}), then we obtain   
\[
\Bigl[{\rm tr}\hspace{0.03cm}
\bigl(t^{\,e}t^{\,s}t^{\,d}\hspace{0.03cm}{\mathfrak n}_{{\bf p}}\hspace{0.03cm}\bigr)
-
{\rm tr}\hspace{0.03cm}
\bigl(t^{\,s}t^{\,d}\hspace{0.03cm}t^{\,e}\hspace{0.03cm}
{\mathfrak n}_{{\bf p}}\hspace{0.03cm}\bigr)\hspace{0.03cm}\Bigr]
f^{\hspace{0.03cm}c\hspace{0.03cm}e\hspace{0.03cm}d}
\]
\[
=-\frac{1}{2}\,i\hspace{0.03cm}
\Bigl[\hspace{0.03cm}{\rm tr}\hspace{0.03cm}
\bigl(T^{\,c}\hspace{0.03cm}T^{\,s}D^{\,\kappa}\hspace{0.03cm}\bigr)
-
{\rm tr}\hspace{0.03cm}
\bigl(T^{\,c}\hspace{0.03cm}T^{\,s}\hspace{0.03cm}T^{\,\kappa}\hspace{0.03cm}\bigr)
\hspace{0.03cm}\Bigr]
n_{\bf p}\hspace{0.03cm}
\bigl\langle\hspace{0.01cm}\mathcal{Q}^{\hspace{0.03cm}\kappa}\hspace{0.03cm}\bigr\rangle
+
\frac{1}{2}\,i\hspace{0.03cm}N_{c}\hspace{0.03cm}
\bigl(d^{\hspace{0.03cm}s\hspace{0.02cm}c\hspace{0.03cm}\kappa}
+
i\hspace{0.03cm}
f^{\hspace{0.03cm}s\hspace{0.02cm}c\hspace{0.03cm}\kappa}
\hspace{0.03cm}\bigr)
n_{\bf p}\hspace{0.03cm}
\bigl\langle\hspace{0.01cm}\mathcal{Q}^{\hspace{0.03cm}\kappa}
\hspace{0.03cm}\bigr\rangle
\]
\[
=
\frac{1}{4}\,i\hspace{0.03cm}N_{c}\hspace{0.03cm}
\bigl(d^{\hspace{0.03cm}s\hspace{0.02cm}c\hspace{0.03cm}\kappa}
+
i\hspace{0.03cm}
f^{\hspace{0.03cm}s\hspace{0.02cm}c\hspace{0.03cm}\kappa}
\hspace{0.03cm}\bigr)
n_{\bf p}\hspace{0.03cm}
\bigl\langle\hspace{0.01cm}\mathcal{Q}^{\hspace{0.03cm}\kappa}
\hspace{0.03cm}\bigr\rangle.
\]
The next step is to contract the above expression with the color charge $\bigl\langle\hspace{0.01cm}\mathcal{Q}^{\hspace{0.03cm}c}\hspace{0.03cm}\bigr\rangle$, as is the case of the term in (\ref{eq:11t}), mixed by the functions $W^{\,l}_{{\bf k}}$ and $N^{\,l}_{{\bf k}}$. Then, the contribution of this term takes the final form
\[
-\frac{1}{8}\,n_{\bf p}\hspace{0.03cm}N^{\hspace{0.03cm}2}_{c}\hspace{0.03cm}
d^{\hspace{0.04cm}s\hspace{0.02cm}c\hspace{0.03cm}\kappa}
\bigl\langle\hspace{0.01cm}\mathcal{Q}^{\hspace{0.03cm}c}\hspace{0.03cm}\bigr\rangle
\bigl\langle\hspace{0.01cm}\mathcal{Q}^{\hspace{0.03cm}\kappa}\hspace{0.03cm}\bigr\rangle\hspace{0.03cm}
\bigl(W^{\,l}_{{\bf k}_{1}}N^{\,l}_{{\bf k}_{2}} - N^{\,l}_{{\bf k}_{1}}W^{\,l}_{{\bf k}_{2}}\bigr).
\]
\indent Let us consider the remaining term in (\ref{eq:11t}), proportional to the product of $W^{\,l}_{{\bf k}_{1}}W^{\,l}_{{\bf k}_{2}}$. With the use of the trace difference (\ref{eq:11y}), it can be represented in a somewhat cumbersome form:
\begin{equation}
\biggl\{
\frac{1}{2}\,\Bigl[\bigl(T^{\,s}D^{\,\kappa}\bigr)^{e\hspace{0.03cm}d}
-\bigl(T^{\,s}\hspace{0.03cm}T^{\,\kappa}\bigr)^{e\hspace{0.03cm}d}\hspace{0.03cm}\Bigr]
+
\frac{1}{2}\,if^{\hspace{0.03cm}e\hspace{0.03cm}d\hspace{0.03cm}\lambda}
\bigl(d^{\hspace{0.03cm}s\hspace{0.02cm}\lambda\hspace{0.03cm}\kappa}
+
i\hspace{0.03cm}
f^{\hspace{0.03cm}s\hspace{0.02cm}\lambda\hspace{0.03cm}\kappa}
\hspace{0.03cm}\bigr)\!\biggr\}
\label{eq:11u}
\end{equation}
\[
\times\hspace{0.03cm}
\Bigl(\delta^{\hspace{0.02cm}c\hspace{0.02cm}e}\delta^{\hspace{0.02cm}\rho\hspace{0.02cm} d}
+
\delta^{\hspace{0.02cm}c\hspace{0.02cm}d}\delta^{\hspace{0.02cm}e\hspace{0.02cm}\rho}
+
\frac{1}{4}\,N_{c}\hspace{0.02cm}\Bigl[\bigl\{D^{\,c},
D^{\,\rho}\bigr\}^{e\hspace{0.02cm}d}
-
d^{\hspace{0.02cm}c\hspace{0.02cm}\rho\lambda}
\bigl(D^{\,\lambda}\bigr)^{e\hspace{0.02cm}d}
\Bigr]\Bigr)\hspace{0.03cm}
n_{\bf p}\hspace{0.03cm}
\bigl\langle\hspace{0.01cm}\mathcal{Q}^{\hspace{0.03cm}\kappa}\hspace{0.03cm}\bigr\rangle
\bigl\langle\hspace{0.01cm}\mathcal{Q}^{\hspace{0.03cm}c}\hspace{0.03cm}\bigr\rangle
\bigl\langle\hspace{0.01cm}\mathcal{Q}^{\hspace{0.03cm}\rho}\hspace{0.03cm}\bigr\rangle
W^{\,l}_{{\bf k}_{1}}W^{\,l}_{{\bf k}_{2}}
\]
\[
=\frac{1}{2}\,\biggl\{
\Bigl(-\Bigl[\bigl(T^{\,c}D^{\,\rho}\bigr)^{s\hspace{0.03cm}\kappa}
+
\bigl(T^{\,\rho}D^{\,c}\bigr)^{s\hspace{0.03cm}\kappa}\Bigr]
+
\frac{1}{4}\,N_{c}\hspace{0.02cm}\Bigl[\hspace{0.03cm}
{\rm tr}\hspace{0.04cm}\bigl(T^{\,s}\hspace{0.02cm} D^{\,\kappa}\hspace{0.02cm}\bigl\{D^{\,c},D^{\,\rho}\bigr\}\bigr)
-\,
d^{\hspace{0.03cm}c\hspace{0.02cm}\rho\lambda}\hspace{0.03cm}
{\rm tr}\hspace{0.03cm}\bigl(T^{\,s}\hspace{0.02cm} D^{\,\kappa}D^{\,\lambda}\bigr)
\Bigr]\Bigr)
\]
\[
-\hspace{0.03cm}  \Bigl(\bigl\{T^{\,c},T^{\,\rho}\bigr\}^{s\hspace{0.02cm}\kappa}
+
\frac{1}{4}\,N_{c}\hspace{0.02cm}\Bigl[{\rm tr}\hspace{0.04cm}\bigl(T^{\,s}\hspace{0.02cm} T^{\,\kappa}\hspace{0.02cm}\bigl\{D^{\,c},D^{\,\rho}\bigr\}\!\bigr)
-\,
d^{\hspace{0.03cm}c\hspace{0.02cm}\rho\lambda}\hspace{0.03cm}
{\rm tr}\hspace{0.03cm}\bigl(T^{\,s}\hspace{0.02cm} T^{\,\kappa}D^{\,\lambda}\bigr)
\Bigr]\Bigr)\biggr\}
n_{\bf p}\hspace{0.03cm}
\bigl\langle\hspace{0.01cm}\mathcal{Q}^{\hspace{0.03cm}\kappa}\hspace{0.03cm}\bigr\rangle
\hspace{0.03cm}
\bigl\langle\hspace{0.01cm}\mathcal{Q}^{\hspace{0.03cm}c}\hspace{0.03cm}\bigr\rangle
\hspace{0.03cm}
\bigl\langle\hspace{0.01cm}\mathcal{Q}^{\hspace{0.03cm}\rho}\hspace{0.03cm}\bigr\rangle
W^{\,l}_{{\bf k}_{1}}\!W^{\,l}_{{\bf k}_{2}}.
\]
The expression in parentheses in the last line is exactly the same expression that we obtained in analyzing the fifth-order trace in section \ref{section_9}, Eqs.\,(\ref{eq:9t}) and (\ref{eq:9y}). There, it was shown that this expression vanishes. Let us consider the expression in parentheses in the next-to-last line. We write out this expression once more, setting by virtue of (\ref{ap:C5}) 
\[
{\rm tr}\hspace{0.03cm}\bigl(T^{\,s}\hspace{0.02cm} D^{\,\kappa}D^{\,\lambda}\bigr) 
= 
i\hspace{0.01cm}\left(\displaystyle\frac{N^{\hspace{0.03cm}2}_{c} - 4}{2\hspace{0.03cm}N_{c}}\right)\! f^{\hspace{0.03cm}s\hspace{0.02cm}\kappa\hspace{0.03cm}\lambda},
\] 
then
\begin{equation}
-\Bigl[\bigl(T^{\,c}D^{\,\rho}\bigr)^{s\hspace{0.03cm}\kappa}
+
\bigl(T^{\,\rho}D^{\,c}\bigr)^{s\hspace{0.03cm}\kappa}\Bigr]
+
\frac{1}{4}\,N_{c}\hspace{0.02cm}\Bigl[\hspace{0.03cm}
{\rm tr}\hspace{0.04cm}\bigl(T^{\,s}\hspace{0.02cm} D^{\,\kappa}\hspace{0.02cm}\bigl\{D^{\,c},D^{\,\rho}\bigr\}\bigr)
-\,
i\hspace{0.02cm}
\left(\displaystyle\frac{N^{\hspace{0.03cm}2}_{c} - 4}{2\hspace{0.02cm}N_{c}}\right)\! f^{\hspace{0.03cm}s\hspace{0.02cm}\kappa\hspace{0.03cm}\lambda}
\hspace{0.03cm}d^{\,c\hspace{0.02cm}\rho\lambda}
\Bigr].
\label{eq:11i}
\end{equation}
We calculate the fourth-order trace,  using the representation (\ref{ap:C8}). It takes the form  
\[
{\rm tr}\hspace{0.04cm}\bigl(T^{\,s}\hspace{0.02cm} D^{\,\kappa}\hspace{0.02cm}\bigl\{D^{\,c},D^{\,\rho}\bigr\}\bigr)
=
i\hspace{0.04cm}\biggl(\frac{N^{\hspace{0.02cm}2}_{c} - 12}{2\hspace{0.02cm}N_{c}}\biggr)
f^{\hspace{0.03cm}s\hspace{0.03cm}\kappa\hspace{0.03cm}\lambda}
\hspace{0.03cm}d^{\,c\hspace{0.03cm}\rho\hspace{0.03cm}\lambda}.
\]
According to (\ref{eq:11u}), the expression (\ref{eq:11i}) should be contracted with 
$\bigl\langle\hspace{0.01cm}\mathcal{Q}^{\hspace{0.03cm}\kappa}
\hspace{0.03cm}\bigr\rangle
\hspace{0.03cm}
\bigl\langle\hspace{0.01cm}\mathcal{Q}^{\hspace{0.03cm}c}\hspace{0.03cm}\bigr\rangle
\hspace{0.03cm}
\bigl\langle\hspace{0.01cm}\mathcal{Q}^{\hspace{0.03cm}\rho}
\hspace{0.03cm}\bigr\rangle$. As a result, we have 
\[
\frac{1}{2}\,
\Bigl[\bigl(T^{\,e}\bigr)^{s\hspace{0.03cm}c}\bigl(D^{\,e}\bigr)^{\rho\hspace{0.03cm}\kappa}
+
\bigl(T^{\,e}\bigr)^{s\hspace{0.03cm}\rho}\bigl(D^{\,e}\bigr)^{c\hspace{0.03cm}\kappa}
+
\bigl(T^{\,e}\bigr)^{s\hspace{0.03cm}\kappa}\bigl(D^{\,e}\bigr)^{c\hspace{0.03cm}\rho}
\Bigr]
\bigl\langle\hspace{0.01cm}\mathcal{Q}^{\hspace{0.03cm}\kappa}
\hspace{0.03cm}\bigr\rangle
\hspace{0.03cm}
\bigl\langle\hspace{0.01cm}\mathcal{Q}^{\hspace{0.03cm}c}\hspace{0.03cm}\bigr\rangle
\hspace{0.03cm}
\bigl\langle\hspace{0.01cm}\mathcal{Q}^{\hspace{0.03cm}\rho}
\hspace{0.03cm}\bigr\rangle.
\]
The color structure in the square brackets is zero. It can be easily verified by rewriting it in the following form: 
\[
\bigl(T^{\,e}\bigr)^{s\hspace{0.03cm}c}\bigl(D^{\,e}\bigr)^{\rho\hspace{0.03cm}\kappa}
+
\bigl(T^{\,e}\bigr)^{s\hspace{0.03cm}\rho}\bigl(D^{\,e}\bigr)^{c\hspace{0.03cm}\kappa}
+
\bigl(T^{\,e}\bigr)^{s\hspace{0.03cm}\kappa}\bigl(D^{\,e}\bigr)^{c\hspace{0.03cm}\rho}
\equiv
\bigl[\hspace{0.03cm}T^{\,s},D^{\,\rho}\hspace{0.03cm}\bigr]^{c\kappa} -i\hspace{0.03cm}
f^{\hspace{0.03cm}s\hspace{0.02cm}\rho\hspace{0.03cm}e}
\hspace{0.01cm}\bigl(D^{\,e}\bigr)^{c\hspace{0.03cm}\kappa}
\]
and making use of the second relation in (\ref{ap:C3}) from the Appendix \ref{appendix_C}. Thus, the contribution proportional to the product $W^{\,l}_{{\bf k}_{1}}W^{\,l}_{{\bf k}_{2}}$ completely drops out of consideration. Collecting all the calculated expressions, instead of (\ref{eq:11t}), we finally find 
\begin{equation}
	\Bigl[{\rm tr}\hspace{0.03cm}
	\bigl(t^{\,e}\hspace{0.03cm}t^{\,s}\hspace{0.03cm}t^{\,d}\hspace{0.03cm}
	{\mathfrak n}_{{\bf p}}\hspace{0.03cm}\bigr)
	-
	{\rm tr}\hspace{0.03cm}
	\bigl(t^{\,s}\hspace{0.03cm}t^{\,d}\hspace{0.03cm}t^{\,e}\hspace{0.03cm}
	{\mathfrak n}_{{\bf p}}\hspace{0.03cm}\bigr)\hspace{0.03cm}\Bigr]
	{\rm tr}\hspace{0.03cm}\bigl(T^{\,d}{\mathcal N}_{{\bf k}_{1}} T^{\,e} {\mathcal N}_{{\bf k}_{2}} \hspace{0.03cm}\bigr)
	\label{eq:11o}
\end{equation}
\[
=
-\frac{1}{2}\,n_{\bf p}\hspace{0.03cm}N^{\hspace{0.03cm}2}_{c}\hspace{0.03cm} \bigl\langle\hspace{0.01cm}\mathcal{Q}^{\hspace{0.03cm}s}\hspace{0.03cm}
\bigr\rangle\hspace{0.03cm} N_{{\bf k}_{1}}N_{{\bf k}_{2}}
-
\frac{1}{8}\,n_{\bf p}\hspace{0.03cm}N^{\hspace{0.03cm}2}_{c}\hspace{0.03cm}
d^{\hspace{0.04cm}s\hspace{0.02cm}c\hspace{0.03cm}\kappa}
\bigl\langle\hspace{0.01cm}\mathcal{Q}^{\hspace{0.03cm}c}\hspace{0.03cm}\bigr\rangle
\hspace{0.03cm}
\bigl\langle\hspace{0.01cm}\mathcal{Q}^{\hspace{0.03cm}\kappa}\hspace{0.03cm}\bigr\rangle\hspace{0.03cm}
\bigl(W^{\,l}_{{\bf k}_{1}}N^{\,l}_{{\bf k}_{2}} - N^{\,l}_{{\bf k}_{1}}W^{\,l}_{{\bf k}_{2}}\bigr).
\]
\indent We proceed now to the consideration of the other expression in the second term in (\ref{eq:11w}), with a different color structure. This expression in view of the decomposition (\ref{eq:8r}) and the traces (\ref{ap:C4}) and (\ref{ap:C5}), can be represented as follows
\begin{equation}
{\rm tr}\hspace{0.03cm}
\bigl(t^{\,s}\hspace{0.03cm}t^{\,d}\hspace{0.03cm} {\mathfrak n}_{{\bf p}}t^{\,e}\hspace{0.03cm} {\mathfrak n}_{{\bf p}}\bigr)
\Bigl[\,{\rm tr}\hspace{0.03cm}
\bigl(T^{\,d}\hspace{0.03cm}T^{\,e} {\mathcal N}_{{\bf k}_{2}} \hspace{0.03cm}\bigr)
-
{\rm tr}\hspace{0.03cm}\bigl(T^{\,e}\hspace{0.03cm}T^{\,d}
{\mathcal N}_{{\bf k}_{1}}\hspace{0.01cm}\bigr)\Bigr]
\label{eq:11p}
\end{equation}
\[
=
{\rm tr}\hspace{0.03cm}
\bigl(t^{\,s}\hspace{0.03cm}t^{\,d}\hspace{0.03cm} {\mathfrak n}_{{\bf p}}t^{\,e}\hspace{0.03cm} {\mathfrak n}_{{\bf p}}\bigr)
\Bigl[\hspace{0.03cm} \delta^{\,d\hspace{0.03cm}e}\hspace{0.03cm}N_{c}\hspace{0.03cm}
\bigl(N^{\,l}_{{\bf k}_{2}} - N^{\,l}_{{\bf k}_{1}}\hspace{0.01cm}\bigr)
+
\frac{1}{2}\,
i\hspace{0.03cm}N_{c}\hspace{0.03cm} f^{\,d\hspace{0.02cm}e\hspace{0.03cm}c}
\hspace{0.03cm}\bigl(W^{\,l}_{{\bf k}_{2}} + 
W^{\,l}_{{\bf k}_{1}}\hspace{0.01cm}\bigr) 
\bigl\langle\hspace{0.01cm}\mathcal{Q}^{\hspace{0.03cm}c}\hspace{0.03cm}\bigr\rangle\Bigr].
\]
As usual, the first step is to analyze the contribution proportional to the trivial color structure $\delta^{\hspace{0.03cm}d\hspace{0.03cm}e}$. Taking into account the relations (\ref{ap:B5}) and (\ref{ap:B7}), we have the following chain of transformations:
\begin{equation}
{\rm tr}\hspace{0.03cm}
\bigl(t^{\,s}\hspace{0.03cm}t^{\,e}\hspace{0.03cm} {\mathfrak n}_{{\bf p}}\hspace{0.03cm}t^{\,e}\hspace{0.03cm}{\mathfrak n}_{{\bf p}}\bigr)
=
\frac{1}{2}\,{\rm tr}\hspace{0.03cm}\bigl(t^{\,s}\hspace{0.03cm}{\mathfrak n}_{{\bf p}}\bigr)\hspace{0.03cm}
{\rm tr}\hspace{0.03cm}\bigl({\mathfrak n}_{{\bf p}}\bigr)
-
\frac{1}{2\hspace{0.02cm}N_{c}}\,{\rm tr}\hspace{0.03cm}\bigl({\mathfrak n}_{{\bf p}}\hspace{0.03cm}t^{\,s}\hspace{0.03cm}{\mathfrak n}_{{\bf p}}\bigr)
\label{eq:11a}
\end{equation}
\[
=
\frac{1}{2}\,
(n_{\bf p})^{2}\hspace{0.03cm}
\hspace{0.03cm}\biggl\{
\bigl\langle\hspace{0.01cm}\mathcal{Q}^{\hspace{0.03cm}s}\hspace{0.03cm}\bigr\rangle
\hspace{0.03cm}\bigl\langle\hspace{0.01cm}\mathcal{Q}\hspace{0.03cm}\bigr\rangle
-
\frac{1}{2\hspace{0.02cm}N_{c}}\,
\biggl(
\frac{4}{N_{c}}\,
\bigl\langle\hspace{0.01cm}\mathcal{Q}^{\hspace{0.03cm}s}\hspace{0.03cm}
\bigr\rangle
\hspace{0.03cm}\bigl\langle\hspace{0.01cm}\mathcal{Q}\hspace{0.03cm}\bigr
\rangle
+
2\hspace{0.04cm}d^{\,s\hspace{0.03cm}d\hspace{0.03cm}e}
\hspace{0.03cm}
\bigl\langle\hspace{0.01cm}\mathcal{Q}^{\hspace{0.03cm}d}\hspace{0.03cm}
\bigr\rangle
\hspace{0.03cm}\bigl\langle\hspace{0.01cm}\mathcal{Q}^{\hspace{0.03cm}
e}\hspace{0.03cm}\bigr\rangle
\biggr)\biggr\}
\]
\[
=
\frac{1}{2}\,
(n_{\bf p})^{2}\hspace{0.03cm}
\hspace{0.03cm}\biggl\{
\left(\displaystyle\frac{N^{\hspace{0.03cm}2}_{c} - 2} {N^{\hspace{0.03cm}2}_{c}}\right)\!
\bigl\langle\hspace{0.01cm}\mathcal{Q}^{\hspace{0.03cm}s}\hspace{0.03cm}
\bigr\rangle
\hspace{0.03cm}\bigl\langle\hspace{0.01cm}\mathcal{Q}\hspace{0.03cm}\bigr
\rangle
-
\frac{1}{N_{c}}\,
d^{\,s\hspace{0.03cm}d\hspace{0.03cm}e}
\hspace{0.03cm}
\bigl\langle\hspace{0.01cm}\mathcal{Q}^{\hspace{0.03cm}d}\hspace{0.03cm}
\bigr\rangle
\hspace{0.03cm}\bigl\langle\hspace{0.01cm}\mathcal{Q}^{\hspace{0.03cm}
e}\hspace{0.03cm}\bigr\rangle
\biggr\}.
\]
Our next task is to consider the term in (\ref{eq:11p}) with the antisymmetric structure constants $f^{\hspace{0.03cm}d\hspace{0.02cm}e\hspace{0.03cm}c}$. Here, we need the relation (\ref{ap:B6_a}). Then, by the use of (\ref{eq:10e}) and (\ref{eq:11e}), we find
\[
{\rm tr}\hspace{0.03cm}
\bigl(t^{\,d}\hspace{0.03cm}t^{\,s}\hspace{0.03cm} {\mathfrak n}_{{\bf p}}\hspace{0.03cm}t^{\,e}\hspace{0.03cm}{\mathfrak n}_{{\bf p}}\bigr)
f^{\hspace{0.03cm}d\hspace{0.02cm}e\hspace{0.03cm}c}
=
-\hspace{0.03cm}\frac{i}{2}\,\Bigl\{{\rm tr}\hspace{0.03cm}\bigl(t^{\,s}\hspace{0.03cm}
{\mathfrak n}_{{\bf p}}\bigr)\hspace{0.03cm}
{\rm tr}\hspace{0.03cm}\bigl(t^{\,c}\hspace{0.03cm}
{\mathfrak n}_{{\bf p}}\bigr)
-
{\rm tr}\hspace{0.03cm}\bigl(t^{\,s}\hspace{0.03cm}t^{\,c}\hspace{0.03cm}{\mathfrak n}_{{\bf p}}\bigr)
{\rm tr}\hspace{0.03cm}\bigl({\mathfrak n}_{{\bf p}}\bigr)\Bigr\}
\]
\[
=
-\hspace{0.03cm}\frac{i}{2}\,
(n_{\bf p})^{2}\hspace{0.03cm}
\hspace{0.03cm}\biggl\{
\bigl\langle\hspace{0.01cm}\mathcal{Q}^{\hspace{0.03cm}s}\hspace{0.03cm}
\bigr\rangle
\hspace{0.03cm}\bigl\langle\hspace{0.01cm}\mathcal{Q}^{\hspace{0.03cm}c}
\hspace{0.03cm}\bigr\rangle
-
\biggl(
\frac{1}{2\hspace{0.03cm}N_{c}}\,\delta^{\hspace{0.03cm}s\hspace{0.03cm}c}
\hspace{0.03cm}\bigl\langle\hspace{0.01cm}{\mathcal Q}\hspace{0.03cm}
\bigr\rangle
+
\frac{1}{2}\,\bigl(d^{\hspace{0.03cm}s\hspace{0.02cm}c\hspace{0.03cm}e}
+
i\hspace{0.03cm}
f^{\hspace{0.03cm}s\hspace{0.02cm}c\hspace{0.03cm}e}\hspace{0.03cm}\bigr)
\hspace{0.03cm}\bigl\langle\hspace{0.01cm}{\mathcal Q}^{\hspace{0.03cm}e}\hspace{0.03cm}
\bigr\rangle
\biggr)\hspace{0.03cm}\bigl\langle\hspace{0.01cm}{\mathcal Q}\hspace{0.03cm}
\bigr\rangle\biggr\}.
\]
By contracting the obtained expression with $\frac{1}{2}\,
i\hspace{0.03cm}N_{c}\hspace{0.03cm}\hspace{0.03cm}\bigl\langle\hspace{0.01cm}\mathcal{Q}^{\hspace{0.03cm}c}
\hspace{0.03cm}\bigr\rangle$ and adding to (\ref{eq:11a}), we finally obtain, instead of (\ref{eq:11p}),
\begin{equation}
	{\rm tr}\hspace{0.03cm}
	\bigl(t^{\,s}\hspace{0.03cm}t^{\,d}\hspace{0.03cm} {\mathfrak n}_{{\bf p}}t^{\,e}\hspace{0.03cm} {\mathfrak n}_{{\bf p}}\bigr)
	\Bigl[\,{\rm tr}\hspace{0.03cm}
	\bigl(T^{\,d}T^{\,e} {\mathcal N}_{{\bf k}_{2}} \hspace{0.03cm}\bigr)
	-
	{\rm tr}\hspace{0.03cm}\bigl(T^{\,e}T^{\,d}{\mathcal N}_{{\bf k}_{1}}\hspace{0.01cm}\bigr)\Bigr]
	\label{eq:11s}
\end{equation}
\[
=
\frac{1}{2}\,N_{c}\hspace{0.03cm}
(n_{\bf p})^{2}\hspace{0.03cm}\biggl[
\hspace{0.03cm}\biggl\{\!\hspace{0.03cm}
\left(\displaystyle\frac{N^{\hspace{0.03cm}2}_{c} - 2} {N^{\hspace{0.03cm}2}_{c}}\right)\!
\bigl\langle\hspace{0.01cm}\mathcal{Q}^{\hspace{0.03cm}s}\hspace{0.03cm}
\bigr\rangle
\hspace{0.03cm}\bigl\langle\hspace{0.01cm}\mathcal{Q}\hspace{0.03cm}\bigr
\rangle
-
\frac{1}{N_{c}}\,
d^{\,s\hspace{0.03cm}d\hspace{0.03cm}e}
\hspace{0.03cm}
\bigl\langle\hspace{0.01cm}\mathcal{Q}^{\hspace{0.03cm}d}\hspace{0.03cm}
\bigr\rangle
\hspace{0.03cm}\bigl\langle\hspace{0.01cm}\mathcal{Q}^{\hspace{0.03cm}
e}\hspace{0.03cm}\bigr\rangle
\biggr\}
\bigl(N^{\,l}_{{\bf k}_{2}} - N^{\,l}_{{\bf k}_{1}}\hspace{0.01cm}\bigr)
\]
\[
+\,
\frac{1}{2}\,
\hspace{0.03cm}\biggl\{
\bigl\langle\hspace{0.01cm}\mathcal{Q}^{\hspace{0.03cm}s}\hspace{0.03cm}
\bigr\rangle
\hspace{0.03cm}\bigl\langle\hspace{0.01cm}\mathcal{Q}^{\hspace{0.03cm}e}
\hspace{0.03cm}\bigr\rangle
\hspace{0.03cm}\bigl\langle\hspace{0.01cm}\mathcal{Q}^{\hspace{0.03cm}e}
\hspace{0.03cm}\bigr\rangle
-
\frac{1}{2}\,\biggl(
\frac{1}{N_{c}}\,\hspace{0.03cm}\bigl\langle\hspace{0.01cm}
\mathcal{Q}^{\hspace{0.03cm}s}
\hspace{0.03cm}\bigr\rangle
\hspace{0.03cm}\bigl\langle\hspace{0.01cm}{\mathcal Q}\hspace{0.03cm}
\bigr\rangle
+
d^{\,s\hspace{0.02cm}d\hspace{0.03cm}e}
\hspace{0.03cm}\bigl\langle\hspace{0.01cm}\mathcal{Q}^{\hspace{0.03cm}d}
\hspace{0.03cm}\bigr\rangle
\hspace{0.03cm}\bigl\langle\hspace{0.01cm}{\mathcal Q}^{\hspace{0.03cm}e}\hspace{0.03cm}
\bigr\rangle
\biggr)\hspace{0.03cm}\bigl\langle\hspace{0.01cm}{\mathcal Q}\hspace{0.03cm}
\bigr\rangle\biggr\}
\hspace{0.03cm}\bigl(W^{\,l}_{{\bf k}_{2}} + 
W^{\,l}_{{\bf k}_{1}}\hspace{0.01cm}\bigr) 
\biggr].
\]
\indent It remains for us to compute the remaining expressions with traces on the right side of equation (\ref{eq:11w}), namely
\[
\Bigl[{\rm tr}\hspace{0.03cm}
\bigl(t^{\,e}\hspace{0.03cm}t^{\,s}\hspace{0.03cm}t^{\,d}\hspace{0.03cm}{\mathfrak n}_{{\bf p}}\hspace{0.03cm}\bigr)
-
{\rm tr}\hspace{0.03cm}
\bigl(t^{\,d}\hspace{0.03cm}t^{\,e}\hspace{0.03cm}t^{\,s}\hspace{0.03cm}{\mathfrak n}_{{\bf p}}\hspace{0.03cm}\bigr)\hspace{0.03cm}\Bigr]
{\rm tr}\hspace{0.03cm}\bigl(T^{\,d}{\mathcal N}_{{\bf k}_{1}} T^{\,e} {\mathcal N}_{{\bf k}_{2}} \hspace{0.03cm}\bigr)
\]
and
\[
{\rm tr}\hspace{0.03cm}
\bigl(\hspace{0.03cm}t^{\,e}\hspace{0.03cm}t^{\,s}\hspace{0.03cm} {\mathfrak n}_{{\bf p}}\hspace{0.03cm}t^{\,d}\hspace{0.03cm} {\mathfrak n}_{{\bf p}}\bigr)
\Bigl[\,{\rm tr}\hspace{0.03cm}
\bigl(T^{\,d}T^{\,e} {\mathcal N}_{{\bf k}_{2}} \hspace{0.03cm}\bigr)
\!-\,
{\rm tr}\hspace{0.03cm}\bigl(T^{\,e}T^{\,d}{\mathcal N}_{{\bf k}_{1}}\hspace{0.03cm}\bigr)\Bigr].
\]
The calculation of the former gives us the expression (\ref{eq:11o}), while for the latter we have (\ref{eq:11s}). Taking into account all the above calculations, using Sokhotsky's formula (\ref{eq:6y}), instead of (\ref{eq:11w}), we get the following equation for the averaged color charge  $\bigl\langle\hspace{0.01cm}\mathcal{Q}^{\hspace{0.03cm}s}\hspace{0.03cm}
\bigr\rangle$:
\begin{equation}
n_{\bf p}\hspace{0.03cm}
\frac{d\hspace{0.04cm} \langle\hspace{0.01cm}\mathcal{Q}^{\hspace{0.03cm}s}\hspace{0.03cm}\rangle}{d\hspace{0.03cm}t}
=
n_{\bf p}\!\int\!d\hspace{0.02cm}{\bf k}\, 
{\rm Im}\hspace{0.03cm}\mathscr{T}^{\,(2,\hspace{0.03cm}{\mathcal A})}_{{\bf k},\,{\bf k}}({\bf v})
\hspace{0.03cm}W^{\hspace{0.03cm}l}_{\bf k}\,
\Bigl\{
\bigl\langle\hspace{0.01cm}\mathcal{Q}\hspace{0.03cm}\bigr\rangle
\hspace{0.03cm}
\bigl\langle\hspace{0.01cm}\mathcal{Q}^{\hspace{0.03cm}s}\hspace{0.03cm}
\bigr\rangle
+
N_{c}\hspace{0.04cm}d^{\hspace{0.03cm}s\hspace{0.02cm}d\hspace{0.03cm}e}
\hspace{0.02cm}
\bigl\langle\hspace{0.01cm}\mathcal{Q}^{\hspace{0.03cm}d}\hspace{0.03cm}
\bigr\rangle
\hspace{0.03cm}
\bigl\langle\hspace{0.01cm}\mathcal{Q}^{\hspace{0.03cm}e}\hspace{0.03cm}
\bigr\rangle\Bigr\}
\label{eq:11d}
\end{equation}
\[
+\hspace{0.03cm}\frac{1}{2}\,N^{\hspace{0.02cm}2}_{c}\hspace{0.04cm}
n_{\bf p}\!\int\!d{\bf k}_{1}\hspace{0.03cm}d{\bf k}_{2}\,
\bigl|\hspace{0.02cm}\mathscr{T}^{\,(2,\hspace{0.03cm}{\mathcal A})}_{{\bf k}_{1},\,{\bf k}_{2}}({\bf v})\hspace{0.02cm}\bigr|^{\hspace{0.03cm}2}
\hspace{0.03cm}
\Bigl\{
\hspace{0.03cm} \bigl\langle\hspace{0.01cm}\mathcal{Q}^{\hspace{0.03cm}s}\hspace{0.03cm}
\bigr\rangle\hspace{0.03cm} N_{{\bf k}_{1}}N_{{\bf k}_{2}}
+
\frac{1}{4}\,
d^{\,s\hspace{0.02cm}d\hspace{0.03cm}e}
\bigl\langle\hspace{0.01cm}\mathcal{Q}^{\hspace{0.03cm}d}\hspace{0.03cm}
\bigr\rangle
\hspace{0.03cm}
\bigl\langle\hspace{0.01cm}\mathcal{Q}^{\hspace{0.03cm}e}\hspace{0.03cm}
\bigr\rangle\hspace{0.03cm}
\bigl(W^{\,l}_{{\bf k}_{1}}N^{\,l}_{{\bf k}_{2}} - N^{\,l}_{{\bf k}_{1}}W^{\,l}_{{\bf k}_{2}}\bigr)\Bigr\}
\]
\[
\times\,
\hspace{0.03cm}
(2\pi)\,\delta(\omega^{l}_{{\bf k}_{1}} - \omega^{l}_{{\bf k}_{2}} - {\mathbf v}\cdot({\bf k}_{1} - {\bf k}_{2}))
\vspace{0.2cm}
\]
\[
\hspace{2cm}
-\,\frac{1}{2}\,N_{c}\hspace{0.03cm}
(n_{\bf p})^{2}\!\hspace{0.03cm}
\!\int\!d{\bf k}_{1}\hspace{0.03cm} d{\bf k}_{2}\,
\bigl|\hspace{0.02cm}\mathscr{T}^{\,(2,\hspace{0.03cm}{\mathcal A})}_{{\bf k}_{1},\,{\bf k}_{2}}({\bf v})\hspace{0.02cm}\bigr|^{\hspace{0.03cm}2}
\hspace{0.03cm}
(2\pi)\,\delta(\omega^{l}_{{\bf k}_{1}} - \omega^{l}_{{\bf k}_{2}} - {\mathbf v}\cdot({\bf k}_{1} - {\bf k}_{2}))
\]
\[
\times\, 
\biggl[
\hspace{0.03cm}\biggl\{\!\hspace{0.03cm}
\left(\displaystyle\frac{N^{\hspace{0.03cm}2}_{c} - 2} {N^{\hspace{0.03cm}2}_{c}}\right)\!
\bigl\langle\hspace{0.01cm}\mathcal{Q}^{\hspace{0.03cm}s}\hspace{0.03cm}
\bigr\rangle
\hspace{0.03cm}\bigl\langle\hspace{0.01cm}\mathcal{Q}\hspace{0.03cm}\bigr
\rangle
-
\frac{1}{N_{c}}\,
d^{\,s\hspace{0.03cm}d\hspace{0.03cm}e}
\hspace{0.03cm}
\bigl\langle\hspace{0.01cm}\mathcal{Q}^{\hspace{0.03cm}d}\hspace{0.03cm}
\bigr\rangle\hspace{0.03cm}
\bigl\langle\hspace{0.01cm}\mathcal{Q}^{\hspace{0.03cm}
e}\hspace{0.03cm}\bigr\rangle
\biggr\}
\bigl(N^{\,l}_{{\bf k}_{1}} - N^{\,l}_{{\bf k}_{2}}\hspace{0.01cm}\bigr)
\vspace{0.1cm}
\]
\[
-\,
\frac{1}{2}\,
\hspace{0.03cm}\biggl\{
\bigl\langle\hspace{0.01cm}\mathcal{Q}^{\hspace{0.03cm}s}\hspace{0.03cm}
\bigr\rangle
\hspace{0.03cm}\bigl\langle\hspace{0.01cm}\mathcal{Q}^{\hspace{0.03cm}e}
\hspace{0.03cm}\bigr\rangle
\hspace{0.03cm}\bigl\langle\hspace{0.01cm}\mathcal{Q}^{\hspace{0.03cm}e}
\hspace{0.03cm}\bigr\rangle
-
\frac{1}{2}\,\biggl(
\frac{1}{N_{c}}\,\hspace{0.03cm}\bigl\langle\hspace{0.01cm}
\mathcal{Q}^{\hspace{0.03cm}s}
\hspace{0.03cm}\bigr\rangle
\hspace{0.03cm}\bigl\langle\hspace{0.01cm}{\mathcal Q}\hspace{0.03cm}
\bigr\rangle
+
d^{\hspace{0.03cm}s\hspace{0.02cm}d\hspace{0.03cm}e}
\hspace{0.03cm}\bigl\langle\hspace{0.01cm}\mathcal{Q}^{\hspace{0.03cm}d}
\hspace{0.03cm}\bigr\rangle
\hspace{0.03cm}\bigl\langle\hspace{0.01cm}{\mathcal Q}^{\hspace{0.03cm}e}\hspace{0.03cm}
\bigr\rangle
\biggr)\hspace{0.03cm}\bigl\langle\hspace{0.01cm}{\mathcal Q}\hspace{0.03cm}
\bigr\rangle\biggr\}
\hspace{0.03cm}\bigl(W^{\,l}_{{\bf k}_{1}} + 
W^{\,l}_{{\bf k}_{2}}\hspace{0.01cm}\bigr) 
\biggr].
\]
By virtue of the same reasoning we used after equation (\ref{eq:10p}) describing the time evolution of the colorless charge $\bigl\langle\hspace{0.01cm}{\mathcal Q}\hspace{0.03cm}
\bigr\rangle$, we can discard the contributions on the right-hand side of (\ref{eq:11d}) containing the differences $\bigl(W_{{\bf k}_{1}}N_{{\bf k}_{2}} - N_{{\bf k}_{1}}W_{{\bf k}_{2}}\bigr)$ and $\bigl(N_{{\bf k}_{1}} - N_{{\bf k}_{2}}\hspace{0.01cm}\bigr)$ in the integrands. In addition, we multiply the left and right-hand sides by ${\bf p}^{2}$ and then integrate over $|{\bf p}|$ with the normalization
\begin{equation}
\biggl(\int\!n_{\bf p}\,{\bf p}^{2}\hspace{0.03cm}
d\hspace{0.02cm}|{\bf p}|\biggr) = 1.
\label{eq:11f}
\end{equation}
As a result, we are left with the following evolution equation, instead of  (\ref{eq:11d}):
\begin{equation}
	\frac{d\hspace{0.04cm} \langle\hspace{0.01cm}\mathcal{Q}^{\hspace{0.03cm}s}\hspace{0.03cm}\rangle}{d\hspace{0.03cm}t}
	=
	\!\int\!d\hspace{0.02cm}{\bf k}\, 
	{\rm Im}\hspace{0.03cm}\mathscr{T}^{\,(2,\hspace{0.03cm}{\mathcal A})}_{{\bf k},\,{\bf k}}({\bf v})
	\hspace{0.03cm}W^{\hspace{0.03cm}l}_{\bf k}\,
	\Bigl\{
	\bigl\langle\hspace{0.01cm}\mathcal{Q}\hspace{0.03cm}\bigr\rangle
	\hspace{0.03cm}
	\bigl\langle\hspace{0.01cm}\mathcal{Q}^{\hspace{0.03cm}s}\hspace{0.03cm}\bigr\rangle
	+
	N_{c}\hspace{0.04cm}d^{\hspace{0.03cm}s\hspace{0.02cm}d\hspace{0.03cm}e}
	\hspace{0.02cm}
	\bigl\langle\hspace{0.01cm}\mathcal{Q}^{\hspace{0.03cm}d}\hspace{0.03cm}
	\bigr\rangle
	\hspace{0.03cm}
	\bigl\langle\hspace{0.01cm}\mathcal{Q}^{\hspace{0.03cm}e}\hspace{0.03cm}
	\bigr\rangle\Bigr\}
\vspace{-0.5cm}
\label{eq:11g}
\end{equation}
\begin{align}
&+\frac{1}{2}\,N^{\hspace{0.02cm}2}_{c}\!
\!\int\!d{\bf k}_{1}\hspace{0.03cm} d{\bf k}_{2}\,
\bigl|\hspace{0.02cm}\mathscr{T}^{\,(2,\hspace{0.03cm}{\mathcal A})}_{{\bf k}_{1},\,{\bf k}_{2}}({\bf v})\hspace{0.02cm}\bigr|^{\hspace{0.03cm}2}
\hspace{0.03cm}
\bigl\langle\hspace{0.01cm}\mathcal{Q}^{\hspace{0.03cm}s}\hspace{0.03cm}
\bigr\rangle\hspace{0.03cm} N_{{\bf k}_{1}}N_{{\bf k}_{2}}
\hspace{0.03cm}
(2\pi)\,\delta(\omega^{l}_{{\bf k}_{1}} - \omega^{l}_{{\bf k}_{2}} - {\mathbf v}\cdot({\bf k}_{1} - {\bf k}_{2}))
\notag\\[1.5ex]
&+\frac{1}{4}\,N_{c}\hspace{0.01cm}
\biggl(\int\!n^{2}_{\bf p}\,{\bf p}^{2}\hspace{0.03cm}
d\hspace{0.02cm}|{\bf p}|\biggr)\!\hspace{0.02cm}
\!\int\!d{\bf k}_{1}\hspace{0.03cm}d{\bf k}_{2}\,
\bigl|\hspace{0.02cm}\mathscr{T}^{\,(2,\hspace{0.03cm}{\mathcal A})}_{{\bf k}_{1},\,{\bf k}_{2}}({\bf v})\hspace{0.02cm}\bigr|^{\hspace{0.03cm}2}
\hspace{0.03cm}
(2\pi)\,\delta(\omega^{l}_{{\bf k}_{1}} - \omega^{l}_{{\bf k}_{2}} - {\mathbf v}\cdot({\bf k}_{1} - {\bf k}_{2}))
\notag
\end{align}
\[
\times\, 
\biggl\{
\bigl\langle\hspace{0.01cm}\mathcal{Q}^{\hspace{0.03cm}s}\hspace{0.03cm}
\bigr\rangle
\hspace{0.03cm}\bigl\langle\hspace{0.01cm}\mathcal{Q}^{\hspace{0.03cm}e}
\hspace{0.03cm}\bigr\rangle
\hspace{0.03cm}\bigl\langle\hspace{0.01cm}\mathcal{Q}^{\hspace{0.03cm}e}
\hspace{0.03cm}\bigr\rangle
-
\frac{1}{2}\,\biggl(
\frac{1}{N_{c}}\,\hspace{0.03cm}\bigl\langle\hspace{0.01cm}
\mathcal{Q}^{\hspace{0.03cm}s}
\hspace{0.03cm}\bigr\rangle
\hspace{0.03cm}\bigl\langle\hspace{0.01cm}{\mathcal Q}\hspace{0.03cm}
\bigr\rangle
+
d^{\,s\hspace{0.02cm}d\hspace{0.03cm}e}
\hspace{0.03cm}\bigl\langle\hspace{0.01cm}\mathcal{Q}^{\hspace{0.03cm}d}
\hspace{0.03cm}\bigr\rangle
\hspace{0.03cm}\bigl\langle\hspace{0.01cm}{\mathcal Q}^{\hspace{0.03cm}e}\hspace{0.03cm}
\bigr\rangle
\biggr)\hspace{0.03cm}\bigl\langle\hspace{0.01cm}{\mathcal Q}\hspace{0.03cm}
\bigr\rangle\biggr\}
\hspace{0.03cm}\bigl(W^{\,l}_{{\bf k}_{1}} + 
W^{\,l}_{{\bf k}_{2}}\hspace{0.01cm}\bigr)
\]
with the initial condition
\[
\langle\hspace{0.01cm}{\mathcal Q}^{\hspace{0.03cm}s}\hspace{0.03cm}
\bigr\rangle|_{t\hspace{0.02cm}=\hspace{0.02cm}t_{0}}
=
{\mathcal Q}^{\hspace{0.03cm}s}_{\hspace{0.03cm}0},
\]
where ${\mathcal Q}^{\hspace{0.03cm}a}_{\hspace{0.03cm}0}$ is some fixed (non-random) vector of color charge that a high-energy particle possessed at the initial moment of time $t_{0}$.\\
\indent We are interested in the time dependence of the quadratic combination of the color charge ${\mathfrak q}_{2}(t)$, as it defined by the expression (\ref{eq:10a}). By virtue of equation (\ref{eq:11g}) we easily find 
\begin{equation}
\frac{d\hspace{0.04cm}{\mathfrak q}_{2}(t)}{d\hspace{0.03cm}t}
=
2\hspace{0.03cm}\!\!\int\!d\hspace{0.02cm}{\bf k}\, 
{\rm Im}\hspace{0.03cm}\mathscr{T}^{\,(2,\hspace{0.03cm}{\mathcal A})}_{{\bf k},\,{\bf k}}({\bf v})
\hspace{0.03cm}W^{\hspace{0.03cm}l}_{\bf k}\,
\Bigl\{
\bigl\langle\hspace{0.01cm}\mathcal{Q}\hspace{0.03cm}\bigr\rangle
\hspace{0.04cm}{\mathfrak q}_{2}(t)
+
N_{c\,}{\mathfrak q}_{3}(t)\Bigr\}	
\label{eq:11h}
\end{equation}
\[
+\hspace{0.03cm}
N^{\hspace{0.02cm}2}_{c}\hspace{0.03cm}{\mathfrak q}_{2}(t)\hspace{0.03cm}
\!\!\int\!d{\bf k}_{1}\hspace{0.03cm}d{\bf k}_{2}\,
\bigl|\hspace{0.02cm}\mathscr{T}^{\,(2,\hspace{0.03cm}{\mathcal A})}_{{\bf k}_{1},\,{\bf k}_{2}}({\bf v})\hspace{0.02cm}\bigr|^{2}
\hspace{0.03cm} N_{{\bf k}_{1}}N_{{\bf k}_{2}}
\hspace{0.03cm}
(2\pi)\,\delta(\omega^{l}_{{\bf k}_{1}} - \omega^{l}_{{\bf k}_{2}} - {\mathbf v}\cdot({\bf k}_{1} - {\bf k}_{2}))
\vspace{0.2cm}
\]
\[
+\,\frac{1}{2}\,N_{c}\hspace{0.02cm}\biggl(\int\!n^{2}_{\bf p}\,{\bf p}^{2}\hspace{0.03cm}
d\hspace{0.02cm}|{\bf p}|\biggr)\!\hspace{0.02cm}
\!\int\!d{\bf k}_{1}\hspace{0.03cm}d{\bf k}_{2}\,
\bigl|\hspace{0.02cm}\mathscr{T}^{\,(2,\hspace{0.03cm}{\mathcal A})}_{{\bf k}_{1},\,{\bf k}_{2}}({\bf v})\hspace{0.02cm}\bigr|^{\hspace{0.03cm}2}
\hspace{0.03cm}
(2\pi)\,\delta(\omega^{l}_{{\bf k}_{1}} - \omega^{l}_{{\bf k}_{2}} - {\mathbf v}\cdot({\bf k}_{1} - {\bf k}_{2}))
\]
\[
\times\hspace{0.03cm}
\biggl\{
\bigl({\mathfrak q}_{2}(t)\bigr)^{\!2}
-
\frac{1}{2}\,\biggl(
\frac{1}{N_{c}}\,\hspace{0.03cm}
{\mathfrak q}_{2}(t)
\hspace{0.03cm}\bigl\langle\hspace{0.01cm}{\mathcal Q}\hspace{0.03cm}
\bigr\rangle
+
{\mathfrak q}_{3}(t)
\biggr)\hspace{0.03cm}\bigl\langle\hspace{0.01cm}{\mathcal Q}\hspace{0.03cm}
\bigr\rangle\biggr\}
\hspace{0.03cm}\bigl(W^{\,l}_{{\bf k}_{1}} + W^{\,l}_{{\bf k}_{2}}
\hspace{0.01cm}\bigr).
\]
Here, we have introduced the notation for the second colorless combination of the third order in the averaged color charge
\[
{\mathfrak q}_{3}(t) \equiv d^{\hspace{0.04cm}a\hspace{0.02cm}b\hspace{0.03cm}c}
\bigl\langle\hspace{0.01cm}\mathcal{Q}^{\hspace{0.03cm}a}\hspace{0.03cm}\bigr\rangle
\hspace{0.03cm}
\bigl\langle\hspace{0.01cm}\mathcal{Q}^{\hspace{0.03cm}b}\hspace{0.03cm}\bigr\rangle
\hspace{0.03cm}
\bigl\langle\hspace{0.01cm}\mathcal{Q}^{\hspace{0.03cm}c}\hspace{0.03cm}\bigr\rangle.
\]
%
To close equation (\ref{eq:11h}) we also deduce an equation for the function ${\mathfrak q}_{3}(t)$:
\begin{equation}
	\frac{d\hspace{0.04cm}{\mathfrak q}_{3}(t)}{d\hspace{0.03cm}t}
	=
	3\hspace{0.03cm}\!\!\int\!d\hspace{0.02cm}{\bf k}\, 
	{\rm Im}\hspace{0.03cm}\mathscr{T}^{\,(2,\hspace{0.03cm}{\mathcal A})}_{{\bf k},\,{\bf k}}({\bf v})
	\hspace{0.03cm}W^{\hspace{0.03cm}l}_{\bf k}\,
	\Bigl\{
	\bigl\langle\hspace{0.01cm}\mathcal{Q}\hspace{0.03cm}\bigr\rangle
	\hspace{0.04cm}{\mathfrak q}_{3}(t)
	+
	N_{c}\hspace{0.04cm}{\mathfrak q}_{4}(t)\Bigr\}	
\label{eq:11k}
\end{equation}
\[	
+\,\frac{3}{2}\,
\hspace{0.03cm}N^{\hspace{0.02cm}2}_{c}\hspace{0.04cm}{\mathfrak q}_{3}(t)
	\!\int\!d{\bf k}_{1}\hspace{0.03cm}d{\bf k}_{2}\,
	\bigl|\hspace{0.02cm}\mathscr{T}^{\,(2,\hspace{0.03cm}{\mathcal A})}_{{\bf k}_{1},\,{\bf k}_{2}}({\bf v})\hspace{0.02cm}\bigr|^{\hspace{0.02cm}2}
\hspace{0.03cm} N_{{\bf k}_{1}}N_{{\bf k}_{2}}
\hspace{0.03cm}
(2\pi)\,\delta(\omega^{l}_{{\bf k}_{1}} - \omega^{l}_{{\bf k}_{2}} - {\mathbf v}\cdot({\bf k}_{1} - {\bf k}_{2}))
\vspace{0.2cm}
\]
\[
\hspace{0.5cm}
+\,\frac{3}{4}\,N_{c}\hspace{0.02cm}\biggl(\int\!n^{2}_{\bf p}\,{\bf p}^{2}\hspace{0.03cm}
d\hspace{0.02cm}|{\bf p}|\biggr)\!\hspace{0.02cm}
\!\int\!d{\bf k}_{1}\hspace{0.03cm}d{\bf k}_{2}\,
\bigl|\hspace{0.02cm}\mathscr{T}^{\,(2,\hspace{0.03cm}{\mathcal A})}_{{\bf k}_{1},\,{\bf k}_{2}}({\bf v})\hspace{0.02cm}\bigr|^{\hspace{0.03cm}2}
\hspace{0.03cm}
(2\pi)\,\delta(\omega^{l}_{{\bf k}_{1}} - \omega^{l}_{{\bf k}_{2}} - {\mathbf v}\cdot({\bf k}_{1} - {\bf k}_{2}))
\]
\[
\times
\hspace{0.03cm}
\biggl\{
{\mathfrak q}_{\hspace{0.02cm}3}(t)
{\mathfrak q}_{\hspace{0.02cm}2}(t)
-
\frac{1}{2}\,\biggl(
\frac{1}{N_{c}}\,\hspace{0.03cm}
{\mathfrak q}_{\hspace{0.02cm}3}(t)
\hspace{0.03cm}\bigl\langle\hspace{0.01cm}{\mathcal Q}\hspace{0.03cm}
\bigr\rangle
+
{\mathfrak q}_{\hspace{0.02cm}4}(t)
\biggr)\hspace{0.03cm}\bigl\langle\hspace{0.01cm}{\mathcal Q}\hspace{0.03cm}
\bigr\rangle\biggr\}
\hspace{0.03cm}\bigl(W^{\,l}_{{\bf k}_{1}} + W^{\,l}_{{\bf k}_{2}}
\hspace{0.01cm}\bigr).
\]
However, on the right-hand side of this equation, a colorless combination of higher fourth order 
\begin{equation}
{\mathfrak q}_{\hspace{0.02cm}4}(t) =
{\mathfrak q}^{\hspace{0.03cm}a}_{\hspace{0.02cm}2}(t)\hspace{0.03cm}
{\mathfrak q}^{\hspace{0.03cm}a}_{\hspace{0.02cm}2}(t)
\label{eq:11l}
\end{equation}
appears, where
\[
{\mathfrak q}^{\hspace{0.03cm}a}_{\hspace{0.02cm}2}(t) 
\equiv
d^{\hspace{0.03cm}a\hspace{0.03cm}b\hspace{0.03cm}c}
\langle\hspace{0.01cm}{\mathcal Q}^{\,b}\hspace{0.03cm}\bigr\rangle
\hspace{0.03cm}
\langle\hspace{0.01cm}{\mathcal Q}^{\,c}\hspace{0.03cm}\bigr\rangle.
\]
It is clear that an attempt to write the equation for ${\mathfrak q}_{\hspace{0.02cm}4}(t)$ will in turn lead to more complicated colorless structures. A coupled chain of equations can be truncated at the first two combinations ${\mathfrak q}_{2}(t)$ and ${\mathfrak q}_{3}(t)$ for the particular Lie algebra $\mathfrak{su}(3_{c})$ (except for the ``trivial'' case $\mathfrak{su}(2_{\hspace{0.02cm}c})$). By virtue of the second relation in (\ref{ap:C15}), the following representation for (\ref{eq:11l}) is valid:
\begin{equation*}
{\mathfrak q}_{4}(t) = \frac{1}{3}\,\bigl({\mathfrak q}_{2}(t)\bigr)^{\!2}.
\end{equation*}
This allows us to completely close the system of three equations for the colorless charge $\langle\hspace{0.01cm}{\mathcal Q}\hspace{0.03cm}\bigr\rangle$, Eq.\,(\ref{eq:10g}), and equations for the colorless combinations ${\mathfrak q}_{2}(t)$ and ${\mathfrak q}_{3}(t)$, Eqs.\,(\ref{eq:11h}) and (\ref{eq:11k}), respectively.\\
\indent The equations (\ref{eq:11h}) and (\ref{eq:11k}) are presented in the most general form, which makes them quite complicated. Let us simplify them. As a first step, we take into account that due to the absence of linear Landau damping, it is necessary to put
\[
{\rm Im}\hspace{0.03cm}\mathscr{T}^{\,(2,\hspace{0.03cm}{\mathcal A})}_{{\bf k},\,{\bf k}}({\bf v}) = 0.
\]
We have already discussed this at the end of the previous section. Next, by virtue of (\ref{eq:10g}), the ``colorless'' charge $\bigl\langle\hspace{0.01cm}{\mathcal Q}\hspace{0.03cm}
\bigr\rangle$ must be assumed to be a constant value. For the sake of simplicity, we set this constant to zero
\[
\bigl\langle\hspace{0.01cm}{\mathcal Q}\hspace{0.03cm}\bigr\rangle \equiv 0.
\]
Thus, instead of the evolution equations (\ref{eq:11h}) and (\ref{eq:11k}), we now get
\begin{align}
	&\frac{d\hspace{0.04cm}{\mathfrak q}_{2}(t)}{d\hspace{0.03cm}t}
	=
	A(t)\hspace{0.03cm}{\mathfrak q}_{2}(t) + B(t)\hspace{0.02cm}\bigl({\mathfrak q}_{2}(t)\bigr)^{\!\hspace{0.03cm}2},
	\!\qquad
	{\mathfrak q}_{2}(t)|_{t\hspace{0.02cm}=\hspace{0.02cm}t_{0}}
	=
	{\mathfrak q}_{\hspace{0.02 cm}2}^{\hspace{0.01cm}0},
	\label{eq:11b}\\[1ex]
	&\frac{d\hspace{0.04cm}{\mathfrak q}_{3}(t)}{d\hspace{0.03cm}t}
	=
	\frac{3}{2}\,\bigl\{\!\hspace{0.03cm}A(t)
	- 
	B(t)\hspace{0.03cm}{\mathfrak q}_{2}(t)\bigr\}\hspace{0.03cm}
	{\mathfrak q}_{3}(t),
	\quad
	{\mathfrak q}_{3}(t)|_{t\hspace{0.02cm}=\hspace{0.02cm}t_{0}}
	=
	{\mathfrak q}_{\hspace{0.02 cm}3}^{\hspace{0.01cm}0},
	\label{eq:11n}
\end{align}
where we have introduced the notations
\begin{align}
	&A(t) \equiv 
	N^{\hspace{0.02cm}2}_{c} \!\int\!d\hspace{0.03cm}{\bf k}_{1}\hspace{0.03cm} d\hspace{0.03cm}{\bf k}_{2}\,
	\bigl|\hspace{0.02cm}\mathscr{T}^{\,(2,\hspace{0.03cm}{\mathcal A})}_{{\bf k}_{1},\,{\bf k}_{2}}({\bf v})\hspace{0.02cm}\bigr|^{\hspace{0.03cm}2}
	\hspace{0.03cm} N^{\hspace{0.03cm}l}_{{\bf k}_{1}}N^{\hspace{0.03cm}l}_{{\bf k}_{2}}\hspace{0.03cm}
	(2\pi)\hspace{0.03cm}\delta(\omega^{l}_{{\bf k}} - \omega^{l}_{{\bf k}_{1}} - {\mathbf v}\cdot({\bf k} - {\bf k}_{1})),
	\label{eq:11v}\\[1.5ex]
	&B(t) \equiv
	\frac{1}{2}\,
	N_{c}\hspace{0.01cm}
	\biggl(\int\!n^{2}_{\bf p}\,{\bf p}^{2}\hspace{0.03cm}
	d\hspace{0.02cm}|{\bf p}|\biggr)\!
	\!\int\!d{\bf k}_{1}\hspace{0.03cm} d{\bf k}_{2}\,
	\bigl|\hspace{0.02cm}\mathscr{T}^{\,(2,\hspace{0.03cm}{\mathcal A})}_{{\bf k}_{1},\,{\bf k}_{2}}({\bf v})\hspace{0.02cm}\bigr|^{\hspace{0.03cm}2}
	\hspace{0.03cm}
	\bigl(W^{\,l}_{{\bf k}_{1}}\! + W^{\,l}_{{\bf k}_{2}}
	\hspace{0.01cm}\bigr) 
	\hspace{0.03cm}
	(2\pi)\,\delta(\omega^{l}_{{\bf k}_{1}}\! - \omega^{l}_{{\bf k}_{2}}\! - {\mathbf v}\cdot({\bf k}_{1} - {\bf k}_{2})).
	\notag
\end{align}
The initial values ${\mathfrak q}_{
\hspace{0.02cm}2}^{\hspace{0.01cm}0}$ and ${\mathfrak q}_{
\hspace{0.02cm}3}^{\hspace{0.01cm}0}$ are defined as
\[
{\mathfrak q}_{\hspace{0.02 cm}2}^{\hspace{0.01cm}0}
=
{\mathcal Q}^{\hspace{0.03cm}a}_{0}\hspace{0.03cm}
{\mathcal Q}^{\hspace{0.03cm}a}_{0},
\qquad
{\mathfrak q}_{\hspace{0.02 cm}3}^{\hspace{0.01cm}0}
=
d^{\hspace{0.04cm}a\hspace{0.02cm}b\hspace{0.03cm}c}
\hspace{0.03cm}\mathcal{Q}^{\hspace{0.03cm}a}_{0}\hspace{0.03cm}
\mathcal{Q}^{\hspace{0.03cm}b}_{0}
\hspace{0.03cm}\mathcal{Q}^{\hspace{0.03cm}c}_{0}.
\]
The equation (\ref{eq:11b}) is a special case of the Bernoulli equation and, therefore, we can immediately write out its solution \cite{kamke_1977}  
\begin{equation}
{\mathfrak q}_{\hspace{0.02cm}2}(t) 
= 
{\mathfrak q}_{\hspace{0.02 cm}2}^{\hspace{0.01cm}0}\, 
\frac{\exp\left\{\,\displaystyle\int^{t}_{t_{0}}\!\!A(\tau)\hspace{0.03cm} d\tau\right\}}
{1 - {\mathfrak q}_{\hspace{0.02 cm}2}^{\hspace{0.01cm}0}
\displaystyle\int^{t}_{t_{0}}\!\!B(\tau)\hspace{0.03cm}
\exp\left\{\,\displaystyle\int^{\tau}_{t_{0}}\!
A(\tau^{\prime})\hspace{0.03cm} d\tau^{\prime}\right\} d\tau},
\label{eq:11m}
\end{equation}
which is qualitatively different from the solution we obtained in \cite{markov_I_2024}. The second colorless combination ${\mathfrak q}_{\hspace{0.02cm}3}(t)$ is trivially determined from the second equation (\ref{eq:11n}). For physical reasons, we consider that the plasmon number density $N^{\,l}_{\bf k}$ is a positive function that, by virtue of the definitions (\ref{eq:11v}), leads in turn to the inequality
\[
A(t) \geq 0.
\] 
Because of this, the exponential function in the solution (\ref{eq:11m}) is an increasing function in time. On the other hand, the color part $W^{\,l}_{\bf k}$ of the plasmon number density is, in general, indefinite and, as a consequence, the function $B(t)$ can be either positive or negative. However, the solution (\ref{eq:11m}) may nevertheless remain a finite value which is physically more reasonable.

\section{Conclusion}
\label{section_17}
\setcounter{equation}{0}

In this paper we have demonstrated in detail that the Hamiltonian formalism proposed in \cite{markov_2023} to describe the nonlinear dynamics of {\it only soft} Fermi- and Bose-excitations contains much more information
about the medium under consideration than was originally assumed. It turned out to be also very suitable for describing another range of physical phenomena, namely the processes of the scattering of colorless plasmons off {\it hard} thermal (or external) color-charged particles moving in a high-temperature quark-gluon plasma. The methodology developed in this paper allowed us to somewhat justify and define more exactly the formalism we proposed within the framework of heuristic approach in \cite{markov_I_2024}. In particular, this is reflected in the appearance of new contributions to both the kinetic equation for color part of the plasmon number density and the evolution equation (\ref{eq:11g}) for the mean value of the color charge  $\bigl\langle\hspace{0.01cm}\mathcal{Q}^{\hspace{0.03cm}a}\hspace{0.03cm}
\bigr\rangle$. The appearance of a new contribution to (\ref{eq:11g}) could drastically change the dynamics of the color charge evolution in contrast to the conclusion of the paper \cite{markov_I_2024}.\\
\indent In this paper we have restricted ourselves to the detailed consideration of only the simplest process of the interaction of soft and hard modes in a quark-gluon plasma: the elastic scattering of plasmon off hard particle occurring without change of statistics of soft and hard excitations. At least for the weakly-excited system corresponding to the level of thermal fluctuations, this process is dominant.
\vspace{-0.5cm}
%
%
\section*{\bf Acknowledgment}
The work was carried out at the expense of the state assignment under the theme ``Development of analytical and numerical methods of description in problems of mathematical physics, continuum mechanics, quantum field theory and nuclear physics'' (no. of state registration: 126021217175-3).

\begin{appendices}
\numberwithin{equation}{section}
\section{Effective three-plasmon vertices}
\numberwithin{equation}{section}
\label{appendix_A}

In this appendix we present an explicit form of the effective three-plasmon vertex functions  ${\mathcal V}_{\, {\bf k},\, {\bf k}_{1},\, {\bf k}_{2}}$ and ${\mathcal U}_{\, {\bf k},\, {\bf k}_{1},\, {\bf k}_{2}}$. They were obtained earlier in \cite{markov_2020} when constructing the Hamiltonian formalism for soft Bose excitations in a hot gluon plasma. These vertices 
read
\begin{equation}
	{\mathcal V}_{\, {\bf k},\, {\bf k}_{1},\, {\bf k}_{2}} = 
	\frac{1}{2^{3/4}}\,g\hspace{0.03cm}
	\left(\frac{{\rm Z}_l({\bf k})}{2\omega_{\bf k}^l}\right)^{\!\!1/2}\!\!\!
\frac{\tilde{u}_{\mu}(k)}{\sqrt{\bar{u}^2(k)}}
\prod\limits_{i = 1}^{2}
	\left(\frac{{\rm Z}_l({\bf k}_{i})}{2\omega_{\bf k}^l}\right)^{\!\!1/2}\!\!\!
\frac{\tilde{u}_{\mu_{i}}(k_{i})}{\sqrt{\bar{u}^2(k_{i})}}
	\,^{\ast}\Gamma^{\mu\mu_1\mu_2}(k,- k_{1},- k_{2})\Bigr|_{\rm \,on-shell}
	\hspace{0.4cm} 
	\label{ap:A1}
\end{equation}
and
\begin{equation}
	{\mathcal U}_{\, {\bf k},\, {\bf k}_{1},\, {\bf k}_{2}} =
	\frac{1}{2^{3/4}}\,g\hspace{0.03cm}
	\left(\frac{{\rm Z}_l({\bf k})}{2\omega_{\bf k}^l}\right)^{\!\!1/2}\!\!\!
	\frac{\tilde{u}_{\mu}(k)}{\sqrt{\bar{u}^2(k)}}
	\prod\limits_{i = 1}^{2}
	\left(\frac{{\rm Z}_l({\bf k}_{i})}{2\omega_{\bf k}^l}\right)^{\!\!1/2}\!\!\!
	\frac{\tilde{u}_{\mu_{i}}(k_{i})}{\sqrt{\bar{u}^2(k_{i})}} 
	\,^{\ast}\Gamma^{\mu\mu_1\mu_2}(- k,- k_{1},- k_{2})\Bigr|_{\rm \,on-shell}.
	\label{ap:A2}
\end{equation}
Two four-vectors 
\begin{equation}
\tilde{u}_{\mu} (k) = \frac{k^2}{(k\cdot u)}\ \! \Bigl(k_{\mu} - u_{\mu}(k\cdot u)\Bigr)
\quad \mbox{and} \quad
\bar{u}_{\mu} (k) = k^2 u_{\mu} - k_{\mu}(k\cdot u)
\label{ap:A3}
\end{equation}
are the projectors onto the longitudinal direction of wavevector ${\bf k}$, written in the Lorentz-covariant form in the Hamilton and Lorentz gauges, respectively. Here, $u^{\mu}$ is the four-velocity of the medium, which in the rest system is $u^{\mu}=(1,0,0,0)$. The explicit form of the effective three-gluon vertex $\,^{\ast}\Gamma^{\mu\mu_1\mu_2}(k, k_{1}, k_{2})$ on the right-hand side of (\ref{ap:A1}) and (\ref{ap:A2}) is defined by formulae (\ref{ap:A4})\,--\,(\ref{ap:A6}) below.\\ 
\indent Effective three-gluon vertex in the hard thermal loop (HTL) approximation has the following form \cite{blaizot_2002, ghiglieri_2020, braaten_1990}
\begin{equation}
	\,^{\ast} \Gamma^{\mu\hspace{0.02cm} \nu  \rho}(k, k_{1}, k_{2}) \equiv
	\Gamma^{\mu\hspace{0.02cm} \nu  \rho}(k, k_{1}, k_{2}) +
	\delta\hspace{0.025cm} \Gamma^{\mu\hspace{0.02cm} \nu  \rho}(k, k_{1}, k_{2}),
	\label{ap:A4}
\end{equation}
where the first term is bare three-gluon vertex
\begin{equation}
	\Gamma^{\mu\hspace{0.02cm} \nu  \rho}(k, k_{1}, k_{2}) =
	g^{\mu\hspace{0.02cm} \nu } (k - k_{1})^{\rho} + g^{\nu \rho} (k_{1} - k_{2})^{\mu} +
	g^{\mu \rho} (k_{2} - k)^{\nu}
	\label{ap:A5}
\end{equation}
and the second one is the corresponding HTL-correction
\begin{equation}
	\delta\hspace{0.025cm} \Gamma^{\mu\hspace{0.02cm} \nu  \rho}(k, k_{1}, k_{2}) =
	3\hspace{0.035cm}\omega^{2}_{\rm pl}\!\int\!\frac{d\hspace{0.035cm}\Omega}{4 \pi} \,
	\frac{v^{\mu}\hspace{0.02cm} v^{\nu} v^{\rho}}{v\cdot k + i\hspace{0.025cm}\epsilon} \,
	\Biggl(\frac{\omega_{2}}{v\cdot k_{2} - i\epsilon} -
	\frac{\omega_1}{v\cdot k_{1} - i\epsilon}\Biggr),
	\quad \epsilon\rightarrow +\hspace{0.02cm}0.
	\label{ap:A6}
\end{equation}
Here $v^{\hspace{0.03cm}\mu} = (1,{\bf {\bf v}})$, $k^{\hspace{0.03cm}\mu} = (\omega, {\bf k})$ is a gluon four-momentum with $k  + k_{1} + k_{2} = 0$, $d\hspace{0.035cm}\Omega$ is a differential solid angle and $\omega_{\rm pl}^2 = g^2(2N_c+N_f)T^2/18$ is plasma frequency squared.


\numberwithin{equation}{section}
\section{Relations and traces for generators in the defining representation of the $SU(N_{c})$ color group}
\numberwithin{equation}{section}
\label{appendix_B}

Let $t^{\hspace{0.03cm}a},\, a = 1,\ldots,\,N_{c}^{\hspace{0.02cm}2} - 1$ be the $SU(N_{c})$ generators in the fundamental representations, then  
\begin{equation}
t^{\hspace{0.03cm}a}t^{\hspace{0.03cm}b} = \frac{1}{2\hspace{0.02cm}N_{c}}\,\delta^{\hspace{0.03cm}a\hspace{0.03cm}b}
\mathbbm{1}
	+
	\frac{1}{2}\,\bigl(d^{\hspace{0.03cm}a\hspace{0.02cm}b\hspace{0.03cm}c}
	+
	i\hspace{0.03cm}
	f^{\hspace{0.03cm}a\hspace{0.02cm}b\hspace{0.03cm}c}\hspace{0.03cm}\bigr)\hspace{0.03cm}t^{c}	
\label{ap:B1}
\end{equation}
and, as a consequence, one has
\begin{equation}
t^{\hspace{0.03cm}a}t^{\hspace{0.03cm}a} = \biggl(\frac{N^{\hspace{0.02cm}2}_{c} - 1}{2\hspace{0.02cm}N_{c}}\biggr)
\mathbbm{1},
\qquad
t^{\hspace{0.03cm}b}t^{\hspace{0.03cm}a}t^{\hspace{0.03cm}b}
=
-\hspace{0.03cm}\frac{1}{2\hspace{0.02cm}N_{c}}\,t^{\hspace{0.03cm}b}.	
\label{ap:B2}
\end{equation} 
Further, the Fierz identities for the $t^{\,a}$ matrices are
\begin{subequations} 
	\label{ap:B3}
	\begin{align}
		&(t^{\hspace{0.03cm}a})^{\hspace{0.03cm}i_{1}\hspace{0.03cm}j_{2}}
		\hspace{0.02cm}
		(t^{\hspace{0.03cm}a}\hspace{0.03cm})^{\hspace{0.03cm}j_{1}
			\hspace{0.03cm}i_{2}}
		= 
		\frac{1}{2}\,\delta^{\hspace{0.03cm}i_{1}\hspace{0.03cm}i_{2}}
		\hspace{0.03cm}\delta^{\hspace{0.03cm}j_{1}\hspace{0.03cm}j_{2}}	
		-
		\frac{1}{2\hspace{0.02cm}N_{c}}\,\delta^{\hspace{0.03cm}i_{1}
			\hspace{0.03cm}j_{2}}\hspace{0.02cm}
		\delta^{\hspace{0.03cm}j_{1}\hspace{0.03cm}i_{2}},
		\label{ap:B3_a}\\[1ex]
		&(t^{\hspace{0.03cm}a})^{\hspace{0.03cm}i_{1}\hspace{0.03cm}j_{2}}
		(t^{\hspace{0.03cm}a}\hspace{0.03cm})^{\hspace{0.03cm}j_{1}
			\hspace{0.03cm}i_{2}}
		= 
		\biggl(\frac{N^{\hspace{0.02cm}2}_{c} - 1}{2\hspace{0.02cm}N^{\hspace{0.02cm}2}_{c}}\biggr) \delta^{\hspace{0.03cm}i_{1}\hspace{0.03cm}i_{2}}
		\hspace{0.02cm}
		\delta^{\hspace{0.03cm}j_{1}\hspace{0.03cm}j_{2}}		
		-
		\frac{1}{N_{c}}\,
		(t^{\,a})^{i_{1}\hspace{0.03cm}i_{2}}
		\hspace{0.02cm}
		(t^{\,a})^{j_{1}\hspace{0.01cm}j_{2}}.
		\label{ap:B3_b}
	\end{align}
\end{subequations}
A trivial consequence of the first relation is the useful identity
\begin{equation}
\delta^{\hspace{0.03cm}i_{1}\hspace{0.03cm}j_{2}}
\hspace{0.02cm}
\delta^{\hspace{0.03cm}j_{1}\hspace{0.03cm}i_{2}}
=
\frac{1}{N_{c}}\,\delta^{\hspace{0.03cm}i_{1}\hspace{0.03cm}i_{2}}
\hspace{0.02cm}
\delta^{\hspace{0.03cm}j_{1}\hspace{0.03cm}j_{2}}
+
2\hspace{0.04cm}
(t^{\,a})^{i_{1}\hspace{0.03cm}i_{2}}
\hspace{0.02cm}
(t^{\,a})^{j_{1}\hspace{0.01cm}j_{2}}.
\label{ap:B4}
\end{equation}
Next, the other consequence of (\ref{ap:B3_a}) is the relation for the trace of the following form:
\begin{equation}
{\rm tr}\hspace{0.03cm}\bigl(A\hspace{0.03cm}t^{\,a\!}B\hspace{0.03cm}t^{\,a}
\bigr) 
=
\frac{1}{2}\,{\rm tr}\hspace{0.03cm}\bigl(A\bigr){\rm tr}\hspace{0.03cm}\bigl(B\bigr)	
-
\frac{1}{2\hspace{0.02cm}N_{c}}\,{\rm tr}\hspace{0.03cm}\bigl(AB\bigr).
\label{ap:B5}
\end{equation}
In addition, if we consider the following representations for the structure constants 
\[
f^{\hspace{0.03cm}a\hspace{0.02cm}b\hspace{0.03cm}c}
=
-2\hspace{0.03cm}i\hspace{0.03cm}{\rm tr}\hspace{0.03cm}\bigl(\bigl[\hspace{0.03cm}t^{\,a\!},
\hspace{0.03cm}t^{\,b}\bigr]\hspace{0.03cm}t^{\,c}\bigr),
\qquad
d^{\,a\hspace{0.02cm}b\hspace{0.03cm}c}
=
2\hspace{0.03cm}{\rm tr}\hspace{0.03cm}\bigl(\bigl\{\hspace{0.03cm}t^{\,a\!},
\hspace{0.03cm}t^{\,b}\bigr\}\hspace{0.03cm}t^{\,c}\bigr), 
\]
then, from (\ref{ap:B3_a}) and (\ref{ap:B4}), it also follows that 
\begin{subequations} 
\label{ap:B6}
\begin{align}
	&f^{\hspace{0.03cm}a\hspace{0.02cm}b\hspace{0.03cm}c}
	(t^{\hspace{0.03cm}b}\hspace{0.02cm})^{\hspace{0.03cm}i_{1}\hspace{0.03cm}j_{2}}
	(t^{\hspace{0.03cm}c}\hspace{0.03cm})^{\hspace{0.03cm}j_{1}
	\hspace{0.03cm}i_{2}} 
	= 
	\frac{i}{2}\,\Bigl\{\delta^{\hspace{0.03cm}j_{1}\hspace{0.03cm}j_2}
	(t^{\,a})^{i_{1}\hspace{0.03cm}i_{2}}
	-
	\delta^{\hspace{0.03cm}i_{1}\hspace{0.03cm}i_2}
	(t^{\,a})^{j_{1}\hspace{0.03cm}j_{2}}\Bigr\},	
	\label{ap:B6_a}\\[1ex]
	&d^{\,a\hspace{0.02cm}b\hspace{0.03cm}c}
	(t^{\hspace{0.03cm}b})^{\hspace{0.03cm}i_{1}\hspace{0.03cm}j_{2}}
	(t^{\hspace{0.03cm}c}\hspace{0.03cm})^{\hspace{0.03cm}j_{1}\hspace{0.03cm}i_{2}} 
	= 
\biggl(\frac{N^{\hspace{0.02cm}2}_{c} - 4} {2\hspace{0.02cm}N^{\hspace{0.02cm}2}_{c}}\biggr)\hspace{0.03cm}
\Bigl\{\delta^{\hspace{0.03cm}j_{1}\hspace{0.03cm}j_2}
	(t^{\,a})^{i_{1}\hspace{0.03cm}i_{2}}
	+
	\delta^{\hspace{0.03cm}i_{1}\hspace{0.03cm}i_2}
	(t^{\,a})^{j_{1}\hspace{0.03cm}j_{2}}\Bigr\}	
	-
	\frac{2}{N_{c}}\,
	d^{\,a\hspace{0.02cm}b\hspace{0.03cm}c}
	(t^{\,b}\hspace{0.02cm})^{i_{1}\hspace{0.03cm}i_{2}}
	\hspace{0.02cm}
	(t^{\,c})^{j_{1}\hspace{0.01cm}j_{2}}.
\label{ap:B6_b}
\end{align}
\end{subequations}
In deriving the last identity, we have used the relation for the sum
\begin{equation}
\delta^{\hspace{0.03cm}i_{1}\hspace{0.03cm}j_2}\hspace{0.03cm}
(t^{\,a})^{j_{1}\hspace{0.03cm}i_{2}}
+
\delta^{\hspace{0.03cm}j_{1}\hspace{0.03cm}i_2}\hspace{0.03cm}
(t^{\,a})^{i_{1}\hspace{0.03cm}j_{2}}
=
\frac{2}{N_{c}}\,\bigl[
\delta^{\hspace{0.03cm}j_{1}\hspace{0.03cm}j_2}
(t^{\,a})^{i_{1}\hspace{0.03cm}i_{2}}
+
\delta^{\hspace{0.03cm}i_{1}\hspace{0.03cm}i_2}
(t^{\,a})^{j_{1}\hspace{0.03cm}j_{2}}
\bigr]
+
2\hspace{0.03cm}d^{\hspace{0.04cm}a\hspace{0.02cm}b\hspace{0.03cm}c}
(t^{\,b})^{i_{1}\hspace{0.03cm}i_{2}}
\hspace{0.02cm}
(t^{\,c})^{j_{1}\hspace{0.01cm}j_{2}},	
\label{ap:B7}
\end{equation}
which is a consequence of (\ref{ap:B4}) and (\ref{ap:B1}). A similar relation for the difference trivially follows from (\ref{ap:B6_a}). Further, a useful consequence is also the relation 
\[
	\bigl(T^{\,a}\hspace{0.03cm}T^{\,b}\bigr)^{c\hspace{0.03cm}d}
	(t^{\hspace{0.03cm}a})^{\hspace{0.03cm}i_{1}\hspace{0.03cm}j_{2}}
	(t^{\hspace{0.03cm}b}\hspace{0.03cm})^{\hspace{0.03cm}j_{1}\hspace{0.03cm}i_{2}} 
\]
\[	
	=
	\frac{1}{2}\,\Bigl[(t^{\,c}\hspace{0.03cm}t^{\,d}\hspace{0.03cm})^{\hspace{0.03cm}i_{1}\hspace{0.03cm}i_{2}}\hspace{0.03cm}
	\delta^{\hspace{0.03cm}j_{1}\hspace{0.03cm}j_{2}}
	+
(t^{\,d}\hspace{0.03cm}t^{\,c}\hspace{0.03cm})^{\hspace{0.03cm}j_{1}
	\hspace{0.03cm}j_{2}}\hspace{0.03cm}
	\delta^{\hspace{0.03cm}i_{1}\hspace{0.03cm}i_{2}}\Bigr]
	-\,
	\frac{1}{2}\,\Bigl[(t^{\,c})^{i_{1}\hspace{0.03cm}i_{2}}
	(t^{\,d}\hspace{0.03cm})^{j_{1}\hspace{0.03cm}j_{2}}
	+
	(t^{\,c})^{\hspace{0.03cm}j_{1}\hspace{0.03cm}j_{2}}
(t^{\,d}\hspace{0.03cm})^{\hspace{0.03cm}i_{1}\hspace{0.03cm}i_{2}}\Bigr].
\]
In section \ref{section_9} we require a special consequence of the previous expression, namely
\begin{equation}
	\bigl\{T^{\,a},\hspace{0.03cm}T^{\,b}\bigr\}^{c\hspace{0.03cm}d}
	\hspace{0.02cm}
	(t^{\hspace{0.03cm}a})^{\hspace{0.03cm}i_{1}\hspace{0.03cm}j_{2}}
	(t^{\hspace{0.03cm}i}\hspace{0.03cm})^{\hspace{0.03cm}j_{1}
	\hspace{0.03cm}i_{2}} 
	=
	\frac{1}{N_{c}}\,
	\delta^{\hspace{0.03cm}c\hspace{0.03cm}d}\hspace{0.03cm}
	\delta^{\hspace{0.03cm}i_{1}\hspace{0.03cm}i_2}\hspace{0.03cm}
	\delta^{\hspace{0.03cm}j_{1}\hspace{0.03cm}j_2}
	\label{ap:B8}
\end{equation}
\[	
+\,
\frac{1}{2}\,\bigl(D^{\,\lambda}\bigr)^{c\hspace{0.03cm}d}
\Bigl[(t^{\,\lambda}\hspace{0.02cm})^{\hspace{0.03cm}i_{1}
	\hspace{0.03cm}i_{2}}\hspace{0.03cm}
\delta^{\hspace{0.03cm}j_{1}\hspace{0.03cm}j_{2}}
+
(t^{\,\lambda}\hspace{0.02cm})^{\hspace{0.03cm}j_{1}
	\hspace{0.03cm}j_{2}}\hspace{0.03cm}
\delta^{\hspace{0.03cm}i_{1}\hspace{0.03cm}i_{2}}\Bigr]
-
\Bigl[(t^{\,c}\hspace{0.02cm})^{i_{1}\hspace{0.03cm}i_{2}}
(t^{\,d}\hspace{0.03cm})^{j_{1}\hspace{0.03cm}j_{2}}
+
(t^{\,c}\hspace{0.02cm})^{\hspace{0.03cm}j_{1}\hspace{0.03cm}j_{2}}
(t^{\,d}\hspace{0.03cm})^{\hspace{0.03cm}i_{1}\hspace{0.03cm}i_{2}}\Bigr].
\]
Finally, we can write down an additional identity for the special case $N_{c} = 3$:
\begin{equation}
(t^{\hspace{0.03cm}a})^{\hspace{0.03cm}i_{1}\hspace{0.03cm}j_{2}}
(t^{\hspace{0.03cm}b}\hspace{0.03cm})^{\hspace{0.03cm}j_{1}
\hspace{0.03cm}i_{2}}
+
(t^{\hspace{0.03cm}b})^{\hspace{0.03cm}i_{1}\hspace{0.03cm}j_{2}}
(t^{\hspace{0.03cm}a}\hspace{0.03cm})^{\hspace{0.03cm}j_{1}\hspace{0.03cm}
i_{2}}
=
(t^{\hspace{0.03cm}a})^{\hspace{0.03cm}i_{1}\hspace{0.03cm}i_{2}}
(t^{\hspace{0.03cm}b}\hspace{0.03cm})^{\hspace{0.03cm}j_{1}
	\hspace{0.03cm}j_{2}}
+
(t^{\hspace{0.03cm}b})^{\hspace{0.03cm}i_{1}\hspace{0.03cm}i_{2}}
(t^{\hspace{0.03cm}a}\hspace{0.03cm})^{\hspace{0.03cm}j_{1}\hspace{0.03cm}
j_{2}}
\label{ap:B9}
\end{equation}
\[
+\;
\delta^{\hspace{0.03cm}a\hspace{0.03cm}b}\hspace{0.04cm}
\biggl\{\frac{1}{9}\,
\delta^{\hspace{0.03cm}i_{1}\hspace{0.03cm}i_{2}}
\delta^{\hspace{0.03cm}j_{1}\hspace{0.03cm}j_{2}}	
-
\frac{1}{3}\,
(t^{\hspace{0.03cm}e})^{\hspace{0.03cm}i_{1}\hspace{0.03cm}i_{2}}
(t^{\hspace{0.03cm}e}\hspace{0.03cm})^{\hspace{0.03cm}j_{1}
	\hspace{0.03cm}j_{2}}\!
\bigg\}
\]
\[
+\,\frac{1}{3}\,
\bigl(D^{\,\lambda}\bigr)^{a\hspace{0.03cm}b}
\Bigl[(t^{\,\lambda}\hspace{0.02cm})^{\hspace{0.03cm}i_{1}
	\hspace{0.03cm}i_{2}}\hspace{0.03cm}
\delta^{\hspace{0.03cm}j_{1}\hspace{0.03cm}j_{2}}
+
(t^{\,\lambda}\hspace{0.02cm})^{\hspace{0.03cm}j_{1}
	\hspace{0.03cm}j_{2}}\hspace{0.03cm}
\delta^{\hspace{0.03cm}i_{1}\hspace{0.03cm}i_{2}}\Bigr]
-
2\hspace{0.03cm}\bigl(D^{\,\lambda}\bigr)^{a\hspace{0.03cm}b}
	d^{\,\lambda\hspace{0.02cm}\kappa\hspace{0.03cm}\rho}
(t^{\,\kappa}\hspace{0.02cm})^{i_{1}\hspace{0.03cm}i_{2}}
\hspace{0.02cm}
(t^{\,\rho})^{j_{1}\hspace{0.01cm}j_{2}}.
\]
This relation can be easily obtained if we first rewrite the left-hand side as 
\[
(t^{\hspace{0.03cm}a})^{\hspace{0.03cm}i_{1}\hspace{0.03cm}j_{2}}
(t^{\hspace{0.03cm}b}\hspace{0.03cm})^{\hspace{0.03cm}j_{1}
	\hspace{0.03cm}i_{2}}
+
(t^{\hspace{0.03cm}b}\hspace{0.03cm})^{\hspace{0.03cm}i_{1}\hspace{0.03cm}j_{2}}
(t^{\hspace{0.03cm}a}\hspace{0.01cm})^{\hspace{0.03cm}j_{1}\hspace{0.03cm}
	i_{2}}
=
\bigl(
\delta^{\hspace{0.02cm}a\hspace{0.02cm}d}\delta^{\hspace{0.02cm}b
\hspace{0.03cm}c}
+
\delta^{\hspace{0.02cm}a\hspace{0.02cm}c}\delta^{\hspace{0.02cm}b
\hspace{0.03cm}d}\hspace{0.03cm}\bigr)
(t^{\hspace{0.03cm}d}\hspace{0.03cm})^{\hspace{0.03cm}i_{1}\hspace{0.03cm}j_{2}}
(t^{\hspace{0.03cm}c}\hspace{0.03cm})^{\hspace{0.03cm}j_{1}
	\hspace{0.03cm}i_{2}},
\]
and then for the color structure $\bigl(
\delta^{\hspace{0.02cm}a\hspace{0.02cm}d}\delta^{\hspace{0.02cm}b
\hspace{0.02cm} c} +\hspace{0.03cm}
\delta^{\hspace{0.02cm}a\hspace{0.02cm}c}\delta^{\hspace{0.02cm}i\hspace{0.03cm}d}\hspace{0.03cm}\bigr)$ we use the first relation in (\ref{ap:C15}) from Appendix \ref{appendix_C} below and further employ the identities (\ref{ap:B3_b}), (\ref{ap:B6_b}) and (\ref{ap:B8}). When we contract (\ref{ap:B9}) with $\delta^{\hspace{0.02cm}a\hspace{0.02cm}b}$ and consider (\ref{ap:C2}), we reproduce the identity (\ref{ap:B3_b}) for $N_{c} = 3$, as it should be. Unfortunately, the relation (\ref{ap:B9}) is not valid for arbitrary $N_{c}$. Indeed, if we use the general relation (\ref{ap:C10}) for the color structure 
$\bigl(\delta^{\hspace{0.02cm}a\hspace{0.02cm}d}\delta^{\hspace{0.02cm}
b\hspace{0.02cm} c} + \delta^{\hspace{0.02cm}a\hspace{0.02cm}c}\delta^{\hspace{0.02cm}b
\hspace{0.03cm}d}\hspace{0.03cm}\bigr)$, then, taking into account (\ref{ap:B4}) and (\ref{ap:B7}), by virtue of the relation
\[
	\bigl\{D^{\,a},\hspace{0.03cm}D^{\,b}\bigr\}^{c\hspace{0.03cm}d}
	\hspace{0.02cm}
	(t^{\hspace{0.03cm}a})^{\hspace{0.03cm}i_{1}\hspace{0.03cm}j_{2}}
	(t^{\hspace{0.03cm}b}\hspace{0.03cm})^{\hspace{0.03cm}j_{1}
		\hspace{0.03cm}i_{2}} 
	=
	\delta^{\hspace{0.03cm}c\hspace{0.03cm}d}\hspace{0.03cm}
	\biggl\{\biggl(\frac{N^{\hspace{0.02cm}2}_{c} - 2} {N^{\hspace{0.02cm}3}_{c}}\biggr)\hspace{0.03cm}
	\delta^{\hspace{0.03cm}i_{1}\hspace{0.03cm}i_2}\hspace{0.03cm}
	\delta^{\hspace{0.03cm}j_{1}\hspace{0.03cm}j_2}
	-
	\frac{4}{N^{\hspace{0.03cm}2}_{c}}\,
(t^{\hspace{0.03cm}e})^{\hspace{0.03cm}i_{1}\hspace{0.03cm}i_{2}}
(t^{\hspace{0.03cm}e}\hspace{0.03cm})^{\hspace{0.03cm}j_{1}
\hspace{0.03cm}j_{2}}\!
\bigg\}
\]
\[	
+\,
\biggl(\frac{N^{\hspace{0.02cm}2}_{c} - 8} {2\hspace{0.02cm}N^{\hspace{0.02cm}2}_{c}}\biggr)\hspace{0.03cm}\,
\bigl(D^{\,\lambda}\bigr)^{c\hspace{0.03cm}d}
\Bigl[(t^{\,\lambda}\hspace{0.02cm})^{\hspace{0.03cm}i_{1}
	\hspace{0.03cm}i_{2}}\hspace{0.03cm}
\delta^{\hspace{0.03cm}j_{1}\hspace{0.03cm}j_{2}}
+
(t^{\,\lambda}\hspace{0.02cm})^{\hspace{0.03cm}j_{1}
	\hspace{0.03cm}j_{2}}\hspace{0.03cm}
\delta^{\hspace{0.03cm}i_{1}\hspace{0.03cm}i_{2}}\Bigr]
+
\Bigl[(t^{\,c}\hspace{0.02cm})^{i_{1}\hspace{0.03cm}i_{2}}
(t^{\,d}\hspace{0.03cm})^{j_{1}\hspace{0.03cm}j_{2}}
+
(t^{\,c}\hspace{0.02cm})^{\hspace{0.03cm}j_{1}\hspace{0.03cm}j_{2}}
(t^{\,d}\hspace{0.03cm})^{\hspace{0.03cm}i_{1}\hspace{0.03cm}i_{2}}\Bigr].
\]
\[	
-\,
\frac{4}{N_{c}}\,\bigl(D^{\,\lambda}\bigr)^{c\hspace{0.03cm}d}
d^{\,\lambda\hspace{0.02cm}\kappa\hspace{0.03cm}\rho}
(t^{\,\kappa}\hspace{0.02cm})^{i_{1}\hspace{0.03cm}i_{2}}
\hspace{0.02cm}
(t^{\,\rho})^{j_{1}\hspace{0.01cm}j_{2}}
-\,
\frac{2}{N_{c}}\,
\Bigl[(t^{\,c}\hspace{0.02cm})^{i_{1}\hspace{0.03cm}j_{2}}
(t^{\,d}\hspace{0.03cm})^{j_{1}\hspace{0.03cm}i_{2}}
+
(t^{\,c}\hspace{0.02cm})^{\hspace{0.03cm}j_{1}\hspace{0.03cm}i_{2}}
(t^{\,d}\hspace{0.03cm})^{\hspace{0.03cm}i_{1}\hspace{0.03cm}j_{2}}\Bigr]
\]
we arrive only at the identity.


\section{Traces for generators in the adjoint repre\-sentation of the  $SU(N_{c})$ color group}
\numberwithin{equation}{section}
\label{appendix_C}

In this Appendix, we have provided an explicit form for the traces of adjoint representation matrices, which we use throughout our work. An extensive list of various traces, relations and identities for color matrices in the adjoint representation can be found in \cite{kaplan_1967, macfarlane_1968, azcarraga_1998, fadin_2005, nikolaev_2005, haber_2021}. Initial definitions of the matrices $T^{\,a}$ and $D^{\,a}$ are
\[
\bigl(T^{\,a}\bigr)^{\hspace{0.01cm}b\hspace{0.03cm}c} 
\equiv 
-i\hspace{0.03cm}f^{\hspace{0.03cm}a\hspace{0.02cm}b\hspace{0.03cm}c},
\qquad
\bigl(D^{\,a}\bigr)^{\hspace{0.01cm}b\hspace{0.03cm}c} 
\equiv 
d^{\hspace{0.03cm}a\hspace{0.02cm}b\hspace{0.03cm}c},
\]
where $f^{\hspace{0.03cm}a\hspace{0.02cm}b\hspace{0.03cm}c}$ and $d^{\hspace{0.03cm}a\hspace{0.02cm}b\hspace{0.03cm}c}$ are the totally antisymmetric and symmetric structure constants for the $SU(N_{c})$ group, respectively. These matrices are traceless, i.e.
\begin{equation}
{\rm tr}\,T^{\,a} = 0, 
\qquad 
{\rm tr}\,D^{\,a} = 0
\label{ap:C2}
\end{equation}
and satisfy the following commutation relations
\begin{equation}
\bigl[\hspace{0.02cm}T^{\,a},T^{\,b}\hspace{0.02cm}\bigr] = i\hspace{0.02cm}f^{\hspace{0.03cm}a\hspace{0.02cm}b\hspace{0.03cm}c}
\hspace{0.03cm}T^{\,c},
\qquad
\bigl[\hspace{0.02cm}T^{\,a},D^{\,b}\hspace{0.02cm}\bigr] = i\hspace{0.02cm}f^{\hspace{0.03cm}a\hspace{0.02cm}b\hspace{0.03cm}c}
\hspace{0.02cm} D^{\,c}.
\label{ap:C3}
\end{equation}
For completeness, we also provide the commutator for the $D^{\,a}$ matrices
\[
\hspace{0.6cm}
\bigl[\hspace{0.02cm}D^{\,a},D^{\,b}\hspace{0.02cm}\bigr]^{c\hspace{0.02cm} d} = i\hspace{0.02cm}f^{\hspace{0.03cm}a\hspace{0.02cm}b\hspace{0.03cm}e}\hspace{0.02cm} \bigl(T^{\,e}\bigr)^{cd}
+
\frac{2}{N_{c}}\,\bigl(\delta^{\hspace{0.02cm}a\hspace{0.02cm}d}\delta^{\hspace{0.02cm}b\hspace{0.02cm} c}
-
\delta^{\hspace{0.02cm}a\hspace{0.02cm}c}\delta^{\hspace{0.02cm}b\hspace{0.02cm}d}\bigr).
\] 
The traces of two generators are given by 
\begin{equation}
{\rm tr}\hspace{0.03cm}
\bigl(T^{\,a}\hspace{0.03cm}T^{\,b}\bigr) = N_{c}\hspace{0.03cm}\delta^{\hspace{0.03cm}a\hspace{0.03cm}b},
\quad
{\rm tr}\hspace{0.03cm}
\bigl(D^{\,a}\hspace{0.02cm}D^{\,b}\bigr) = \left(\frac{N^{\hspace{0.03cm}2}_{c} - 4}{N_{c}}\right)\delta^{\hspace{0.04cm}a\hspace{0.03cm}b},
\quad
{\rm tr}\hspace{0.03cm}
\bigl(T^{\,a}\hspace{0.02cm}D^{\,b}\bigr) = 0,
\label{ap:C4}
\end{equation}
and for the traces of three generators, we have, in turn,
\begin{equation}
\begin{array}{ll}
{\rm tr}\hspace{0.03cm}
\bigl(T^{\,a}\hspace{0.03cm}T^{\,b}\hspace{0.03cm}T^{\,c}\bigr) = \displaystyle\frac{i}{2}\,N_{c}\hspace{0.03cm}
f^{\hspace{0.03cm}a\hspace{0.02cm}b\hspace{0.03cm}c},\\[4ex]
\vspace{0.03cm}
{\rm tr}\hspace{0.03cm}
\bigl(D^{\,a}\hspace{0.03cm}T^{\,b}\hspace{0.03cm}T^{\,c}\bigr) = \displaystyle\frac{1}{2}\,N_{c}\hspace{0.04cm}d^{\hspace{0.03cm}a
\hspace{0.02cm}b\hspace{0.03cm}c},
\end{array}
\qquad
\begin{array}{ll}
\vspace{0.2cm}
{\rm tr}\hspace{0.03cm}
\bigl(D^{\,a}\hspace{0.03cm}D^{\,b}\hspace{0.03cm}T^{\,c}\bigr) = i\hspace{0.05cm}\left(\displaystyle\frac{N^{\hspace{0.03cm}2}_{c} - 4}{2\hspace{0.03cm}N_{c}}\right) f^{\hspace{0.03cm}a\hspace{0.02cm}b\hspace{0.03cm}c},\\[2ex]
\vspace{0.03cm}
{\rm tr}\hspace{0.03cm}
\bigl(D^{\,a}\hspace{0.03cm}D^{\,b}\hspace{0.03cm}D^{\,c}\bigr) = \left(\displaystyle\frac{N^{\hspace{0.03cm}2}_{c} - 12}{2\hspace{0.03cm}N_{c}}\right) d^{\hspace{0.03cm}a\hspace{0.02cm}b\hspace{0.03cm}c}.
\end{array}
\label{ap:C5}
\end{equation}
The traces of four generators are
\begin{align}
&{\rm tr}\hspace{0.03cm}\bigl(T^{\,a}\hspace{0.02cm} T^{\,b}\hspace{0.02cm} T^{\,c}\hspace{0.02cm}T^{\,d}\hspace{0.03cm}\bigr)
=
\delta^{\hspace{0.02cm}a\hspace{0.02cm}b}\hspace{0.03cm}\delta^{\hspace{0.02cm}c\hspace{0.03cm}d}
+
\delta^{\hspace{0.02cm}a\hspace{0.02cm}d}\hspace{0.03cm}\delta^{\hspace{0.02cm}c\hspace{0.03cm}b}
+
\frac{1}{4}\,N_{c}\hspace{0.02cm}\Bigl[\bigl\{D^{\hspace{0.03cm}a},D^{\,c}\bigr\}^{b\hspace{0.03cm}d}
-
d^{\,a\hspace{0.02cm}c\lambda}\hspace{0.03cm}\bigl(D^{\hspace{0.03cm}\lambda}\bigr)^{b\hspace{0.03cm}d}\hspace{0.04cm}\Bigr],
\label{ap:C6} \\[1ex]
&{\rm tr}\hspace{0.03cm}\bigl(T^{\,a}\hspace{0.02cm} T^{\,b}\hspace{0.02cm} D^{\,c}\hspace{0.02cm}D^{\,d}\hspace{0.03cm}\bigr)
=
\biggl(\frac{N^{\hspace{0.02cm}2}_{c} - 4}{N^{\hspace{0.02cm}2}_{c}}\biggr)
\bigl(\delta^{\hspace{0.02cm}a\hspace{0.02cm}b}\delta^{\hspace{0.02cm}
c\hspace{0.02cm}d}
\!-
\delta^{\hspace{0.02cm}a\hspace{0.02cm}c}\delta^{\hspace{0.02cm}b\hspace{0.02cm}d}\bigr)
+\,
\biggl(\frac{N^{\hspace{0.02cm}2}_{c} - 8}{4\hspace{0.02cm}N_{c}}\biggr)
\bigl(d^{\,a\hspace{0.03cm}b\hspace{0.03cm}e}
\hspace{0.03cm}d^{\,c\hspace{0.03cm}d\hspace{0.03cm}e}
\!-
d^{\,a\hspace{0.03cm}c\hspace{0.03cm}e}
\hspace{0.03cm}d^{\,b\hspace{0.03cm}d\hspace{0.03cm}e}\bigr)
\label{ap:C7}\\[1ex] 
&\hspace{6cm}+\,
\frac{1}{4}\,N_{c}\hspace{0.03cm}
d^{\,a\hspace{0.03cm}d\hspace{0.03cm}e}
\hspace{0.03cm}d^{\,b\hspace{0.03cm}c\hspace{0.03cm}e},
\notag\\[1ex]
&{\rm tr}\hspace{0.03cm}\bigl(T^{\,a} D^{\,b}\hspace{0.02cm} D^{\,c}\hspace{0.02cm}D^{\,d}\hspace{0.03cm}\bigr)
=
i\hspace{0.04cm}\biggl(\frac{N^{\hspace{0.02cm}2}_{c} - 12}{4\hspace{0.02cm}N_{c}}\biggr)
f^{\hspace{0.03cm}a\hspace{0.03cm}b\hspace{0.03cm}e}
\hspace{0.03cm}d^{\,c\hspace{0.03cm}d\hspace{0.03cm}e}
+
\frac{i}{N_{c}}\,
\bigl(f^{\hspace{0.03cm}a\hspace{0.03cm}d\hspace{0.03cm}e}
\hspace{0.03cm}d^{\,b\hspace{0.03cm}c\hspace{0.03cm}e}
-
f^{\hspace{0.03cm}a\hspace{0.03cm}c\hspace{0.03cm}e}
\hspace{0.03cm}d^{\,b\hspace{0.03cm}d\hspace{0.03cm}e}\bigr)
\!+
\frac{1}{4}\,i\hspace{0.03cm}N_{c}\hspace{0.03cm}
d^{\,a\hspace{0.03cm}b\hspace{0.03cm}e}
\hspace{0.01cm}f^{\hspace{0.03cm}c\hspace{0.03cm}d\hspace{0.03cm}e}.
\label{ap:C8} 
\end{align}
The representation (\ref{ap:C8}) is convenient because it clearly shows the symmetry of the first term on the right-hand side and the antisymmetry of the second and third terms with respect to the permutation of indices $c$ and $d$. We employ this fact in the section \ref{section_11}. Further, the trace (\ref{ap:C6}) is written in such a way that makes its symmetry with respect to the permutation of indices $a$ and $c$, as well as with respect to the indices $b$ and $d$, immediately apparent, i.e., 
\begin{equation}
	{\rm tr}\hspace{0.03cm}\bigl(T^{\,a}\hspace{0.02cm} T^{\,b}\hspace{0.02cm} T^{\,c}\hspace{0.02cm} 
	T^{\,d}\hspace{0.03cm}\bigr)	
	=
	{\rm tr}\hspace{0.03cm}\bigl(T^{\,c}\hspace{0.02cm} T^{\,b}\hspace{0.02cm} T^{\,a}\hspace{0.02cm} 
	T^{\,d}\hspace{0.03cm}\bigr).
\label{ap:C9}
\end{equation} 
If we use the anticommutation relation 
\begin{equation}
\bigl\{\hspace{0.02cm}T^{\,a},T^{\,b}\hspace{0.02cm}\bigr\}^{c\hspace{0.02cm} d}
+
\bigl\{\hspace{0.02cm}D^{\,a},D^{\,b}\hspace{0.02cm}\bigr\}^{c\hspace{0.02cm} d} 
= 
\frac{4}{N_{c}}\,\delta^{\hspace{0.02cm}a\hspace{0.02cm}b}\delta^{\hspace{0.02cm}c\hspace{0.02cm} d} 
+
2\hspace{0.02cm}d^{\hspace{0.03cm}a\hspace{0.02cm}b\hspace{0.03cm}e}\hspace{0.02cm} \bigl(D^{\,e}\bigr)^{cd}
-
\frac{2}{N_{c}}\,\bigl(\delta^{\hspace{0.02cm}a\hspace{0.02cm}d}\delta^{\hspace{0.02cm}b\hspace{0.02cm}c}
+
\delta^{\hspace{0.02cm}a\hspace{0.02cm}c}\delta^{\hspace{0.02cm}
b\hspace{0.02cm}d}\bigr),
\label{ap:C10}
\end{equation}
then the trace (\ref{ap:C6}) can also be represented in a slightly different form
\begin{equation}
{\rm tr}\hspace{0.03cm}\bigl(T^{\,a}\hspace{0.02cm}T^{\,b}\hspace{0.02cm} T^{\,c}\hspace{0.02cm}T^{\,d}\hspace{0.03cm}\bigr)
=
\delta^{\hspace{0.02cm}a\hspace{0.02cm}c}\hspace{0.03cm}
\delta^{\hspace{0.02cm}b\hspace{0.03cm}d}
+
\frac{1}{2}\,\bigl(
\delta^{\hspace{0.02cm}a\hspace{0.02cm}b}\hspace{0.03cm}
\delta^{\hspace{0.02cm}c\hspace{0.03cm}d}
+
\delta^{\hspace{0.03cm}a\hspace{0.02cm}d}\hspace{0.03cm}
\delta^{\hspace{0.03cm}c\hspace{0.03cm}b}
\bigr)
-
\frac{1}{4}\,N_{c}\hspace{0.02cm}\Bigl[\bigl\{T^{\,a},T^{\,c}\bigr\}^{b\hspace{0.03cm}d}
-
d^{\hspace{0.03cm}a\hspace{0.02cm}c\lambda}\hspace{0.03cm}\bigl(D^{\,\lambda}\bigr)^{b\hspace{0.03cm}d}
\hspace{0.04cm}\Bigr].
\label{ap:C11}
\end{equation}
\indent The trace of five generators $T^{\,a}$ can be presented as a linear combination of the traces of four generators  \cite{ritbergen_1999}\footnote{\hspace{0.03cm}In the paper  \cite{ritbergen_1999} in the formula (45) for the trace of five generators in one of the terms on the right-hand side, two indices are incorrectly placed.}
\begin{equation}
{\rm tr}\hspace{0.03cm}\bigl(T^{\,a_{1}}\hspace{0.02cm} T^{\,a_{2}}\hspace{0.02cm} T^{\,a_{3}}\hspace{0.02cm} 
T^{\,a_{4}}\hspace{0.03cm}T^{\,a_{5}}\hspace{0.02cm}\bigr)
\label{ap:C12}
\end{equation}
\[
\begin{split}
=
-\hspace{0.03cm}\frac{i}{2}\,\Bigl\{ 
&f^{\hspace{0.03cm}a_{3}\hspace{0.02cm}a_{2}\hspace{0.02cm}b}\hspace{0.04cm}
{\rm tr}\hspace{0.03cm}\bigl(T^{\,a_{1}}\hspace{0.02cm} T^{\,b}\hspace{0.02cm}T^{\,a_{5}}\hspace{0.02cm} 
T^{\,a_{4}}\hspace{0.03cm}\bigr)
+
f^{\hspace{0.03cm}a_{5}\hspace{0.02cm}a_{4}\hspace{0.02cm}b}\hspace{0.04cm}
{\rm tr}\hspace{0.03cm}\bigl(T^{\,a_{1}}\hspace{0.02cm} T^{\,a_{2}}\hspace{0.02cm} 
T^{\,a_{3}}\hspace{0.02cm}T^{\,b}\hspace{0.03cm}\bigr),\\[0.7ex]
+\,
&f^{\hspace{0.03cm}a_{3}\hspace{0.02cm}a_{1}\hspace{0.02cm}b}\hspace{0.04cm}
{\rm tr}\hspace{0.03cm}\bigl(T^{\,b}\hspace{0.02cm}T^{\,a_{2}}\hspace{0.02cm} T^{\,a_{5}}\hspace{0.02cm} 
T^{\,a_{4}}\hspace{0.03cm}\bigr)
+
f^{\hspace{0.03cm}a_{2}\hspace{0.02cm}a_{1}\hspace{0.02cm}b}\hspace{0.04cm}
{\rm tr}\hspace{0.03cm}\bigl(T^{\,a_{3}}\hspace{0.02cm} T^{\,b}\hspace{0.02cm} T^{\,a_{5}}\hspace{0.02cm} 
T^{\,a_{4}}\hspace{0.03cm}\bigr)\Bigr\}.
\end{split}
\]
This expression is a consequence of the sign reversal property of permutation of matrices $T^{\,a}$ under the trace sign in reverse order
\[
{\rm tr}\hspace{0.03cm}\bigl(T^{\,a_{1}}\hspace{0.02cm} T^{\,a_{2}}\hspace{0.02cm} T^{\,a_{3}}\hspace{0.02cm} 
T^{\,a_{4}}\hspace{0.03cm}T^{\,a_{5}}\hspace{0.02cm}\bigr)
=
-{\rm tr}\hspace{0.03cm}\bigl(T^{\,a_{5}}\hspace{0.02cm} T^{\,a_{4}}\hspace{0.02cm} T^{\,a_{3}}\hspace{0.02cm} 
T^{\,a_{2}}\hspace{0.03cm}T^{\,a_{1}}\hspace{0.02cm}\bigr),
\]
which in turn is a trivial consequence of the identity
\[
{\rm tr}\hspace{0.03cm}\bigl(T^{\,a_{1}}\hspace{0.02cm} T^{\,a_{2}}\hspace{0.02cm} T^{\,a_{3}}\hspace{0.02cm} 
T^{\,a_{4}}\hspace{0.03cm}T^{\,a_{5}}\hspace{0.02cm}\bigr)
=
-\hspace{0.03cm}2\hspace{0.03cm}
{\rm tr}\hspace{0.03cm}\bigl(t^{\,b}\hspace{0.02cm}\bigl[\hspace{0.03cm}t^{\,a_{1}},\bigl[\hspace{0.02cm} t^{\,a_{2}},\bigl[\hspace{0.02cm} t^{\,a_{3}},\bigl[\hspace{0.02cm}t^{\,a_{4}},\bigl[\hspace{0.03cm}t^{\,a_{5}},t^{\,b}\bigr]\bigr]\bigr]\bigr]\bigr]\hspace{0.02cm}\bigr).
\]
\indent The second-order Casimiris are
\begin{equation*}
T^{\,a}T^{\,a} = N_{c}\hspace{0.03cm}I, 
\qquad 
D^{\,a}D^{\,a} =  \left(\frac{N^{\hspace{0.03cm}2}_{c} - 4}{N_{c}}\right)\! I,
\end{equation*}
where $I$ is the $\bigl(N^{2}_{c} - 1\bigr)\!\times\!\bigl(N^{2}_{c} - 1\bigr)$ unit matrix. Also it is useful the following formula
\begin{equation}
T^{\,a}\hspace{0.02cm}T^{\,b}\hspace{0.03cm}T^{\,a} = \frac{1}{2}\,N_{c}\hspace{0.03cm}T^{\,b}.
\label{ap:C14}
\end{equation}
\indent In addition, there are two additional identities for the special case $N_{c} = 3$ \cite{macfarlane_1968, haber_2021}, which we use in the text of this article:
\begin{equation}
\begin{split}
	&\bigl\{\hspace{0.02cm}T^{\,a},T^{\,b}\hspace{0.02cm}\bigr\}^{c\hspace{0.02cm} d}
		= 
	3\hspace{0.03cm}d^{\,a\hspace{0.02cm}b\hspace{0.03cm}e}\hspace{0.02cm} \bigl(D^{\,e}\bigr)^{cd}
	+
	\delta^{\hspace{0.02cm}a\hspace{0.03cm}b}
	\delta^{\hspace{0.02cm}c\hspace{0.03cm}d} 
	-
	\delta^{\hspace{0.02cm}a\hspace{0.03cm}d}
	\delta^{\hspace{0.02cm}b\hspace{0.03cm}c}
	-
	\delta^{\hspace{0.02cm}a\hspace{0.03cm}c}
	\delta^{\hspace{0.02cm}b\hspace{0.03cm}d},
	\\[1ex]
	&\bigl\{\hspace{0.02cm}D^{\,a},D^{\,b}\hspace{0.02cm}\bigr\}^{c\hspace{0.02cm} d}
= 
-\hspace{0.03cm}d^{\,a\hspace{0.02cm}b\hspace{0.03cm}e}\hspace{0.02cm} \bigl(D^{\,e}\bigr)^{cd}
+
\frac{1}{3}\,\bigl(
\delta^{\hspace{0.02cm}a\hspace{0.03cm}b}
\delta^{\hspace{0.02cm}c\hspace{0.03cm}d} 
+
\delta^{\hspace{0.02cm}a\hspace{0.03cm}d}
\delta^{\hspace{0.02cm}b\hspace{0.03cm}c}
+
\delta^{\hspace{0.02cm}a\hspace{0.03cm}c}
\delta^{\hspace{0.02cm}b\hspace{0.03cm}d}\bigr).
\end{split}	
\label{ap:C15}
\end{equation}

\end{appendices}

\newpage

\end{document}